\documentclass[12pt]{report}
\usepackage[width=6.1in, height=8.0in, top=1.0in]{geometry}
\usepackage[pdftex]{graphicx}
\usepackage{amsmath, amsfonts, amssymb}
\usepackage{mathtools}
\usepackage{mathrsfs}
\usepackage{yhmath}
\usepackage{amssymb}
\usepackage{tensor}
\usepackage{accents}
\usepackage{tipa}
\usepackage{textcomp}
\usepackage{fancyhdr}
\usepackage{tikz}
\usepackage{hyperref}

\hypersetup{%
  colorlinks=true,
  linkcolor=purple,
  linkbordercolor={0 0 1}
}

\usepackage{mlmodern}
\usepackage[T1]{fontenc}
\usepackage{dsfont}
\usepackage{enumerate}
\usepackage{braket}
\usepackage{stackengine}
\usepackage[english]{babel}
\usepackage{amsthm}
\newtheorem{theorem}{Theorem}[chapter]
\newtheorem{lemma}{Lemma}[chapter]
\newtheorem{definition}{Definition}[chapter]

\stackMath

\def\gtlt{\mathrel{%
  \stackon[1pt]{\scriptscriptstyle\wedge}{\scriptscriptstyle\vee}}}

\usepackage[framemethod=TikZ]{mdframed}
\mdfdefinestyle{testframe}{topline=false,rightline=false,bottomline=false,%
innerleftmargin=1em,linecolor=white,%
tikzsetting={draw=black,line width=.5pt,dashed,dash pattern= on 1pt off 3pt}, splittopskip=2 }

\allowdisplaybreaks

\fancypagestyle{plain}{%
\fancyhead{}

}
\begin{document}
%
\chapter*{\center Quantum Gravity as a Theory of Connections}
\section*{\center Hassan Mehmood \\ ID: 23100127 }
\section*{\center Supervisor: Dr. Syed Moeez Hassan}
\section*{\center BS Thesis \\ Department of Physics \\ Lahore University of Management Sciences}
\newpage
\clearpage
\begin{center}
    \thispagestyle{empty}
    \vspace*{\fill}
    \textit{To Mamoon, \\ for his unconditional love, \\ and to Ami, \\ for keeping me sane.}
    \vspace*{\fill}
\end{center}
\clearpage

\newpage
\begin{abstract}
    Consider the interior of a black hole or the very early universe: matter is so densely localized that neither the effects of gravity nor those of quantum theory can be ignored. But this entails that neither general relativity nor quantum theory on its own can fully describe such a situation, for some of the most fundamental principles inhering in these two theories are haunted by the specter of incompatibility. Quantum gravity is the name for the bewildering penumbra of theories that seek to exorcise this demon. But it turns out that the metrical variables of general relativity constitute a lamp too narrow to bottle the phantom, and loop quantum gravity is a fascinating enterprise that seeks the Aladdin who does possess the required lamp. This is achieved by recasting general relativity as a theory of connections, rather than that of metrics. This shift of emphasis allows one to use a number of mathematical tools that make it possible to arrive at a fully consistent, almost background-independent theory of quantum gravity. This thesis endeavours to probe these ideas in detail.  
\end{abstract}

\tableofcontents
%
%

\chapter*{\center Acknowledgements}
\addcontentsline{toc}{chapter}{\protect\numberline{}Acknowledgements}
This thesis is by no means an individual endeavour. While my debts are too great to recount fully, I would be remiss not to mention them at all. 

I would like to begin by thanking my supervisor, Dr. Moeez. I cannot imagine myself going through a tragedy as colossal as his and still finding the strength to even fully smile ever again, let alone instruct courses and supervise theses. It is a testament to his courage that he supervised and saw to fruition my immensely painstaking efforts to thoroughly grasp the subtleties of loop quantum gravity. I have lost count of the number of times that feelings of insignificance tormented me while I would be stuck understanding some obscure and technical point, only to be washed away by a rekindled ambition to complete my project in the aftermath of an engaging discussion with him. 

I also wish to offer my deep gratitude to Dr. Adam. His role in my journey as an aspiring theoretical physicist has been nothing short of foundational. His pedagogic prowess and mastery of physics have left an indelible mark on my understanding of all the subjects he has instructed me in -- the most notable being quantum mechanics. Moreover, his immensely supportive character in a world that cherishes a toxic and competitive academic environment has been absolutely essential in pulling me through the vicissitudes of my undergraduate career. 

My family's constant love and support have also been truly decisive in all of my successes. Ami, Appa Khala, Asma Phupo, Bhaya, Faisal Mamoon, Shazi Mamoon, Qamar -- I could not have done any of this without any of you. You are all shining stars in my sky. 

Finally, it would be a sin if I were not to record my debts to my second family: the amazing, wonderful, and wholesome people I found in the physics department at LUMS. It would be an equally great misfortune to not mention them by name and recognise them for the unique individuals they are. Therefore, I wish to express my very deep gratitude to each of these people individually. 

To Alina, for being the loveliest twin I could have ever hoped for, for her unswerving honesty which I cherish so much, for being the friend who would know from a hundred miles away that another friend needs her, and for giving me the confidence to talk in baby talk. 

To Azka, for blessing me with the warmth of her friendship when I needed it the most, and for teaching me that one can colour one's hearts with colours that they like.  

To Bismah, for teaching me what it means to be truly happy, and what it means to love, for being the person whose smiles and wellbeing give incalculable meaning to my life, for being the person whose very existence inspires me to get up from my bed on days when one wishes one did not exist, and for being the biggest contributor to my collection of cute WhatsApp stickers. 

To Hammas, for being the friend I always dreamt of finding, for being the source of my comfort and solace in the most dreary storms of my life, for being my moral compass when terrible circumstances convulsed my own conscience, and for rivalling me in my love for Noam Chomsky.

To Irfan, for taking care of me like a lovely elder brother, for blessing my life with his sheer brilliance, and for making me confident in my obsession with perfection. 

To Mahnoor, for always allowing me to unwaveringly vent in front of her in the face of academic burnouts and work-induced bouts of despair, for constantly uplifting my downtrodden spirits, and for engaging in our endless laptop battles (mine is still better!). 

To Manahil, for being the person whose mere existence inspires me to continuously become a better version of myself, for helping me discover unexplored treasures within myself, for being a queen of empathy and compassion, and for convincing me of the heavenly merits of coffee. 

To Muzammil, for taking care of me like another lovely elder brother, for teaching me not to doubt my own abilities, and for simply blessing my life with his intrinsic buddy-ness. 

To Salman, for being another friend that I always dreamt of finding, for being the crucible of relatibility in which the unbreakable bonds of our friendship were forged, for being my profoundest corner of safety, calm and warmth, for our immensely engaging philosophical discussions, and for sprinkling my sense of humour with let's-not-say-what. 

To Abdullah, Diya, Fariha, Huzaifa, Shirin and Wahaj, people who I could not discover as deeply as others, but who, along with all the others, form the sun in my sky. Here is a little poem for you all. 
\newpage

\begin{verse}
Everyday, I walk the treacherous path that in front of me lies,\\
Riding upon the horse of gravity, which I strive to quantise!

But the rope of hope loosens, and I heave countless sighs\\
As my spirits this mighty task threatens to pulverise! 

But then come these lovely souls, embracing me in their care;\\
Joy incarnate are the numberless moments that we share.

They offer me these crowns, for they say I make them proud;\\
Hope dances into my life again, and meaning to it is endowed. 

They take my hand, lest on this pesky road I fall;\\
Oh I wish I could tell, how much I love you all!
\end{verse}

\chapter*{\center Preface}
\addcontentsline{toc}{chapter}{\protect\numberline{}Preface}

The term `quantum gravity' carries a number of connotations. To a student of physics, it seems to convey the idea of a mathematical framework that describes gravity with recourse to quantum theory. But such a characterisation begets more questions than it answers. What is quantum theory? Is it the quantum mechanics of non-relativistic particles? Or is it the quantum theory of relativistic fields? Or is it yet some other physical theory whose name contains the word `quantum'? In any case, are there some salient `quantum principles' that undergird all these different incarnations of quantum theory? If so, does gravity as we understand it lack those principles, so that we are interested in garbing it in quantum-theoretical clothes? But what is gravity in the first place? Is there something inherently and conceptually non-quantum about it or is its non-quantum nature simply an artefact of how we typically describe it mathematically? 

Let us make an attempt to broach these questions. Consider the second end of the devil's horn, namely gravity. Since Einstein's general relativity, we have a fairly precise idea of what gravity is. We imagine the universe around us as a four-dimensional manifold called spacetime, and gravity is understood as a set of fields that characterise the geometry of this manifold. Thus, as these fields change, the geometry of spacetime changes, which in turn influences the dynamics of matter contained in the universe. Furthermore, this influence is not a totalitarian rule where only gravitational fields determine the fate of matter. It is a democratic calculus: matter, which can also be described via fields on spacetimes, also monitors the behaviour of gravity. This two-way street is paved with the Einstein field equations,
\begin{equation*}
    G_{\alpha\beta} = 8\pi T_{\alpha\beta},
\end{equation*}
which are dynamical equations with spatial and temporal derivatives of the gravitational fields on the left-hand side, and those of the matter fields on the right-hand side. We refer to this duel between geometry and matter as \textit{background independence}, which is meant to convey the message that there is no background arena in nature determined by fixed gravitational fields that are inert to the machinations of matter fields. In the time-honoured words of John Wheeler, ``Spacetime tells matter how to move; matter tells spacetime how to curve.'' 

Thus, in a nutshell, gravity could be envisioned as a background-independent field. This characterisation lets us confront the remaining end in the devil's horn of `quantum gravity'. Since the modern understanding of gravity is in terms of a field theory, to make an appropriate comparison with quantum theory, we must invoke relativistic quantum field theories, such as Klein-Gordon scalar field theory, quantum electrodynamics, and so on. What are the features of these field theories that make them fundamentally different from general relativity? The first point to notice is that in relativistic quantum field theories, the relevant fields have a very different structure from the fields in general relativity. Unlike gravitational fields, they are represented as operators with possibly non-vanishing commutators. It is this property of `quantum' fields that differentiates them from classical fields such as gravity in general relativity, and it has nontrivial empirical consequences, such as the manifestation of the uncertainty principle, probabilistic amplitudes, and so on. Indeed, this signals the need to modify general relativity, for it assumes matter fields to be non-quantum along with gravitational fields, whereas the stunning empirical achievements of particle physics tell us that matter is undoubtedly quantum.

The other salient feature of relativistic quantum field theories is that they are typically formulated as dynamical theories of matter fields living on the fixed background of Minwkowskian spacetime -- hence the name relativistic. Thus, they are not background independent. In fact, since the geometry of Minkowski spacetime is flat, gravity is altogether absent from these theories. From an empirical standpoint, this is justified, for general relativity dictates that small masses localised in small regions of spacetime are immune to the effects of the geometry of spacetime at large, and relativistic quantum field theories are used to understand the microscopic world of particle physics, which deals with small masses localised at laboratory scales. 

However, there might be realistic scenarios in which huge amounts of masses may be localised in small regions of spacetime, such as for instance, the interior of a black hole or the very early universe. In such circumstances, the effects of spacetime geometry become non-negligible. Thus, it would be incorrect to describe the behaviour of particles through quantum fields on an inert background. Here one could smuggle gravity into the the theory by changing the background spacetime from Minwkowskian to one with nontrivial geometry. This is the paradigm of \textit{quantum field theory on curved spacetime}, and has been successful, for example, in formulating the inflationary model of the expansion of the universe. However, this is still a background-dependent scheme, for the background geometry, despite being nontrivial, is still fixed; quantum fields are affected by it, but do not in turn influence it. To really develop a background-independent theory of quantum fields, one will have to somehow construct a framework in which \textit{quantum} matter fields influence gravitational fields. Thus, no matter how we contrast quantum field theory and general relativity, we arrive at the need to develop a background-independent theory of gravity in which matter is necessarily quantum, a task beyond the conceptual framework of both these theories. This is how the term `quantum gravity' is interpreted in this thesis.

Broadly speaking, there are two ways to realise such a vision of quantum gravity. One way is to keep treating gravity as a classical field, while somehow trying to describe how quantum matter fields may be able to influence the classical gravitational fields. This approach is called \textit{semiclassical gravity}, and involves a penumbra of schemes, ranging from modifying the Einstein field equations to employing path-integral methods. But we will be interested in the other second approach to our problem. It consists in treating gravity on an equal footing with matter. All fields, whether matter or gravitational, are considered quantum, and thus represented as operators on a Hilbert space. In this thesis, we will delve into the gravitational aspect of this approach, namely the description of gravity in vacuum in terms of quantum fields.   

How are we to proceed with such a task? Typically, quantum theories are constructed from classical theories via a specific scheme of quantisation, such as canonical quantisation, path-integral methods, and so forth. In the case of gravity, this observation translates to the quantisation of general relativity, which is a classical field theory of gravity. In the canonical quantisation of a field theory, one begins by passing to the Hamiltonian formulation of the classical theory. One then applies a well-defined procedure to convert the dynamical fields and observables involved in the Hamiltonian formulation into operators on a Hilbert space. In contrast, path-integral quantisation proceeds from the Lagrangian formulation of a classical field theory, and directly obtains amplitudes related to quantum fields by integrating over all classical histories. This thesis will only explore only the canonical quantisation of gravity.    

Regardless of which scheme of quantisation one wishes to use, in general, there are two ways to quantise a classical field theory: perturbative and non-perturbative. Perturbative methods are motivated from standard relativistic quantum field theory with interactions, in which we describe interacting fields as perturbations whose effects manifest in amplitudes and scattering cross sections. This method applied to gravity entails that we split the classical gravitational fields into two parts, one of which is considered as a fixed background upon which the second part is a perturbation; then we subsequently quantise this perturbation. However, this is anathema to the spirit of background-independence, for the non-perturbed part of the gravitational fields furnishes an inert background. Indeed, perturbative quantum gravity along the lines of perturbative quantum field theory is known to be non-renormalisable, i.e. riddled with incurable divergences, an indication, perhaps of our carelessness concerning background independence. However, this does not mean that there are no viable perturbative frameworks of quantum gravity -- the most famous example is string theory. We will, however, insist upon background-independence and accordingly, focus on non-perturbative methods of quantisation, which seek to fully quantise classical gravitational fields. 

Thus, our central aim in this thesis is to study non-perturbative canonical quantisation of general relativity. As we shall repeatedly see, walking down this road forces us to forgo some of our most cherished ideas about gravity and spacetime. Chapter 1 will spell trouble for the idea of quantising gravity using a mathematical framework in which general relativity is traditionally conceived, i.e. the description of gravitational fields via a metric on a Lorentzian spacetime. We will formulate a Hamiltonian theory of metrical variables, and find that an attempt at quantising it faces a plethora of insurmountable difficulties. Notice, however, that we were decisively vague when we described gravity above as a theory of fields that characterise the geometry of spacetime. This suggests that perhaps, the metric is not the only field that accomplishes this task. Indeed, Chapter 2 will be devoted to elucidating this point. We shall formulate alternative, equivalent versions of canonical general relativity, which demote the metric to a secondary character, according primary importance, instead, to connections on local sections of the tangent bundle of spacetime. This is where the name of the thesis comes from, and this approach of sacralising connections in favour of the metric is the starting point of one of the most refined and sophisticated non-perturbative frameworks of quantum gravity -- loop quantum gravity. Chapter 3 -- the longest one -- shall delve into this vast theory and develop a rigorous framework to quantise the connection formulations of canonical general relativity at the kinematic level. Chapter 4 will elaborate some tantalising predictions of this new framework, such as the discreteness of spatial geometry, which casts a shadow of doubt on the time-honoured understanding of spacetime as a continuum. Finally, Chapter 5 shall be concerned with implementing the dynamics of gravity at the quantum level. 

It is worth pointing out that we will explore loop quantum gravity in its canonical framework. There also exist path-integral versions of loop quantum gravity, but we will not explore them. They are subsumed under the umbrella of spinfoam gravity, also known as covariant loop quantum gravity \cite{rovelli 4}. Moreover, connections will be at the forefront of all our constructions. Accordingly, we shall focus on the connection-space representation of loop quantum gravity. An alternative formulation is in terms of the so-called loop variables \cite{rovelli 3}. Both formulations are completely equivalent \cite{ashtekar 5}. For a comprehensive overview of the various threads interspersed in the fabric of loop quantum gravity, we refer the reader to Ref. \cite{ashtekar 10}.

\chapter*{\center Note to the Reader}
\addcontentsline{toc}{chapter}{\protect\numberline{}Note to the Reader}
This thesis is written with an advanced undergraduate audience in mind. It assumes a background in general relativity and quantum field theory at the level of the first part of Wald \cite{wald} and the first ten chapters of Srednicki \cite{srednicki}. In practice, this translates to a thorough mastery over index gymnastics and familiarity with basic field-theoretic and quantum-theoretical concepts such as Poisson brackets, canonical commutation relations, functional derivatives, a Hilbert space, and so on. With such a background, most of the contents of this thesis can be grasped through a patient perusal, provided the reader possesses some mathematical sophistication to make sense of some technical mathematical concepts introduced here and there. In fact, some of this mathematical experience is certainly gained in one's undergraduate journey in theoretical physics. On this account, we assume at least a nodding acquaintance with some mathematical terms that we do not define, such as isomorphism, diffeomorphism, algebra, group, and so forth. 

There is an exception to this general theme: Chapter 3 demands a fair amount of prior mathematical expertise to be thoroughly understood. A grounding in undergraduate analysis would certainly help, as would some basic familiarity with concepts from measure theory and topology. Where possible, we have given references to books and papers where the relevant concepts can be studied in more detail, and have intuitively explained the import of technical theorems and results that we state without proof. Unfortunately, it is not possible to do justice to the concepts involved in loop quantum gravity without recourse to these mathematical preliminaries. In fact, contrary to appearances, Chapter 3 still severely lacks in mathematical rigour, and brushes many technical nuances under the carpet. Our aim there is to help the reader gain at least some appreciation for the very beautiful mathematical structures and notions that undergird the field of loop quantum gravity, without introducing so much technical baggage as to scare even the most serious reader away. Such an approach is bound to make both experts and non-experts unhappy, but even if one person found our coddiwomple through loop quantum gravity helpful in understanding the often severely dense foundational papers in the field, our efforts would not have been in vain. To both the timid novice and the haughty expert, we apologise in advance.

\chapter{Spectres That Haunt ADM}
One of the very successful ways of quantising classical systems is to pass to their Hamiltonian formulation, and then promote functions that are observables in the classical phase space to self-adjoint operators in a Hilbert space by promoting the Poisson brackets between those functions to commutators between operators. Since there exists a well-worked-out Hamiltonian formulation of general relativity, namely, the ADM formalism (due to Arnowitt, Deser and Misner), it seems natural to ask whether it can be subjected to the methods of Dirac quantisation. If such a task can be accomplished, one will have a fully consistent non-perturbative theory of quantum gravity. It would be an immensely satisfactory marriage of the principles of quantum mechanics and of general relativity, for gravity would have become a quantum phenomenon that respects background independence, which is a cornerstone of general relativity. However, as we shall see, a straightforward quantisation of the ADM formalism leads to a morass of intractable difficulties. To step into these murky waters, we shall first have to come to grips with the ADM formalism itself. 

\section{The ADM Formalism of General Relativity}
The Hamiltonian formulation of a field theory proceeds with the choice of a configuration variable and its conjugate momentum, which involves a time derivative. Thus one has to choose a time coordinate before one can even begin to write down a Hamiltonian formulation. It seems rather perverse to do such a thing in the context of a theory that boasts of general diffeomorphism invariance. The remedy, of course, is to demonstrate that the formulation so constructed will be independent of the particular choice of the time coordinate.

\begin{figure}[h!]
    \centering
    \tikzset{every picture/.style={line width=0.75pt}} 

\begin{tikzpicture}[x=0.75pt,y=0.75pt,yscale=-1,xscale=1]

\draw   (272,185.8) .. controls (272,189) and (284.31,191.6) .. (299.5,191.6) .. controls (314.69,191.6) and (327,189) .. (327,185.8) .. controls (327,182.6) and (339.31,180) .. (354.5,180) .. controls (369.69,180) and (382,182.6) .. (382,185.8) -- (382,232.2) .. controls (382,229) and (369.69,226.4) .. (354.5,226.4) .. controls (339.31,226.4) and (327,229) .. (327,232.2) .. controls (327,235.4) and (314.69,238) .. (299.5,238) .. controls (284.31,238) and (272,235.4) .. (272,232.2) -- cycle ;
\draw   (281,169.3) .. controls (281,172.78) and (294.21,175.6) .. (310.5,175.6) .. controls (326.79,175.6) and (340,172.78) .. (340,169.3) .. controls (340,165.82) and (353.21,163) .. (369.5,163) .. controls (385.79,163) and (399,165.82) .. (399,169.3) -- (399,219.7) .. controls (399,216.22) and (385.79,213.4) .. (369.5,213.4) .. controls (353.21,213.4) and (340,216.22) .. (340,219.7) .. controls (340,223.18) and (326.79,226) .. (310.5,226) .. controls (294.21,226) and (281,223.18) .. (281,219.7) -- cycle ;
\draw   (289.5,154.4) .. controls (289.5,157.93) and (304.05,160.8) .. (322,160.8) .. controls (339.95,160.8) and (354.5,157.93) .. (354.5,154.4) .. controls (354.5,150.87) and (369.05,148) .. (387,148) .. controls (404.95,148) and (419.5,150.87) .. (419.5,154.4) -- (419.5,205.6) .. controls (419.5,202.07) and (404.95,199.2) .. (387,199.2) .. controls (369.05,199.2) and (354.5,202.07) .. (354.5,205.6) .. controls (354.5,209.13) and (339.95,212) .. (322,212) .. controls (304.05,212) and (289.5,209.13) .. (289.5,205.6) -- cycle ;

\draw (368,234.4) node [anchor=north west][inner sep=0.75pt]    {$t=0$};
\draw (402,213.4) node [anchor=north west][inner sep=0.75pt]    {$t=1$};
\draw (422,197.4) node [anchor=north west][inner sep=0.75pt]    {$t=2$};
\draw (339,186.7) node [anchor=north west][inner sep=0.75pt]    {$\Sigma $};

\end{tikzpicture}

    \caption{Foliation of spacetime into a set of spacelike hypersurfaces}
    \label{fig1.1}
\end{figure}
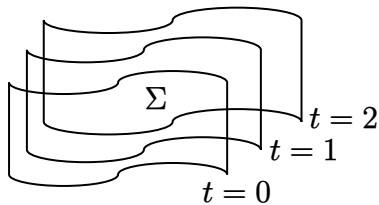

The spacetime manifold $M$ is locally diffeomorphic to $\mathbb{R}^4$, which in turn can be described by a set of four coordinates. Reserving one of these coordinates for time is thus equivalent to finding a diffeomorphism $\phi: M \rightarrow \mathbb{R} \times \Sigma$, where $\Sigma$ is a three-dimensional manifold and $\mathbb{R}$ is the abode of our time coordinate. $\Sigma$ can be thought of as a hypersurface in $M$. The picture that emerges is then of the spacetime being sliced into three-dimensional hypersurfaces, where each hypersurface is the three-dimensional space at a particular time $t$ (Fig. \ref{fig1.1}). Now, any dynamical theory must have a well-defined initial-value formulation, in the sense that specifying the initial values of the dynamical fields must uniquely determine their values at all future times, the future values being given by the relevant dynamical equations. In Hamiltonian general relativity, this amounts to there being a hypersurface at some initial time $t = t_o$ such that specifying the values of the dynamical fields on that hypersurface uniquely determines a solution of Hamilton's equations of motion. But to achieve this, the hypersurface must be special: it must be \textit{spacelike and Cauchy}. A spacelike hypersurface is one that has an everywhere nonvanishing timelike normal vector field. Intuitively, this means that locally, one cannot join any points on the surface with a timelike or null path, and this makes sense with regard to the initial-value formulation being well-defined because the existence of nonspacelike paths between two points makes them causally connected, curtailing our freedom to freely specify initial values. However, the existence spacelike hypersurfaces is not sufficient, for although all points on them are \textit{locally} causally disconnected, they may not be so \textit{globally}\footnote{Imagine a spacelike hypersurface that spirals round and round. Then for small enough regions, every two points are causally disconnected, but two points in different rings of the spiral can be connected by a timelike path.}. Thus we further require the hypersurfaces to be Cauchy, which is to demand that every two points on the surface are \textit{achronal}, i.e. do not lie in each other's causal future\footnote{Our account of the initial-value formulation of general relativity and the attendant concepts of the causal structure of spacetime, such as spacelike Cauchy hypersurfaces, is quite heuristic. For a more thorough treatment, see \cite{wald, hawking}. Also, for an illuminating introduction to causal structure, see \cite{witten}.}. Intuitive though they seem, these conditions are highly non-trivial -- only globally hyperbolic spacetimes admit such a foliation into spacelike Cauchy hypersurfaces\footnote{Most physical spacetimes considered in relativity are globally hyperbolic, so this is not a particularly damning restriction.}.

With these provisos, it should be clear that the Hamiltonian formulation of general relativity will be a prescription that specifies how a particular hypersurface changes in time, i.e. given the values of the configuration variable and its conjugate momentum, both defined on $\Sigma$, at an initial time $t=t_o$, what will be their values at a later time. Thus the first task we shall set ourselves up to is to define on $\Sigma$ the relevant structures and quantities that will realise this picture of evolution. To that end, we first observe that since the dynamical quantity in relativity is the metric $g_{\alpha\beta}$, the configuration variable in the Hamiltonian formulation will be this metric restricted to $\Sigma$, i.e. the pullback of $g$ by $\phi$, $\phi^* g$. We shall call this the \textit{induced metric}, denoted by\footnote{Greek indices ($\alpha$, $\beta$, etc.) refer to coordinates in spacetime and thus range from 0 to 3, whereas Latin indices ($a$, $b$, etc.) refer to the intrinsic coordinates on $\Sigma$ and so range from 1 to 3.} $q_{ab}$. The first virtue possessed by this induced metric is that it is Riemannian. That is, for any nonzero vector $v \in T_p\Sigma$, the tangent space of $\Sigma$ at the point\footnote{A point of clarification: since $v$ lives on $\Sigma$, we should have used Latin indices to represent its components. However, given the diffeomorphism $\phi$, we can always transport objects in $\Sigma$ to their counterparts in $M$ and vice versa, via pullbacks and pushforwards. Intuitively, this freedom stems from the fact that $\Sigma$ is a surface embedded in $M$, and thus anything that lives in the former lives, a fortiori, in the latter as well. We shall exploit this freedom at whim.} $p$, 
\begin{equation}
    g_{\alpha\beta}v^\alpha v^\beta > 0.
\end{equation}\label{eq1.1}
Thus the configuration space of Hamiltonian general relativity is the space of all Riemannian metrics $q_{ab}$ on $\Sigma$; it is denoted by Met$(\Sigma)$ and called the \textit{superspace}. 

Next, since $\Sigma$ is everywhere spacelike, we can find a timelike unit-vector field $n$ that is everywhere orthogonal to it, i.e. 
\begin{equation}
    g_{\alpha\beta}n^\alpha n^\beta = -1, \qquad g_{\alpha\beta}n^\alpha v^\beta = 0, \label{eq1.2}
\end{equation}
for all $v \in T_p\Sigma$. With the help of $n$, we can decompose any vector $u \in T_pM$ into a component normal and a component tangent to $\Sigma$:
\begin{equation}
    u^\alpha = -g_{\mu\nu}u^\mu n^\nu n^\alpha + (u^\alpha + g_{\mu\nu}u^\mu n^\nu n^\alpha). 
\end{equation}\label{eq1.3}
In particular, we can decompose the metric as
\begin{align*}
    g_{\alpha\beta} &= (g^{\mu\nu}g^{\rho\lambda}g_{\mu\rho}n_\nu n_\lambda)n^\alpha n^\beta + (g_{\alpha\beta} - (g^{\mu\nu}g^{\rho\lambda}g_{\mu\rho}n_\nu n_\lambda)n_\alpha n_\beta)\\
    &= -n_\alpha n_\beta + (g_{\alpha\beta} + n_\alpha n_\beta),
\end{align*}
which allows us to write the metric induced on $\Sigma$ as
\begin{equation}
    q_{\alpha\beta} = g_{\alpha\beta} + n_\alpha n_\beta. \label{eq1.4}
\end{equation}
This induced metric allows us to project objects in $M$ onto $\Sigma$, for given $v \in M$, $n_\alpha\tensor{q}{^\alpha_\beta}v^\beta = n_\alpha\delta^\alpha_\beta v^\beta + n_\alpha n^\alpha n_\beta v^\beta = n_\beta v^\beta - n_\beta v^\beta = 0$. Also, the induced metric can be used to raise and lower indices on objects in $\Sigma$.

\begin{figure}[h!]
    \centering
   \tikzset{every picture/.style={line width=0.75pt}} 

\begin{tikzpicture}[x=0.75pt,y=0.75pt,yscale=-1,xscale=1]

\draw    (157,148) .. controls (322,74) and (341,197) .. (440,142) ;
\draw    (193,133) -- (236.1,118.63) ;
\draw [shift={(238,118)}, rotate = 161.57] [color={rgb, 255:red, 0; green, 0; blue, 0 }  ][line width=0.75]    (10.93,-3.29) .. controls (6.95,-1.4) and (3.31,-0.3) .. (0,0) .. controls (3.31,0.3) and (6.95,1.4) .. (10.93,3.29)   ;
\draw    (277,125) -- (302.13,127.84) -- (328.01,130.78) ;
\draw [shift={(330,131)}, rotate = 186.46] [color={rgb, 255:red, 0; green, 0; blue, 0 }  ][line width=0.75]    (10.93,-3.29) .. controls (6.95,-1.4) and (3.31,-0.3) .. (0,0) .. controls (3.31,0.3) and (6.95,1.4) .. (10.93,3.29)   ;
\draw    (193,133) -- (182.62,100.9) ;
\draw [shift={(182,99)}, rotate = 72.07] [color={rgb, 255:red, 0; green, 0; blue, 0 }  ][line width=0.75]    (10.93,-3.29) .. controls (6.95,-1.4) and (3.31,-0.3) .. (0,0) .. controls (3.31,0.3) and (6.95,1.4) .. (10.93,3.29)   ;
\draw    (277,125) -- (280.79,89.99) ;
\draw [shift={(281,88)}, rotate = 96.17] [color={rgb, 255:red, 0; green, 0; blue, 0 }  ][line width=0.75]    (10.93,-3.29) .. controls (6.95,-1.4) and (3.31,-0.3) .. (0,0) .. controls (3.31,0.3) and (6.95,1.4) .. (10.93,3.29)   ;

\draw (442,145.4) node [anchor=north west][inner sep=0.75pt]    {$\Sigma $};
\draw (209,128.4) node [anchor=north west][inner sep=0.75pt]    {$v^{\alpha }$};
\draw (157,91.4) node [anchor=north west][inner sep=0.75pt]    {$n^{\alpha }$};
\end{tikzpicture}
    \caption{The effect of parallel-transporting $n^\alpha$ along $u^\alpha$.}
    \label{fig1.2}
\end{figure}
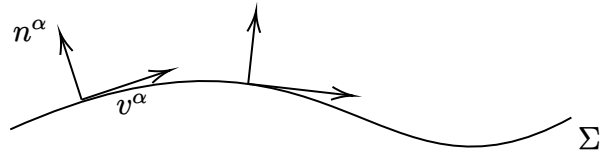

We have already identified the induced metric as the configuration variable of the Hamiltonian formulation. What will be the momentum conjugate to it? To answer this question, it is worth observing that $\Sigma$ is a hypersurface embedded in $M$. Thus to get a complete picture of how $\Sigma$ evolves in time, it is necessary to not only know its intrinsic properties, which are encoded in the induced metric, but also how it is embedded in $M$ – its extrinsic properties. Therefore, we seek an object that describes how $\Sigma$ is curved in the embedding manifold $M$. It then seems reasonable to expect that the evolution of the induced metric will effect this object, which will thus be related to the momentum conjugate to the induced metric. To look for this object, we take two vectors $u$, $v \in T_p\Sigma$ and using Eq~(\ref{eq1.3}), see how the parallel transport of one along the other will decompose into normal and tangential parts:
\begin{equation}
    \nabla_u v^\alpha = -(\nabla_u v_\beta) n^\beta n^\alpha + (\nabla_u v^\alpha + (\nabla_u v_\beta) n^\beta n^\alpha). \label{eq1.5}
\end{equation}
Consider the first term in this equation. Using the product rule and the fact that $v_\beta n^\beta = 0$ (since $v^\beta$ is tangent to $\Sigma$), we get $n^\alpha(\nabla_u n_\beta)v^\beta = n^\alpha u^\gamma (\nabla_\gamma n_\beta)v^\beta$, which, being the projection onto $v^\beta$ of the result of parallel-transporting $n^\beta$ along $u^\alpha$, describes how much $n^\beta$ rotates in the direction of $v^\beta$ when it is parallel-transported along $u^\alpha$ – gauging $n^\beta$ along $\Sigma$ gives us a measure of how $\Sigma$ is curved in $M$ (Fig.~\ref{fig1.2}). This suggests that we should define a quantity $\nabla_\alpha n_\beta$, which would describe the so-called \textit{extrinsic curvature} of $\Sigma$. However, $\nabla_\alpha n_\beta$ evidently has components both in $M$ and $\Sigma$, and we had promised initially to define quantities living solely in $\Sigma$. Thus we use the induced metric to project $\nabla_\alpha n_\beta$ onto $\Sigma$ and thereby define the extrinsic curvature:
\begin{equation}
    \begin{split}
         K_{\alpha\beta} &\equiv \tensor{q}{^\gamma_\alpha}\tensor{q}{^\lambda_\beta}\nabla_\gamma n_\lambda \\
         &= \tensor{q}{^\gamma_\alpha}\nabla_\gamma n_\beta. 
    \end{split}
\end{equation}\label{eq1.6}

To flesh out the relationship between $K_{\alpha\beta}$ and the momentum conjugate to $q_{\alpha\beta}$, we now hark back to $\mathbb{R}$ in the splitting of spacetime into space and time, $\phi: M \rightarrow \mathbb{R}\times{\Sigma}$. The pullback of the time coordinate $\tau \in \mathbb{R}$ will give rise to a time coordinate $t$ in $M$, i.e. $t = \phi^* \tau$. Similarly, the pushforward of the vector field $\partial_\tau$ will give rise to a vector field $t^\alpha \equiv \phi_*\partial_\tau$. This vector field `points forwards in time' in $M$, and it is the Lie derivative along this vector field that shall be associated with the time derivative in $M$. Like any object of interest, we first decompose $t^\alpha$ into normal and tangential components:
\begin{equation}
    t^\alpha = Nn^\alpha + N^\alpha, \label{eq1.7}
\end{equation}
where, in accordance with Eq~(\ref{eq1.3}),
\begin{align}
    N &= -g_{\alpha\beta}t^\alpha n^\beta \label{eq1.8}\\
    N^\alpha &= t^\alpha + (g_{\mu\nu}t^\mu n^\nu)n^\alpha. \label{eq1.9} 
\end{align}
$N$ is called the lapse and $N^\alpha$ is called the shift vector. Their physical interpretation and importance will become clear later. 

We will now show that $K_{\alpha\beta}$ is a function of the Lie derivative of the induced metric. To this end, we first prove that $K_{\alpha\beta}$ is symmetric. Take $u, v \in T_p\Sigma$. Since the spatial projection of the parallel transport of one vector along the other should result in a vector that is still tangent to $\Sigma$, we have that
\begin{align*}
    0 &= n_\alpha (\tensor{q}{^\gamma_\beta}u^\beta\nabla_\gamma v^\alpha - \tensor{q}{^\gamma_\beta}v^\beta\nabla_\gamma u^\alpha)\\
    &= -v^\alpha u^\beta  \tensor{q}{^\gamma_\beta}\nabla_\gamma n_\alpha + u^\alpha v^\beta \tensor{q}{^\gamma_\beta} \nabla_\gamma n_\alpha \\
    &= u^\beta v^\alpha (\tensor{q}{^\gamma_\alpha}\nabla_\gamma n_\beta - \tensor{q}{^\gamma_\beta}\nabla_\gamma n_\alpha), 
\end{align*}
which establishes the symmetry of $K_{\alpha\beta}$. Next, we observe that
\begin{equation}
    \begin{split}
        \mathcal{L}_n q_{\alpha\beta} &= n^\gamma\nabla_\gamma q_{\alpha\beta} + q_{\gamma\beta}\nabla_\alpha n^\gamma + q_{\alpha\gamma}\nabla_\beta n^\gamma\\
        &= n^\gamma\nabla_\gamma(g_{\alpha\beta} + n_\alpha n_\beta) + (g_{\gamma\beta} + n_\gamma n_\beta)\nabla_\alpha n^\gamma + (g_{\alpha\gamma} + n_\alpha n_\gamma)\nabla_\beta n^\gamma\\
        &= n^\gamma\nabla_\gamma(n_\alpha n_\beta) + \nabla_\alpha n_\beta + \nabla_\beta n_\alpha + n_\gamma n_\beta \nabla_\alpha n^\gamma + n_\alpha n_\gamma \nabla_\beta n^\gamma \\
        &= n^\gamma\nabla_\gamma(n_\alpha n_\beta) + \nabla_\alpha n_\beta + \nabla_\beta n_\alpha\\
        &= (\delta^\gamma_\alpha - n_\alpha n^\gamma)\nabla_\gamma n_\beta + (\delta^\gamma_\beta - n_\beta n^\gamma)\nabla_\gamma n_\alpha\\
        &= 2K_{\alpha\beta}.
    \end{split}\label{eq1.10}
\end{equation}
In the last step above, we used the symmetry of $K_{\alpha\beta}$. Next, we use Eq~(\ref{eq1.7}) in Eq~(\ref{eq1.10}) to conclude that
\begin{align*}
    K_{\alpha\beta} = \frac{1}{2N}\mathcal{L}_{t-N}q_{\alpha\beta} = \frac{1}{2N}(\mathcal{L}_t q_{\alpha\beta} - \mathcal{L}_N q_{\alpha\beta}). 
\end{align*}
Since $K_{\alpha\beta}$ already lives in $\Sigma$, we can contract it with $\tensor{q}{^\alpha_\beta}$ as many times as we like. This allows us to recast the preceding equation as
\begin{equation}
    K_{\alpha\beta} = \tensor{q}{^\gamma_\alpha}\tensor{q}{^\lambda_\beta}K_{\gamma\lambda} = \frac{1}{2N}(\tensor{q}{^\gamma_\alpha}\tensor{q}{^\lambda_\beta}\mathcal{L}_t q_{\gamma\lambda} - \tensor{q}{^\gamma_\alpha}\tensor{q}{^\lambda_\beta}\mathcal{L}_N q_{\gamma\lambda}) \label{eq1.11}
\end{equation}
We define the first term to be the time derivative of the induced metric, $\Dot{q}_{\alpha\beta}$. Consider the second term. We get
\begin{equation}
    \begin{split}
     \tensor{q}{^\gamma_\alpha}\tensor{q}{^\lambda_\beta}\mathcal{L}_N q_{\gamma\lambda} &= \tensor{q}{^\gamma_\alpha}q_{\mu\beta}\nabla_\gamma N^\mu + q_{\mu\alpha}\tensor{q}{^\lambda_\beta}\nabla_\lambda N^\mu + \tensor{q}{^\gamma_\alpha}\tensor{q}{^\lambda_\beta}N^{\mu}\nabla_{\mu}q_{\gamma\lambda}\\
    &= \tensor{q}{^\gamma_\alpha}q_{\mu\beta}\nabla_\gamma N^\mu + q_{\mu\alpha}\tensor{q}{^\lambda_\beta}\nabla_\lambda N^\mu,
    \end{split} \label{eq1.12}
\end{equation}
where the third term in the first line becomes zero because 
\begin{equation}
    \begin{split}
        \tensor{q}{^\gamma_\alpha}\tensor{q}{^\lambda_\beta}\nabla_\mu q_{\gamma\lambda} &= \tensor{q}{^\gamma_\alpha}\tensor{q}{^\lambda_\beta}\nabla_\mu g_{\gamma\beta} + \tensor{q}{^\gamma_\alpha}\tensor{q}{^\lambda_\beta}\nabla_\mu (n_\gamma n_\lambda) \\
        &= \tensor{q}{^\gamma_\alpha}\tensor{q}{^\lambda_\beta}\nabla_\mu (n_\gamma n_\beta)\\
        &= \tensor{q}{^\gamma_\alpha}\nabla_\mu(\tensor{q}{^\lambda_\beta} n_\gamma n_\lambda) - \tensor{q}{^\gamma_\alpha}n_\gamma n_\lambda \nabla_\mu \tensor{q}{^\lambda_\beta} = 0.
    \end{split} \label{eq1.13}
\end{equation}
Incidentally, this last calculation shows that the covariant derivative of the induced metric projected onto $\Sigma$ is compatible with the induced metric. That is, we are naturally led to define a covariant derivative on $\Sigma$:
\begin{equation}
    D_{\alpha}v^{\beta} \equiv \tensor{q}{^\mu_\alpha}\tensor{q}{^\beta_\nu}\nabla_{\mu}v^{\nu} \label{eq1.14}
\end{equation}
for all $v \in T_p\Sigma$. Eqs~(\ref{eq1.12}--\ref{eq1.14}) now make us able to write Eq~(\ref{eq1.11}) as
\begin{equation}
    K_{\alpha\beta} = \frac{1}{2N}(\Dot{q}_{\alpha\beta} - D_\alpha N_\beta - D_\beta N_\alpha). \label{eq1.15}
\end{equation}
As advertised, $K_{\alpha\beta}$ is a function of the time derivative of the induced metric. 

The covariant derivative on $\Sigma$ immediately gives rise to a Riemann curvature tensor on $\Sigma$:
\begin{equation}
    \tensor{^{(3)}R}{^\alpha_{\beta\mu\nu}}V_\alpha \equiv [D_\beta, D_\mu]V_\nu. 
\end{equation}

We have set up all the mathematical machinery necessary to perform a Legendre transform of the Einstein-Hilbert action, 
\begin{equation}
    S = \int d^4x \sqrt{g}R, \label{eq1.17}
\end{equation}
whose variation with respect to the metric yields the Einstein field equations. We seek to cast Eq~(\ref{eq1.17}) in the form $\int dt \int d^3x (p^{ab}\Dot{q}_{ab} - H))$, where $p^{ab}$ is the momentum conjugate to $q_{ab}$ and $H$ is the Hamiltonian. To that end, it is necessary to decompose $\sqrt{g}R$ into objects living solely in $\Sigma$. We shall now perform this so-called 3+1 decomposition.

Since our final expressions will be manifestly coordinate invariant, it will be harmless to work in a convenient set of local coordinates. To this end, in a neighbourhood of any $p \in \Sigma$, we choose local coordinates $x^\alpha$ such that $x^0 = \phi^*\tau$, $\partial_0 = t^\alpha\nabla_\alpha = \nabla_{\phi_*\partial_\tau}$ and the spatial vector fields $\partial_a$ are tangent to $\Sigma$ at $p$\footnote{Choosing such local coordinates is always possible. Indeed, since the diffeomorphism $\phi$ is arbitrary, we are free to set $x^0 = \phi^*\tau$. Further, to get the desired $\partial_a$, we can use Eq~(\ref{eq1.3}) to extract the tangential components of the vector fields $\partial_x$, $\partial_y$, $\partial_z$ of coordinates $(x,y,z)$ in $\Sigma$.}. These virtuous coordinates afford the simplification that only the spatial components of tensors in $M$ that are tangent to $\Sigma$ are nonzero. 

Let us begin with the decomposition of $\sqrt{g}$. Recalling that $t^\alpha$ points in the direction of time, we write $dx^\alpha = t^\alpha dt + \tensor{q}{^\alpha_\beta} dx^\beta$ and thus, using Eq~(\ref{eq1.7}),
\begin{align*}
    ds^2 &= g_{\alpha\beta}[(Ndt)n^\alpha + (N^\alpha dt + \tensor{q}{^\alpha_\gamma}dx^\gamma)][(Ndt)n^\alpha + (N^\beta dt + \tensor{q}{^\beta_\lambda}dx^\lambda)]\\
    &= -N^2dt^2 + g_{\alpha\beta}N^\alpha N^\beta dt^2 + g_{\alpha\beta}N^\alpha \tensor{q}{^\beta_\lambda}dx^\lambda + g_{\alpha\beta}N^\beta \tensor{q}{^\alpha_\gamma}dx^\gamma + g_{\alpha\beta}\tensor{q}{^\alpha_\gamma}\tensor{q}{^\beta_\lambda}dx^\gamma dx^\lambda,
\end{align*}
where use was made of the orthogonality of $n$ to any objects living on $\Sigma$. Now, by virtue of our chosen coordinates, except $g_{\alpha\beta}$, all the tensorial terms above have nonzero components only for spatial indices, for which the metric $g$ is equivalent to the induced metric. Therefore,  we can recast the preceding equation as
\begin{equation}
    ds^2 = -N^2 dt^2 + q_{ab}(dx^a + N^a dt)(dx^b + N^b dt), \label{eq1.18}
\end{equation}
which admits the following matrix interpretation: 
\begin{equation}
   g_{\alpha\beta} = \left(\begin{array}{c|c c} 
	-N^2 + N^aN_a & N_b\\ 
	\hline 
	N^a & q_{ab}  
\end{array}\right). \label{eq1.19}
\end{equation}
Now suppose $t^\alpha = n^\alpha$, so that $N^\alpha = 0$. Then the equation above guarantees
\begin{equation}
    g = -N^2 q, \label{eq1.20}
\end{equation}
which, being coordinate invariant, holds for arbitrary $t^\alpha$ as well.

Eq~(\ref{eq1.18}) furnishes the physical interpretation of the lapse and the shift, namely, that displacements in spacetime are determined by the metric induced on the spatial slice one happens to be on, as well as by the deformations of neighbouring slices, encoded in $N$ and $N^a$, with respect to each other. In particular, under time evolution, the lapse measures how much a slice $\Sigma$ is pushed in the normal direction, while the shift specifies a push in the tangent direction. 

Now, let us decompose $R$. For this, we need to find the normal and tangential projections of the Riemann curvature tensor. These are given by the \textit{Gauss-Codazzi} equations:
\begin{subequations}
\label{eq1.21}
\begin{align}
    \tensor{q}{^\gamma_\alpha}\tensor{q}{^\lambda_\beta}\tensor{q}{^\mu_\rho}\tensor{q}{^\nu_\sigma}\tensor{R}{^\sigma_{\gamma\lambda\mu}} = \tensor{^{(3)}R}{^\nu_{\alpha\beta\rho}} + K_{\alpha\rho}\tensor{K}{^\nu_\beta} - K_{\beta\rho}\tensor{K}{^\nu_\alpha},
\end{align}
\begin{align}
    \tensor{q}{^\alpha_\mu}\tensor{q}{^\beta_\nu}\tensor{q}{^\gamma_\lambda}R_{\alpha\beta\gamma\rho}n^\rho = D_\mu K_{\nu\lambda} - D_\nu K_{\mu\lambda},
\end{align}
\begin{align}
    R_{\alpha\beta\gamma\lambda}n^\beta n^\lambda = -\mathcal{L}_nK_{\alpha\beta} + K_{\alpha\gamma}\tensor{K}{^\gamma_\beta} + D_{(\alpha}n^\gamma \nabla_{|\gamma|} n_{\beta)} + (n^\gamma\nabla_\gamma n_\alpha)(n^\lambda\nabla_\lambda n_\beta)
\end{align}
\end{subequations}
Using them, we find (up to total derivatives\footnote{Since our purpose here is more illustrative than pedantic, we shall content ourselves with ignoring any boundary terms in the action.})
\begin{equation}
\begin{split}
    R &= \tensor{^{(3)}R}{} + K^{\alpha\beta}K_{\alpha\beta} - (\tensor{K}{^\alpha_\alpha})^2,\\
    &= \tensor{^{(3)}R}{} + K^{ab}K_{ab} - (\tensor{K}{^a_a})^2,
    \\
    &= \tensor{^{(3)}R}{} + \text{tr}(K^2) - (\text{tr}K)^2.
\end{split} \label{eq1.22}
\end{equation}
Note that in the second line above, we switched the spacetime indices to purely spatial indices, since the extrinsic curvature has no components normal to $\Sigma$ and thus, in our chosen coordinates, its temporal components are zero. We are now ready to perform a Legendre transform on the action (Eq~(\ref{eq1.17})), which now becomes
\begin{equation}
    S = \int dt\int_\Sigma d^3x \,\mathcal{L}(p^{ab}, q_{ab}), \label{eq1.23}
\end{equation}
where
\begin{equation}
    \mathcal{L} = \sqrt{q}N(\tensor{^{(3)}R}{} + \text{tr}(K^2) - (\text{tr}K)^2) \label{eq1.24}
\end{equation}
is the Lagrangian density, now canonically decomposed. We immediately identify the momentum conjugate to $q_{ab}$:
\begin{equation}
    \Tilde{p}^{ab} := \frac{\partial\mathcal{L}}{\partial \Dot{q}_{ab}} = \sqrt{q}(K^{ab} - \text{tr}(K)q^{ab}), \label{eq1.25}
\end{equation}
where the tilde over $p$ indicates the fact that the conjugate momentum is a tensor density of weight 1, i.e. comes with one factor of the square-root of the metric. Using the preceding equation, we can recast $\mathcal{L}$ in terms of $\Tilde{p}^{ab}$ and then finally perform the Legendre transform to find the Hamiltonian of general relativity:
\begin{equation}
    \begin{split}
        H &:= \int_\Sigma d^3x (\Tilde{p}^{ab}\Dot{q}_{ab}-\mathcal{L}),\\
        &= \int_\Sigma d^3x \sqrt{q}(NC + N^a C_a),
    \end{split} \label{eq1.26}
\end{equation}
where
\begin{equation}
    C = -^{(3)}R + \frac{1}{q}\left(\text{tr}(\Tilde{p}^2)-\frac{1}{2}\text{tr}(\Tilde{p})^2\right) \label{eq1.27}
\end{equation}
and
\begin{equation}
    C_a = -2D^b\left(\frac{1}{\sqrt{q}}\Tilde{p}_{ab}\right). \label{eq1.28}
\end{equation}
We pause here to point out a peculiar feature of this Hamiltonian. Using the Gauss-Codazzi equations, it is possible to show that
\begin{equation}
    C = -2G_{\alpha\beta}n^\alpha n^\beta, \quad C_a = -2G_{\alpha a}n^\alpha. \label{eq1.29}
\end{equation}
But this, via the vacuum Einstein equations, implies that $C$ and $C_a$ vanish, and thus the Hamiltonian of general relativity is zero! This seems to be a puzzling fact, for it would appear that the dynamics of general relativity is trivial. However, one can explicitly work out the (rather complicated) Poisson brackets of $\Tilde{p}^{ab}$ and $q_{ab}$ with $H$ to confirm that they do evolve nontrivially in time. In fact, the vanishing of the Hamiltonian in a generally diffeomorphism invariant theory like general relativity is to be expected, for otherwise, one would be left with the unsatisfactory conclusion that there is an absolute time with respect to which a non-vanishing Hamiltonian generates evolution. Therefore, as promised, the canonical decomposition performed above is independent of the choice of the diffeomorphism $\phi$. It is worth mentioning here that standard quantum mechanics is not invariant with respect to reparametrisations of its time parameter. This fundamental disjunction between how these two theories treat time is the origin of the infamous \textit{problem of time}, which we shall return to later. 

The vanishing of $C$ and $C_a$ is a pleasing conclusion from another standpoint as well, since it lends credence to the fact that the initial-value formulation of general relativity is well posed. This is so because explicitly writing out Eqs~(\ref{eq1.29}) shows that they contain, at most, the first time derivatives of the metric -- thus, they encapsulate nothing but constraints on the initial values of the metric and its first-order time derivatives, and this is to be expected of the initial-value formulation of a theory described by a system of second-order partial differential equations, as general relativity is.

But what is the physical meaning of the constraints? It turns out that there is a precise sense in which the constraints can be thought of as generating deformations of the spatial slice $\Sigma$ in time. To understand this geometric meaning of the constraints, it is essential to study the Poisson-bracket structure of general relativity. 


\section{The Poisson-bracket Structure of Hamiltonian General Relativity}

Let us begin by recalling the formal definition of the Poisson bracket between any two functions $f(q,\Tilde{p})$ and $g(q,\Tilde{p})$ on the phase space:
\begin{equation}
    \left\{f, g\right\} = \int_\Sigma \sqrt{q}\,d^3x\left(\frac{\delta f}{\delta \Tilde{p}^{ab}(x)}\frac{\delta g}{\delta q_{ab}(x)} - \frac{\delta f}{\delta q
    _{ab}(x)}\frac{\delta g}{\delta \Tilde{p}^{ab}(x)} \right) \label{eq1.30} 
\end{equation}
It is straightforward to verify that Poisson brackets are Lie brackets. 

Eq~(\ref{eq1.30}) readily yields the Poisson brackets between the canonical variables:
\begin{subequations}
\label{eq1.31}
\begin{align}
    \left\{\Tilde{p}^{ab}(x), q_{cd}(y)\right\} &= \delta^{(a}_c\delta^{b)}_d\delta^3(x-y)\\ 
    \left\{\Tilde{p}^{ab}(x), \Tilde{p}^{cd}(y)\right\} &= 0 \\
     \left\{q_{ab}(x), \Tilde{p}_{cd}(y)\right\} &= 0 
\end{align}\textbf{}
\end{subequations}
Next, as indicated earlier, one could evaluate the brackets $\left\{H, q_{ab}\right\}:=\Dot{q}_{ab}$ and $\left\{H, \Tilde{p}^{ab}\right\}:=\Dot{\Tilde{p}}^{ab}$ to find how the canonical variables evolve in time, but neither are their complicated expressions particularly illuminating, nor will we need them in the subsequent discussion. Our essential motive in using Poisson brackets is to obtain the brackets between the canonical variables and analyse the nature of the constraints so as to be able to understand their geometric meaning. We focus on this second task now. 

To start with, as noted earlier, the lapse $N$ and the shift $N^a$ are a measure of how much time evolution pushes the surface $\Sigma$ in the normal direction and the tangent direction, respectively. Now, in the Hamiltonian, the former occurs with $C$ and the latter with $C_a$. Therefore, we expect that the part of the Hamiltonian containing\footnote{Eq~(\ref{eq1.32}) is what is called the \textit{smeared} version of the constraint $C$. Smearing the constraints is important on two accounts: (1) it reveals a clear physical interpretation of the constraints, and (2) the Poisson brackets between unsmeared constraints are ill-defined, involving Dirac deltas.} $C$,
\begin{equation}
    C_N := \int_\Sigma d^3x \sqrt{q} NC, \label{eq1.32}
\end{equation}
generates time evolution in a manner that corresponds to nudging $\Sigma$ in the normal direction. On the other hand, the part of the Hamiltonian containing $C_a$,
\begin{equation}
    C_{\Vec{N}} := \int_\Sigma d^3x \sqrt{q} N^a C_a, \label{eq1.33} 
\end{equation}
should generate time evolution that corresponds to deformations of $\Sigma$ tangent to itself. Both of these expectations can be made precise. In particular, we can show that for any phase-space function $f(q, \Tilde{p})$, 
\begin{subequations}
\label{eq1.33*}
    \begin{align}
        \{f, C_{\Vec{N}}\} &= \mathcal{L}_{\Vec{N}}f \\
        \{f, C_{N}\} &= (\cdots) + \mathcal{L}_{Nn} f 
    \end{align}
\end{subequations}
where $(\cdots)$ indicates terms proportional to $C_{\Vec{N}}$ and the Einstein field equations, and $n$ is the unit normal to $\Sigma$. In other words, the flow of $N^a$, which spans $\Sigma$, is a one-parameter family of diffeomorphisms of $\Sigma$; thus, according to Eq~(\ref{eq1.33*}a), $C_{\Vec{N}}$ generates transformations of the physical phase space (the subspace on which $C_a = 0$) corresponding to diffeomorphisms of $\Sigma$. For this reason, $C_{\Vec{N}}$ (or $C_a$) is called the \textit{diffeomorphism constraint} of general relativity. Similarly, provided that the Einstein equations and the diffeomorphism constraint are in force, Eq~(\ref{eq1.33*}b) tells us that $C_N$, called the \textit{Hamiltonian constraint}, generates transformations of $\Sigma$ in the normal direction. We will derive Eq~(\ref{eq1.34}a) below; Eq~(\ref{eq1.34}b) requires a fairly involved calculation, for which we refer to Ref. \cite{thiemann}. 

It is worth emphasising that owing to the presence of Lie derivatives in Eq~(\ref{eq1.33*}), we interpret $C_N$ and $C_{\Vec{N}}$ as generators of \textit{infinitesimal} transformations. To generate \textit{finite} transformations, one will have to exponentiate these constraints. In general, this is only possible for the diffeomorphism constraint, since the Hamiltonian constraint involves the nontrivial $(\cdots)$ terms as well. For the diffeomorphism constraint, exponentiating yields the flow generated by $\Vec{N}$, i.e. the integral curves $\phi_t : \mathbb{R}\to\Sigma$ of the differential equation
\begin{align}
        \frac{d\phi_t(x)}{dt} &= \Vec{N}(\phi_t(x)).   \label{eq1.33**}
\end{align}
Note that the one-parameter family of maps $\phi_t$ forms a group under the operation $\phi_s \cdot \phi_t = \phi_{s+t}$. It is called the \textit{spatial diffeomorphism group.}  

Having understood the physical meaning of the constraints, we now derive the Poisson brackets between them. We find:
\begin{subequations}
\label{eq1.34}
\begin{align}
    \left\{C_{\Vec{M}}, C_{\Vec{N}}\right\} &= C_{\mathcal{L}_{\Vec{M}}\Vec{N}},\\
    \left\{C_{\Vec{N}}, C_N \right\} &= C_{\mathcal{L}_{\Vec{N}}N},\\
    \{C_M, C_N \} &= C_{q^{ab}(M\partial_bN-N\partial_bM)}.
\end{align}
\end{subequations}
A number of comments are in order here. First, the constraints are closed under taking Poisson brackets -- the bracket of two constraints is again a constraint. Such constraints are said to be \textit{first class}, and Eqs~(\ref{eq1.34}) form what is called a \textit{Dirac algebra}. As we shall see, the first-class nature of the constraints is important in finding physical quantum states in Dirac quantisation. Second, due to the presence of $q^{ab}$ in Eq~(\ref{eq1.34}a), the Dirac algebra is not a Lie algebra. If it were, the task of quantisation would be significantly simpler, for one would then simply have to look for the representations of the group of which the Dirac algebra would be a Lie algebra. 

\begin{mdframed}[style=testframe]
    \textbf{Derivation -- Eqs~(\ref{eq1.33*}a) and (\ref{eq1.34})}
    \begin{enumerate}[(i)]
        \item We head straight into the calculation:
        \begin{align*}
            \{C_{\Vec{M}}, C_{\Vec{N}}\} &= \int d^3z \sqrt{q(z)} \left[\frac{\delta C_{\Vec{M}}}{\delta\Tilde{p}^{ab}(z)}\frac{\delta C_{\Vec{N}}}{\delta q_{ab}(z)} - (\Vec{M} \leftrightarrow \Vec{N}) \right]\\
            &= \int d^3z\sqrt{q(z)}\frac{\delta}{\delta\Tilde{p}^{ab}(z)}\left[\int d^3x\sqrt{q(x)}M^c C_c(x) \right]\\
            &\quad\times\frac{\delta}{\delta q_{ab}}\left[ \int d^3y\sqrt{q(y)}N^dC_d(y)\right] - (\Vec{M}\leftrightarrow\Vec{N})\\
            &= \frac{1}{2}\int d^3xd^3y\sqrt{q(x)q(y)}M^c\frac{\delta C_c(x)}{\delta\Tilde{p}^{ab}(y)}q^{ab}N^dC_d(y)\\
            &\quad + \int d^3zd^3xd^3y\sqrt{q(x)q(y)q(z)}M^c\frac{\delta C_c(x)}{\delta\Tilde{p}^{ab}(y)}N^d\frac{\delta C_d(y)}{\delta q_{ab}(z)} \\
            &\quad - (\Vec{M}\leftrightarrow\Vec{N})
        \end{align*}
        In the fourth line, we applied the product rule on the derivatve in the third line and used the identity $\delta q(x)/\delta q_{ab}(y) = q(x)q^{ab}(x)\delta^3(x-y)$. The term in the second-last line above cancels the corresponding term in $(\Vec{M}\leftrightarrow\Vec{N})$. On the other hand, ignoring total derivatives,
        \begin{align*}
            N_d(y)\frac{\delta C^c(x)}{\delta\Tilde{p}^{ab}(y)}q^{ab}(y) &= \left[-\frac{2}{\sqrt{q(x)}} N^d(y)\frac{\delta}{\delta\Tilde{p}^{ab}(y)}D^{(x)}_l\Tilde{p}^{cl}(x) \right]q^{ab}(y)\\
            &= \frac{2}{\sqrt{q(x)}}(D_l N_d)\delta^{(c}_a\delta^{l)}_bq^{ab}(y)\delta^3(x-y)\\
            &= \frac{2}{\sqrt{q(x)}}(D^c N_d)\delta^3(x-y),
        \end{align*}
        which upon substitution into its parent equation yields
        \begin{align*}
            \{C_{\Vec{M}}, C_{\Vec{N}}\} = \int d^3x\sqrt{q}[M_c(D^cN_d)C^d - (\Vec{M}\leftrightarrow\Vec{N})] = \int d^3x\sqrt{q}\mathcal{L}_{\Vec{M}}N_d C^d.
        \end{align*}

        \item We begin by writing the diffeomorphism constraint in the following way \cite{minahil}. 
        \begin{align*}
            C_{\Vec{N}} = -2\int d^3x\,N^bD^a\Tilde{p}_{ab} = 2\int d^3x \,\Tilde{p}_{ab}D^aN^b
            &= 2\int d^3x\, \Tilde{p}_{ab}D^{(a}N^{b)}\\
            &= \int d^3x\, \Tilde{p}_{ab}\,\mathcal{L}_{\Vec{N}}q^{ab}\\
            &= -\int d^3x\,q^{ab}\,\mathcal{L}_{\Vec{N}}\Tilde{p}^{ab}
        \end{align*}
        It is then easy to see that
        \begin{align*}
            \frac{\delta C_{\Vec{N}}}{\delta\Tilde{p}^{ab}} &= \mathcal{L}_{\Vec{N}}q_{ab},\\
            \frac{\delta C_{\Vec{N}}}{\delta q_{ab}} &= -\mathcal{L}_{\Vec{N}}\Tilde{p}^{ab},
        \end{align*}
        which at once yields validates Eq~(\ref{eq1.33*}a). It is also not difficult to find that
        \begin{equation*}
             \frac{\delta C_M}{\delta\Tilde{p}^{ab}} = q^{-1/2}M[2\Tilde{p}_{ab} - q_{ab}\text{tr}(\Tilde{p})].
        \end{equation*}
        Next, consider the identity 
        \begin{equation*}
            \frac{\delta R(x)}{\delta q_{ab}(y)} = -R^{ab}(x)\delta^3(x-y) + q^{ab}D^kD_k\delta^3(x-y) - D^{(a}D^{b)}\delta^3(x-y),
        \end{equation*}
        which, via two successive integrations-by-parts, entails that
        \begin{align*}
            \int d^3x\,N \frac{\delta R(x)}{\delta q_{ab}(y)} = -R^{ab} - D^{(a}D^{b)}N + q^{ab}D^kD_kN. 
        \end{align*}
        This, along with the fact that traces are metric-dependent and $\delta q(x)/\delta q_{ab}(y) = q(x)q^{ab}(x)\delta^3(x-y)$, yields
        \begin{align*}
            \frac{\delta C_N}{\delta q_{ab}} &= \int d^3x\,N\frac{\delta}{\delta q_{ab}(y)}\left[-q^{1/2}R(x) + q^{-1/2}\left( \text{tr}(\Tilde{p}^2) - \frac{1}{2}\,\text{tr}(\Tilde{p}) \right) \right]\\
            &= Nq^{1/2}\left[R^{ab}-\frac{q^{ab}R}{2}\right] + \left[D^{(a}D^{b)}N - q^{ab}D^kD_kN\right] \\
            &\quad + 2Nq^{-1/2}\left[\tensor{\Tilde{p}}{^{(a}_n}\tensor{\Tilde{p}}{^{b)n}} - \frac{1}{2}\,\text{tr}(\Tilde{p})\,\Tilde{p}^{ab}\right] - \frac{Nq^{-1/2}}{2}q^{ab}\left[\text{tr}(\Tilde{p}^2) - \frac{1}{2}\,\text{tr}(\Tilde{p})\right]. 
        \end{align*}
        Finally, then, some integrations-by-parts and rearrangements yield the desired result. 

        \item Using the derivatives of $C_N$ as found in (ii), we get
        \begin{align*}
            \{C_M, C_N\} &= \int d^3z\sqrt{q(z)}\left[ \frac{\delta C_M}{\delta\Tilde{p}^{ab}(z)}\frac{\delta C_N}{\delta q_{ab}(z)} - (M\leftrightarrow N) \right]\\
            &= \int MN(\cdots) \\
            &\quad +\int d^3x\,q^{-1/2}M[2\Tilde{p}_{ab}-q_{ab}\text{tr}(\Tilde{p})][D^{(a}D^{b)}N - q^{ab}D^kD_kN]\\
            &\quad - (M\leftrightarrow N)\\
            &= \int d^3x\,q^{-1/2}[2M\Tilde{p}_{ab}D^aD^bN - 2M\,\text{tr}(\Tilde{p})D_kD^kN\\ 
            &\quad - \text{tr}(\Tilde{p})D_kD^kN + 3\,\text{tr}(\Tilde{p})D_kD^kN] - (M\leftrightarrow N)\\
            &= 2\int d^3x\,q^{-1/2}M\Tilde{p}_{ab}D^aD^bN - (M\leftrightarrow N)\\
            &= -2\int d^3x\,q^{-1/2}(D^aM)(D^bN)\Tilde{p}_{ab} - 2\int d^3x\,q^{-1/2}MD^bN D^a\Tilde{p}_{ab}\\ 
            &\quad -(M\leftrightarrow N)\\
            &= -2\int d^3x\,q^{-1/2}(MD^bN-ND^bM)D^a\Tilde{p}_{ab},
        \end{align*}
        which is the desired result. 
    \end{enumerate}
    
\end{mdframed}

\section{An Attempt at Quantisation}
We will now attempt to canonically quantise general relativity. Rather than give a general algorithm for quantising any constrained system and subsequently apply it to general relativity, we shall take the less abstract route of building such an algorithm by way of quantising general relativity. It is worth emphasising that our algorithm will be one among many nonequivalent ones that can be used to quantise different constrained systems. Our choice is motivated by its transparency in highlighting the fundamental issues in quantising general relativity -- in particular the ADM formulation of the theory -- and by its success in being applied to loop quantum gravity. For details regarding different quantisation schemes and their application to several examples, the reader is referred to Ref. \cite{ashtekar 1, ashtekar 5}.  

\subsection{A home for general quantum states}
To begin with, as in usual quantum mechanics, our first task is to find a representation of the Poisson algebra between the canonical variables (Eqs~(\ref{eq1.31})) on some complex vector space $\mathcal{V}$, which shall be the home of our quantum states. In other words, we wish to promote $q_{ab}$ and $\Tilde{p}^{ab}$ to operators\footnote{In a field theory, variables defined at a single point are singular, as evidenced by the appearance of the Dirac delta in Eqs~(\ref{eq1.31}) and (\ref{eq1.35}). This leads to infinities in the quantum theory. To remove these, one should instead work with smeared versions of the classical variables, i.e. integrating these variables using some test functions that probe the structure of the variables in some local region of the underlying manifold (Cf. Note 7 above). This is not difficult to do in any field theory and sheds no light on the difficulties that a quantisation of the ADM variables faces, and thus we will not pursue this issue here. However, in the interests of rigour, we will return to this point when formulating loop quantum gravity proper (Cf. Section 3.2).} on $\mathcal{V}$ that satisfy the following commutation relations.
\begin{subequations}
\label{eq1.35}
\begin{align}
    [\hat{\Tilde{p}}^{ab}(x), \hat{q}_{cd}(y)] &= -i\delta^{(a}_c\delta^{b)}_d\delta^3(x-y),\\
     [\hat{\Tilde{p}}^{ab}(x), \hat{\Tilde{p}}^{cd}(y)] &= 0,\\
     [\hat{q}_{ab}(x), \hat{q}_{cd}(y)] &= 0. 
\end{align}
\end{subequations}
To accomplish this task, we first define $\mathcal{V}$ to be the space of all complex-valued smooth scalar functions on Met$(\Sigma)$. Then at every $x \in \Sigma$ we define the operators $\hat{q}_{ab}(x)$ and $\hat{p}^{ab}(x)$ to be such that for all $\psi \in \mathcal{V}$ and for all $q \in \text{Met}(\Sigma)$,
\begin{subequations}
\label{eq1.36}
\begin{align}
    (\hat{q}_{ab}(x)\psi)(q) &= g_{ab}(x)\psi(q),\\
    (\Tilde{p}^{ab}(x)\psi)(q) &= -i\frac{\delta}{\delta q_{ab}(x)}\psi(q),
\end{align}
\end{subequations}
where $g$ is a three-metric in $\Sigma$. As can be verified, the operators so defined satisfy Eqs~(\ref{eq1.35}).

Now, we must recall that $(q, {\Tilde{p}})$ are not the only objects to be promoted to operators on $\mathcal{V}$: we have in the classical theory various observables each of which ought to have an unambiguous quantum analogue. In general, these observables are functions of the phase-space variables. Therefore, in order to consistently promote all the observables to operators, we need a set $\mathcal{S}$ of elementary classical variables composed of the phase-space variables; promotion of these elementary variables to operators should automatically guarantee the promotion of all suitable functions of the phase-space variables to operators. These considerations require that $\mathcal{S}$ is in general a subspace of the vector space of all smooth, complex-valued functions on the phase space such that
\begin{enumerate}[(a)]
    \item it is sufficiently large, so that any function on the phase space can be expressed in terms of the sums of products of the elementary variables;
    \item the elementary variables are closed under certain operations, such as Poisson brackets and complex conjugation, i.e. $\forall F,\, G \in \mathcal{S}, \{F, G\} \in \mathcal{S}$ and $F^* \in \mathcal{S}$. 
\end{enumerate}
For instance, consider a particle moving in three dimensions. Its phase space is coordinatised by its position $(x, y, z)$ and momentum $(p_x, p_y, p_z)$. $\mathcal{S}$ can be identified with complex linear combinations of these six variables. 

Once $\mathcal{S}$ is identified, we need to do two things. First, associate with each $F \in \mathcal{S}$ an abstract operator $\hat{F}$ by constructing a free associative algebra $\mathcal{B}$ on $\mathcal{S}$. This can be done by imposing (1) canonical commutation relations $[\hat{F}, \hat{G}] = i\{\widehat{F, G}\}$, and (2) (possibly though not necessarily) anti-commutation relations that capture any non-trivial algebraic relations between the elementary variables. Second, construct a $\star$-algebra $\mathcal{B}^{\star}$, where $\star$ denotes an involution\footnote{An involution is a map $\star: \mathcal{B}\to\mathcal{B}$ that satisfies the following conditions. \begin{enumerate}[(i)]
    \item anti-linearity: $(\alpha A + \beta B)^{\star} = \alpha^*A^\star + \beta^*B^\star$ for all $\alpha,\,\beta\in\mathbb{C}$ and $A,\,B\in\mathcal{B}$.
    \item $(AB)^{\star} = B^\star A^\star$.
    \item self-inverse: $(A^\star)^\star = A$. 
\end{enumerate}}, on $\mathcal{B}$ by requiring that if two classical variables are related by $F^* = B$, then $\hat{F}^\star = \hat{B}$. From this and the properties of the involution, it follows that the $\star$-algebra serves to mimic Hermitian conjugation. In fact, if $\mathcal{V}$ were a Hilbert space, we could find a linear representation of $\mathcal{B}^{\star}$ on $\mathcal{V}$ via linear operators, i.e. a map $R: \mathcal{B}^{\star}\to\mathcal{L}(\mathcal{V})$, $\mathcal{L}(\mathcal{V)}$ being the set of all linear operators on $\mathcal{V}$, such that $R(\hat{A}^{\star}) = R(\hat{A})^\dagger$ for all $\hat{A}$ in $\mathcal{B}^{\star}$, where $\dagger$ stands for Hermitian conjugation with respect to the inner product on $\mathcal{V}$ \footnote{This round-about way of arriving at Hermitian conjugation via $\star$-algebras is not an exercise in pedantry, since we shall find it necessary for a rigorous construction of the Hilbert space of loop quantum gravity.}.  

\subsection{A home for physical states}
The next step in the process is to realise that not all states $\psi \in \mathcal{V}$ are physical. The reason is simple: in the classical theory, only those points in the phase space $(q, \Tilde{p})$ that satisfy the constraints $C_a$ and $C$ are physically allowed. Therefore, these constraints need to be somehow incorporated in the quantum theory. There are at least three (non-equivalent) ways of doing this \cite{rovelli 3}. We describe each of them below.

\subsubsection{1. Dirac quantisation}
Since physical states ought to be quantum, and $\mathcal{V}$ is the home of general quantum states, in this framework, we demand that physical states are those that lie in the null space of all the constraint operators, i.e. $\mathcal{V}_{phy} = \{\psi \in \mathcal{H}: \hat{C}\psi = \hat{C}_a\psi = 0 \}$. Thus two tasks need to be performed: (1) promote the classical constraints to operators on $\mathcal{V}$, and (2) find the null space of these operators. However, there are serious obstructions to carrying out both these tasks in the ADM formalism of general relativity.

To start with, a glance at Eqs~(\ref{eq1.34}) will shed light on the enormity of the task of promoting the constraints to operators. These constraints are highly non-polynomial functions of the canonical variables. In particular, it is not even clear how to define a ``square-root'' operator corresponding to the square-root of the metric determinant. Is a regularisation procedure required? If so, will it be compatible with the regularisation of the other terms in a particular constraint? 

Moreover, any operator representation of the constraints will necessarily have a non-unique commutator algebra. This is so because the commutator between the canonical variables (Eq~(\ref{eq1.35})a) is non-trivial, and thus different ways of ordering these variables in the classical constraint functions will yield different commutators between the constraint operators. This is a particularly severe problem, since consistency requires that the constraint operators be \textit{weakly} closed under commutation, which is to say that the commutator between any constraint operators yields an operator that has a constraint on the extreme right, so that the action of the commutator on states in $\mathcal{V}_{phy}$ is again zero; otherwise, on the one hand, $C_i\psi = 0$ would imply $[C_i, C_j]\psi = 0$, and on the other hand, $[C_i, C_j] = (\cdots)f(q,p)$ for some $f(p,q) \neq C_i$ would imply the exact opposite.

Finally, it is worth noting that this approach works only if the constraints are first class, for otherwise, weak closure could not be achieved even at the classical level. This point highlights the importance of always working with a formulation of general relativity that has first-class constraints, because in their absence, one would not even know where to begin. That is, given first-class constraints, we can at least hope to find an operator ordering that yields commutation relations that are analogous to the Dirac algebra:
\begin{subequations}
\label{eq1.37}
\begin{align}
    [\hat{C}_{\Vec{M}}, \hat{C}_{\Vec{N}}] &= -i\hat{C}_{\mathcal{L}_{\Vec{M}}\Vec{N}},\\
    [\hat{C}_{\Vec{N}}, C_N] &= -i\hat{C}_{\mathcal{L}_{\Vec{N}}N},\\
    [\hat{C}_M, \hat{C}_N] &= -i\hat{C}_{q^{ab}(M\partial_b N - N\partial_b M)}.
\end{align}
\end{subequations}
One could then invoke Occam's razor and pretend to brush other operator orderings under the carpet. Thus the lessons to be learnt from the failure of this approach to quantising the constraints are the intractability of non-polynomial constraints and the importance of working with first-class constraints only. As to the second point, we shall keep it in mind in formulating alternative Hamiltonian formulations of general relativity in the next chapter. As to the first point, we will see that these alternative formulations have constraints that are, at most, quadratic polynomials in the canonical variables (or derivatives thereof). 

Since, as the discussion above makes clear, there is no satisfactory way of promoting the constraints to operators in the ADM variables, the question of solving these constraints becomes moot. 

\subsubsection{2. 
Reduced phase space quantisation}
In this scheme, one solves the constraints at the classical level, reducing the unconstrained phase space to its physical subspace on which the constraints vanish. This is done by finding classical functions of the phase space variables that have vanishing Poisson brackets with the constraints either identically or only on the constraint surface (i.e. the Poisson brackets are again constraints). In the former case, the functions are called \textit{strong observables} and in the latter, \textit{weak observables}. The name observable comes from the fact that since the constraints of general relativity encode the diffeomorphism invariance of the theory, these functions are the only diffeomorphism-invariant and hence, physically observable quantities. 

Now, we need not find all such observables, but only a representative set of elementary observables such that any function on the reduced phase phase space can be expressed as (possibly a limit of) a sum of products of the elementary functions. These elementary observables can then be promoted to operators in a consistent manner, so that all functions on the reduced phase space are also automatically promoted to operators. 

In this scheme, it is necessary that the constraints be first class, for otherwise no unique physical subspace of the unconstrained phase space will be picked out by the constraints, and the notion of observables as functions with vanishing Poisson brackets with the constraints would become ambiguous. Thus once again, that the constraints be first class is a crucial factor in successfully quantising any Hamiltonian formulation of general relativity. If there are second-class constraints, they must be solved before proceeding with any quantisation scheme. 

Reduced space quantisation is difficult to implement in any Hamiltonian formulation of general relativity. This is because, owing to the highly complicated nature of the constraints (the Hamiltonian constraint, for instance, involves the Ricci scalar, an object composed of second-order partial derivatives of the metric), the task of finding functions that have vanishing Poisson brackets with the constraints is highly non-trivial. In fact, no such functions are known in the case of vacuum general relativity on a spatially compact spacetime. When matter is included, only certain diffeomorphism invariant functions describing matter are known, but no such functions for the gravitational field are known. It is only for asymptotically flat spacetimes that 10 such functions are known; these are the ADM charges given by the generators of Poincare transformations at spatial infinity \cite{dittrich 1}. But these, of course, will not be enough obtain all the observables of the theory. 

\subsubsection{3. A group-theoretic approach}
As explained above, the constraints of general relativity generate the full diffeomorphism group of the theory. Thus one natural avenue for dealing with constraints would be to find a representation of the diffeomorphism group on $\mathcal{V}$ and then identify the quantum constraints as generators of this representation. Physical states would then be those left invariant under the action of the elements of the chosen representation. 

This method of solving the constraints requires a reasonable degree of control on the constraint algebra. The first issue is, again, the highly non-polynomial nature of the constraints. Thus, once more, the ADM variables turn out to be ill-suited for quantisation, and the polynomiality of the new formalism that will developed in the next chapter will enable us to deal with at least the diffeomorphism constraint in this group-theoretic framework.

However, the Hamiltonian constraint, whether it is polynomial or not, poses problems in any formalism. This is because the appearance of the metric in the Poisson algebra (Eq~\ref{eq1.34}c) of the Hamiltonian constraint makes it difficult to find its representations in a background-independent way. Thus, while some progress has been made in taming this constraint in loop quantum gravity, it is still an open problem. 

\subsection{Going physical: the need for a Hilbert space}
Even if the foregoing problems that a quantisation of ADM formalism faces are solved by some miracle, the quantum theory will be incapable of making any physical predictions. For this task, one would have to have a finite inner product on the space $\mathcal{V}_{phy}$ of physical states obtained by solving the constraints (by any one of the above three methods or a combination thereof). Classical observables will then be represented as operators on $\mathcal{H}_{phy}$ that are self-adjoint with respect to this inner product.

The central concern here, then, is to convert $\mathcal{V}_{phy}$ into a Hilbert space $\mathcal{H}_{phy}$. There are two ways to proceed here. The first is to begin with the space $\mathcal{V}$ of general quantum states and seek to introduce an inner product on it. In the Dirac method of solving the constraints, the physical space $\mathcal{V}_{phy}$, being the null space of the constraint operators, turns out to be a subspace of $\mathcal{V}$. One would thus expect the inner product on $\mathcal{V}$ to project down to an inner product on $\mathcal{V}_{phy}$. However, there is no guarantee that this will be the case. Moreover, depending on the method used to solve the constraints, $\mathcal{V}_{phy}$ might not even be a subspace of $\mathcal{V}$. Thus, the second approach, which seeks to directly endow $\mathcal{V}_{phy}$ with a Hilbert space structure seems better equipped to deal with the problem at hand. 

In any case, it is nonetheless true that $\mathcal{V}_{phy}$ will in some sense be constructed from $\mathcal{V}$\footnote{See Chapter 3 for a detailed account of how this works in loop quantum gravity.}. Now recall that $\mathcal{V}$ is the space of complex-valued functions on the configuration space Met($\Sigma$) of the ADM formalism. Thus it is necessary to develop a theory of integration on the configuration space. Unfortunately, there is no known way of doing that in Met($\Sigma$), which is infinite-dimensional. This is in contrast to finite-dimensional configuration spaces, which usually come equipped with Lebesgue measures, which can be used to define finite $L^2$ norms on $\mathcal{V}$ and thus convert it into a Hilbert space $\mathcal{H}$. For example, the configuration space of a particle moving in three dimensions is $\mathbb{R}^3$, which has measure $d^3x$ -- thus $\mathcal{H} = L^2(\mathbb{R}^3, d^3x)$. 

This does not mean that infinite-dimensional configuration spaces cannot have a well-defined analogue of a Lebesgue measure. One routinely encounters the problem of defining precisely such measures in, for instance, the path-integral formulation of quantum field theory. One can construct well-defined Borel measures in such situations using a variety of different techniques. However, there is no general recipe for doing this; one needs to proceed on a case-by-case basis to see what kind of a measure a particular configuration space admits (see \cite{ashtekar 3} and Chapter 3 for details). For example, in the Klein-Gordon scalar field theory, the configuration space is the space of all scalar field configurations, $\{\phi(x)\}$, on spacetime. Here, one can extend the Gaussian measure on finite-dimensional subspaces of $\{\phi(x)\}$ to its topological dual. As another example, consider Yang-Mills theory, where the configuration space is the space of all smooth $SU(N)$ connections on $\Sigma$. In this case, the natural Haar measure on $SU(N)$ can serve as a candidate measure. However, it is not at all clear how to carry over these techniques to Met($\Sigma$).

The absence of a well-defined inner product is perhaps the most unsalvageable lacuna that riddles the quantisation of the ADM formalism, and thus the need for an alternative framework becomes indispensable. As we shall see in the following chapters, loop quantum gravity provides a viable solution to fill this huge gap.  

\subsection{The problem of time}
Unfortunately, even solving all the manifestations of the inner product problem does not exorcise all the spectres that haunt canonical quantum gravity, because there is one -- perhaps the most troubling -- problem that remains. Recall that the Hamiltonian vanishes on $\mathcal{H}_{phy}$. Therefore, any operator on $\mathcal{H}_{phy}$ trivially commutes with the Hamiltonian, and thus does not change in time:
\begin{equation}
    \frac{d\hat{A}}{dt} = i[\hat{H}, \hat{A}] = 0.
\end{equation}
Time has disappeared from the theory! We stumble upon the notorious \textit{problem of time}. One might retort that the disappearance of the dynamics of the classical theory is not a problem, because unlike $(q_{ab}, \tilde{p}^{ab})$, which describe the states of classical gravity at a particular time, the states of $\mathcal{V}_{phy}$ just encode the information about the state of a system that is invariant under all spacetime diffeomorphisms. However, the observables that we know of in the classical theory do not exhibit this peculiar behaviour, and we are thus left clueless as to what candidate observables ought to be represented as self-adjoint operators on $\mathcal{H}_{phy}$.

The problem of time persists in loop quantum gravity as well. However, some of the solutions suggested to it in the traditional canonical framework, such as the introduction of a universal scalar field that keeps track of time, work equally well in loop quantum gravity too. Furthermore, there is a rather satisfactory solution to the problem in the linearised theory that results from taking the weak-field limit of loop quantum gravity \cite{ashtekar 1}.  This, combined with the fact that loop quantum gravity almost solves the other three problems that riddle the traditional framework, is reason enough to embark upon a careful study of the former. 

\chapter{Virtues of Gauge}
In the preceding chapter, we saw the ADM formalism of general relativity and surveyed the problems that occur when one attempts to quantise it. It is quite evident that the most troubling of these problems are the non-polynomiality of the ADM constraints, and the absence of control over the infinite-dimensional space of all three-metrics. Therefore, in this chapter, we shall consider the possibility of reformulating general relativity with a view to eliminating these two problems. We will find that each of these problems can be gotten rid of, but only at the expense of the other one. 

\section{The Tetrad Formulation of General Relativity}
\subsection{The spirit of Lorentz lives on}
Let us start with the simple observation that the spacetime manifold $M$ is locally diffeomorphic to $\mathbb{R}^4$. From this it follows that at every $x\in M$, there exists a basis of vectors $e_I$ in which the metric $g_{\alpha\beta}$ on $M$ becomes the Minkowski metric:
\begin{equation}
    g_{\alpha\beta}\tensor{e}{^\alpha_I}\tensor{e}{^\beta_J} = \eta_{IJ}, \label{eq2.1}
\end{equation}
where $\tensor{e}{^\alpha_I}$ are the components of $e_I$, called \textit{tetrads} or \textit{frame fields}. We will begin by studying the structure of these triads. 

Consider the trivial vector bundle $M \times \mathbb{R}^4$. Since $M$ is locally diffeomorphic to $\mathbb{R}^4$, there exists a local trivialisation of its tangent bundle $TM$. That is, for every $x \in M$, there exists an open set $U$ containing $x$ and a local vector bundle isomorphism 
\begin{equation*}
    e: U \times \mathbb{R}^4 \to \left. TM \right|_U 
\end{equation*}
that takes each fiber $\{x\} \times \mathbb{R}^4$ of $M \times \mathbb{R}^4$ to the tangent space $T_xM$ at $x$ \footnote{For simplicity, henceforth, we will take $M$ to mean a small enough subset of $M$ such that the tangent bundle restricted to the subset, $\left.TM\right|_U$, is trivial in the above sense.}. Since each fibre $\{x\}\times\mathbb{R}^4$ can be thought of as a function from $M$ to $\mathbb{R}^4$, i.e. a section, $e$ can be conceived of as mapping sections in $M\times\mathbb{R}^4$ to vectors in $TM$. Conversely, $e^{-1}$ is a map from vectors in $TM$ to sections in $M\times\mathbb{R}^4$. 

The idea of the tetrad formalism is to use $M\times\mathbb{R}^4$ as a substitute for $M$, and this is possible because we can go back and forth between the two using $e$. Let us see how this can be precisely done. 

To begin with, we equip $M\times\mathbb{R}^4$ with the Minkowski metric. That is, we define the inner product between two sections $s$ and $r$ of $M\times\mathbb{R}^4$ by
\begin{equation*}
    s\cdot r = \eta_{IJ}s^Ir^J. 
\end{equation*}
With this, the isomorphism $e$ becomes precisely the transformation described in Eq~(\ref{eq2.1}). That is, given vectors $v$ and $w \in T_pM$ and sections $s$ and $t$ at $\{p\}\times\mathbb{R}^4$, $g(v,w) = \eta(e^{-1}(v), e^{-1}(w))$ and $\eta(s,t) = g(e(s), e(w))$.

$\mathbb{R}^4$ is called the \textit{internal space}, which we distinguish from $M$ by the use of \textit{internal indices} $I, J, K,$ etc., which can be raised and lowered using the Minkowski metric.  We can now raise and lower indices on objects with mixed indices as well, e.g. $\tensor{e}{^I_\alpha}$. For instance, we define the \textit{cotetrad}\footnote{Formally, we have $e^{-1}(\partial_\alpha) = \tensor{e}{^I_\alpha}\lambda_I$, where $\partial_\alpha$ are elements of the coordinate basis of vectors at every point in $M$ and $\lambda_I = e^{-1}(e_I)$.}
\begin{equation}
    \tensor{e}{^I_\alpha} = \eta^{IJ}g_{\alpha\beta}\tensor{e}{^\beta_J}.  \label{eq2.2}
\end{equation}
Multiplying this equation on both sides by $\tensor{e}{^\alpha_K}$ and using Eq~(\ref{eq2.1}), we get \begin{equation}
    \tensor{e}{^\alpha_I}\tensor{e}{^\beta_J} = \delta^I_J, \label{eq2.3}
\end{equation}
which in turn allows us to express the metric in $M$ in terms of the Minkowski metric:
\begin{equation}
    g_{\alpha\beta} = \eta_{IJ}\tensor{e}{^I_\alpha}\tensor{e}{^J_\beta}. \label{eq2.4}
\end{equation}
Next, just as in $M$, we need a connection on $M\times\mathbb{R}^4$ to parallel-transport sections on it. We thus define
\begin{equation}
    D_\alpha s^I = \partial_\alpha s^I + \tensor{A}{_\alpha^I_J}s^J. \label{eq2.5}
\end{equation}
In analogy with the Levi-Civita connection on $M$, we demand that $A$ be metric-preserving, i.e. $D_\alpha \eta_{IJ} = 0$, which entails that
\begin{equation}
    \tensor{A}{_\alpha^{IJ}} = -\tensor{A}{_\alpha^{JI}}. \label{eq2.6}
\end{equation}
Consider $\Lambda = e^{ia^\alpha \tensor{A}{_\alpha^I_J}}$. To first order in $A$, we find, using Eq~(2.6), that $\Lambda^T\eta\Lambda = \eta$. Thus the connection lives in the Lie algebra $\mathfrak{so}(3,1)$ of the Lorentz group \footnote{There is a fancier way to see this in the language of fibre bundles. Since $M$ is Lorentzian, it can be thought of as the associated bundle to an SO(3,1)-principal bundle, and connections on the latter are $\mathfrak{so}(3,1)$-valued.}. We call $A$ a \textit{Lorentz connection}. As we shall soon see, the Lorentz connection will be one of the configuration variables for gravity in the tetrad formulation. We thus see here the beginnings of an analogy with gauge theories of particle physics, where the configuration variables are $\mathfrak{g}(N)$-valued, $\mathfrak{g}(N)$ being the Lie algebra of the gauge group of the theory. In this sense, the gauge group of general relativity is seen to be $SO(3,1)$. However, as opposed to the gauge theories of particle physics, such as the Yang-Mills theory, where the gauge group is compact, $SO(3,1)$ is not a compact group. The difficulty of developing an integration theory on the configuration space of gravity can be traced back to this crucial difference: loosely speaking, while we know how to integrate functions on compact spaces, the same is not true of noncompact ones. 

With this important digression in mind, let us continue to build the tetrad formalism. Just as in $M$, we can define a curvature corresponding to the Lorentz connection:
\begin{equation}
    \tensor{\Omega}{^{IJ}_{\alpha\beta}} = 2\partial_{[\alpha}\tensor{A}{_{\beta]}^{IJ}} + [A_\alpha, A_\beta]^{IJ}, \label{eq2.7}
\end{equation}
which is also antisymmetric in the internal indices. The appearance of partial derivatives in this equation means that the Lorentz connection must be smooth (i.e. $\tensor{A}{_\alpha^{IJ}}(x)$ are smooth functions of $x\in M$). 

\subsection{Lagrangian formulation}
We are now in a position to recast general relativity in the tetrad formalism. We define the \textit{Palatini action}:
\begin{equation}
    S(A, e) = \int_M d^4x(e)\tensor{e}{^\alpha_I}\tensor{e}{^\beta_J}\tensor{\Omega}{^{IJ}_{\alpha\beta}}. \label{eq2.8}
\end{equation}
Here, $e$ is the determinant of the triad. As we will soon see, the variation of this action with respect to the tetrad and the Lorentz connection yields the Einstein field equations. Therefore, the Palatini action is completely equivalent to the Einstein-Hilbert action. The novelty in the tetrad formalism is that the dynamical variables are the tetrad and the Lorentz connection, rather than the metric, which is to emerge secondarily via Eq~(\ref{eq2.4}) once tetrads and Lorentz connections that solve the equations of motion are found. We say that the emphasis has shifted from \textit{geometrodynamics} to \textit{connection dynamics}. 

What is the motivation behind the definition of the Palatini action? Recall that in ordinary general relativity, the metric is the basic ingredient that defines all the relevant geometric objects in the theory, such the Levi-Civita connection, the Riemann curvature tensor, the Einstein equations, and so on. In the tetrad formalism, we proceeded by trading the metric for the tetrads. Thus it is only natural to expect them to enter into the action of the theory. However, these tetrads have no relation to the Lorentz connection, which is independently introduced as the Lorentz-metric-preserving connection. Thus it should be incorporated into the action independently of the tetrads.

We will now show that the variation of the Palatini action furnishes the Einstein field equations. To that end, we first establish a correspondence between the structures that enter the Palatini action and those that enter the Einstein-Hilbert action. First, using tetrads and cotetrads enables us to transport the Lorentz connection from $M\times\mathbb{R}^4$ to the tangent bundle $TM$. We define
\begin{equation}
    D_\alpha v^\beta = \partial_\alpha v^\beta + C^\beta_{\alpha\gamma}v^\gamma, \label{eq2.9}
\end{equation}
where 
\begin{equation}
    C^\beta_{\alpha\gamma} = \tensor{A}{_\alpha^J_I}\tensor{e}{^I_\gamma}{e}{^\beta_J}. \label{eq2.10}
\end{equation}
Note that this allows us to parallel-transport objects with both spacetime and internal indices. For instance, for $V^I_\alpha$, we extend the definition of the covariant derivative $D$ as follows.
\begin{equation}
    D_\alpha V^I_\beta := \partial_\alpha V^I_\beta + \tensor{A}{_\alpha^I_J}V^J_\beta - C^\gamma_{\alpha\beta}V^I_\gamma. \label{eq2.10*}
\end{equation}
Next, we define a curvature tensor on $M$:
\begin{equation}
    \tensor{R}{^\alpha_{\beta\gamma\lambda}} = \tensor{F}{^{IJ}_{\beta\gamma}}\tensor{e}{^\alpha_I}e_{\lambda J}, \label{eq2.11}
\end{equation}
which, as can be readily verified, is the curvature tensor of the connection in Eq~(\ref{eq2.10}). With these definitions at hand, if one varies the Palatini action with respect to the tetrad, one finds \cite{baez 1, ashtekar 1}
\begin{equation}
    R_{\alpha\beta} - \frac{1}{2}R g_{\alpha\beta} = 0, \label{eq2.12}
\end{equation}
which are almost the Einstein equations, except that the Ricci tensor and scalar appearing above come from a curvature tensor that is not necessarily the Riemann curvature tensor of the metric $g$. However, variation of the Palatini action with respect to the Lorentz connection yields \cite{baez 1, ashtekar 1}
\begin{equation}
    C^\gamma_{\alpha\beta} = \frac{1}{2}g^{\gamma\lambda}(\partial_\alpha g_{\beta\lambda} + \partial_\beta g_{\alpha\lambda} - \partial_\lambda g_{\alpha\beta}), \label{eq2.13}
\end{equation}
which imples that the connection in Eq~(\ref{eq2.10}) is metric compatible and thus, the curvature in Eq~(\ref{eq2.11}) is nothing but the Riemann curvature tensor of the metric $g$. Therefore, the Palatini formalism is completely equivalent to general relativity. 

\subsection{Hamiltonian formulation}
Now that we are in possession of an alternative Lagrangian formulation of general relativity, the next natural vocation is to study the Hamiltonian theory that results from it. In this section, we will perform a Legendre transform of the Palatini action and analyse the phase-space structure of the resultant theory. 

As explained in chapter 1 in detail, we start by foliating the spacetime into spacelike hypersurfaces ($\phi: M \to \mathbb{R}\times\Sigma$) each of which is intersected by the integral curves of the timelike vector field $t^\alpha = \phi_*\partial_\tau$, which can be decomposed into components tangent and normal to each hypersurface via
\begin{equation*}
    t^\alpha = Nn^\alpha + N^\alpha,
\end{equation*}
where $n^\alpha$ is everywhere orthogonal to $\Sigma$, $N^\alpha$ is the shift vector, which is tangent to $\Sigma$ and $N$ is the lapse. Let $n_I = n_\alpha e^\alpha_I$, and consider the spatial projection $E^I_\alpha = e^\beta_I(g^\alpha_\beta + n^\alpha n_\beta)$ of the tetrad. Using these and the fact that $e = EN$, where $E$ is the determinant of $E^\alpha_I$, we can write
\begin{align}
    S &= \int_M d^4x (e)e^\alpha_I e^\beta_J \Omega^{IJ}_{\alpha\beta} \nonumber \\
    &= \int d^4x EN E^\alpha_I E^\beta_J \Omega^{IJ}_{\alpha\beta}  -2\int d^4x EN E^\alpha_I n_J n^\beta \Omega^{IJ}_{\alpha\beta} + \int d^4x EN n_I n_J n^a n^b \Omega^{IJ}_{\alpha\beta} \label{eq2.1.3.1}.
\end{align}
The last term above vanishes by virtue of the antisymmetry of $\Omega$ and the symmetry of $n_I n_J$. Consider the second term. We get
\begin{align*}
    \int d^4x E^\alpha_{I} n_{J} (Nn_\beta) \Omega^{IJ}_{\alpha\beta}
    &= \int d^4xEE^\alpha_{[I}n_{J]}t^\beta\Omega^{IJ}_{\alpha\beta} - \int d^4xEE^\alpha_{[I}n_{J]}N^\beta\Omega^{IJ}_{\alpha\beta} \\
    &= \int d^4x EE^\alpha_{[I} n_{J]}\left[ t^\beta\partial_\alpha A_\beta^{IJ} - t^\beta\partial_\beta A_\alpha^{IJ} + t^\beta[A_\alpha, A_\beta]^{IJ}  \right]\\
    &\qquad\qquad\qquad\qquad- \int d^4x EE^\alpha_{[I}n_{J]} N^\beta \Omega^{IJ}_{\alpha\beta} \\
    &= \int d^4x EE^\alpha_{[I} n_{J]}\left[- A_\beta^{IJ}\partial_\alpha t^\beta - t^\beta\partial_\beta A_\alpha^{IJ} + \partial_\alpha(t^\beta A_\beta^{IJ}) + t^\beta[A_\alpha, A_\beta]^{IJ} \right]\\ 
    &\qquad\qquad\qquad\qquad - \int d^4x EE^\alpha_{[I}n_{J]}N^\beta \Omega^{IJ}_{\alpha\beta} \\
    &= \int d^4x EE^\alpha_{[I} n_{J]}\left[ -\mathcal{L}_tA_\alpha^{IJ} + D_\alpha(t^\beta A_\beta^{IJ}) \right] - \int d^4x EE^\alpha_{[I}n_{J]} N^\beta \Omega^{IJ}_{\alpha\beta}.
\end{align*}
The action thus becomes
\begin{align*}
    S &= \int d^4x E\left[NE^\alpha_IE^\beta_J\Omega^{IJ}_{\alpha\beta} - 2n_{[I}E^\alpha_{J]}D_\alpha(A_\beta^{IJ}t^\beta) + 2n_{[I}E^\alpha_{J]}\mathcal{L}_tA^{IJ}_\alpha + 2N^\alpha n_{[I}E^\beta_{J]}\Omega^{IJ}_{\alpha\beta} \right]\\
    &= \int dt \int_\Sigma d^3x E\left [ NE^a_IE^b_J\Omega^{IJ}_{ab} - 2(A_\beta^{IJ}t^\beta)D_a(E^a_{[I}n_{J]}) - 2E^a_{[I}n_{J]}\mathcal{L}_tA^{IJ}_a + 2N^a n_{[I}E^b_{J]}\Omega^{IJ}_{ab} \right].
\end{align*}
In the second line, we have relabelled some indices and replaced all the terms (except one) in the first line with their pullbacks to $\Sigma$, since each term was a contraction with purely spatial objects. Let $\undertilde{N} = E^{-1}N$ and
\begin{subequations}
\label{eq2.14}
\begin{align}
    \tilde{E}^a_I &= E E^\alpha_I,\\
    \tilde{\alpha}^a_{IJ} &= \tilde{E}^a_{[I}n_{J]}.
\end{align}
\end{subequations}
We immediately have $-4\tilde{\alpha}^a_{IK}\tilde{\alpha}^{bK}_J\Omega^{IJ}_{ab} = \tilde{E}^a_I\tilde{E}^b_J\Omega^{IJ}_{ab}$. Hence, finally, the action canonically decomposes as
\begin{equation}
    S(A, \tilde{\alpha}) = 2\int dt \int_\Sigma d^3x\, \text{tr}\left(-2\undertilde{N}\tilde{\alpha}^a\tilde{\alpha}^b\Omega_{ab} - (t^\alpha A_\alpha)D_a\tilde{\alpha}^a + N^a\tilde{\alpha}^b\Omega_{ab} - \tilde{\alpha}^a\mathcal{L}_tA_a \right), \label{eq2.15}
\end{equation}
where the trace is over the internal indices. 

Since the action above is now written in the form $\int (p\Dot{q}-H)$, we readily notice a number of features. First, $\tilde{\alpha}^a_{IJ}$ is the momentum conjugate to $A^{IJ}_a$, whence
\begin{equation}
    \{A^{IJ}_a(x), \tilde{\alpha}^b_{MN}(y) \} = \delta^b_a\delta^I_{[M}\delta^J_{N]}\delta^3(x,y). \label{eq2.16}
\end{equation}
Second, Eq~(\ref{eq2.15}) does not contain the time derivatives of $\undertilde{N},\, N^\alpha, \, (t^\alpha A^{IJ}_\alpha)$, which, therefore are Lagrange multipliers, variation with respect to which yields the constraints of the theory:
\begin{subequations}
\label{eq2.17}
\begin{align}
    &S := \text{tr}\, \tilde{\alpha}^a\tilde{\alpha}^b\Omega_{ab} \approx 0, \\
    &V_a := \text{tr}\,\tilde{\alpha}^b\Omega_{ab} \approx 0,\\
    &G_{IJ} := D_a\tilde{\alpha}^a_{IJ} \approx 0.
\end{align}
\end{subequations}
The Hamiltonian, being the sum of all constraints, is thus given by
\begin{equation}
    H = 2\int_\Sigma d^3x \left( 2\undertilde{N}S - N^aV_a + \text{tr}\,(t^\alpha A_\alpha)G \right). 
\end{equation}
A number of comments are in order. It seems that the constraints of general relativity in the Palatini framework are polynomial. However, this is illusory, since Eqs~(\ref{eq2.17}) are not all the constraints. We are ignoring the fact that we cannot choose the conjugate momentum $\tilde{\alpha}^a_{IJ}$ arbitrarily; its form is constrained by (Eq~(\ref{eq2.14}a)). It can be shown \cite{ashtekar 1} that this constraint on the form of the conjugate momentum is equivalent to the constraint
\begin{equation}
    \phi^{ab} := \epsilon^{IJKL}\Tilde{\alpha}^a_{IJ}\Tilde{\alpha}^b_{KL} = 0, \label{eq2.1.3.2}
\end{equation}
and this constraint is needed to show that $\{S(x), S(y)\}$ vanishes on the constraint surface. Thus the constraints in Eq~(\ref{eq2.17}) do not form a closed system by themselves. Furthermore, the Poisson brackets between $S(x)$ and $\phi^{ab}(y)$ fail to vanish, giving rise to a second-class constraint. It can be shown \cite{ashtekar 1} that when these two (i.e. $\phi^{ab}$ and $\{S(x),\phi^{ab}(y)\}$) second class constraints are solved, the remaining first class constraints reduce to the old ADM constraints. This fact, combined with the observation that Lorentz connection in the tetrad formalism belongs to the Lie algebra of a noncompact group, makes this formalism an unsuitable alternative to the ADM formalism. However, familiarity with the phase-space structure of the tetrad formalism paves the way towards the introduction of other strategies that may alleviate one of the two foregoing problems. It is the task of subsequent sections to explore these different strategies. 

\section{The Ashtekar Variables}
In ordinary general relativity, the spacetime manifold $M$ is locally diffeomorphic to $\mathbb{R}^4$. On the other hand, to introduce the Ashtekar variables, one needs to work with complex general relativity, wherein $M$ is locally diffeomorphic to $\mathbb{C}^4$. Is this not departing from the general relativity that is known to be experimentally correct? The answer is no, for at least classically, one can always impose suitable reality conditions on one's basic variables to recover real general relativity. In the quantum theory, dealing with the reality conditions is a lot more complicated, but more on that later. 

Let us spell out the changes that going to complex general relativity incurs in the mathematical structures introduced so far. We define the complexified tangent bundle, $\mathbb{C}TM$, of $M$ to be the vector bundle whose fiber at each $x \in M$ is the vector space $\mathbb{C}\otimes T_xM$. Similarly, the trivial bundle $M\times\mathbb{R}^4$ of the Palatini formalism becomes $M\times\mathbb{C}^4$, and the tetrads become vector bundle isomporhisms from the latter to the complexified tangent bundle of $M$, i.e. $e: M\times\mathbb{C}^4 \to \mathbb{C}TM$. The Lorentz connection now becomes an $\mathfrak{so}(3,1)\otimes\mathbb{C}$-valued 1-form on $M$. This is itself a Lie algebra\footnote{Given a Lie algebra $\mathfrak{g}$, we define its complexification to be the vector space $\mathfrak{g}\otimes\mathbb{R}$. As is easily verified, this vector is itself a Lie algebra, provided we define the Lie bracket of $x\otimes\alpha, y \otimes\beta\in \mathfrak{g}\otimes\mathbb{C}$ by $[x\otimes\alpha, y\otimes\beta)] = [x,y]\otimes\alpha\beta$}.

Since the Lorentz connection is a 2-form with respect to internal indices, we define an internal \textit{Hodge star operator}, which is a map from 2-forms to 2-forms, given by
\begin{equation}
    \star T^{IJ} = \frac{1}{2}\tensor{\epsilon}{^{IJ}_{KL}}T^{KL}.\label{eq2.19}
\end{equation}
We say that a 2-form is \textit{self-dual} if $\star T^{IJ} = iT^{IJ}$ and \textit{anti-self-dual} if $\star T^{IJ} = -iT^{IJ}$. Any 2-form can be decomposed into its self-dual and anti-self-dual parts, i.e.
\begin{align*}
    T^{IJ} &= ^+T^{IJ} + ^-T^{IJ},\\
    ^{\pm}T^{IJ} = \frac{1}{2}(T^{IJ} \mp& i\star T^{IJ}), \quad \star ^{\pm}T^{IJ} = \pm i^{\pm}T^{IJ}.
\end{align*}

We are now in a position to reformulate general relativity in terms of the Ashtekar variables. As in the Palatini case, there are two dynamical variables, namely, the tetrads and the Lorentz connection. However, now, instead of using any Lorentz connection, we demand that it be self-dual. The rest of the framework is entirely similar. That is, we write the action as
\begin{equation}
    S(A,e) = \int_M d^4x (e) e^\alpha_I e^\beta_I F^{IJ}_{\alpha\beta}, \label{eq2.20}
\end{equation}
where
\begin{equation}
    F^{IJ}_{\alpha\beta} = 2\partial_{[\alpha}A^{IJ}_{\beta]} + [A_\alpha, A_\beta]^{IJ} \label{eq2.21}
\end{equation}
is the curvature of the self-dual Lorentz connection $A$, and, as can at once be seen from its explicit form, is itself self-dual (self-dual 2-forms are closed under Lie brackets). 

Once again, variation of the self-dual action with respect to the connection and the tetrad yields the Einstein field equations (see \cite{baez 1, ashtekar 1} for a derivation), but for complex general relativity. To recover the real theory, as remarked earlier, one needs to impose some reality conditions, which we shall discuss in the context of the Hamiltonian theory of the self-dual formulation.

It is quite natural to inquire at this point why the Asthekar variables work at all? First, using only the self-dual part of the curvature tensor in the action (Eq~\ref{eq2.21}) seems a little ad hoc. Second, the success of the new strategy in reproducing the Einstein equations seems to rely on the curious coincidence that the curvature tensor, and hence the Einstein equations resulting from it, are all self-dual. Is it possible to decipher some more fundamental principle behind these two points? The answer is in the affirmative, as we now demonstrate \cite{baez 1}.

Given a real Lie algebra $\mathfrak{g}$, we can decompose its complexification $\mathfrak{g}\otimes\mathbb{C}$ into a direct sum of two of its Lie subalgebras:
\begin{subequations}
\label{eq2.24*}
    \begin{equation}
        \mathfrak{g}\otimes\mathbb{C} = \mathfrak{g}_+ \oplus \mathfrak{g}_-, 
    \end{equation}
    where
    \begin{equation}
        \mathfrak{g}_{\pm} = \{x\otimes 1 \pm ix\otimes i: x\in\mathfrak{g}\}.
    \end{equation}
\end{subequations}
We can show further that each of $\mathfrak{g}_{\pm}$ is isomorphic as a Lie algebra to $\mathfrak{g}$. Now recall that a complexified Lorentz connection in the Asthekar formalism lies in $\mathfrak{so}(3,1)\otimes\mathbb{C}$. On the other hand, we know that the space $SL(2,\mathbb{C})$ of complex $2\times2$ matrices with unit determinant is a double cover of (the identity component of) $SO(3,1)$, whence $\mathfrak{so}(3,1)\otimes\mathbb{C}$ is isomorphic to $\mathfrak{sl}(2,\mathbb{C})\otimes\mathbb{C}$. But by Eq~(\ref{eq2.24*}),  $\mathfrak{sl}(2,\mathbb{C})\otimes\mathbb{C}$ can be decomposed into two Lie subalgebras, each of which is isomorphic to $\mathfrak{sl}(2,\mathbb{C})$. Objects living in these two subalgebras are precisely self-dual and anti-self-dual forms, respectively. It thus follows that self-dual (anti-self-dual) Lorentz connections lie in the self-dual (anti-self-dual) subalgebra of $\mathfrak{sl}(2,\mathbb{C})\otimes\mathbb{C}$ or equivalently in $\mathfrak{sl}(2,\mathbb{C})$. Thus, in the Ashtekar formulation, $\mathfrak{sl}(2,\mathbb{C})$ emerges as the gauge group for gravity. It is still a noncompact group, and so offers no advantage over the Palatini formalism. However, it turns out that in the Ashtekar formulation, the constraints of general relativity are polynomial in canonical variables.

To see this, let us pass to the Hamiltonian formulation of the self-dual theory. The Legendre transform can be performed quite analogously to what was done in the Palatini framework. However, here we can additionally make use of the self-duality of the curvature 2-form; that changes the form of the canonical momentum. More precisely, in the second term in Eq~(\ref{eq2.1.3.1}), use the fact that $F^{IJ}_{\alpha\beta}$ is self-dual in the internal indices to write $F^{IJ}_{\alpha\beta} = -\frac{i}{2}\tensor{\epsilon}{^{IJ}_{KL}}F^{KL}_{\alpha\beta}$ and then repeat exactly the same subsequent steps, except for the anti-symmetrisation over the internal indices in $E^\alpha_In_J$. We then find the action in Eq~(\ref{eq2.20}) to reduce to that in Eq~(\ref{eq2.15}), except that each occurrence of the term $\Tilde{E}^a_{[I}n_{J]}$ is replaced with $\tilde{P}^a_{IJ} := -i\tilde{E}^a_M n_N\tensor{\epsilon}{^{MN}_{IJ}}$. Explicitly,
\begin{equation}
    S = 2\int dt \int_\Sigma d^3x\, \text{tr}\, \left( -\tilde{P}^a\mathcal{L}_tA_a + N^a\tilde{P}^bF_{ab} - (A^\alpha t_\alpha)D_a\tilde{P}^a - \undertilde{N}\tilde{P}^a\tilde{P}^bF_{ab} \right), \label{eq2.22}
\end{equation}
where now the momentum conjugate to the self-dual connection $A_a^{IJ}$ is $\Tilde{P}^a_{IJ}$. A direct calculation reveals that $\Tilde{P}^a_{IJ}$ is self-dual, whence
\begin{equation}
    \left\{A^{IJ}_a(x), \tilde{P}^b_{MN}(y) \right\} = \delta^b_a\delta^3(x,y)\left[\delta^I_{[M}\delta^J_{N]} - \frac{i}{2}\tensor{\epsilon}{_{MN}^{IJ}}\right]. \label{eq2.23}
\end{equation}
This might suggest that $A^{IJ}_a$ and $\Tilde{P}^b_{MN}$ are not ``true'' canonically conjugate variables. But we shall soon recast them in a way that makes the Poisson brackets between them canonical.

The Hamiltonian is just a sum of the constraints, which, as in the Palatini case are
\begin{subequations}
\label{eq2.24}
\begin{align}
     &S := \text{tr}\, \tilde{P}^a\tilde{P}^bF_{ab} \approx 0, \\
    &V_a := \text{tr}\,\tilde{P}^bF_{ab} \approx 0,\\
    &G_{IJ} := D_a\tilde{P}^a_{IJ} \approx 0.
\end{align}
\end{subequations}
However, unlike the Palatini case, there are no constraints on the form of the canonical momentum; it can be any arbitrary self-dual one-form (self-duality is not a constraint, but a defining assumption of the Ashtekar formalism). Therefore, Eqs~(\ref{eq2.24}) are the only constraints, and can explicitly be checked to be first class \cite{ashtekar 1}. Thus, the Ashtekar constraints are polynomial in the basic variables, as we had promised. The first constraint is called the \textit{Gauss constraint}. This is a new constraint in this formulation, and arises from the use of tetrads, instead of metrics, as our dynamical variables, because the former have four more independent components than the latter, resulting in our freedom to perform four rotations in the internal space; the Gauss constraint describes this rotational freedom. The remaining two constraints are equivalent to the diffeomorphism and Hamiltonian constraints of the ADM variables, respectively. 

To confirm that we are still dealing with complex general relativity, let us count the number of degrees of freedom. Each canonical variable, being self-dual, has 3 spatial and 3 internal degrees of freedom, giving a total of 9 degrees of freedom. The constraints in the preceding equations are $1+3+3 =  7$ in number. Therefore, the physical degrees of freedom reduce to 2, which are the degrees of freedom of general relativity. 

This is all very good, but we must not lose sight of the fact that we are doing complex general relativity. To make contact with physics, we must specify a way to recover the real theory. This can be done by relating the Ashtekar variables to the old ADM variables, which will now be complex. Then we impose certain reality conditions on the Ashtekar variables so that the ADM variables obtained from them are real. Before we describe this, let us first recast the canonical variables in a simpler and more convenient form. To this end, note that our canonical variables $A_a^{IJ}$ and $\Tilde{P}^a_{IJ}$ are both antisymmetric and self-dual in the internal indices. These conditions entail that for each value of $a$, these variables have only 3 independent components. In other words, we can replace the internal indices $I, J$ with a single index $i$ that runs from 1 to 3. More precisely, let $T_i^{IJ}\in \mathfrak{sl}(2,\mathbb{C})$ be such that \cite{ashtekar 1}:
\begin{subequations}
\label{eq2.25}
\begin{align}
    [T_i, T_j] &= \tensor{\epsilon}{_{ij}^k}T_k\\
    \text{tr}\,T_i T_j &= -\delta_{ij}\\
    T_i^{MN}T^i_{IJ} &= \frac{1}{2}\left(\delta^M_{[I}\delta^N_{J]} - \frac{i}{2}\tensor{\epsilon}{_{IJ}^{MN}} \right)
\end{align}
\end{subequations}
These equations ensure that any such $T_i^{MN}$ form a basis of the self-dual subalgebra of $\mathfrak{so}(3,1)\otimes\mathbb{C}$, and are also antisymmetric in the internal indices. Thus they span the phase space. Explicitly, any self-dual and antisymmetric $S^{IJ}$ can be written as
\begin{equation}
    S^{IJ} = (-\text{tr}\,ST^i)T_i^{IJ}. \label{eq2.26}
\end{equation}
This then lets us define
\begin{subequations}
\label{eq2.27}
\begin{align}
    A^i_a &:= -\text{tr}\,A_aT^i,\\
    F^i_{ab} &:= -\text{tr}\,F_{ab}T^i = 2\partial_{[a}A^i_{b]} + \epsilon^{ijk}A_{bj}A_{ak},\\
    \tilde{E}^a_{i} &:= -i\text{tr}\,\tilde{P}^aT_i, 
\end{align}
\end{subequations}
where $A_a,\, F_{ab}, \, \tilde{P}^a$ are the self-dual connection, its curvature and its conjugate momentum, respectively, and the right-hand side of the second equation follows if we choose a locally coordinate-independent basis, i.e. $\partial_a T_i^{IJ} = 0$, which is always possible. What we have essentially done is to exploit the fact that the self-dual subalgebra of $\mathfrak{so}(3,1)\otimes\mathbb{C}$ is isomorphic to $\mathfrak{sl}(2,\mathbb{C})$ to write the canonical variables as matrix components of $\mathfrak{sl}(2,\mathbb{C})$ matrices. In the literature on loop quantum gravity, it is the pair $(A^i_a, \tilde{E}^a_i)$ that is more often called the Ashtekar variables. In terms of these, the Poisson brackets and constraints can be written as\footnote{The $i$ in Eq~(\ref{eq2.28}a) comes from Eq~(\ref{eq2.27}c). This is done just for conformity with the standard literature. Note that this choice forces us to include a factor of $-i$ in the definition of the Poisson bracket between two functions of $A^i_A$ and $\Tilde{E}^b_j$. }
\begin{subequations}
\label{eq2.28}
\begin{align}
    \{\tilde{E}^a_i(x), A^i_b(y)\} &= -i\delta^a_b\delta^j_i\delta^3(x,y),\\
    \{\tilde{E}^a_i(x), \tilde{E}^b_j(y)\} = 0, &\quad \{A^i_a(x), A^j_b(y)\} = 0. 
\end{align}
\end{subequations}

\begin{subequations}
\label{eq2.29}
\begin{align}
    S &= \epsilon^{ijk}\tilde{E}^a_i\tilde{E}^b_jF_{abk} = 0,\\
    V_b &= \tilde{E}^a_iF^i_{ab} = 0,\\
    G_i &= D_a\tilde{E}^a_i = 0. 
\end{align}
\end{subequations}

Notice the striking resemblance to the gauge theories of particle physics that this version of the constraints of general relativity affords. In Yang-Mills theory, for instance, the configuration variable is a vector potential which is an $\mathfrak{su}(N)$-valued connection, its `electric field' being the conjugate momentum and satisfying the so-called Gauss constraint, which looks exactly like Eq~(\ref{eq2.29}c). In the Ashtekar formulation of general relativity, the configuration variable is an $\mathfrak{sl}(2,\mathbb{C})$-valued connection and its conjugate momentum also satisfies a Gauss constraint. Moreover, in Yang-Mills theory, the Gauss constraint generates internal $SU(N)$ transformations. Similarly, here the Gauss constraint generates internal rotations on the triads $E^a_i$. This is to be expected, because the Ashtekar variables can be conceived of as an enlargement of the ADM phase space in the following sense. The three-metric $q_{ab}$ has six independent components, whereas the spatial triad $E^a_i$ has nine independent components; the Gauss constraint reflects our freedom in rotating the extra three components in the latter. Since we know how to quantise Yang-Mills theory, which contains the Gauss constraint, perhaps unsurprisingly, we will see that this constraint is the easiest to quantise in loop quantum gravity. 

However, this is as far as the analogy goes. There are no analogs of Eqs~(\ref{eq2.29}a) and (\ref{eq2.29}b) in Yang-Mills theory. As explained in Chapter 1, these equations encode the background-independence of general relativity. They are, of course, not present in Yang-Mills theory, which is formulated on flat Minkowski spacetime and hence, is not, like general relativity, background independent. Moreover, it is also worth emphasising once more that $sl(2,\mathbb{C})$ is a noncompact group, whereas the $SU(N)$, the gauge group of Yang-Mills theory, is compact. These considerations mean that general relativity is a gauge theory of a very different kind than Yang-Mills theory. 

After this important digression, we can now make contact between the Ashtekar variables and the ADM variables. Recall that the tetrads can be used to define the metric on $M$ (Eq~(\ref{eq2.4})). Thus it is natural that the three-metric of the ADM variables depends on the triads:
\begin{equation}
    (q)q^{ab} = \tilde{E}^a_i\tilde{E}^{bi}. \label{eq2.30}
\end{equation}
As for the canonical momentum in the ADM variables, we require more work. Start with transforming a spatial index of the extrinsic curvature $K_{ab}$ to an internal index:
\begin{equation}
    K^i_a := K_{ab}E^b_k\delta^{ik} \label{eq2.31},
\end{equation}
where $E^b_k := q^{-1/2}\tilde{E}^a_i$ is the dedensitised triad. Next, recall that a Lorentz connection $\omega_\alpha^{IJ}$ on $M\times\mathbb{C}^4$ induces a connection on $M$ (Eq~(\ref{eq2.10})) that is the Levi-Civita connection when the equations of motion hold -- in particular, when $D_\alpha e^\beta_I = 0$, where $D$ is the covariant derivative corresponding to $\omega_\alpha^{IJ}$. If we now choose internal coordinates such that
\begin{equation}
    e^\alpha_0 = -n^\alpha, \label{eq2.35*}
\end{equation}
where $n^\alpha$, recall, is the unit normal to the spatial slice $\Sigma$, it follows that the restriction to $\Sigma$ of the Lorentz connection is compatible\footnote{$D_\alpha e^\beta_I = 0$ implies that $D_a E^b_I = 0$. And Eq~(\ref{eq2.35*}) means that $E^b_0 = 0$ (the spatial projection of the unit normal is obviously zero), whence $D_aE^b_0 = 0$.} with the triads, i.e. $D_aE^b_j = 0$, which allows us to write it in terms of the Christoffel symbols:
\begin{equation}
    \tensor{\omega}{_a^i_j} = -E^b_j(\partial_ae^i_b - \Gamma^c_{ab}E^i_c). \label{eq2.32}
\end{equation}
The choice in Eq~(\ref{eq2.35*}) is harmless since we are always free to perform rotations on the internal indices without changing any physics; this specific choice is called the \textit{time gauge}, and geometrically amounts to our arranging for the spatial components $e^\alpha_i$ of the tetrads to span the tangent space of $\Sigma$ at each point. Now comes the punchline: the self-dual connection $A^i_a$ is related to the ADM variables via the following canonical transformation.
\begin{equation}
    A^i_a := \omega^i_a - iK^i_a, \label{eq2.33}
\end{equation}
where the identification $\omega^i_a \cong \tensor{\omega}{_a^i_j}$ makes sense because the latter has only three independent components (due to antisymmetry). Thus the perhaps elusive self-dual connection is nothing but a consequence of a canonical transformation on the old ADM variables. 

Eqs~(\ref{eq2.30}) and (\ref{eq2.33}) shed light on the reality conditions required to recover real general relativity from the self-dual framework. Since the three-metric and the extrinsic curvature are both real in the real theory, we require that
\begin{equation}
     \tilde{E}^a_i \, \text{be real$\quad$and} \quad i(A^i_a - \omega^i_a) \, \text{be imaginary}. \label{eq2.34} 
\end{equation}
It turns out that classically, these are not hard to implement. But in the quantum theory, their implementation is highly non-trivial and has not yet been achieved. 
\newline

Eqs~(\ref{eq2.30}--\ref{eq2.34}) specify how to recover the \textit{real} ADM variables from the Ashtekar variables. Incidentally, these equations also suggest a way to go in the opposite direction \cite{barbero, immirzi, samuel}. We begin with the \textit{real} ADM variables $K_{ab}$ and $q_{ab}$. We then introduce the triads $e^\alpha_I$ and the Lorentz connection $\omega_\alpha^{IJ}$, and project these onto the spatial slice $\Sigma$, whence we obtain $e^a_I$ and $\omega_a^{IJ}$. From here, there are two ways to proceed. One way is to pick the time gauge and then use Eqs~(\ref{eq2.30}) and (\ref{eq2.31}) to replace $K_{ab}$ and $q_{ab}$ with $\Tilde{E}^a_i$ and $K^i_a$, which are themselves canonical phase-space variables, since 
\begin{equation}
    \{K^i_a(x), \Tilde{E}^b_j(y)\} = \delta^b_a\delta^i_j\delta^3(x,y). \label{eq2.39*}
\end{equation}
Substituting these new phase-space variables into the ADM constraints (Eqs~(\ref{eq1.27}) and (\ref{eq1.28})), we can express these constraints in terms of the new variables:
\begin{subequations}
\label{eq2.40*}
    \begin{equation}
    C_a = \Tilde{E}^b_i\partial_{[a}K^i_{b]} = 0
    \end{equation}
    \begin{equation}
        C = \sqrt{q}R + \frac{2}{\sqrt{q}}\Tilde{E}^{[a}_i\Tilde{E}^{b]}_jK^i_a K^j_b = 0.
    \end{equation}
    But now these are not all the constraints. Notice that $K^i_a$ has nine components in total, whereas $K_{ab}$, owing to its symmetry, has six independent components only. This imposes a further constraint on $K^i_a$ and $\Tilde{E}^a_i$, namely
    \begin{equation}
        K_{i[a}\Tilde{E}^i_{c]} = 0,
    \end{equation}
\end{subequations}
which follows from making use of Eqs~(\ref{eq2.30}) and \ref{eq2.31} in $K_{[ab]} = 0$. Thus, in going from $(q^{ab}, K_{cd})$ to $(K^i_a, \Tilde{E}^b_j)$, we have enlarged the ADM phase space the enlargement being reflected in the internal gauge freedom provided by Eq~(\ref{eq2.40*}). Finally, we introduce the Ashtekar connection via the canonical transformation in Eq~(\ref{eq2.33}), where $\omega^i_a$, again, is given by Eq~(\ref{eq2.32}), which follows from $D_aE^b_j = 0$, which in turn is a consequence of using the time gauge. The factor of $i$ forces us into the arena of complex general relativity. Substituting the Ashtekar connection into Eq~(\ref{eq2.40*}) yields the polynomial constraints in the Ashtekar formulation (Eq~(\ref{eq2.29})).

But, as we mentioned above, there is also another way to proceed, and it avoids any internal gauge fixing. This uses the fact that since we wish to introduce self-dual connections and self-dual `triads', all we need to do is to project the spatial projections of the Lorentz connection and the tetrads to the self-dual subalgebra of $\mathfrak{so}(3,1)\otimes\mathbb{C}$. This can be achieved via \cite{immirzi}
\begin{subequations}
    \label{eq2.41*}
    \begin{equation}
        A^i_a := T^i_{IJ}\omega^{IJ}_a,
    \end{equation}
    \begin{equation}
        E^{ai} = -\epsilon^{abc}T^i_{IJ}e^I_b e^J_c, 
    \end{equation}
\end{subequations}
where $T^i_{IJ}$ is the basis given by Eq~(\ref{eq2.27}), and the second equation above represents a self-dual product of two tetrads. 
\newline

Let us summarise our achievements so far. Up until now, we have seen two alternative formulations of general relativity, namely the Palatini and Ashtekar formulations. In the first case, the constraints turn out to be non-polynomial as in the ADM case, whereas in the second case, the constraints turn out to be polynomial. However, the Ashtekar variables describe complex general relativity, and the reality conditions required to recover the real theory seem too complicated to be implemented at the quantum level. On the other hand, the polynomiality of constraints in the Ashtekar formulation seems to arise from self-duality (since otherwise, as in the Palatini case, the conjugate momentum is constrained, giving rise to second-class, non-polynomial constraints), and self-duality in spacetimes requires the use of complex variables by definition. Thus it seems necessary that we consider complex general relativity to achieve polynomial constraints.

\section{Barbero's Insight}
In either the Palatini or the Ashtekar formulation, the configuration variables are found to lie in the Lie algebra of a noncompact group. In other words, the gauge group of gravity is noncompact. In this section, we consider the possibility of altering this state of affairs. On the face of it, this seems to be a silly question to entertain, for the Palatini framework has already established the noncompactness of the gauge group of general relativity. But as we will see, there is a certain loose sense in which we could regard the configuration variables for gravity as coming from the Lie algebra of a compact group. 

The ADM-Ashtekar canonical transformation (Eq~\ref{eq2.33}) naturally prompts one to consider more general canonical transformations of the same form. That is, rather than $i$, we introduce a one-parameter family of canonical transformations,
\begin{equation}
    ^{(\beta)}A^i_a = \omega^i_a + \beta K^i_a, \label{eq2.3.1}
\end{equation}
where $\beta$ is an arbitrary parameter of our choosing; it is called the \textit{Barbero-Immirzi} parameter \cite{barbero, immirzi}. Specifically, $\beta =\pm i$ yields the Ashtekar variables, sending one to complex general relativity. However, if $\beta$ is kept real, we remain within the confines of real general relativity. We shall call the real-valued connection obtained in this way the \textit{Barbero connection}. 

There are both advantages and disadvantages of using real $\beta$. Let us probe the demerits first. The most obvious problem is that $\beta$ is an arbitrary parameter in our theory. At the classical level, there is no way to distinguish between different values of $\beta$, since each one yields a theory equivalent to general relativity. At the quantum level, each value of $\beta$ may make unique predictions at the quantum level, which can only be confirmed if one has access either to some independent means of arriving at some $\beta$-independent version of the same predictions or to some experiments. There are evidently no quantum-gravity experiments, so we are left with the former possibility only. For instance, the Barbero-Immirzi parameter is a multiplicative factor in the formula for the blackhole entropy derived via loop quantum gravity. One could thus determine the value of $\beta$ by comparing this formula with the already-known one derived through semiclassical methods. But this is not very neat, since the whole point of arriving at the blackhole entropy via loop quantum gravity is because one may not trust the semiclassical result in the first place. Thus, the ambiguity introduced into the theory by the use of $\beta$ remains. We shall invoke simplicity to set $\beta = 1$ if we wish to work with real canonical variables and $\beta = -i$ if we wish to work with the complex Ashtekar variables. In either case, we shall denote the transformed connection as $A^i_a$, the context making it clear which variable we are referring to. 

The second disadvantage of introducing the Barbero-Immirzi parameter concerns the fact that the constraints are polynomial only when $\beta =\pm i$. This deserves elaboration. Recall from the end of the previous section that to apply transformations of the form (\ref{eq2.3.1}), one first needs to enlarge the ADM phase space by going from $(q^{ab}, K_{cd})$ to $(K^i_a, \Tilde{E}^b_j)$. This enlargement leads to the constraints in Eq~(\ref{eq2.40*}). Now if we perform the canonical transformation (\ref{eq2.3.1}), keeping $\beta$ arbitrary, we find, after some algebra \cite{barbero, immirzi}, that the constraints (\ref{eq2.40*}) become
\begin{subequations}
    \label{eq2.3.2}
    \begin{equation}
        S = \frac{\epsilon^{ijk}}{\Tilde{E}}\Tilde{E}^a_i\Tilde{E}^b_j\, \tensor{^{(\beta)}F}{_{abk}} - \frac{2(\beta^2 + 1)}{\beta^2\Tilde{E}}\Tilde{E}^a_{[i}\Tilde{E}^b_{j]}(^{(\beta)}A^i_a - \omega^i_a)(^{(\beta)}A^j_b - \omega^j_b) = 0,
    \end{equation}
    \begin{equation}
        V_b = \Tilde{E}^a_i\tensor{^{(\beta)}F}{^i_{ab}} = 0
    \end{equation}
    \begin{equation}
        G_i = D_a\Tilde{E}^a_i = 0,
    \end{equation}
\end{subequations}
where $\Tilde{E} = E^2 = q$ is the determinant of the (densitised) triads. As can be seen, these constraints are polynomial\footnote{It might appear that the constraint obtained thus is different from Eq~(\ref{eq2.29}a). The determinant of the triad does not appear in Eq~(\ref{eq2.29}a) because we had already absorbed it in the lapse function above Eq~(\ref{eq2.14}a); we can do the same thing here as well, recovering the Ashtekar constraint.} only when $\beta = \pm i$, i.e. for the Ashtekar variables. Keeping $\beta$ real, therefore, adds a non-polynomial term to the Hamiltonian constraint. Once again, it seems that polynomiality of constraints is a special feature of self-dual \textit{complex} general relativity. Note, however, that the Hamiltonian constraint above is still a little better than the ADM Hamiltonian constraint, which involves a factor of the square-root of the three-metric, one of the most troublesome objects to quantise. 

What are the advantages of keeping $\beta$ real? One good thing about this choice is that we have a formulation of Hamiltonian general relativity that prioritises connections as the configuration variables over three-metrics, and that has first-class constraints which are perhaps easier to quantise than the ADM constraints. But the real advantage of this new scheme lies in the fact that we can now regard the configuration space of general relativity as the Lie algebra of a compact group. To see how this comes about, notice that in enlarging the ADM phase space to arrive at the new variables for\footnote{For $\beta = \pm i$, we always have the choice of using Eq~(\ref{eq2.41*}) directly.} $\beta \neq \pm i$, we have to work in the time gauge. Since this gauge fixes the time component of the tetrads, which encode our freedom to perform internal $SO(3,1)$ transformations in real general relativity, for real values of $\beta$, we are effectively left with the freedom to perform internal $SO(3)$ transformations only. This means that $^{(\beta)}A^i_a$ can now be regarded as taking values in $\mathfrak{so}(3)$. Since $SO(3)$ is a compact group, we may take this to mean that the ``gauge group'' of gravity is compact in this new formulation. 

But this is a heuristic argument at best. In fact, as shown by Samuel \cite{samuel}, for real values of $\beta$, $^{(\beta)}A^i_a$ cannot be regarded as the pullback to $\Sigma$ of the spacetime Lorentz connection. Since it is the latter that determines the gauge group of general relativity, the former does not belong to the Lie algebra of the gauge group of general relativity. This is in striking contrast to the Ashtekar connection, which, in view of Eq~(\ref{eq2.41*}), \textit{can} be regarded as the pullback to $\Sigma$ of (the self-dual part of) the spacetime Lorentz connection. What are we to make of the argument in the preceding paragraph, then? The answer, again, is provided by Samuel \cite{samuel}. Recall that we arrived at the Ashtekar phase-space variables from a manifestly covariant Lagrangian formulation. Incidentally, such a passage to phase-space variables with arbitrary values of $\beta$ is also available. This is provided by Holst's \cite{holst} modification to the Palatini action (\ref{eq2.8}), namely
\begin{equation}
    S = \frac{1}{2}\int_M d^4x(e)e^\alpha_I e^\beta_J \left(F_{\alpha\beta}^{IJ} - \frac{1}{2\beta}\tensor{\epsilon}{^{IJ}_{MN}}F_{\alpha\beta}^{MN} \right). \label{eq2.4.3}
\end{equation}
As can be readily verified, setting $\beta = \pm i$ recovers Ashtekar's self-dual action (\ref{eq2.23}). If one follows Holst's derivation \cite{holst, samuel} of the Hamiltonian theory corresponding to this action for arbitrary $\beta$, one finds that one has to ``gauge away'' part of the spacetime connection $F^{IJ}_{\alpha\beta}$ to derive the Barbero connection. In other words, the Barbero connection is identified with certain components of the spacetime Lorentz connection, and is not arrived at by pulling the latter back to $\Sigma$. 

One would do well to keep the foregoing subtleties in mind. But it is also true that the Barbero connection and $\Tilde{E}^a_i$ have canonical Poisson brackets among them. Thus, in so far as these variables can be regarded as the phase-space variables for general relativity, we are justified in using them to develop quantum general relativity. This is highly desirable, since now the configuration space of the theory will be the Lie algebra of a compact group, upon which, as hinted in Chapter 1, one can formulate a viable integration theory -- a precursor to constructing the Hilbert spaces of quantum theory. On this account, we shall restrict our attention to real general relativity as defined by the Barbero connection. 

\section{A Convenient Recasting of the New Variables}
In this section, we will see that the variables $(A^i_a, \Tilde{E}^b_j)$ of the previous two sections can be recast as $SU(2)$ spinors, which are very convenient for calculations. For simplicity, we shall restrict our attention to real connections and triads; analogous results hold for the complex case, since $\mathfrak{sl}(2,\mathbb{C})$ is isomorphic to $\mathfrak{su}(2)\otimes\mathbb{C}$.

Recall that $A^i_a$ can be thought of as components of matrices in the Lie algebra of $SO(3)$. Now, this Lie algebra is isomorphic to $\mathfrak{su}(2)$, which, essentially, is the space $\mathcal{M}$ of traceless, anti-Hermitian $2
\times2$ matrices, which are spanned by $i\tau_i$, $\tau_i$ being the Pauli matrices: 
\begin{equation}
  \tau_1 = \begin{pmatrix}
0 & 1\\ 
1 & 0
\end{pmatrix}, \quad
\tau_2 = \begin{pmatrix}
0 & -i\\ 
i & 0
\end{pmatrix}, \quad 
\tau_3 = \begin{pmatrix}
1 & 0\\ 
0 & -1
\end{pmatrix}. \label{eq2.36}
\end{equation}
We can thus introduce an isomporphism between the space of connections and $\mathcal{M}$. Most generally, we can set 
\begin{equation*}
    \tensor{A}{_{aA}^B} = ikA^i_a\tensor{\tau}{_{iA}^B}
\end{equation*}
 for some constant $k$; the indices $A$, $B$ refer to the component of a matrix in row $A$ and column $B$, and are called \textit{spinor} indices. Analogously, we define $\tensor{\Tilde{E}}{^a_A^B} = ik\Tilde{E}^a_i\tensor{\tau}{^i_A^B}$. That these two equations define an isomorphism is fairly obvious. In fact, by construction, distinct choices of $A^i_a$ will contract with $\tau_i$ to give distinct matrices in $\mathcal{M}$, and so the map is injective. Thus the map restricted to its range is invertible. Since addition and scalar multiplication are trivially preserved under the map, it is an isomorphism. 
 
 We can fix the value of $k$ by requiring that the isomorphism preserve the curvature of the connection. That is, first, we define the curvature of the connection $\tensor{A}{_{aA}^B}$:
 \begin{equation}
     \tensor{F}{_{abA}^B} = 2\partial_{[a}\tensor{A}{_{b]A}^B} + \tensor{[A_a, A_b]}{_A^B}. \label{eq2.37}
 \end{equation}
We then require that $\tensor{F}{_{abA}^B} = ik F^i_{ab}\tensor{\tau}{_{iA}^B}$, where 
 \begin{equation*}
     F^i_{ab} = 2\partial_{[a}A^i_{b]} + \tensor{\epsilon}{^i_{jk}}A^j_aA^k_b,
 \end{equation*}
 is the curvature of $A^i_a$. Substituting the expressions for the two $F$'s in $\tensor{F}{_{abA}^B} = ik F^i_{ab}\tensor{\tau}{_{iA}^B}$ and using the fact that
 \begin{equation}
     \tensor{\tau}{_{iA}^C}\tensor{\tau}{_{jC}^B} = i\tensor{\epsilon}{_{ij}^k}\tensor{\tau}{_{kA}^B} + \delta_{ij}\delta_A^B, \label{eq2.38}
 \end{equation}
 we obtain $k = -1/2$. Hence, we get the modified variables:
 \begin{subequations}
 \label{eqs2.39}
     \begin{align}
         \tensor{A}{_{aA}^B} &= -\frac{i}{2}A^i_a\tensor{\tau}{_{iA}^B}\\
         \tensor{\Tilde{E}}{^a_A^B} &= -\frac{i}{2}\Tilde{E}^a_i\tensor{\tau}{^i_A^B}
     \end{align}
 \end{subequations}
 
Since $\Tilde{E}^a_i$ are vector fields on the spatial manifold $\Sigma$, $\tensor{\Tilde{E}}{^a_A^B}$ is an isomorphism between the tangent space at a point in $\Sigma$ and $\mathcal{M}$. A fancy way of describing this is to say that the objects $\tensor{\Tilde{E}}{^a_A^B}$ ``solder'' $\mathcal{M}$ to the tangent space at every point. Accordingly, they are called \textit{soldering forms}. They contain as much information about the geometry of $\Sigma$ as does the spatial metric $q_{ab}$. In fact, Eq~(\ref{eq2.38}) implies that
\begin{equation}
    \text{tr}\,\Tilde{E}^a\Tilde{E}^b = \tensor{\Tilde{E}}{^a_A^B}\tensor{\Tilde{E}}{^b_B^A} = -2qq^{ab}. \label{eq2.40}
\end{equation}
Thus the modified momenta are sufficient to reconstruct the spatial metric. 

Furthermore, we can evaluate the Poisson brackets between the modified variables. We find that all the brackets vanish except (derivation below)
\begin{equation}
    \{\tensor{\Tilde{E}}{^a_{CD}}(x), \tensor{A}{_b^{AB}}(y)\} = -i\delta^a_b\delta_C^{(A}\delta_D^{B)}\delta^3(x,y). \label{eq2.41}
\end{equation}
Therefore, the spinorial variables form a faithful set of canonical variables, which can be used in place of the non-spinorial variables to describe canonical general relativity. That they also preserve the particular advantages of the Ashtekar formulation is further confirmed by the fact that they do not change the form of the constraints in Eq~(\ref{eq2.29}), which now become
\begin{subequations}
\label{eq2.42}
    \begin{align}
        S &= \text{tr}\,(\Tilde{E}^a\Tilde{E}^bF_{ab}) = 0,\\
        V_b &= \text{tr}\,(\Tilde{E}^aF_{ab}) = 0,\\
        G_i &= D_a\tensor{\Tilde{E}}{^a_A^B} = 0. 
    \end{align}
\end{subequations}

\begin{mdframed}[style=testframe]
    \textbf{Derivation -- Eq~(\ref{eq2.42})}
    
    We have defined the spinorial variables with one index downstairs and one upstairs, whereas Eq~(\ref{eq2.41}) contains spinors with both indices upstairs and downstairs. Therefore, we first need a way to raise and lower spinor indices. To this end, we look for an object that is invariant under $SU(2)$ transformations. This is analogy to the Minkowski metric, which is invariant under Lorentz transformations, being used for index raising and lowering in Minkowski space. Now, for any $2\times2$ matrix $\tensor{M}{_A^B}$, we have that
    \begin{equation}
        \text{det}\,(\tensor{M}{_A^B}) = \frac{1}{2}\epsilon_{AB}\epsilon^{CD}\tensor{M}{_C^A}\tensor{M}{_D^B}. \label{eq2.43} 
    \end{equation}
    where
    \begin{equation}
        \epsilon_{AB} = \epsilon^{AB} = \begin{pmatrix}
            0 & -1\\
            1 & 0
        \end{pmatrix} \label{eq2.44}
    \end{equation}
    is the two-dimensional Levi-Civita tensor, which satisfies 
    \begin{equation}
        \epsilon_{AB}\epsilon^{CD} = 2\delta_A^{(C}\delta_B^{D)}. \label{eq2.45}
    \end{equation}
    Since $SU(2)$ matrices have determinant 1 and are Hermitian, Eqs~(\ref{eq2.43}) and (\ref{eq2.45}) entail that for an $SU(2)$ matrix $\tensor{U}{_A^B}$,
    \begin{equation}
        \tensor{U}{_C^A}\epsilon_{AB}\tensor{U}{^B_D} = 1. \label{eq2.46}
    \end{equation}
    Thus the required object is $\epsilon_{AB}$. Since it is antisymmetric, we need to fix a convention for index positioning when raising and lowering indices. We choose
    \begin{equation}
        L^A = \epsilon^{AB}L_B, \quad L_B = L^A\epsilon_{AB},
        \label{eq2.47}
    \end{equation}
    which means that index raising (lowering) occurs by multiplication with $\epsilon^{AB}$ ($\epsilon_{AB}$) from the left (right). 
    
    With this, we are ready to evaluate the Poisson bracket between $\Tilde{E}^a$ and $A_b$. Using Eqs~(\ref{eq2.28}) and (\ref{eqs2.39}), we obtain
    \begin{equation}
        \{\tensor{\Tilde{E}}{^a_{CD}}(x), \tensor{A}{_b^{AB}}(y)\} = \frac{i}{4}\delta^a_b\delta^3(x,y)\tensor{\tau}{^i_{CD}}\tensor{\tau}{_i^{AB}}. \label{eq2.48}
    \end{equation}
    Next, from the completeness of the Pauli matrices, i.e. 
    \begin{equation}
        \tensor{\tau}{^i_A^B}\tensor{\tau}{_i_C^D} = 2\delta_A^D\delta^B_C - \delta_{AC}\delta^{BD},
    \end{equation}
    and Eqs~(\ref{eq2.47}) and (\ref{eq2.45}), it follows that
    \begin{align}
        \tensor{\tau}{^i_{CD}}\tensor{\tau}{_i^{AB}} &= \epsilon_{ED}\epsilon^{AF}\tensor{\tau}{^i_C^E}\tensor{\tau}{_i_F^B}\\
        &= 2\delta_C^B\epsilon_{FD}\epsilon^{AF} - \epsilon_{CD}\epsilon^{AB}\\
        &= -2\delta_C^{(A}\delta_D^{B)},
    \end{align}
    which upon substitution into Eq~(\ref{eq2.48}) yields the desired result. 
\end{mdframed}

A useful remark is in order. The symmetrisation in Eq~(\ref{eq2.41}) suggests that in index gymnastics, one can take the connection and the associated momentum to be symmetric in spinor indices if they are both upstairs or downstairs. This is indeed the case, as we now verify. Consider any traceless, anti-Hermitian matrix $\tensor{M}{_A^B}$. It can be written as
\begin{equation}
    \tensor{M}{_A^B} = i\begin{pmatrix}
      a & b\\
      b^* & -a
    \end{pmatrix}, \label{eq2.53}
\end{equation}
which implies that 
\begin{subequations}
\label{eq2.54}
    \begin{equation}
    M_{AB} = \tensor{M}{_A^C}\epsilon_{CB} = -i\begin{pmatrix}
        -b^* & a\\
        a & b
    \end{pmatrix}
\end{equation}
and that
\begin{equation}
    M^{AB} = \epsilon^{AC}\tensor{M}{_C^B} = i\begin{pmatrix}
        -b^* & a \\
        a & b
    \end{pmatrix}.
\end{equation}
\end{subequations}

\chapter{Connections Go Quantum}
In this chapter, we will embark upon a rigorous quantisation of gravity written in the new (Barbero) variables. We will construct an the algebra of observables that are to be promoted to operators during quantisation. Then we will see in detail how to construct a quantum configuration space, and understand how it is different from the classical configuration space. This shall allow us to obtain a Hilbert space of quantum states satisfying the Gauss constraint; the other two constraints will be dealt with in the next chapter. Finally, we will also derive an explicit orthonormal basis for this Hilbert space. 

\section{A Brief Review of Canonical Quantisation}
Let us begin by outlining a general algorithm for the quantisation of a fully constrained Haimltonian theory, as exemplified in Section 1.3. 
\begin{enumerate}
    \item Let $\Gamma$ be the classical phase space, coordinatised by configuration and momentum variables. Choose a set $\mathcal{S}$ of elementary classical variables that consists of complex-valued functions $f: \Gamma \to \Gamma$ subject to the following consistency conditions:
    \begin{enumerate}[(a)]
        \item $\mathcal{S}$ should be sufficiently large, so that any function on the phase space can be expressed as a sum of products of the elementary variables.
        \item Closure under Poisson brackets, i.e. for any $F, G \in \mathcal{S}$, $\{F, G\} \in \mathcal{S}$.
        \item Closure under complex conjugation, i.e. $F^* \in \mathcal{S}$ for all $F \in \mathcal{S}$. 
    \end{enumerate}
    \item Promote all $F \in \mathcal{S}$ to operators $\hat{F}$ that satisfy the free associative algebra $\mathcal{B}$ that results from imposing (1) canonical commutation relations $[\hat{F}, \hat{G}] = \{\widehat{F, G}\}$, and (2) any anticommutation relations that capture the algebraic relations between the elements of $\mathcal{S}$. 
    \item Via an involution $\star$, introduce a $\star$-algebra $\mathcal{B}^{(\star)}$ on $\mathcal{B}$ such that for any $F$, $G \in \mathcal{S}$, if $F^* = G$, then $\hat{F}^\star = \hat{G}$. 
    \item Construct a Hilbert space $\mathcal{H}$ that is in some sense an $L^2$-normed vector space of complex-valued functions on the classical configuration space (or an enlargement thereof\footnote{See below for elaboration of this point.}). 
    \item Find a representation of $\mathcal{B}^{(\star)}$ on the space $\mathcal{L(H)}$ of (bounded) linear operators on $\mathcal{H}$ via
    \begin{equation}
        R(A^\star) = R(A)^\dagger, \label{eq3.1*}
    \end{equation}
    where $\dagger$ denotes Hermitian conjugation.
    \item Via step 1 and the properties of the involution, the constraints $C_i$ of the classical theory are now represented as self-adjoint operators $\hat{C}_i$ on $\mathcal{H}$. Find the space $\mathcal{V}_{phy}$ of solutions to these constraints (this step, as we will see below, may require us to represent the exponentiated version of $C_i$ as unitary operators $\hat{U}_i$ on $\mathcal{H}$).
    \item Induce a Hilbert space structure on $\mathcal{V}_{phy}$, obtaining the physical Hilbert space $\mathcal{H}_{phy}$.
\end{enumerate}

In the subsequent sections, we shall see in detail how all these steps can be (almost) implemented in loop quantum gravity. However, before proceeding, let us address a possible technical objection to the steps above. We have required in step four that the vector space of complex-valued functions on the configuration space be a Hilbert space. One might retort that this is an unnecessary step, since for physical predictions, all we require is the existence of a finite inner product on the physical Hilbert space. This objection is certainly valid, but irrelevant. This is so because, as will be recalled from Section 1.3, $\mathcal{H}$ arose in the context of defining a measure on the configuration space. Since, as we will see, $\mathcal{H}_{phy}$ has to be in some sense constructed from $\mathcal{H}$, the definition of an inner product on the former at least requires a knowledge of how to perform integrations on the latter, which thus needs to have a well-defined measure on it. Thus, even though the relevant space for the introduction of inner products is $\mathcal{H}_{phy}$, in the framework that we shall use, constructing $\mathcal{H}$ can be conceived of as an intermediate step in that very direction. 

\section{Loops, Holonomies, and All That}
As outlined in the preceding section, the first task in the quantisation program is to pick a space $\mathcal{S}$ of elementary classical variables that need to be quantised. Now, our phase space is coordinatised by $(A^i_a, \Tilde{E}^b_j)$. Which functions of these variables ought to constitute $\mathcal{S}$? Two considerations are germane to an answer to this question. First, in view of rigour, whatever functions are used, they must smear $(A^i_a, \Tilde{E}^b_j)$, and, in the interests of consistency, do so in a background-independent manner. Second, as explained in the first chapter, we need a way to integrate functions on the phase space, and this requires the construction of well-defined measures on the domain space. In the following sections, we shall keep both these points in mind. 

\subsection{Configuration variables}
We shall start with $A^i_a$, which, it will be recalled, is an $SU(2)$ connection on the principal bundle associated to the tangent bundle of a spacelike hypersurface $\Sigma$ in $M$. Smearing a variable involves some kind of its integration. One natural avenue to search for a suitable ``integrating procedure" is the equation for the parallel transport of sections in $\Sigma\times\mathbb{R}^3$, since it is a differential equation involving the connection $A^i_a$. This strategy also promises to cater to our measure-theoretic concerns. Recall that, as mentioned in Chapter 1, one of the advantages of reformulating gravity as an $SU(2)$ gauge theory is that there exists the natural Haar measure on $SU(2)$ (or on any compact Lie group, for that matter), which can be used to construct a well-defined measure on the configuration space. Notice, however, that the configuration space is the space of all $SU(2)$ connections, which take values in the Lie algebra of $SU(2)$ and not, directly, in $SU(2)$. It thus seems natural to transport these connections to $SU(2)$ in a background-independent manner. Now, the parallel-transport of a vector along curves in $\Sigma$ involves the connection and is a way of moving across the tangent spaces of $\Sigma$ at different points. On the other hand, in view of the tetrad formalism, vectors at different points in (sufficiently small regions of) $\Sigma$ are related via $SU(2)$ transformations. Thus there is a one-one correspondence between connections and $SU(2)$ elements via parallel transport. 

More precisely, let $\gamma: [r, s] \to \Sigma$ be an analytic\footnote{Analytic here means the existence of derivatives of all orders. This condition is required for purely technical reasons pertaining to the application of standard measure-theoretic results in subsequent sections.} curve in $\Sigma$. We say that a section $v$ on $\Sigma\times\mathbb{R}^3$ is parallel-transported along $\gamma$ if for all $t\in [r,s]$,
\begin{equation}
    D_{\dot{\gamma}(t)}v(t) := \dot{v}(t) + A(\dot{\gamma}(t))v(t) = 0, \label{eq3.2}
\end{equation}
where the over-dot indicates a $t$-derivative, $A(\dot{\gamma}(t))v(t) = A_a(t)\dot{\gamma}^a(t)v(t)
$, and $v(t) := v(\gamma(t))$ and $A_a(t) := A_a(\gamma(t))$. Integrating this equation, we obtain
\begin{equation*}
    v(t) = v(0) - \int_r^s dt_1 A(\dot{\gamma}(t_1))v(t_1),
\end{equation*}
which, when substituted recursively into itself yields
\begin{equation}
    v(t) = \sum_{n=0}^{\infty} \left[ (-1)^n\int_{s \geq t_1 \geq \cdots t_{n-1} \geq r} dt_n\cdots dt_1 A(\dot{\gamma}(t_1))\cdots A(\dot{\gamma}(t_n))  \right]v(0) \label{eq3.3}
\end{equation}
This sum can be shown to converge and be differentiable \cite{baez 1}. Thus we are in safe territory. To recast the preceding equation in a more palatable form, we define a permutation $\rho(i), \, i \in \{1,\cdots,n\}$ such that $t_{\rho(1)}\geq \cdots \geq t_{\rho(n)}$. Then, given a set $\{K(t_i): i \in \{1,\cdots,n\}\}$, the \textit{path-ordered product} of the $K(t_i)$ is defined as
\begin{equation*}
    \mathcal{P} K(t_1)\cdots K(t_n) := K(t_{\rho(1)})\cdots K(t_{\rho(n)}).
\end{equation*}
The integral in Eq~(\ref{eq3.3}) can then be written as 
\begin{align*}
    \int_{s \geq t_1 \geq \cdots t_{n-1} \geq r} dt_n\cdots dt_1 &A(\dot{\gamma}(t_1))\cdots A(\dot{\gamma}(t_n))\\ 
    &= \frac{1}{n!}\int_{t_i\in [r,s]}dt_n\cdots dt_1 \mathcal{P}A(\dot{\gamma}(t_1))\cdots A(\dot{\gamma}(t_n))\\
    &= \frac{1}{n!}\left(\int_r^s dt A(\dot{\gamma}(t)) \right)^n,
\end{align*}
which allows us to rewrite Eq~(\ref{eq3.3}) as
\begin{equation}
    v(s) = \mathcal{P}\exp\left({-\int_r^s dtA(\dot{\gamma}(t))}\right) v(r), \label{eq3.4}
\end{equation}
where
\begin{equation}
     \mathcal{P}\exp\left({-\int_r^s dtA(\dot{\gamma}(t))}\right) := \sum_{n=0}^{\infty}\frac{(-1)^n}{n!}\left(\int_r^s ds A(\dot{\gamma}(s)) \right)^n \label{eq3.5}
\end{equation}
is a \textit{path-ordered exponential}. It can also be thought of as a linear map from $T_p\Sigma$ to $T_q\Sigma$, $p = \gamma(r)$ and $q = \gamma(s)$ being two points in $\Sigma$. We call this map the \textit{parallel propagator} around $\gamma$ and label as
\begin{equation*}
    U_{\gamma(t)}(s, r)[A],
\end{equation*}
where $t$ is the parameter along the curve. Note that the propagator acts from right to left, i.e. given $\gamma: [t_1, t_2] \to \Sigma$,
\begin{equation}
    v(t_2) = U_{\gamma(s)}(t_2, t_1)v(t_1). \label{eq3.6}
\end{equation}
We began with an analytic curve, but we can be more general. Consider a piecewise-analytic path $\gamma: [r,s]$. We can break it up into maximal\footnote{That is, you break the path exactly at points of non-analyticity.} analytic pieces $\gamma_i: [t_i, t_{i+1}] \to \Sigma$, where $1 \leq i \leq n$ and $t_1 = r$, $t_n = s$. Then it is evident that we can parallel-transport a vector along $\gamma$ by transporting it along one analytic piece at a time:
\begin{equation}
    U_{\gamma}(s, r) = U_{\gamma_n}(s, t_{n-1})\cdots U_{\gamma_i}(t_{i+1}, t_{i})\cdots U_{\gamma_1}(t_1, r). \label{eq3.7}
\end{equation}

If the path $\gamma$ returns to itself, it is called a \textit{loop}. In that case, the parallel propagator is called a \textit{holonomy}, and we will denote it as
\begin{equation*}
    U_{\gamma(s)}(t, t)[A] = U_{\gamma(s)}(t)[A],
\end{equation*}
$\gamma(t)$ being the point where the loop starts and ends. We can also think of a loop as a map $\gamma: S^1 \to \Sigma$, where $S^1$ is the boundary of the unit circle. For each loop $\gamma$ and connection $A$, we define the so-called \textit{Wilson loop}
\begin{equation}
    T^0_{\gamma(s)}[A] = \text{tr}\,U_{\gamma(s)}[A], \label{eq3.8}
\end{equation}
where the trace is evaluated in the fundamental representation of $SU(2)$, namely the algebra of the Pauli matrices. For a reason that will become clear below, we shall identify the new configuration variables with the traces $T^0_{\gamma}$ of holonomies of connections around loops.   

Since exponentiation of Lie-algebra elements generates Lie-group elements, as promised, parallel propagators are elements of $SU(2)$. 

There are a number of other interesting properties that parallel propagators satisfy. First, they require no background metric to be defined. Second, a substitution of variables proves that they are reparametrisation invariant. That is, given a curve $\gamma$ parametrised by $s$, let $t = f(s)$ for some smooth function $f$. Then it is easily seen that $U_{\gamma(t)} = U_{\gamma(s)}$. Thus one might as well write $U_{\gamma}$, dropping the parameter label, without loss of generality. Third, a parallel propagator along a composition of smooth paths that are joined end-to-end splits naturally into parallel propagators along each separate, smooth path. In other words, let $\gamma: [0, s] \to \Sigma$ be a path from $p$ to $q$, and let $\lambda: [0,t] \to \Sigma$ be a path from $q$ to $r$. Then we define the product $\lambda\gamma: [0, s+t] \to \Sigma$ of the two paths by
\begin{equation*}
    (\lambda\gamma)(u) := \begin{cases}
 \gamma(u)& \text{ if } 0\leq u\leq s \\
 \lambda(u-s)& \text{ if } s \leq u \leq s+t. 
\end{cases}
\end{equation*}
A direct substitution of this equation into Eq~(\ref{eq3.5}) reveals that 
\begin{subequations}
    \label{eq3.9}
    \begin{align}
        U_{\lambda\gamma}(0, t) = U_\lambda(0,s)U_\gamma(s, t).
    \end{align}
    Fourth, given a path $\gamma: [s,t] \to \Sigma$ from $p$ to $q$, there is an inverse path $\gamma^{-1}$ from $q$ to $p$ given by $\gamma^{-1}(u) = \gamma(t-u)$. Moreover, for every point $p$ we have an \textit{identity loop} $1_p: [s,t]\to \Sigma$, which is a path that stays at $p$, i.e. $1_p(u) = p$ $\forall u \in [s,t]$. Again, a direct substitution of these definitions into Eq~(\ref{eq3.5}) yields the following identities.
    \begin{align}
        U_{\gamma}^{IJ}(s) &= -U_{\gamma^{-1}}^{JI}(s)\\
        U_{1_p\gamma} &= U_{\gamma 1_p} = U_\gamma \\
        U_{1_p} &= \mathds{1}_{SU(2)}.
    \end{align}
    In the first equation, the superscripts label the matrix elements of a holonomy. Fifth, parallel propagators transform in a simple manner under a gauge transformation. Given $\Lambda \in SU(2)$ (or whatever the relevant gauge group) and a connection $A$, we find
    \begin{align}
        U_{\gamma}(s,t)[A'] = \Lambda(\gamma(s))U_\gamma(s,t)[A]\Lambda(\gamma(t))^{-1},
    \end{align}
    where $A'$ is the gauge-transformed connection. This can be verified as follows. Begin with the parallel-transport equation. For a $u(x) \in T_{\gamma(x)}\Sigma$, we write
    \begin{equation*}
        \dot{u}(x) = -A(\dot{\gamma}(x))u(x) = - \dot{\gamma}^a(x)A_a(\gamma(x)) u(x).
    \end{equation*}
    Under a guage transformation $\Lambda$, we have $u(x) \to \Lambda(\gamma(x))u(x) := v(x)$. Now observe that
    \begin{align*}
        \dot{v}(x) &= \dot{\Lambda}(\gamma(x))u(x) + \Lambda(\gamma(x))\dot{u}(x)\\
        &= \dot{\gamma}^a(x)(\partial_a\Lambda)\Lambda^{-1}v(x) - \dot{\gamma}^a(x)A_a u\Lambda^{-1}v(x) \\
        &= -\dot{\gamma}^a(x)(\Lambda\partial_a\Lambda^{-1} + \Lambda A_a \Lambda^{-1})v(x),
    \end{align*}
    where the last line follows from $(\partial_a\Lambda) = \partial_a(\Lambda\Lambda^{-1}) - \Lambda\partial_a\Lambda^{-1} = - \Lambda\partial_a\Lambda^{-1}$. The term in the parentheses in the last line above is nothing but the gauge transformation of the connection. Therefore, under a gauge transformation, $D_{\dot{\gamma}(x)}u(x) \to D'_{\dot{\gamma}(x)}v(x) = 0$, where $D'$ is the covariant derivative associated with $A' = \Lambda\partial_a\Lambda^{-1} + \Lambda A_a \Lambda^{-1}$. Since $U_\gamma[A]$ maps $u(s)$ to $u(t)$ and $U_\gamma[A']$ maps $v(s) = \Lambda(\gamma(s))u(s)$ to $v(t) = \Lambda(\gamma(t))u(t)$, Eq~(\ref{eq3.9}d) follows.    

    Fifth, Eq~(\ref{eq3.9}e) unravels a peculiar characteristic of holonomies. Since a loop $\alpha$ starting at $p$ will eventually return to $p$, under a gauge transformation, $U_\alpha[A'] = \Lambda(p)U_\alpha[ A]\Lambda(p)^{-1}$, which, via the cyclic invariance of the trace operation, implies that
    \begin{equation}
        \text{tr}\,U_\alpha[A'] = \text{tr}\,U_\alpha[A],
    \end{equation}
    In other words, Eq~(\ref{eq3.9}f) tells us that the trace of a holonomy is gauge invariant! Thus holonomies furnish a natural way to extract information about a system that is in principle physically observable. For example, consider quantum electrodynamics, in which the vector potential is a $U(1)$ connection, for which the path-ordered exponential in Eq~(\ref{eq3.5}) reduces to an ordinary exponential (since $U(1)$ is abelian), which is nothing but the Aharanov-Bohm phase. 

    Finally, since connections give rise to parallel propagators and holonomies, it is natural to inquire whether the converse is true. That is, using an $SU(2)$-valued function of piecewise-analytic paths $\gamma \subset \Sigma$ that satisfies Eq~(\ref{eq3.9}a), can we construct a connection on $\Sigma\times\mathbb{R}^3$? The answer is yes, provided the function satisfies some more properties. To investigate these properties, let us remind ourselves that a connection $A$ is smooth in the sense that its components $A_a(x)$ are smooth functions of $x\in \Sigma$. This smoothness translates to certain smoothness conditions on the parallel propagators. This can most easily be seen \cite{giles} by comparing parallel propagators for neighbouring paths. Let $x^a(t) \in \Sigma$ be a piecewise-analytic path from $x^a(0) = y^a$ to $x^a(1) = z^a$, and let $u^a(t)$ be a piecewise-analytic path such that $u^a(0) = u^a(1) = 0$. Then $X^a(s,t) = x^a(t) + su^a(t)$ is a one-parameter family of paths $\gamma_s$ with endpoints $y^a$ and $z^a$. It follows from Eq~(\ref{eq3.5}) that as $s\to 0$,
    \begin{equation}
        U_{\gamma_s}(z, y)[A] = U_{\gamma_0}(z, y)[A] + O(s^2), 
    \end{equation}
    where we have indulged in a slight abuse of notation by labelling the arguments of $U$ by points in $\Sigma$ rather than by their corresponding parameter values. Furthermore, if $\alpha$ is a straight-line path from $y^a$ to $z^a$, then as $y^a \to z^a$,
    \begin{equation}
        U_\alpha(z, y)[A] = 1 + (z^a-y^a)A_a(x) + O((z^a-y^a)^2).
    \end{equation}
    In deriving these equations, smoothness of the connections is crucial, for the limits taken would exist only if the connections were bounded, a fact which follows from the smoothness of connections on the continuous image of $[0,1]$ under the paths (recall that a continuous image of compact set is compact, and a smooth function on a compact set is always bounded). We shall interpret these equations as smoothness conditions on the parallel propagators. If an $SU(2)$-valued function of piecewise-analytic paths satisfies Eqs~(\ref{eq3.9}a, g, h), we can use that function to reconstruct the connection $A$. In particular, we have
    \begin{equation}
        A_a(x) = \lim_{\epsilon^a\to 0}\frac{U_\alpha(z+\epsilon, y^a)-1}{\epsilon^a}, 
    \end{equation}
    since in the limit, any path $\alpha$ can be approximated via a straight line, warranting the use of Eq~(\ref{eq3.9}h). In fact, approximating an arbitrary path $\gamma$ from $y$ to $z$ by a large number $N$ of straight-line paths $\gamma_n$, we see that
    \newpage
    \begin{align}
         U_\gamma(z,y) &= \lim_{N\to\infty}U_{\gamma_n}(z, x_{n-1})\cdots U_{\gamma_1}(x_1, y) \nonumber \\
         &= [1 + (z^a-x^a_{n-1})A_a(x)]\cdots[1 + (x^a_1-y^a)A_a(x)] = \mathcal{P}\exp{\left(\int_y^z dx^aA_a\right)},
    \end{align}
    which validates Eq~(\ref{eq3.9}i). Thus there is a one-to-one correspondence\footnote{This correspondence can be made much more rigorous and elegant than our heuristic proof here. See, for instance, the illuminating reconstruction theorems of Barret \cite{barret}, which, among other things, yield a completely novel reformulation of general relativity.} between smooth connections and smooth $SU(2)$-valued functions $U$ of piecewise-analytic paths $\gamma$. This immensely useful result means that instead of the space of smooth connections, one could equivalently regard the space of the maps $U$ as the configuration space for gravity. This fact will become very important when we shall have to obtain a precise characterisation of the configuration space for the quantum theory in Section 3.4.2.
    
\end{subequations}

The gauge-invariance of the traces of holonomies explains our motivation in using them, rather than holonomies directly, as configuration variables. Instead of working in the space $\mathcal{A}$ of connections, we can now work in the space $\mathcal{A/G}$ of connections modulo gauge transformations, which the traces of holonomies, being gauge invariant, project down to\footnote{More precisely, every equivalence class $[A]$ of connections related by a gauge transformation corresponds to a unique function, which is the trace of the holonomy of any representative in $[A]$.}. Therefore, one need not worry about the Gauss constraint, which generates internal $SU(2)$ gauge transformations, which have been factored out in $\mathcal{A/G}$; upon quantisation, only the diffeomorphism and Hamiltonian constraints need to be implemented. As per the typology enunciated in Section 1.3.2, the Gauss constraint has been implemented via reduced phase space quantisation.

Henceforth, the classical configuration space will be $\mathcal{A/G}$. 

\subsection{Momentum variables}
Let us now turn to the momentum variables. To this end, recall that the central motivation for constructing the set $\mathcal{S}$ of elementary variables is for us to be able to write any well-behaved function on the phase space as (possibly a limit of) a sum of products of elements in $\mathcal{S}$. Therefore, it suffices to consider at most linear combinations of suitably smeared phase-space $(A^i_a, \Tilde{E}^b_j)$ variables. Now since the phase space forms a cotangent bundle, via vector fields on the spatial manifold $\Sigma$, the conjugate variable $\Tilde{E}^b_j$ admits a natural smearing that yields variables essentially linear in $\Tilde{E}^b_j$. This is how it works. Let $f$ be a vector field on $\Sigma$ and $\epsilon_{abc}$ be the Levi-Civita tensor. We can smear $\Tilde{E}^b_j$ over an arbitrary surface $S \subset \Sigma$ as 
\begin{equation*}
    \int_S dx^a\wedge dx^b\epsilon_{abc}f^i\Tilde{E}^c_i,
\end{equation*}
where $x^a = (s, t)$ are coordinates on $S$. 

Now, since $A^i_a$ are $\mathfrak{su}(2)$-valued, and $(A^i_a, \Tilde{E}^b_j)$ form a cotangent bundle, $\Tilde{E}^b_j$ must live in the dual of $\mathfrak{su}(2)$. Therefore, the structure of the preceding equation indicates that $f$ must also be $\mathfrak{su}(2)$-valued. Furthermore, a suitable surface $S$ in $\Sigma$ can be foliated by a one-parameter family of loops, i.e. we can think of $S$ as a map $S: (0,1)\times S^1 \to \Sigma$. We call such surfaces \textit{ribbons} or \textit{strips}. Finally, we observe that on each loop in the family is defined a holonomy, which is a function of $\mathfrak{su}(2)$-valued connections, and that since we are now using connections modulo gauge transformations, for consistency, we want the momentum variables to be gauge invariant as well. These considerations allow us to introduce a strip functional:
\begin{subequations}
\label{eq3.10}
    \begin{equation}
    T^{1}_S[A] = \int_S dx^a\wedge dx^b \epsilon_{abc}T^c_{\gamma_s}(t), 
\end{equation}
\begin{equation}
    T^c_{\gamma_s}(t) = \text{tr}\,(U_{\gamma_s}(t)[A]\Tilde{E}^c(s,t)),
\end{equation}
\end{subequations}
where $s$ labels loops within $S$ and $t$ is the starting point of a particular loop $\gamma_s$. In other words, we insert $\Tilde{E}^c$ at the starting point $t$ of a particular loop $\gamma_s$, obtaining $T^c_{\gamma_s}(t)$, which we then integrate over all possible starting points of all loops. Again, to save clutter $\Tilde{E}^c(s,t) := \Tilde{E}^c(\gamma_s(t))$; here we use the spinor representation of the triads, introduced in Section 2.4. As can be readily verified, strip functionals are gauge invariant, and thus well-defined on the cotangent bundle over $\mathcal{A/G}$. They are also linear in $\Tilde{E}^a_i$. Therefore, they are ideal candidates for being the momentum variables. We call $T^1_S[A]$ a \textit{strip functional} or a $T^1$ variable.

In what follows, it will be important to keep in mind certain properties of the $T^c$ variables occurring in the definition of the strip functionals. To start with, they are not, unfortunately, reparametrisation invariant. However, under reparametrisations that preserve the orientation of a loop, the $T^c$ are covariant. To see this, let $\gamma'(t) = \gamma(f(t))$, with $f'(t) > 0$. Then it is easily seen that
\begin{equation}
    T^c_{\gamma(f(t))}(f(s)) = T^c_{\gamma'(t)}(s) \label{eq3.11}
\end{equation}
On the other hand, if the orientation of a loop is changed, then from Eq~(\ref{eq3.7}b) and the symmetry of $\Tilde{E}^a$ in the spinor indices, it follows that the $T^c$ variables change sign. In view of these properties, one concludes that the $T^c$ variables depend on \textit{oriented, unparametrised loops}. This fact will become important later in calculations.  

\subsection{The algebra of \texorpdfstring{$\mathcal{S}$}{TEXT}}
We will now show that the $T^0$ and $T^1$ variables defined above satisfy all the properties for being identified with the set $\mathcal{S}$ of elementary classical variables. 

First, as to the fact that the $T^0$ and $T^1$ variables (almost) span the gauge-invariant subspace of the phase space, we refer the reader to Ref. \cite{ashtekar 1}. However, there seems to be a caveat here. As opposed to Ref. \cite{ashtekar 1, ashtekar 5}, Rovelli and Smolin \cite{rovelli 3} argue that the question of these variables spanning the (gauge-invariant) phase space is still open. I have not been able to confirm which reference is right. However, it is nonetheless agreed that these $T$ variables can be extended to include a more general set of variables (see \cite{rovelli 3}) that do span the whole gauge-invariant phase space. Therefore, for now, we shall rest content with these variables only.

Second, the $T$ variables are closed under complex conjugation. To see this, note that upon complex conjugation, both the connection and its conjugate momentum pick up a minus sign, since they are anti-Hermitian. In the $T^1$ variables, the two minus signs cancel, while in the $T^0$ variables, the minus sign can be absorbed into a reversal of the orientation of the loop around which the holonomy is taken, and we know that the $T^0$ variables are invariant under all loop reparametrisations.

Third, the Poisson brackets of these variables are also closed. This, however, will require significant work to demonstrate. We shall embark upon this task now. 

Let us first describe a general method of performing variational calculations with path-ordered exponentials \cite{rovelli 2}.

\begin{mdframed}[style=testframe]
    \textbf{Variation of Path-ordered Exponentials \cite{rovelli 2}}

    One way to proceed is to try to vary Eq~(\ref{eq3.5}) by brute force with respect to the connection. However, we will not pursue this strategy here. Instead, we start by substituting Eq~(\ref{eq3.5}) into Eq~(\ref{eq3.2}):
    \begin{equation}
        \frac{\partial U_{\gamma(s)}}{\partial s} + \dot{\gamma}^a(s)A_a(s)U_{\gamma(s)} = 0. \label{eq3.12}
    \end{equation}
    Now we vary this equation with respect to the connection. 
    \begin{equation}
        \left[\frac{\partial}{\partial s} + \dot{\gamma}^a(s)A_a(s)\right]\delta U_{\gamma(s)} = -\dot{\gamma}^a(s)(\delta A_a(s))U_{\gamma(s)}. \label{eq3.13}
    \end{equation}
    This can be solved by the ansatz
    \begin{equation}
        \delta U_{\gamma(s)} = U_{\gamma(s)}F(s). \label{eq3.14}
    \end{equation}
    Substituting this equation into Eq~(\ref{eq3.13}) and making use of Eq~(\ref{eq3.12}), we obtain $U_{\gamma(s)}\partial F(s)/\partial s = -\dot{\gamma}^a(s)(\delta A_a(s))U_{\gamma(s)}$, which upon integration yields
    \begin{equation}
        \delta U_{\gamma(s)}(t_2, t_1) = U_{\gamma(s)}(t_2, t_1)\left[c - \int_{t_1}^{t_2} ds\,U_{\gamma(s)}(t_1, s)\dot{\gamma}^a(s)(\delta A_a(s))U_{\gamma(s)}(s, t_1) \right], \label{eq3.15}
    \end{equation}
    where $c$ is a constant of integration, which we can at once set equal to zero, since $\delta U_{\gamma(s)}(t_1, t_1) = 0$. Let us now stop being sloppy and recall that $A_a(s) = A_a(\gamma(s))$. Recall further that 
    \begin{equation}
        \frac{\delta A_a^{AB}(\gamma(s))}{\delta A_b^{CD}(x)} = \delta^b_a\delta^{(A}_C\delta^{B)}_D\delta^3(\gamma(s),x). \label{eq3.16}
    \end{equation}
    It is now prudent to reintroduce spinor indices in Eq~(\ref{eq3.15}). Then upon a substitution of the preceding equation, we have
    \begin{equation}
        \frac{\delta U_{\gamma(s)}^{AB}(t_2, t_1)}{\delta A_a^{CD}(x)} = \int_{t_1}^{t_2}ds\,\delta^3(\gamma(s), x)\dot{\gamma}^a(s)U_\gamma\tensor{(t_2, s)}{^A_{(C}}U_\gamma \tensor{(s, t_1)}{_{D)}^B}, \label{eq3.17}
    \end{equation}
    which in turn yields 
    \begin{equation}
        \frac{\delta\,\text{tr}\, U_\gamma}{\delta A_a^{CD}(x)} = \oint_{t_1}^{t_2}ds\,\delta^3(\gamma(s), x)\dot{\gamma}^a(s)U_{\gamma (CD)}(t_2, t_1), \label{eq3.18}
    \end{equation}
    where, in view of Eq~(\ref{eq3.7}), we have used $U_\gamma\tensor{(t_2, s)}{^A_{(C}}U_\gamma \tensor{(s, t_1)}{_{D)A}} = U_{\gamma (CD)}(t_2, t_1)$. 
\end{mdframed}

We are now ready to evaluate the Poisson brackets between the $T$ variables. The bracket between the $T^0$ variables is trivial, for they do not depend on the triads. Thus, for two loops $\gamma$ and $\eta$,
\begin{equation}
    \{T^0_\gamma, T^0_\eta\} = 0. \label{eq3.19}
\end{equation}

As for $T^0$ and $T^1$, we first find the bracket between $T^0$ and the $T^a$ variables of the previous section.
\begin{align}
    \{T^a_{\gamma}(s), T^0_{\eta}\} &= \{\text{tr}\,(U_{\gamma(s)}\Tilde{E}^a(s)), \text{tr}\,U_{\eta}\}\nonumber\\
    &= U_{\gamma}(s)^{AB}\{\tensor{\Tilde{E}}{^a_{AB}}(\gamma(s)), \text{tr}\,U_{\eta}\} \nonumber\\
    &= -i\int_\Sigma d^3x\, U_{\gamma}(s)^{AB}\frac{\delta \Tilde{E}^a_{AB}(\gamma(s))}{\delta \Tilde{E}^b_{CD}(x)}\frac{\delta\,\text{tr}\,U_{\eta(t)} }{\delta A_b^{CD}(x)} \nonumber \\
    &= -iU_{\gamma}(s)^{AB}\oint dt\, \delta^3(\gamma(s),\eta(t))\dot{\eta}^a(t)U_{\eta(t)(AB)}. \label{eq3.20} 
\end{align}
In the second line, we used $\{fg, h\} = f\{g,h\} + \{f, h\}g$; in the third, we used the definition of the Poisson bracket, and in the last line, we employed Eq~(\ref{eq3.18}). Let us focus our attention on the term $U_{\gamma}(s)^{AB}U_{\eta(t)(AB)}$. We have
\begin{align}
    2U_{\gamma}(s)^{AB}U_{\eta(t)(AB)} &= U_{\gamma}(s)^{AB}U_{\eta(t)AB} + U_{\gamma}(s)^{AB}U_{\eta(t)BA}\nonumber \\
    &= -U_{\gamma}(s)^{AB}U_{\eta^{-1}(t)BA} + U_{\gamma}(s)^{AB}U_{\eta(t)BA}\nonumber \\
    &= -\text{tr}\,(U_{\gamma}U_{\eta^{-1}}) + \text{tr}\,(U_{\gamma}U_{\eta}), \label{eq3.21} 
\end{align}
where in the second line, we used Eq~(\ref{eq3.9}b). Now, the product of the holonomies inside the traces in Eq~(\ref{eq3.21}) can be interpreted as a single holonomy around a loop formed by composing $\gamma$ and $\eta$. On the other hand, due to the presence of the Dirac delta, Eq~(\ref{eq3.20}) is nonzero only if $\gamma(s) = \eta(t)$, i.e. the loops intersect. Therefore, it seems natural to compose the two loops by breaking them at their points of intersection and rejoining in such a way as to form a single loop. Let us pause here to describe how this can be done. 

Let $\alpha$  and $\beta$ be two loops that intersect at finitely many points. Let $x = \alpha(\bar{s}) = \beta(\bar{t})$ be an intersection point. We construct a new loop, denoted as $\alpha\#_x\beta$\footnote{We may frequently indulge in a slight notational abuse by also writing $\alpha\#_{\bar{s}}\beta$ or $\alpha\#_{\bar{t}}\beta$, and by omitting the subscript altogether when the context is clear.}, by starting from $x$, going first around $\alpha$ and then around $\beta$. We now establish a convention for parametrising such a composed loop. First, we will always assume that the parameters of single loops are defined modulo $2\pi$. Second, note that $\alpha(s+\bar{s})$ and $\beta(t+\bar{t})$ are parametrised loops that start and end at the intersection point $x$. Then it is easy to see that 
\begin{equation}
    \alpha\#_x\beta(u) = \left\{\begin{matrix}
\alpha(2u+\bar{s})\quad \text{for} \quad 0 < u < \pi,\\ \beta(2u+\bar{t})\quad \text{for} \quad \pi < u < 2\pi.
\end{matrix}\right. \label{eq3.22}
\end{equation}

With these definitions, we see that $\text{tr}\,(U_{\gamma}(s)U_{\eta(t)}) = T^0_{\gamma\#_s\eta}$ if $\eta$ intersects $\gamma$ at the latter's starting point. Hence we obtain
\begin{equation}
    \{T^a_{\gamma}(s), T^0_{\eta}\} = -i\Delta^a[\gamma, \eta](s)\left[T^0_{\gamma\#\eta} - T^0_{\gamma\#\eta^{-1}}\right], \label{eq3.23} 
\end{equation}
where we have defined
\begin{equation}
    \Delta^a[\gamma, \eta](s) := \frac{1}{2}\oint dt\,\delta^3(\gamma(s), \eta(t))\dot{\eta}^a(t). \label{eq3.24}
\end{equation}

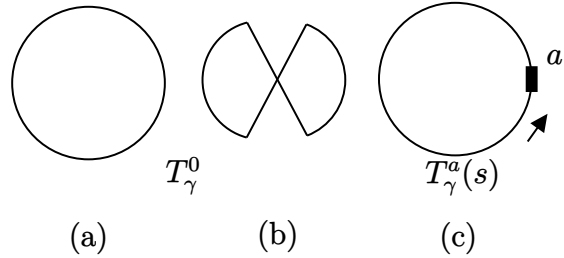
\begin{figure}[ht]
    \centering
    \tikzset{every picture/.style={line width=0.75pt}} 

\tikzset{every picture/.style={line width=0.75pt}} 

\begin{tikzpicture}[x=0.75pt,y=0.75pt,yscale=-1,xscale=1]

\draw   (64,118.5) .. controls (64,97.24) and (81.24,80) .. (102.5,80) .. controls (123.76,80) and (141,97.24) .. (141,118.5) .. controls (141,139.76) and (123.76,157) .. (102.5,157) .. controls (81.24,157) and (64,139.76) .. (64,118.5) -- cycle ;
\draw  [draw opacity=0] (182.27,145.99) .. controls (169.45,142.58) and (160,130.9) .. (160,117) .. controls (160,103.02) and (169.56,91.27) .. (182.5,87.94) -- (190,117) -- cycle ; \draw   (182.27,145.99) .. controls (169.45,142.58) and (160,130.9) .. (160,117) .. controls (160,103.02) and (169.56,91.27) .. (182.5,87.94) ;  
\draw  [draw opacity=0] (212.42,88.86) .. controls (223.85,93.09) and (232,104.1) .. (232,117) .. controls (232,129.77) and (224.02,140.68) .. (212.77,145.01) -- (202,117) -- cycle ; \draw   (212.42,88.86) .. controls (223.85,93.09) and (232,104.1) .. (232,117) .. controls (232,129.77) and (224.02,140.68) .. (212.77,145.01) ;  
\draw    (182.5,87.94) -- (212.77,145.01) ;
\draw    (182.27,145.99) -- (212.42,88.86) ;
\draw   (249,116.5) .. controls (249,95.24) and (266.24,78) .. (287.5,78) .. controls (308.76,78) and (326,95.24) .. (326,116.5) .. controls (326,137.76) and (308.76,155) .. (287.5,155) .. controls (266.24,155) and (249,137.76) .. (249,116.5) -- cycle ;
\draw  [color={rgb, 255:red, 0; green, 0; blue, 0 }  ,draw opacity=1 ][fill={rgb, 255:red, 0; green, 0; blue, 0 }  ,fill opacity=1 ] (323.5,110.5) -- (328.5,110.5) -- (328.5,122.5) -- (323.5,122.5) -- cycle ;
\draw    (324,148) -- (332.26,136.44) ;
\draw [shift={(334,134)}, rotate = 125.54] [fill={rgb, 255:red, 0; green, 0; blue, 0 }  ][line width=0.08]  [draw opacity=0] (8.93,-4.29) -- (0,0) -- (8.93,4.29) -- cycle    ;

\draw (91,187) node [anchor=north west][inner sep=0.75pt]   [align=left] {(a)};
\draw (186,186) node [anchor=north west][inner sep=0.75pt]   [align=left] {(b)};
\draw (140,154.4) node [anchor=north west][inner sep=0.75pt]    {$T_{\gamma }^{0}$};
\draw (271,155.4) node [anchor=north west][inner sep=0.75pt]    {$T_{\gamma }^{a}( s)$};
\draw (332,100.4) node [anchor=north west][inner sep=0.75pt]    {$a$};
\draw (278,187) node [anchor=north west][inner sep=0.75pt]   [align=left] {(c)};

\end{tikzpicture}

    \caption{(a) and (b) denote $T^0_\gamma$; (c) denotes $T^a_{\gamma}(s)$}.
    \label{fig3.1}
\end{figure}

Using our conventions for composing loops, Eq~(\ref{eq3.23}) lends itself to a very useful graphical interpretation, which we will find extremely convenient for subsequent calculations. Recall that the $T^0$ variables depend on unoriented, unparametrised loops, while the $T^a$ variables depend on oriented, unparametrised loops. Thus we shall pictorially identify the former with arbitrary closed curves, and the latter with loops that contain (1) an arrow indicating their orientation and (2) a marker signifying the location of the loop's starting point and hence the point where we have inserted a triad (see Fig~\ref{fig3.1}). We call the marker signifying the insertion of a triad a \textit{hand}.    

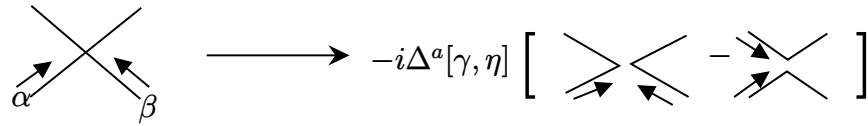
\begin{figure}[ht]
    \centering
    
\tikzset{every picture/.style={line width=0.75pt}} 

\begin{tikzpicture}[x=0.75pt,y=0.75pt,yscale=-1,xscale=1]

\draw    (81,117) -- (137,72) ;
\draw    (81.5,71) -- (136.5,118) ;

\draw    (74,112) -- (89.57,100.76) ;
\draw [shift={(92,99)}, rotate = 144.16] [fill={rgb, 255:red, 0; green, 0; blue, 0 }  ][line width=0.08]  [draw opacity=0] (8.93,-4.29) -- (0,0) -- (8.93,4.29) -- cycle    ;
\draw    (141,112) -- (125.3,98.92) ;
\draw [shift={(123,97)}, rotate = 39.81] [fill={rgb, 255:red, 0; green, 0; blue, 0 }  ][line width=0.08]  [draw opacity=0] (8.93,-4.29) -- (0,0) -- (8.93,4.29) -- cycle    ;
\draw    (170,96) -- (239,96) ;
\draw [shift={(242,96)}, rotate = 180] [fill={rgb, 255:red, 0; green, 0; blue, 0 }  ][line width=0.08]  [draw opacity=0] (10.72,-5.15) -- (0,0) -- (10.72,5.15) -- (7.12,0) -- cycle    ;
\draw   (350,85) -- (378.03,101.27) -- (351,115) ;
\draw   (413,117) -- (384.04,100.35) -- (412,86) ;
\draw   (443.4,117.33) -- (462.62,104.11) -- (483,118) ;
\draw   (483,85) -- (463,98.2) -- (443,84) ;
\draw    (355,118) -- (372.21,111.11) ;
\draw [shift={(375,110)}, rotate = 158.2] [fill={rgb, 255:red, 0; green, 0; blue, 0 }  ][line width=0.08]  [draw opacity=0] (8.93,-4.29) -- (0,0) -- (8.93,4.29) -- cycle    ;
\draw    (405,121) -- (389.62,112.46) ;
\draw [shift={(387,111)}, rotate = 29.05] [fill={rgb, 255:red, 0; green, 0; blue, 0 }  ][line width=0.08]  [draw opacity=0] (8.93,-4.29) -- (0,0) -- (8.93,4.29) -- cycle    ;
\draw    (436,117) -- (451.57,105.76) ;
\draw [shift={(454,104)}, rotate = 144.16] [fill={rgb, 255:red, 0; green, 0; blue, 0 }  ][line width=0.08]  [draw opacity=0] (8.93,-4.29) -- (0,0) -- (8.93,4.29) -- cycle    ;
\draw    (437,87) -- (451.48,96.37) ;
\draw [shift={(454,98)}, rotate = 212.91] [fill={rgb, 255:red, 0; green, 0; blue, 0 }  ][line width=0.08]  [draw opacity=0] (8.93,-4.29) -- (0,0) -- (8.93,4.29) -- cycle    ;

\draw (70,113.4) node [anchor=north west][inner sep=0.75pt]    {$\alpha $};
\draw (134,114.4) node [anchor=north west][inner sep=0.75pt]    {$\beta $};
\draw (251,89.4) node [anchor=north west][inner sep=0.75pt]    {$-i\Delta ^{a}[ \gamma ,\eta ]$};
\draw (326,83) node [anchor=north west][inner sep=0.75pt]   [align=left] {{\Huge [}};
\draw (496,83) node [anchor=north west][inner sep=0.75pt]   [align=left] {{\Huge ]}};
\draw (421,89.4) node [anchor=north west][inner sep=0.75pt]    {$-$};

\end{tikzpicture}
    \caption{Action of the grasp operator.}
    \label{fig3.2}
\end{figure}

Now, Eq~(\ref{eq3.23}) is nonzero only if $\eta$ and $\gamma$ intersect at $s$, i.e. the triad is inserted at the intersection point of the two loops or in other words, the loop $\gamma$ starts right where it intersects $\eta$. If that is so, we say that the loop $\gamma$ \textit{sees} the loop $\eta$, and describe the contents of Eq~(\ref{eq3.23}) as follows. Fix an orientation on the loop $\eta$ ($\gamma$ has already a fixed orientation, since it refers to $T^a_{\gamma}(s)$). We break the loops at the location of the hand, and rejoin each resulting leg of one of the loops with that of the other. There are two ways of doing this, one in which the orientations of the rejoined legs match naturally, and the other in which they clash. In the first case, we obtain the loop $\gamma\#\eta$, and in the second case, we have to flip the orientation on $\eta$, thus obtaining the loop $\gamma\#\eta^{-1}$. Evidently, the first of these loops refers to the variable $T^0_{\gamma\#\eta}$, and the second one refers to $T^0_{\gamma\#\eta^{-1}}$. The Poisson bracket in Eq~(\ref{eq3.24}) is given by the difference between the first loop and the second loop, multiplied by $-i\Delta^a[\gamma, \eta](s)$. The whole process described here is called the \textit{grasp} operation -- we say that the hand on $\gamma$ grasps the loop $\eta$ (Fig~\ref{fig3.2}).

\begin{figure}[ht]
    \centering

\tikzset{every picture/.style={line width=0.75pt}} 

\tikzset{every picture/.style={line width=0.75pt}} 

\begin{tikzpicture}[x=0.75pt,y=0.75pt,yscale=-1,xscale=1]

\draw   (65.71,145.58) .. controls (54.36,130.78) and (55.94,108.53) .. (69.25,95.9) .. controls (82.56,83.27) and (102.55,85.03) .. (113.9,99.84) .. controls (125.25,114.65) and (123.67,136.89) .. (110.36,149.52) .. controls (97.06,162.15) and (77.06,160.39) .. (65.71,145.58) -- cycle ;
\draw  [color={rgb, 255:red, 0; green, 0; blue, 0 }  ,draw opacity=1 ][fill={rgb, 255:red, 0; green, 0; blue, 0 }  ,fill opacity=1 ] (111.84,94.34) -- (115.96,94.34) -- (115.96,105.33) -- (111.84,105.33) -- cycle ;

\draw   (113.9,99.84) .. controls (125.17,84.95) and (145.15,83.05) .. (158.53,95.59) .. controls (171.9,108.14) and (173.61,130.37) .. (162.34,145.26) .. controls (151.07,160.14) and (131.09,162.04) .. (117.71,149.5) .. controls (104.34,136.96) and (102.63,114.72) .. (113.9,99.84) -- cycle ;
\draw    (68,87.46) -- (54.5,100.91) ;
\draw [shift={(52.37,103.03)}, rotate = 315.12] [fill={rgb, 255:red, 0; green, 0; blue, 0 }  ][line width=0.08]  [draw opacity=0] (8.93,-4.29) -- (0,0) -- (8.93,4.29) -- cycle    ;

\draw    (177.35,122.51) -- (210.55,122.51) ;
\draw [shift={(213.55,122.51)}, rotate = 180] [fill={rgb, 255:red, 0; green, 0; blue, 0 }  ][line width=0.08]  [draw opacity=0] (8.93,-4.29) -- (0,0) -- (8.93,4.29) -- cycle    ;
\draw  [draw opacity=0] (299.99,169.1) .. controls (297.08,157.51) and (295.3,142.02) .. (295.3,125) .. controls (295.3,108.12) and (297.05,92.74) .. (299.92,81.18) -- (312.99,125) -- cycle ; \draw   (299.99,169.1) .. controls (297.08,157.51) and (295.3,142.02) .. (295.3,125) .. controls (295.3,108.12) and (297.05,92.74) .. (299.92,81.18) ;  
\draw  [draw opacity=0] (347.67,155.39) .. controls (344,157.05) and (339.99,157.96) .. (335.78,157.96) .. controls (318.29,157.96) and (304.11,142.18) .. (304.11,122.71) .. controls (304.11,103.25) and (318.29,87.46) .. (335.78,87.46) .. controls (339.99,87.46) and (344,88.38) .. (347.67,90.03) -- (335.78,122.71) -- cycle ; \draw   (347.67,155.39) .. controls (344,157.05) and (339.99,157.96) .. (335.78,157.96) .. controls (318.29,157.96) and (304.11,142.18) .. (304.11,122.71) .. controls (304.11,103.25) and (318.29,87.46) .. (335.78,87.46) .. controls (339.99,87.46) and (344,88.38) .. (347.67,90.03) ;  
\draw  [draw opacity=0] (356.79,90.03) .. controls (360.47,88.38) and (364.48,87.46) .. (368.69,87.46) .. controls (386.18,87.46) and (400.36,103.25) .. (400.36,122.71) .. controls (400.36,142.18) and (386.18,157.96) .. (368.69,157.96) .. controls (364.48,157.96) and (360.47,157.05) .. (356.79,155.39) -- (368.69,122.71) -- cycle ; \draw   (356.79,90.03) .. controls (360.47,88.38) and (364.48,87.46) .. (368.69,87.46) .. controls (386.18,87.46) and (400.36,103.25) .. (400.36,122.71) .. controls (400.36,142.18) and (386.18,157.96) .. (368.69,157.96) .. controls (364.48,157.96) and (360.47,157.05) .. (356.79,155.39) ;  
\draw    (346.62,89.58) .. controls (332.08,115.85) and (343.6,148.8) .. (357.85,155.84) ;
\draw    (357.85,89.58) .. controls (368.28,97.54) and (367.45,149.72) .. (346.62,155.84) ;

\draw  [draw opacity=0] (481.77,115.64) .. controls (482.1,117.69) and (482.28,119.81) .. (482.28,121.97) .. controls (482.28,141.43) and (468.1,157.21) .. (450.61,157.21) .. controls (433.11,157.21) and (418.93,141.43) .. (418.93,121.97) .. controls (418.93,102.5) and (433.11,86.72) .. (450.61,86.72) .. controls (455.67,86.72) and (460.45,88.04) .. (464.69,90.39) -- (450.61,121.97) -- cycle ; \draw   (481.77,115.64) .. controls (482.1,117.69) and (482.28,119.81) .. (482.28,121.97) .. controls (482.28,141.43) and (468.1,157.21) .. (450.61,157.21) .. controls (433.11,157.21) and (418.93,141.43) .. (418.93,121.97) .. controls (418.93,102.5) and (433.11,86.72) .. (450.61,86.72) .. controls (455.67,86.72) and (460.45,88.04) .. (464.69,90.39) ;  
\draw  [draw opacity=0] (462.29,90.47) .. controls (466.76,87.78) and (471.89,86.25) .. (477.33,86.25) .. controls (494.82,86.25) and (509,102.03) .. (509,121.5) .. controls (509,140.96) and (494.82,156.74) .. (477.33,156.74) .. controls (459.84,156.74) and (445.66,140.96) .. (445.66,121.5) .. controls (445.66,119.01) and (445.89,116.59) .. (446.33,114.26) -- (477.33,121.5) -- cycle ; \draw   (462.29,90.47) .. controls (466.76,87.78) and (471.89,86.25) .. (477.33,86.25) .. controls (494.82,86.25) and (509,102.03) .. (509,121.5) .. controls (509,140.96) and (494.82,156.74) .. (477.33,156.74) .. controls (459.84,156.74) and (445.66,140.96) .. (445.66,121.5) .. controls (445.66,119.01) and (445.89,116.59) .. (446.33,114.26) ;  
\draw    (446.33,114.26) .. controls (452.86,91.9) and (484.39,101.88) .. (481.77,115.64) ;
\draw  [draw opacity=0] (513.31,80.9) .. controls (516.22,92.49) and (518,107.98) .. (518,125) .. controls (518,141.88) and (516.25,157.26) .. (513.38,168.82) -- (500.31,125) -- cycle ; \draw   (513.31,80.9) .. controls (516.22,92.49) and (518,107.98) .. (518,125) .. controls (518,141.88) and (516.25,157.26) .. (513.38,168.82) ;  

\draw (14.1,111.91) node [anchor=north west][inner sep=0.75pt]    {$\{\ ,\ \} :$};
\draw (216.28,113.16) node [anchor=north west][inner sep=0.75pt]    {$-i\Delta ^{a}[ \gamma ,\eta ]$};
\draw (63.83,158.19) node [anchor=north west][inner sep=0.75pt]    {$\gamma $};
\draw (151.38,157.27) node [anchor=north west][inner sep=0.75pt]    {$\eta $};
\draw (403.41,117.4) node [anchor=north west][inner sep=0.75pt]    {$-$};
\draw (335.27,164.09) node [anchor=north west][inner sep=0.75pt]    {$\gamma \#\eta $};
\draw (445.73,160.34) node [anchor=north west][inner sep=0.75pt]    {$\gamma \#\eta ^{-1}$};

\end{tikzpicture}
    \caption{Poisson bracket of $T^0$ and $T^a$.}
    \label{fig3.3}
\end{figure}
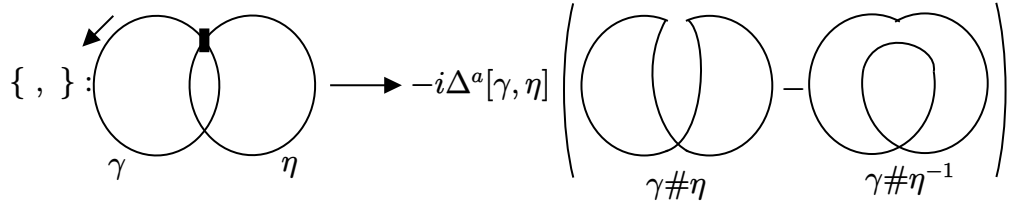

Note that the final result is oblivious to the orientation which we give to $\eta$, since a reversal of the orientation also reverses the direction of the tangent vector $\dot{\eta}^a$, and hence the sign of $\Delta^a$, and this caters for the reversal in the order in which we subtract the composed loops $\gamma\#\eta$ and $\gamma\#\eta^{-1}$. Thus the graphical calculation described in the preceding paragraph can be represented as in Fig~(\ref{fig3.3}).  

In accordance with Fig~(\ref{fig3.2}), we introduce the notation
\begin{equation}
    (\gamma\#\eta)^{><} := \gamma\#\eta, \quad (\gamma\#\eta)^{\gtlt} := \gamma\#\eta^{-1}. \label{eq3.25} 
\end{equation}
Furthermore, we define $|><| = 0$ and $|\gtlt| := 1$, and let the symbol $\lozenge$ take the value $><$ or $\gtlt$. Then Eq~(\ref{eq3.23}) may be concisely written as
\begin{equation}
    \{T^a_{\gamma}(s), T_\eta\} = -i\sum_{\lozenge} (-1)^{|\lozenge|}\Delta^a[\gamma, \eta](s)T^0_{(\gamma\#\eta)^{\lozenge}}. \label{eq3.26}
\end{equation}

With all the clutter out of the way, we are now ready to evaluate the Poisson bracket between $T^0$ and $T^1_S$. To this end, we first observe that with $T^1_S$, we essentially integrate $T^a_{\gamma_s}(t)$ over the surface $S$ formed by a congruence of loops $\gamma_s: [s_1, s_2]\times S^1:\to\Sigma$ (see Eq~(\ref{eq3.10}a)). This surface integral, combined with the line integral in Eq~(\ref{eq3.24}), should kill the delta function. Let us see how this precisely comes about. We begin with
\begin{align}
    \{T^1_S, T^0_{\eta}\} &= -\frac{i}{2}\int_S\oint dudx^adx^b\epsilon_{abc}\delta^3(\gamma_{s}(t),\eta(u))\dot{\eta}^c(u)\sum_{\lozenge}(-1)^{|\lozenge|}T^0_{(\gamma_{s}\#_t\eta)^{\lozenge}} \label{eq3.27}
\end{align}
where $x^a = (s,t) \in [s_1,s_2]\times S^1$. First, the integral over $u$ gives a nonzero contribution only when $\eta$ intersects a curve $\gamma_s$ at its starting point, namely $t$. But as the integrals over $s$ and $t$ probe the surface $S$, every point of each loop becomes its starting point, and we are thus led to consider all possible intersections of $\eta$ with $S$; we denote the composition of the loop $\eta$ with a loop at the $i$th intersection point by $S\circ_i\eta$. Second, notice that the object $dx^adx^b\epsilon_c$ is tangential to the surface $S$, and it is contracted with $\dot{\eta}^c$, the tangent vector of the curve $\eta$. Therefore, if the curve intersects the surface at right angles, the Poisson bracket must be zero, and for non-orthogonal transverse intersections, we should get $\pm 1$, depending on which side of the surface $\dot{\eta}^c$ points to. This fact is captured by the function $\text{sgn}_i(S,\eta)$, which takes values $0, \pm 1$ at each intersection point $i$. Since the delta function is zero except at the intersections, we finally have
\begin{equation}
    \{T^1_S, T^0_{\eta}\} = -i\sum_i\,\text{sgn}_i(S,\eta)\left[T^0_{S\circ_i\eta} - T^0_{S\circ_i\eta^{-1}}\right]. \label{eq3.28}
\end{equation}

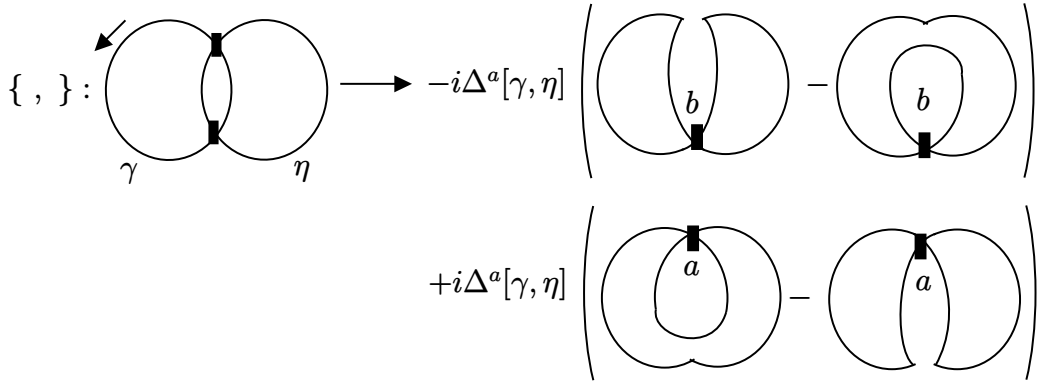
\begin{figure}[ht]
    \centering

\tikzset{every picture/.style={line width=0.75pt}} 

\begin{tikzpicture}[x=0.75pt,y=0.75pt,yscale=-1,xscale=1]

\draw   (70.71,143.58) .. controls (59.36,128.78) and (60.94,106.53) .. (74.25,93.9) .. controls (87.56,81.27) and (107.55,83.03) .. (118.9,97.84) .. controls (130.25,112.65) and (128.67,134.89) .. (115.36,147.52) .. controls (102.06,160.15) and (82.06,158.39) .. (70.71,143.58) -- cycle ;
\draw   (118.9,97.84) .. controls (130.17,82.95) and (150.15,81.05) .. (163.53,93.59) .. controls (176.9,106.14) and (178.61,128.37) .. (167.34,143.26) .. controls (156.07,158.14) and (136.09,160.04) .. (122.71,147.5) .. controls (109.34,134.96) and (107.63,112.72) .. (118.9,97.84) -- cycle ;

\draw  [color={rgb, 255:red, 0; green, 0; blue, 0 }  ,draw opacity=1 ][fill={rgb, 255:red, 0; green, 0; blue, 0 }  ,fill opacity=1 ] (115.36,136.54) -- (119.48,136.54) -- (119.48,147.52) -- (115.36,147.52) -- cycle ;

\draw  [color={rgb, 255:red, 0; green, 0; blue, 0 }  ,draw opacity=1 ][fill={rgb, 255:red, 0; green, 0; blue, 0 }  ,fill opacity=1 ] (116.84,92.34) -- (120.96,92.34) -- (120.96,103.33) -- (116.84,103.33) -- cycle ;
\draw    (73,85.46) -- (59.5,98.91) ;
\draw [shift={(57.37,101.03)}, rotate = 315.12] [fill={rgb, 255:red, 0; green, 0; blue, 0 }  ][line width=0.08]  [draw opacity=0] (8.93,-4.29) -- (0,0) -- (8.93,4.29) -- cycle    ;

\draw  [draw opacity=0] (354.67,150.39) .. controls (351,152.05) and (346.99,152.96) .. (342.78,152.96) .. controls (325.29,152.96) and (311.11,137.18) .. (311.11,117.71) .. controls (311.11,98.25) and (325.29,82.46) .. (342.78,82.46) .. controls (346.99,82.46) and (351,83.38) .. (354.67,85.03) -- (342.78,117.71) -- cycle ; \draw   (354.67,150.39) .. controls (351,152.05) and (346.99,152.96) .. (342.78,152.96) .. controls (325.29,152.96) and (311.11,137.18) .. (311.11,117.71) .. controls (311.11,98.25) and (325.29,82.46) .. (342.78,82.46) .. controls (346.99,82.46) and (351,83.38) .. (354.67,85.03) ;  
\draw  [draw opacity=0] (363.79,85.03) .. controls (367.47,83.38) and (371.48,82.46) .. (375.69,82.46) .. controls (393.18,82.46) and (407.36,98.25) .. (407.36,117.71) .. controls (407.36,137.18) and (393.18,152.96) .. (375.69,152.96) .. controls (371.48,152.96) and (367.47,152.05) .. (363.79,150.39) -- (375.69,117.71) -- cycle ; \draw   (363.79,85.03) .. controls (367.47,83.38) and (371.48,82.46) .. (375.69,82.46) .. controls (393.18,82.46) and (407.36,98.25) .. (407.36,117.71) .. controls (407.36,137.18) and (393.18,152.96) .. (375.69,152.96) .. controls (371.48,152.96) and (367.47,152.05) .. (363.79,150.39) ;  
\draw    (353.62,84.58) .. controls (339.08,110.85) and (350.6,143.8) .. (364.85,150.84) ;
\draw    (364.85,84.58) .. controls (375.28,92.54) and (374.45,144.72) .. (353.62,150.84) ;

\draw  [color={rgb, 255:red, 0; green, 0; blue, 0 }  ,draw opacity=1 ][fill={rgb, 255:red, 0; green, 0; blue, 0 }  ,fill opacity=1 ] (358.79,138.39) -- (363.79,138.39) -- (363.79,150.39) -- (358.79,150.39) -- cycle ;
\draw  [draw opacity=0] (494.77,112.64) .. controls (495.1,114.69) and (495.28,116.81) .. (495.28,118.97) .. controls (495.28,138.43) and (481.1,154.21) .. (463.61,154.21) .. controls (446.11,154.21) and (431.93,138.43) .. (431.93,118.97) .. controls (431.93,99.5) and (446.11,83.72) .. (463.61,83.72) .. controls (468.67,83.72) and (473.45,85.04) .. (477.69,87.39) -- (463.61,118.97) -- cycle ; \draw   (494.77,112.64) .. controls (495.1,114.69) and (495.28,116.81) .. (495.28,118.97) .. controls (495.28,138.43) and (481.1,154.21) .. (463.61,154.21) .. controls (446.11,154.21) and (431.93,138.43) .. (431.93,118.97) .. controls (431.93,99.5) and (446.11,83.72) .. (463.61,83.72) .. controls (468.67,83.72) and (473.45,85.04) .. (477.69,87.39) ;  
\draw  [draw opacity=0] (475.29,87.47) .. controls (479.76,84.78) and (484.89,83.25) .. (490.33,83.25) .. controls (507.82,83.25) and (522,99.03) .. (522,118.5) .. controls (522,137.96) and (507.82,153.74) .. (490.33,153.74) .. controls (472.84,153.74) and (458.66,137.96) .. (458.66,118.5) .. controls (458.66,116.01) and (458.89,113.59) .. (459.33,111.26) -- (490.33,118.5) -- cycle ; \draw   (475.29,87.47) .. controls (479.76,84.78) and (484.89,83.25) .. (490.33,83.25) .. controls (507.82,83.25) and (522,99.03) .. (522,118.5) .. controls (522,137.96) and (507.82,153.74) .. (490.33,153.74) .. controls (472.84,153.74) and (458.66,137.96) .. (458.66,118.5) .. controls (458.66,116.01) and (458.89,113.59) .. (459.33,111.26) ;  
\draw    (459.33,111.26) .. controls (465.86,88.9) and (497.39,98.88) .. (494.77,112.64) ;
\draw  [color={rgb, 255:red, 0; green, 0; blue, 0 }  ,draw opacity=1 ][fill={rgb, 255:red, 0; green, 0; blue, 0 }  ,fill opacity=1 ] (473.5,142.5) -- (478.5,142.5) -- (478.5,154.5) -- (473.5,154.5) -- cycle ;
\draw  [draw opacity=0] (307.99,165.1) .. controls (305.08,153.51) and (303.3,138.02) .. (303.3,121) .. controls (303.3,104.12) and (305.05,88.74) .. (307.92,77.18) -- (320.99,121) -- cycle ; \draw   (307.99,165.1) .. controls (305.08,153.51) and (303.3,138.02) .. (303.3,121) .. controls (303.3,104.12) and (305.05,88.74) .. (307.92,77.18) ;  
\draw  [draw opacity=0] (526.31,76.9) .. controls (529.22,88.49) and (531,103.98) .. (531,121) .. controls (531,137.88) and (529.25,153.26) .. (526.38,164.82) -- (513.31,121) -- cycle ; \draw   (526.31,76.9) .. controls (529.22,88.49) and (531,103.98) .. (531,121) .. controls (531,137.88) and (529.25,153.26) .. (526.38,164.82) ;  
\draw    (181.35,117.51) -- (214.55,117.51) ;
\draw [shift={(217.55,117.51)}, rotate = 180] [fill={rgb, 255:red, 0; green, 0; blue, 0 }  ][line width=0.08]  [draw opacity=0] (8.93,-4.29) -- (0,0) -- (8.93,4.29) -- cycle    ;
\draw  [draw opacity=0] (480.63,193.61) .. controls (484.3,191.95) and (488.31,191.04) .. (492.52,191.04) .. controls (510.01,191.04) and (524.19,206.82) .. (524.19,226.29) .. controls (524.19,245.75) and (510.01,261.54) .. (492.52,261.54) .. controls (488.31,261.54) and (484.3,260.62) .. (480.63,258.97) -- (492.52,226.29) -- cycle ; \draw   (480.63,193.61) .. controls (484.3,191.95) and (488.31,191.04) .. (492.52,191.04) .. controls (510.01,191.04) and (524.19,206.82) .. (524.19,226.29) .. controls (524.19,245.75) and (510.01,261.54) .. (492.52,261.54) .. controls (488.31,261.54) and (484.3,260.62) .. (480.63,258.97) ;  
\draw  [draw opacity=0] (471.51,258.97) .. controls (467.83,260.62) and (463.82,261.54) .. (459.61,261.54) .. controls (442.12,261.54) and (427.94,245.75) .. (427.94,226.29) .. controls (427.94,206.82) and (442.12,191.04) .. (459.61,191.04) .. controls (463.82,191.04) and (467.83,191.95) .. (471.51,193.61) -- (459.61,226.29) -- cycle ; \draw   (471.51,258.97) .. controls (467.83,260.62) and (463.82,261.54) .. (459.61,261.54) .. controls (442.12,261.54) and (427.94,245.75) .. (427.94,226.29) .. controls (427.94,206.82) and (442.12,191.04) .. (459.61,191.04) .. controls (463.82,191.04) and (467.83,191.95) .. (471.51,193.61) ;  
\draw    (481.68,259.42) .. controls (496.22,233.15) and (484.7,200.2) .. (470.45,193.16) ;
\draw    (470.45,259.42) .. controls (460.02,251.46) and (460.85,199.28) .. (481.68,193.16) ;

\draw  [draw opacity=0] (340.53,231.36) .. controls (340.2,229.31) and (340.02,227.19) .. (340.02,225.03) .. controls (340.02,205.57) and (354.2,189.79) .. (371.69,189.79) .. controls (389.19,189.79) and (403.37,205.57) .. (403.37,225.03) .. controls (403.37,244.5) and (389.19,260.28) .. (371.69,260.28) .. controls (366.63,260.28) and (361.85,258.96) .. (357.61,256.61) -- (371.69,225.03) -- cycle ; \draw   (340.53,231.36) .. controls (340.2,229.31) and (340.02,227.19) .. (340.02,225.03) .. controls (340.02,205.57) and (354.2,189.79) .. (371.69,189.79) .. controls (389.19,189.79) and (403.37,205.57) .. (403.37,225.03) .. controls (403.37,244.5) and (389.19,260.28) .. (371.69,260.28) .. controls (366.63,260.28) and (361.85,258.96) .. (357.61,256.61) ;  
\draw  [draw opacity=0] (360.01,256.53) .. controls (355.54,259.22) and (350.41,260.75) .. (344.97,260.75) .. controls (327.48,260.75) and (313.3,244.97) .. (313.3,225.5) .. controls (313.3,206.04) and (327.48,190.26) .. (344.97,190.26) .. controls (362.46,190.26) and (376.64,206.04) .. (376.64,225.5) .. controls (376.64,227.99) and (376.41,230.41) .. (375.98,232.74) -- (344.97,225.5) -- cycle ; \draw   (360.01,256.53) .. controls (355.54,259.22) and (350.41,260.75) .. (344.97,260.75) .. controls (327.48,260.75) and (313.3,244.97) .. (313.3,225.5) .. controls (313.3,206.04) and (327.48,190.26) .. (344.97,190.26) .. controls (362.46,190.26) and (376.64,206.04) .. (376.64,225.5) .. controls (376.64,227.99) and (376.41,230.41) .. (375.98,232.74) ;  
\draw    (375.98,232.74) .. controls (369.44,255.1) and (337.91,245.12) .. (340.53,231.36) ;
\draw  [color={rgb, 255:red, 0; green, 0; blue, 0 }  ,draw opacity=1 ][fill={rgb, 255:red, 0; green, 0; blue, 0 }  ,fill opacity=1 ] (361.8,201.5) -- (356.8,201.5) -- (356.8,189.5) -- (361.8,189.5) -- cycle ;
\draw  [draw opacity=0] (527.31,178.9) .. controls (530.22,190.49) and (532,205.98) .. (532,223) .. controls (532,239.88) and (530.25,255.26) .. (527.38,266.82) -- (514.31,223) -- cycle ; \draw   (527.31,178.9) .. controls (530.22,190.49) and (532,205.98) .. (532,223) .. controls (532,239.88) and (530.25,255.26) .. (527.38,266.82) ;  
\draw  [draw opacity=0] (308.99,267.1) .. controls (306.08,255.51) and (304.3,240.02) .. (304.3,223) .. controls (304.3,206.12) and (306.05,190.74) .. (308.92,179.18) -- (321.99,223) -- cycle ; \draw   (308.99,267.1) .. controls (306.08,255.51) and (304.3,240.02) .. (304.3,223) .. controls (304.3,206.12) and (306.05,190.74) .. (308.92,179.18) ;  

\draw  [color={rgb, 255:red, 0; green, 0; blue, 0 }  ,draw opacity=1 ][fill={rgb, 255:red, 0; green, 0; blue, 0 }  ,fill opacity=1 ] (476.51,205.61) -- (471.51,205.61) -- (471.51,193.61) -- (476.51,193.61) -- cycle ;

\draw (12,110.91) node [anchor=north west][inner sep=0.75pt]    {$\{\ ,\ \} :$};
\draw (68.83,156.19) node [anchor=north west][inner sep=0.75pt]    {$\gamma $};
\draw (156.38,155.27) node [anchor=north west][inner sep=0.75pt]    {$\eta $};
\draw (223.28,108.16) node [anchor=north west][inner sep=0.75pt]    {$-i\Delta ^{a}[ \gamma ,\eta ]$};
\draw (420.89,232.6) node [anchor=north west][inner sep=0.75pt]  [rotate=-180]  {$-$};
\draw (224.28,209.16) node [anchor=north west][inner sep=0.75pt]    {$+i\Delta ^{a}[ \gamma ,\eta ]$};
\draw (470,213.4) node [anchor=north west][inner sep=0.75pt]    {$a$};
\draw (354,119.4) node [anchor=north west][inner sep=0.75pt]    {$b$};
\draw (353,205.4) node [anchor=north west][inner sep=0.75pt]    {$a$};
\draw (470.51,118.01) node [anchor=north west][inner sep=0.75pt]    {$b$};
\draw (414.41,111.4) node [anchor=north west][inner sep=0.75pt]    {$-$};

\end{tikzpicture}
    \caption{Poisson bracket between two $T^a$'s.}
    \label{fig3.4}
\end{figure}

Finally, we now turn attention towards the Poisson bracket between two strip functionals. Again, as an intermediate step, we first find the bracket between $T^a_{\gamma}(s)$ and $T^b_{\eta}(s)$. Let us first see what we expect from the graphical calculus we have developed. First, we now have two handed loops. In analogy with the bracket of $T^0$ and $T^a$, we expect the bracket between two handed loops to be nonzero only when the hands lie at intersection points of the two loops. Second, we now have to take into account two grasps, one that of the hand of $\gamma$ over $\eta$ and the other that of the hand of $\eta$ over $\gamma$. Each grasp should result in a loop with one hand, i.e. a $T^a$ variable, multiplied by $\Delta^a$. Furthermore, given a fixed orientation on the two loops, it is not difficult to convince oneself that the two grasps should yield results that differ by a sign. That is, if the orientation of a leg of $\gamma$ matches naturally with the orientation of a leg of $\eta$ at the location of one hand, then these orientations clash at the location of the other hand. Thus we expect to have the situation shown in Fig~(\ref{fig3.4}).

Our expectations are indeed borne out, as we now verify \cite{rovelli 3}. 
\begin{align}
    i\{T^a_{\gamma}(s), T^b_{\eta}(t)\} &= U_{\gamma}(s)^{AB}\{\Tilde{E}^a_{AB}(\gamma(s)),U_{\eta}(t)^{CD}\}\Tilde{E}^b_{CD}(\eta(t))\nonumber \\
    &\quad + U_{\eta}(t)^{CD}\{U_{\gamma}(s)^{AB},\Tilde{E}^b_{CD}(\eta(t))\}\Tilde{E}^a_{AB}(\gamma(s)) \nonumber\\
    &= \Delta^a[\gamma, \eta](s)\left( \text{tr}\,\left[U_\gamma(s)U_\eta(t,u)\Tilde{E}^b(\eta(t))U_{\eta}(u,t)\right]\right. \nonumber \\
    &\qquad\qquad\qquad - \left.\text{tr}\,\left[U_{\gamma^{-1}}(t)U_\eta(t,u)\Tilde{E}^b(\eta(t))U_\eta(u,t)\right]  \right) \nonumber \\
    &\quad - \Delta^b[\eta,\gamma](t)\left(\text{tr}\,\left[ U_\eta(t)U_\gamma(s,v)\Tilde{E}^a(\gamma(s))U_\gamma(u,s) \right] \right. \nonumber \\
    &\qquad\qquad\qquad - \left.\text{tr}\,\left[ U_{\eta^{-1}}(t)U_\gamma(s,v)\Tilde{E}^a(\gamma(s))U_{\gamma}(u,s) \right] \right) \nonumber \\
    &= -\sum_{\lozenge}(-1)^{|\lozenge|}\Delta^b[\eta,\gamma](t)T^a_{(\eta\#_t\gamma)^\lozenge}((s-t)/2) \nonumber \\
    &\quad +\sum_{\lozenge}(-1)^{|\lozenge|}\Delta^a[\gamma,\eta](s)T^b_{(\gamma\#_s\eta)^\lozenge}((t-s)/2). \label{eq3.29}
\end{align}

Eq~(\ref{eq3.29}) can now be substituted into $\{T^1_S, T^1_P\}$ to find the Poisson bracket between two strip functionals. However, we will not need it in subsequent sections, so we omit the expression. What is important to note is that the expression is nonzero only at the intersection points of the two surfaces in the definition of the two strip functionals come. This fact, together with Eqs~(\ref{eq3.28}) and (\ref{eq3.19}), implies that the algebra of the loop and strip functionals is closed. Hence, we have successfully constructed the set of elementary variables to be quantised. 

\section{Interlude: Constructive Quantum Field Theory}
We have successfully completed step 1 of the quantisation scheme outlined in Section 3.1. The next step is to promote the loop and strip functionals to operators by constructing a free associative and involutive algebra $\mathcal{B}^{(\star)}$ from the Poisson brackets studied above (see steps 2 and 3 on p. 40 above). While straightforward in principle, this task is complicated by the fact that we eventually seek a representation of $\mathcal{B}^{(\star)}$ on a suitable Hilbert space. Now, in analogy with standard quantum mechanics, the intuitive expectation is that the required Hilbert space be the $L_2$-normed space of complex-valued functions on the classical configuration space, which in this case is the space $\mathcal{A/G}$ of smooth connections modulo gauge transformations. However, since the classical configuration spaces of field theories are infinite-dimensional, the existence of an $L_2$ norm on the space of functions on these configuration spaces is far from guaranteed; indeed, it may not even be possible. The essential difficulty lies in the impossibility of constructing measures on certain infinite-dimensional spaces, and one needs a suitable integration theory to define inner products and thence Hilbert spaces. In this section, we will explore this problem in significant detail. To make the discussion tractable, we shall focus on a similar problem in an unrelated area, namely, constructive quantum field theory of a free scalar field. 

There are many ways of probing the issue of making the standard textbook treatment of quantum field theory mathematically more rigorous \cite{vincent}. A convenient way to begin is to anchor the discussion in the path integral formulation of a quantum theory.  

\subsection{A primer on path integrals}
We will begin by first presenting the standard textbook treatment of path integrals. Then, we will point out the mathematical lucanae in such a treatment and motivate why they must be filled when doing quantum gravity.

\subsubsection{The standard treatment}
\begin{figure}[ht]
    \centering

\tikzset{every picture/.style={line width=0.75pt}} 

\begin{tikzpicture}[x=0.75pt,y=0.75pt,yscale=-1,xscale=1]

\draw    (172,210) -- (360,210) ;
\draw    (172,79) -- (360,79) ;
\draw  [dash pattern={on 4.5pt off 4.5pt}]  (172,191) -- (360,191) ;
\draw  [dash pattern={on 4.5pt off 4.5pt}]  (172,172) -- (360,172) ;
\draw  [dash pattern={on 0.84pt off 2.51pt}]  (260,135) -- (260,159) ;
\draw  [dash pattern={on 4.5pt off 4.5pt}]  (172,100) -- (360,100) ;
\draw  [dash pattern={on 4.5pt off 4.5pt}]  (172,120) -- (360,120) ;
\draw    (213,209) .. controls (240.32,188.51) and (302,202) .. (271.33,187.17) .. controls (240.66,172.33) and (289.32,180.51) .. (302,171) ;
\draw    (259,80) .. controls (244,100) and (333,114) .. (316,132) .. controls (299,150) and (320,157) .. (302,171) ;

\draw    (181,209) .. controls (208.32,188.51) and (233,182) .. (203,181) .. controls (173,180) and (190.32,179.51) .. (203,170) ;
\draw    (224,80) .. controls (209,100) and (237,111) .. (220,129) .. controls (203,147) and (221,156) .. (203,170) ;

\draw    (297,211) .. controls (324.32,190.51) and (372,181) .. (342,180) .. controls (312,179) and (323.32,178.51) .. (336,169) ;
\draw    (341,79) .. controls (307,101) and (330,153) .. (349,136) .. controls (368,119) and (354,155) .. (336,169) ;
\draw    (267,215) -- (267,236) ;
\draw [shift={(267,238)}, rotate = 270] [color={rgb, 255:red, 0; green, 0; blue, 0 }  ][line width=0.75]    (10.93,-3.29) .. controls (6.95,-1.4) and (3.31,-0.3) .. (0,0) .. controls (3.31,0.3) and (6.95,1.4) .. (10.93,3.29)   ;

\draw (151,202.4) node [anchor=north west][inner sep=0.75pt]    {$t_{0}$};
\draw (151,180.4) node [anchor=north west][inner sep=0.75pt]    {$t_{1}$};
\draw (150,68.4) node [anchor=north west][inner sep=0.75pt]    {$t_{N}$};
\draw (150,99.4) node [anchor=north west][inner sep=0.75pt]    {$\epsilon \ \{$};
\draw (369,199.4) node [anchor=north west][inner sep=0.75pt]    {$| q_{0} \rangle \ :=\ |q_{I} \rangle $};
\draw (367,69.4) node [anchor=north west][inner sep=0.75pt]    {$| q_{N} \rangle \ :=\ |q_{F} \rangle $};
\draw (369,178.4) node [anchor=north west][inner sep=0.75pt]    {$| q_{1} \rangle $};
\draw (225,243) node [anchor=north west][inner sep=0.75pt]   [align=left] {The configuration space};

\end{tikzpicture}
    \caption{A sum over histories.}
    \label{fig3.5}
\end{figure}
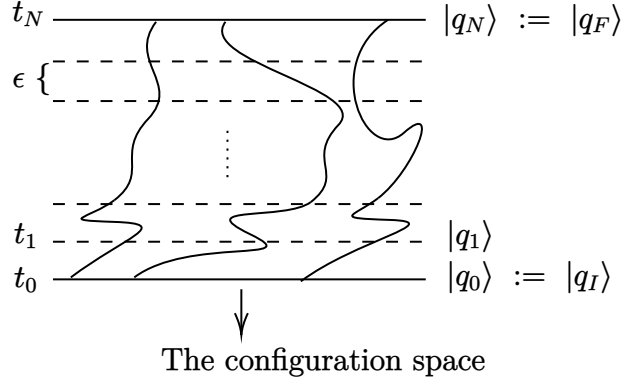

Consider a general quantum system described by $M$ number of canonical position and momentum coordinates $q^i$ and $p_i$, with a Hamiltonian $H(p,q)$. We write $|q_1,\cdots,q_N\rangle := |q\rangle$, and are interested in finding the amplitude of transition corresponding to an initial state $|q_I\rangle$ evolving in a time $T$ to a final state $q_F$:
\begin{equation}
    Z = \langle q_F|e^{-iHT}|q_I\rangle. \label{eq3.30}
\end{equation}
Feynman's nifty idea was to essentially sum over all the possible ways of reaching $|q_F\rangle$ from $|q_I\rangle$. This can be done as follows. Break the time interval $T$ into $N$ equal intervals $[t_n, t_{n+1}]$, $n \in \{0, N-1\}$, each of duration $\epsilon$, and write the state of the system at $t_n$ as $|q_n\rangle$, with $q_0 := q_I$ and $q_N := q_F$. Then we can write 
\begin{equation}
    e^{-iHT} = \underbrace{e^{-iH\epsilon}\cdots e^{-iH\epsilon}}_{N\text{ factors}} \label{eq3.31}
\end{equation}
and insert, in succession, a complete set of position states,
\begin{equation}
    \mathds{1} = \int d^Mq_n |q_n\rangle\langle q_n|, \quad n \in \{1,\cdots,N-1\} \label{eq3.32}  
\end{equation}
between alternating terms in Eq~(\ref{eq3.31}). As a result, we have
\begin{align}
    Z &= \prod_{n=0}^{N-1}\int d^Mq_n \langle q_{n+1}|e^{-iH\epsilon}|q_n\rangle \nonumber \\
    &= \int\prod_{n=0}^{N-1} d^Mq_n \langle q_{n+1}|e^{-iH\epsilon}|q_n\rangle ,\label{eq3.33}
\end{align}
where $d^Mq_n$ is shorthand for $dq^1_ndq^2_n\cdots dq^M_n$. The integral over each $q_n$ encodes all possible positions that the state could be in at time $t_n$, and the "sum over all possible paths" (or histories) from $q_I$ to $q_F$ is obtained in the limit as $N \to \infty$ or $\epsilon \to 0$ (see Fig~(\ref{fig3.5}). 

As is stands, Eq~(\ref{eq3.33}) does not seem particularly revealing. But it can be massaged into a very useful form. In the limit of small $\epsilon$, assuming that $H(p,q) = f(q) + h(p)$ (i.e. the Hamoltonian does not depend on products of positions and momenta)\footnote{The derivation works for more general Hamiltonians as well, as a perusal of any standard textbook on quantum field theory will reveal \cite{srednicki, peskin}. We make this choice so that one does not lose the forest for the trees.}, and inserting a complete set of momentum states, we find 
\begin{align}
    \langle q_{n+1}|e^{-iH\epsilon}|q_n\rangle &= \langle q_{n+1}|(1-iH\epsilon)|q_n\rangle \nonumber \\  
    &= (1-i\epsilon f(q_n))\langle q_{n+1}|q_n\rangle -i\epsilon \langle q_{n+1}|h(p)|q_n\rangle \nonumber \\
    &= \int \frac{d^Mp_n}{2\pi} (1-i\epsilon H(q_n, p_n))e^{ip_n\cdot(q_{n+1} - q_n)} \nonumber \\
    &= \int \frac{d^Mp_n}{2\pi} e^{-i\epsilon H} e^{ip_n\cdot(q_{n+1} - q_n)},\label{eq3.34}
\end{align}
where in the second-last line we used the fact that $\langle q|p\rangle = e^{ip\cdot q}$, with $q\cdot p := q^ip_i$. Substituting this last equation into Eq~(\ref{eq3.33}), we obtain
\begin{align}
   Z &= \int\prod_{n=0}^{N-1}\frac{d^Mp_n}{2\pi}d^Mq_n\exp{\left(i\epsilon\left[p_n\cdot\frac{q_{n+1}-q_n}{\epsilon} - H(q_n, p_n)\right]\right)} \nonumber\\
   &= \int\prod_{n=0}^{N-1}\frac{d^Mp_n}{2\pi}d^Mq_n\prod_{n=0}^{N-1}\exp{\left(i\epsilon\left[p_n\cdot\frac{q_{n+1}-q_n}{\epsilon} - H(q_n, p_n)\right]\right)}\nonumber \\
   &= \int\prod_{n=0}^{N-1}\frac{d^Mp_n}{2\pi}d^Mq_n\exp{\left(i\sum_{n=0}^{N-1}\epsilon\left[p_n\cdot\frac{q_{n+1}-q_n}{\epsilon} - H(q_n, p_n)\right]\right)}. \label{eq3.35}
\end{align}
Now, in the limit that $\epsilon\to 0$ and $N \to \infty$, the sum inside the exponent is replaced by an integral over time, $(q_{n+1}-q_n)/\epsilon \to \dot{q}$. Therefore, 
\begin{equation}
    Z = \int Dp Dq \exp{\left(i\int_{0}^{T}dt[ p(t)\cdot\dot{q}(t) - H(q,p)]\right)}, \label{eq3.36}
\end{equation}
where the functions $q(t)$ are constrained by $q(T) = q_F$ and $q(0) = q_I$ at the endpoints, but the functions $p(t)$ are not. We have suggestively written $DpDq$ in the preceding equation to refer to the measure that, if it exists, arises by taking the formal limit $N \to \infty$. Heuristically speaking, it can understood as the product of the (Lebesgue) measure $dq(t)dp(t)/2\pi$ at each point in time, i.e.
\begin{equation}
    DqDp = \prod_{t\in[0,T]}\frac{d^Mp(t)d^Mq(t)}{2\pi}. \label{eq3.37}
\end{equation}
With this understanding, if we assume that the Hamiltonian is at most quadratic in $p$, then we can complete the square in the exponent of Eq~(\ref{eq3.36}), perform the Gaussian integral that results over $Dp$, and absorb the resulting constant into the definition of $Dq$. When all is said and done, we obtain
\begin{equation}
    Z = \int Dq\, e^{i\int_0^Tdt\, L(q(t),\dot{q}(t))} = \int Dq\, e^{iS}, \label{eq3.38}
\end{equation}
where $L(q(t), \dot{q}(t))$ and $S$, respectively, are the Lagrangian and the action of the theory. 

Let us now narrow the discussion to our main object of interest, namely, a free, massive scalar field $\phi$. The action is given by
\begin{subequations}
\label{eq3.39}
    \begin{align}
        S &= \int_{\mathbb{R}^D}d^Dx\,\mathcal{L},\\
        \mathcal{L} &= \frac{1}{2}\partial_\alpha\phi(x)\partial^\alpha\phi(x) - \frac{m^2}{2}\phi(x), 
    \end{align}
\end{subequations}
where $m$ is the mass of the field and $D$ the dimension of spacetime. The classical dynamics is governed by the equation of motion that results from varying the action with respect to the field,
\begin{equation}
    \partial_\alpha\partial^\alpha\phi(x) = -m^2\phi(x). \label{eq3.40}
\end{equation}
On the other hand, the quantum dynamics inheres in the path integral for the transition of a field configuration from $\phi_I(\mathbf{x},t_I)$ to $\phi_F(\mathbf{x}, t_F)$. In analogy with Eq~(\ref{eq3.38}), we have
\begin{equation}
    Z = \int D\phi\, e^{iS[\phi]}, \label{eq3.41}
\end{equation}
where the integral measure $D\phi$ for now is to be understood as being proportional to the product of "Lebesgue measures", if they exist, on the infinite-dimensional space of all possible field configurations at $\mathbf{x}$,
\begin{equation}
    D\phi \propto \prod_{t\in [t_I, t_F]} \phi(\mathbf{x},t). \label{eq3.42}
\end{equation}

For future reference, we recall here that in quantum field theory, we are also interested in calculating the amplitudes for scattering and collision processes. These amplitudes depend on the correlations between the field operators at different spacetime locations, computed by sandwiching the time-ordered product of those operators at those locations between the vacuum state,
\begin{equation}
    S(x_1, \ldots, x_N) := \langle 0|T[\phi(x_1)\cdots\phi(x_N)]|0\rangle. \label{eq3.43}
\end{equation}
These correlations can be straightforwardly evaluated in the  path-integral formalism. One finds \cite{srednicki, peskin}
\begin{equation}
    S(x_1, \ldots, x_N) = \int \prod_{i=1}^{N}\phi(x_i) e^{iS[\phi]}D\phi. \label{eq3.44}
\end{equation}
Thus, in essence, the path-integral formalism of a quantum field theory consists in integrating functions of a field with respect to the measure
\begin{equation}
    e^{iS[\phi]}D\phi. \label{eq3.45}
\end{equation}

Finally, in anticipation of the rigorous mathematical setting in which we shall consider a scalar field theory below, we will work with a slight modification of the scalar field theory we considered above. In Eqs~(\ref{eq3.39}) through (\ref{eq3.41}), we will replace the time variable $t$ with $-it$, passing from a scalar field on Minkowski space to a scalar field in Euclidean space. This mathematical trick is called analytic continuation, and is performed because we have a better control over the behaviour of the relevant functions in Euclidean field theory. It can be rigorously shown that one can always go back to the Minkowskian theory from its Euclidean counterpart (see \cite{jaffe} for how this can be done in all its gory detail). Notice that in the Euclidean theory, Eqs~(\ref{eq3.40}) and (\ref{eq3.41}) become
\begin{equation}
    \nabla^2\phi(x) = -m^2\phi(x), \quad Z = \int\, D\phi e^{S[\phi]}, \label{eq3.46}
\end{equation}
where $\nabla^2 := \partial_0\partial^0 + \partial_i\partial^i$, in contrast to $\partial_\alpha\partial^\alpha = -\partial_0\partial^0 + \partial_i\partial^i$.

\subsubsection{Path integrals under scrutiny}
Admittedly, the ``derivations'' presented in the preceding pages would torment a thoroughgoing mathematician. First, a fastidious perusal of these pages would have revealed the subtle sleight-of-hand performed in interchanging the product and the integral in coming to Eq~(\ref{eq3.33}), since a limit of a sequence of integrals may very well not be the integral of the limit of a sequence. Second, there is no guarantee that the limits taken in going from Eq~(\ref{eq3.35}) to (\ref{eq3.36}) even exist. Third, the mathematical meaning of the measure in Eq~(\ref{eq3.37}) is entirely unclear. Is it a measure in the rigorous measure-theoretic sense, i.e. is it countably additive? Is it a well-defined procedure to multiply uncountably many Lebesgue measures? The answer to both these questions is no, for an infinite product of Lebesgue measures in $\mathbb{R}$ does not give us a countably additive measure \cite{jose}. Third, is the analogical jump from Eq~(\ref{eq3.38}) to (\ref{eq3.41}) justified? Indeed, something must have gone wrong here, since the space of all field configurations is infinite-dimensional, and there exists a theorem asserting that there exists no non-trivial, locally finite and translation-invariant measure on an infinite-dimensional (separable Banach) space\footnote{Here is the proof. Suppose that there is a locally finite translation-invariant measure on a separable Banach space. Thus there exists an open ball of radius $\epsilon$ that has finite measure. Since the space is infinite dimensional, by Riesz's lemma, there exists a sequence of pairwise-disjoint open balls of radius $\epsilon/4$, all lying in the $\epsilon$-ball. By translation invariance, each of the $\epsilon/4$-balls must have the same measure. Since the space is separable, it can be covered by a countable collection of such $\epsilon/4$-balls, and so by countable additivity, the measure of the union of these balls should be infinite -- a contradiction \cite{hunt}.}, whereas if the measure in Eq~(\ref{eq3.41}) is to be understood as a product of Lebesgue measures, it must be translation invariant!  

However, the kinds of calculations we have presented above are routinely used in the literature on quantum field theory. Indeed, they are physically and conceptually clean, and the motivation of a quantum field theorist in using path integrals is that they enable us to empirically verify our theoretical predictions about the quantum world to an extraordinary level of precision. Thus from an empirical standpoint, it does not seem unjustified to indulge in the kind of mathematical imprecision that plagues the standard treatments of path integrals. 

Nonetheless, theoretical physics proceeds not only by constructing models in the light of empirical results, but also by building careful models that predict phenomena that have not been verified yet. Now, a paucity of experimental input in quantum gravity thoroughly makes it an enterprise of the latter kind, and in such a situation, it appears necessary to be mathematically as consistent as possible. It is on this account that we find ourselves embroiled in the business of understanding the construction of measures on infinite-dimensional configuration spaces, as exemplified in constructive quantum field theorists' attempts to make rigorous sense out of field-theoretic path integrals. 

\subsection{Measures on infinite-dimensional spaces: the problem}
At this point, one might feel that we have come quite far away from our initial goal of illustrating how an infinite-dimensional classical configuration space needs to be enlarged in order to realise quantum states as complex-valued functions on it. However, witnessing the problem of making path integrals in quantum field theory rigorous has introduced us to the question of constructing measures on infinite-dimensional configuration spaces. It is in developing an integration theory on infinite-dimensional configuration spaces that an enlargement of these spaces is required. We will now illustrate this with the example of scalar field theory. The following discussion will have significant input from measure theory on infinite-dimensional spaces, for a review of which we refer the reader to Ref. \cite{jose}. We have based our discussion on Ref. \cite{ashtekar 2, ashtekar 3}. It is worth pointing out that there are several ways of broaching the question of constructing measures on infinite-dimensional spaces. The approach adopted here is motivated by its appeal in illustrating the kind of constructions used in loop quantum gravity. 

Let us begin with a precise characterisation of the classical configuration space of a free scalar field theory. Eq~(\ref{eq3.40}) suggests that our scalar field should be at least a $C^2$ function of $x$, and decrease rapidly at infinity. Thus we identify the classical configuration space with the linear vector space of all $C^2$ functions $\phi$ that rapidly decrease at infinity. We shall call this space $\mathcal{P}$. It is evidently infinite-dimensional. We shall attempt to construct a countably additive measure on this space, and shall find ourselves forced to enlarge it in order to have such a measure on it. 

We wish integrate complex-valued functions on $\mathcal{P}$, which is infinite-dimensional. Now, we know very well how to integrate functions on finite-dimensional spaces, such as $\mathbb{R}^N$. Thus one strategy to pursue for integrating functions on $\mathcal{P}$ is to first transport fields to $\mathbb{R}^N$ and then consider functions on this simpler space. Of course, this will inevitably cause a good deal of information about $\mathcal{P}$ to be lost. Our task then is to ensure that the measures we construct on the simpler structures we consider can be consistently extended to $\mathcal{P}$ in order to include the lost information. We will see that this cannot be done consistently unless we enlarge $\mathcal{P}$. 

Consider then the space of all infinitely differentiable functions on $\mathbb{R}^D$ that fall off sufficiently rapidly at infinity. More precisely, for $f \in \mathcal{J}$,
\begin{equation}
    \sup_{x\in\mathbb{R}^D}\left| x_1^{k_1}\ldots x_D^{k_D}\frac{\partial^{j_1}}{\partial x_1^{j_1}}\cdots\frac{\partial^{j_D}}{\partial x_D^{j_D}}f(x) \right| < \infty, \quad \forall k_1,\ldots,k_D, j_1, \ldots, j_D \in \mathbb{Z}_{+}. \label{eq3.47}
\end{equation}
$\mathcal{J}$ is called the \textit{Schwartz space}. A function $g$ in it can be used to probe the structure of the fields $\phi$ in $\mathcal{P}$ in a neighbourhood of $\mathbb{R}^D$ in which $g$ is nonzero. To see this, for a fixed $g \in \mathcal{J}$, consider a linear function $F_g: \mathcal{P} \to \mathbb{R}$ such that
\begin{equation}
    F_g(\phi) := \int_{\mathbb{R}^D}d^Dx\,\phi(x)g(x). \label{eq3.48}
\end{equation}
 In view of this, then, let $g_1,\dots,g_N$ be a set of linearly independent Schwartz functions. Consider the projection 
 \begin{align}
     p_{g_1,\ldots,g_N}:\, &\mathcal{P} \to \mathbb{R}^N \nonumber \\
     &\phi \mapsto (F_{g_1}(\phi), \ldots, F_{g_N}(\phi)). \label{eq3.49} 
 \end{align}
In this way, we can use finite-dimensional subspaces $V_{g_1,\ldots,g_N}$ of $\mathcal{J}$ that are spanned by a finitely many $\{g_1,\ldots,g_N\}$ Schwartz functions to probe "parts" of $\mathcal{P}$, i.e. subsets of $\mathcal{P}$ of the form
\begin{equation*}
    \{\phi \in \mathcal{P}: (F_{g_1}(\phi), \ldots, F_{g_N}(\phi)) \in \mathbb{R}^N\}.
\end{equation*}
We will first describe a way to integrate complex-valued functions on such "parts" of $\mathcal{P}$. To this end, consider well-behaved functions $f: \mathcal{P} \to \mathbb{C}$ and $\Tilde{f}: \mathbb{R}^N \to \mathbb{C}$ such that
\begin{align}
    f(\phi) &:= p^*_{g_1,\ldots,g_N}\Tilde{f}(\phi) \label{eq3.50}\\ 
    &:= \Tilde{f}(p_{g_1,\ldots,g_N}(\phi)) = \Tilde{f}(F_{g_1}(\phi), \ldots, F_{g_N}(\phi)). \label{eq3.51}
\end{align}
The functions $f$ are called \textit{cylindrical} with respect to $V_{g_1,\ldots, g_N} \subset \mathcal{J}$. Also, the notation $p^*$ denotes the \textit{pullback} of $\Tilde{f}$ by $p$, i.e. in order to define functions on $\mathcal{P}$, we can pullback functions on $\mathbb{R}^N$. 

Since cylindrical functions depend on only a finite number of independent parameters (see Eq~(\ref{eq3.51})), integrating them is not difficult. For every $V_{g_1,\ldots,g_N} \subset \mathcal{J}$, we pick a normalized Borel measure $\mu_N$ on $\mathbb{R}^N$ and define, via Eq~(\ref{eq3.50}), the integral of a cylindrical function $f$ over $\mathcal{P}$ to be its integral over $\mathbb{R}^N$:
\begin{equation}
    \int_{\mathcal{P}}d\mu(\phi)f(\phi) := \int_{\mathbb{R}^N} d\mu_N(\eta_1, \ldots, \eta_N)\Tilde{f}(\eta_1, \ldots, \eta_N), \label{eq3.52}
\end{equation}
where $\eta_1, \ldots, \eta_N$ are coordinates on $\mathbb{R}^N$, given explicitly by Eq~(\ref{eq3.49}). However, as it stands, Eq~(\ref{eq3.52}) may not be well-defined. To see this, notice that the definition of a cylindrical function $f$ with respect to a particular subspace $V \in \mathcal{J}$ depends on the basis $g_1,\ldots, g_N$ chosen to span the subspace (see Eq~\ref{eq3.51}). It may be that a subspace $V_{g_1,
\ldots,g_N}$ is also spanned by another set of linearly independent Schwartz functions $h_1,\ldots, h_N$. We must make sure that the integral of $f$ is independent of the basis. More generally, it is not difficult to see that a function cylindrical with respect to a subspace $V$ is also cylindrical with respect to another subspace $V'$ if $V \subseteq V'$. Therefore, in order for Eq~(\ref{eq3.52}) to be well-defined, the finite-dimensional measures $\mu_N$ ought to satisfy some consistency conditions that ensure that the integral of a cylindrical function is independent of its representation. To flesh out these conditions, let $V_N$ and $V_M$ be $N$- and $M$-dimensional subspaces of $\mathcal{J}$, respectively, with bases $g_1, \ldots, g_N$ and $h_1, \ldots, h_M$. Consider the projection
\begin{equation}
    \pi_{NM}: \mathbb{R}^M \to \mathbb{R}^N. \label{eq3.53}
\end{equation}
If $V_N \subseteq V_M$, we require
\begin{equation}
    \mu_N = (\pi_{NM})_{*}\mu_M := \mu_M\circ\pi^{-1}_{NM}. \label{eq3.54}
\end{equation}
$(\pi_{NM})_{*}$ is called the \textit{pushforward} of the measure $\mu_M$ to $\mu_N$. Loosely speaking, we want the measure on $\mathbb{R}^N$ to be the same as the measure on $\mathbb{R}^M$. 

When the consistency conditions (\ref{eq3.54}) are satisfied, $\mathcal{P}$ is said to be equipped with a \textit{cylindrical measure}, denoted by $d\mu(\phi)$ in Eq~(\ref{eq3.52}). However, this measure is not countably additive, and this brings us to the essential problem: without enlarging $\mathcal{P}$, there is no way to extend a cylindrical measure to a genuine countably additive measure. 

To explore this issue in detail, we need to analyse the support of cylindrical measures as we have defined them. Recall that the measures $\mu_N$ are defined on Borel sets in $\mathbb{R}^N$. For any Borel set $B \subset \mathbb{R}^N$, the preimage of the projections in Eq~(\ref{eq3.49})
\begin{equation}
    p^{-1}_{g_1,\ldots, g_N}(B) = \{\phi \in \mathcal{P}: (F_{g_1}(\phi),\cdots,F_{g_N}(\phi)) \in B \subset \mathbb{R}^N\}, \label{eq3.55}
\end{equation}
yields a set in $\mathcal{P}$. Such a set is called a \textit{cylindrical set}. A cylindrical measure is defined on a family of cylindrical sets. Explicitly, given $V_{g_1,\ldots, g_N}$ and the family $\mathcal{B}$ of Borel sets in $\mathbb{R}^N$, let
\begin{equation}
    \mathcal{C} = \{p^{-1}_{g_1,\ldots, g_N}(B): B \in \mathcal{B}\}  \label{eq3.56}
\end{equation}
be the corresponding family of cylindrical sets in $\mathcal{P}$. Then for every $C \in \mathcal{C}$, the cylindrical measure $\mu$ is given by
\begin{equation}
    \mu(C) = \mu_N(p_{g_1,\ldots, g_N}(C)), \label{eq3.57}
\end{equation}
where $p_{g_1, \ldots, g_N}(C)$ is the image of $C$ under $p_{g_1,\ldots, g_N}$. 

Now recall that in a set $\mathcal{M}$, countably additive measures are defined on a $\sigma$-algebra, which is a collection of subsets of $\mathcal{M}$ that are closed under complements and \textit{countable} unions. Unfortunately, cylindrical sets fail to satisfy this last property, i.e. while a finite union of cylindrical sets is again a cylindrical set, a countably infinite one may not be\footnote{See \href{https://math.stackexchange.com/questions/1279824/show-that-collection-of-finite-dimensional-cylinder-sets-is-an-algebra-but-not?rq=1}{this link} for such a counterexample.}. On the other hand, the cylindrical measure constructed above is defined only on cylindrical sets. This is not a problem in itself, for it could be the case that one could use arbitrary unions of cylindrical sets to construct a family of non-cylindrical sets that forms a $\sigma$-algebra and hence, is measurable. Then one could perhaps extend the cylindrical measure to this family. For instance, if one can find an increasing sequence of cylindrical sets, then the cylindrical measure can be extended to a countably additive measure by the following standard result in measure theory. 

\begin{theorem}
    Let $\mu$ be a countably additive finite measure and $A_1 \subset A_2 \subset \cdots$ be an increasing sequence of measurable sets (i.e. those that belong to a $\sigma$-algebra). Then
    \begin{equation}
        \mu\left(\cup_n A_n\right)=\lim _{n \rightarrow \infty} \mu\left(A_n\right) \label{eq3.58}
    \end{equation}
\end{theorem}

Regrettably, such constructions are, in general, not possible for $\mathcal{P}$. They are, however, possible for the dual space of $\mathcal{P}$, namely, the set of all linear functionals on the Schwartz space $\mathcal{J}$\footnote{There is a potential for confusion here. Strictly speaking, the dual of $\mathcal{P}$ is the set of all linear functionals on $\mathcal{P}$, not on $\mathcal{J}$. However, observe that $\mathcal{J} \subset \mathcal{P}$, which implies that the dual of $\mathcal{P}$ is contained in the dual of $\mathcal{J}$. Thus we can work with the latter without loss of generality.}. Since $\mathcal{P}$ is infinite-dimensional, its dual space is "larger" than it. It is in this sense that an enlargement of $\mathcal{P}$ is required to develop a well-defined integration theory on it.  

There are many ways to understand why dual spaces of infinite-dimensional spaces admit convenient measure-theoretic constructions. One such way is to recognise that these dual spaces can naturally be regarded as the projective limit of a family of their subspaces. A digression on projective limits is thus in order.

\subsection{A digression on projective limits}
Let us begin with some preliminary definitions. 
\begin{definition}
(Partially ordered set) A set $\mathcal{L}$ is said to be partially ordered if it is equipped with a binary relation, denoted $\geq$, such that
\begin{enumerate}[(1)]
    \item (reflexivity) $L \geq L, \forall L \in \mathcal{L}$;
    \item (transitivity) for all $L, L', L'' \in \mathcal{L}$, if $L \geq L'$ and $L' \geq L''$, then $L \geq L''$;
    \item (anti-symmetry) for all $L, L' \in \mathcal{L}$, if $L \geq L'$ and $L' \geq L$, then $L = L'$.
\end{enumerate}
\end{definition}

\begin{definition}
    (Directed set) A partially ordered set $\mathcal{L}$ is called a directed set if for all $L', L'' \in \mathcal{L}$, there is some $L \in \mathcal{L}$ such that $L \geq L'$ and $L \geq L''$.
\end{definition}

\begin{definition}
    (Projective family) Let $\{M_L\}_{L\in\mathcal{L}}$ be a family of sets, where $\mathcal{L}$ is a directed index set. Suppose that for every $L, L' \in \mathcal{L}$ such that $L' \geq L$, the (necessarily surjective) projections
    \begin{equation}
        p_{L,L'}: M_L' \to M_L \label{eq3.59}
    \end{equation}
    satisfy
    \begin{equation}
        p_{L,L'} \circ p_{L', L''} = p_{L,L''}, \text{ for } L'', \geq L' \geq L. \label{eq3.60}
    \end{equation}
    Then the collection $\{M_L, p_{L,L'}\}_{L,L'\in\mathcal{L}}$ of such sets and projections is called a projective family. 
\end{definition}

We will be interested in the Cartesian product of the sets $M_L$,
\begin{equation}
    M_{\mathcal{L}} := \prod_{L\in\mathcal{L}} M_L. \label{eq3.61}
\end{equation}
An element in $M_{\mathcal{L}}$ is denoted as $(x_L)_{L\in\mathcal{L}}$, where $x_L \in M_L$. These definitions enable us to introduce the very useful notion of a projective limit.

\begin{definition}
    (Projective limit) The projective limit of a projective family $\{M_L, p_{L,L'}\}_{L,L'\in\mathcal{L}}$ is the subset of $M_{\mathcal{L}}$ defined as
    \begin{equation}
        M_{\infty} := \left\{(x_L)_{L\in\mathcal{L}}: L'\geq L \Rightarrow p_{L,L'}(x_L') = x_L \right\}. \label{eq3.62}
    \end{equation}
\end{definition}
In words, if $L'\geq L$, then $x_{L'} \in M_{L'}$ is sufficient to determine $x_L \in M_L$. 

We wish to develop an integration theory on the projective limit of a projective family. This requires us to put more structure on the sets $M_L$ of the family. Accordingly, for every $M_L$, we consider a $\sigma$-algebra $\mathcal{B}_L$ of its subsets. This makes $(M_L, \mathcal{B}_L)$ a measurable space for every $L$. Furthermore, recall that given two measurable spaces $X$ and $Y$ with $\sigma$-algebras $\mathcal{B}_Y$ and $\mathcal{B}_X$ respectively, a function $f: X \to Y$ is said to be measurable if the preimage of a measurable set is also measurable, i.e. $f^{-1}(B) \in \mathcal{B}_X$ for every $B \in \mathcal{B}_Y$. This allows us to define a measurable projective family.

\begin{definition}
    (Measurable projective family) A family $\{(M_L, \mathcal{B}_L), p_{L,L'}\}_{L,L'\in\mathcal{L}}$ is said to be a measurable projective family if each $(M_L, \mathcal{B}_L)$ is a measurable space and if $\{M_L, p_{L,L'}\}$ is a projective family such that every projection $p_{L,L'}$ is a measurable function. 
\end{definition}

These definitions make it possible to first turn the product space $M_\mathcal{L}$ in Eq~(\ref{eq3.61}) into a measurable space, and thence, render the projective limit measurable as well. 

To begin with, consider the projections $P_{L,\mathcal{L}}: M_\mathcal{L} \to M_L$ and the collection of subsets of $M_{\mathcal{L}}$ of the form
\begin{equation}
    A := \bigcup_{L\in\mathcal{L}}P^{-1}_{L,\mathcal{L}}\,\mathcal{B}_{L}, \label{eq3.63}
\end{equation}
where $P^{-1}_{L,\mathcal{L}}\,\mathcal{B}_L := \{P^{-1}_{L,\mathcal{L}}(B_L): B_L \in \mathcal{B}_L\}$. It is easily seen that the collection (Eq~(\ref{eq3.63})) is closed under complements and countable unions, and thus forms a $\sigma$-algebra on $M_\mathcal{L}$. Therefore, it is also the case that the projections $P_{L,\mathcal{L}}$ are measurable. We define the smallest $\sigma$-algebra containing $A$ such that all projections are measurable, and denote it as
\begin{equation}
    \mathcal{B}_{\mathcal{L}} := \mathbb{B}\left(\bigcup_{L\in\mathcal{L}}P^{-1}_{L,\mathcal{L}}\,\mathcal{B}_{L}\right) \label{eq3.64}
\end{equation}
It thus follows that $(M_{\mathcal{L}}, \mathcal{B}_{\mathcal{L}})$ is a measurable space. It can be shown \cite{jose} that given normalised, countably additive measures $\mu_L$ in each $(M_L, \mathcal{B}_{\mathcal{L}})$, one can construct a normalised, countably additive measure in $(M_{\mathcal{L}}, \mathcal{B}_{\mathcal{L}})$. However, our main motivation in constructing the foregoing structures is to convert the projective limit $M_{\infty}$ into a measurable space and construct a measure thereon. We now proceed with this task. 

Consider the collection of subsets of $M_{\infty}$ of the form
\begin{equation}
    \mathcal{B}_{\infty} := \mathcal{B}_{\mathcal{L}} \cap M_{\infty} := \{B \cap M_{\infty}: B \in \mathcal{B}_{\mathcal{L}}\}. \label{eq3.65}
\end{equation}
This collection forms a $\sigma$-algebra, as we now verify. 

\begin{lemma}
    The collection $\mathcal{B}_{\infty}$ (Eq~(\ref{eq3.65})) forms a $\sigma$-algebra on $M_{\infty}$.
\end{lemma}
\begin{proof}
    Since for every $B_n \in \mathcal{B}_{\mathcal{L}}, n \in \mathbb{Z}_{+}$,  
\begin{equation*}
    \cup_{n}(B_n \cap M_{\infty}) = (\cup_{n}B_n)\cap M_{\infty},
\end{equation*}
$\mathcal{B}_{\infty}$ is closed under countable unions. Furthermore, $M_{\infty},  B\cap M_{\infty}  \subset M_{\mathcal{L}}$ for any $B \in \mathcal{B}_{\mathcal{L}}$. Since $M_{\mathcal{L}}$ is measurable, we have
\begin{align*}
    M_{\infty}\setminus(B\cap M_{\infty}) &= M_{\infty}\cap(M_{\mathcal{L}}\cap(B\cap M_{\infty})) \\
    &= M_{\infty}\setminus((M_{\mathcal{L}}\setminus B)\cap(M_{\mathcal{L}}\setminus M_{\infty})) \\
    &= M_{\infty} \cap (M_{\mathcal{L}}\setminus B) \in \mathcal{B}_{\infty}.
\end{align*}
Hence, $\mathcal{B}_{\infty}$ is also closed under complements. 
\end{proof}

Thus, $(M_{\infty}, \mathcal{B}_{\infty})$ forms a measurable space -- we christen it as the measurable projective family. In fact, $\mathcal{B}_{\infty}$ is the smallest $\sigma$-algebra on $M_{\infty}$ in a certain sense, which we now express. 

For every projection $P_{L,\mathcal{L}}: M_{\mathcal{L}} \to M_L$, consider its restriction to $M_{\infty}$, given explicitly by
\begin{equation}
    p_{L} := P_{L,\mathcal{L}}\circ i_{\infty}, \label{eq3.66}
\end{equation}
where $i_{\infty}: M_{\infty} \to M_{\mathcal{L}}$ is the inclusion of $M_{\infty}$ in $M_{\mathcal{L}}$ (it is just the identity map on $M_{\infty}$). Since $P_{L,\mathcal{L}}$ and $i_{\infty}$ are both measurable functions, $p_L$ is also\footnote{Note that these projections also allow us to see the projective limit as the ''largest'' subset of $M_{\mathcal{L}}$ in the sense that one can unambigiously project down from it to any set $M_L$ in the projective family. That is, $p_L((x_L')_{L\in\mathcal{L}}) = (x_L)$. Thus the name projective \textit{limit} is quite apt.} measurable for all $L \in \mathcal{L}$. Moreover, the consistency conditions that define $M_{\infty}$ (Eq~(\ref{eq3.62})) can be recast as
\begin{equation}
    p_L = p_{L,L'}\circ p_L', \quad \forall L, L' \in \mathcal{L} \text{ such that } L' \geq L. \label{eq3.67}
\end{equation}
This lets us establish the following lemma.

\begin{lemma}
    The collection of sets
    \begin{equation}
        \mathcal{F}_{\infty} := \bigcup_{L\in\mathcal{L}}p^{-1}_L\mathcal{B}_L, \label{eq3.68}
    \end{equation}
    where $p^{-1}_L\mathcal{B}_L = \{p^{-1}_L(B_L): B_L \in \mathcal{B}_L\}$, forms an algebra. 
\end{lemma}
\begin{proof}
    Closure under complements can be demonstrated in the same way as in Lemma 3.1. As for closure under finite unions, consider $\bigcup_{n=1}^{N}p^{-1}_{L_n}(B_{L_n})$. In view of Definition 3.2, there exists\footnote{Evidently, such an $L$ would not exist if the collection $\{L_n\}$ were countable. This is another way to understand why cylindrical sets are not closed under arbitrary unions.} $L \in \mathcal{L}$ such that $L \geq L_n$ for all $n \in \{1,\cdots, N\}$. From Eq~(\ref{eq3.67}), we then get
    \begin{align*}
        \bigcup_n p^{-1}_{L_n}(B_{L_n}) &= p^{-1}_{L}(p^{-1}_{L_n, L}(B_{L_n}))\\
        &= p^{-1}_L\left( \bigcup_n p^{-1}_{L_n, L}(B_{L_n})\right)
    \end{align*}
    The set in the round brackets in the second line above clearly belongs to $\mathcal{B}_L$.
\end{proof}

The sets $p^{-1}_{L}(B_L)$ are called cylindrical sets (compare with Eq~(\ref{eq3.55})). As we remarked earlier, one can always construct (finitely) additive measures on the collection (Eq~(\ref{eq3.68})) of cylindrical sets. Eqs~(\ref{eq3.54})--(\ref{eq3.57}), \textit{mutatis mutandis}, apply here. That is, suppose we have a collection of measures $\{\mu_L\}_{L\in\mathcal{L}}$ on the projective family, where $\mu_L : \mathcal{B}_L \to \mathbb{R}$. Suppose further that these measures satisfy the self-consistency conditions
\begin{equation}
    \mu_L = (p_{L,L'})_{*}\mu_{L'}, \quad \forall L, L' \text{ such that } L' \geq L. \label{eq3.69}
\end{equation}
Then we get a cylindrical measure $\mu$ on every cylindrical collection $p^{-1}\mathcal{B}_L$ of sets by
\begin{equation}
    \mu = p^*_L \mu_L, \, \text{ i.e. } \mu(p^{-1}(B_L)) = \mu_L(B_L), \quad \forall B_L \in \mathcal{B}_L. \label{eq3.70}
\end{equation}
This measure, as noted in the previous section, lets us integrate cylindrical functions on $M_{\infty}$. In current notation, these functions are pullbacks by $p_L$ of functions on $M_L$ that are integrable with respect to $\mu_L$. More precisely, let $F_L$ be an integrable function on $M_L$. Then 
\begin{equation}
    f := p^*_L F_L = F_L \circ p_L \label{eq3.71}
\end{equation}
is a cylindrical function on $M_{\infty}$, in view of the preceding two equations, can be integrated as 
\begin{align}
    \int_{M_\infty} d\mu f = \int d(p^*_L\mu_L)f = \int d\mu_L (p_L)_*f =  \int_{M_L}d\mu_L F_L.  
     \label{eq3.72} 
\end{align}
The self-consistency conditions (Eq~\ref{eq3.69}) guarantee that we get a well-defined answer above, i.e. the answer does not depend on the choice of $L$, since if $L'\geq L$ and $F_L, F_{L'}$ represent the same cylindrical functions, we have
\begin{equation*}
    \int_{M_L}d\mu_L F_L = \int d((p_{L,L'})_*\mu_{L'})F_L = \int d\mu_{L'} (p^*_{L,L'}F_L) = \int_{M_{L'}}d\mu_{L'} F_{L'},
\end{equation*}
where the last equation follows from the definition of the projective limit.  

But our goal, of course, is to introduce countably additive measures on the $\sigma$-algebra $\mathcal{B}_{\infty}$ that we defined above (Eq~\ref{eq3.65}). The fundamental advantage of working with a projective limit is that one can use arbitrary unions of cylindrical sets to construct $\sigma$-algebras on $M_{\infty}$ -- this was precisely the question posed at the end of the preceding section. In particular, we have the following lemma \cite{jose}.

\begin{lemma}
    $\mathcal{B}_{\infty} := \mathbb{B}(\mathcal{F}_{\infty})$,
    where $\mathbb{B}(\mathcal{F}_{\infty})$ is the smallest $\sigma$-algebra on $M_\infty$ such that all projections $p_L: M_{\infty} \to M_L$ are measurable.
\end{lemma} 

Thus, we have effectively reduced our problem to that of extending cylindrical measures on $\mathcal{F}_{\infty}$ to countably additive measures on $\mathbb{B}(\mathcal{F}_{\infty})$. Of course, the existence of the $\sigma$-algebra $\mathcal{B}_{\infty}$ does not, in itself, guarantee the existence of a measure on $(M_{\infty}, \mathcal{B}_{\infty})$. There does exist a general theorem characterising the extendability of cylindrical measures to non-cylindrical measures in a projective family \cite{yamasaki}, but we shall not need it. For our purposes, it is sufficient to note that under suitable topological conditions on the projective family, extendability is ensured. For example, a regular Borel measure on the projective family formed by compact Hausdorff spaces spaces can be extended to a regular Borel measure on the projective limit. Incidentally, this situation will occur in the enlargement of the configuration space $\mathcal{A/G}$ of connections modulo gauge transformations! We defer a detailed discussion of this important point to later.


    
    

\subsection{Measures on infinite-dimensional spaces: resolution of the problem}
Now that we have understood projective limits and the measure-theoretic gymnastics that they are amenable to, we can return to the problem of constructing a countably additive measure on the configuration space of free scalar field theory. As argued at the end of Section 3.3.2, rather than the space $\mathcal{P}$ of all rapidly-falling-at-infinity $C^2$ functions on $\mathbb{R}^D$, the algebraic dual of $\mathcal{P}$, by virtue of its being a projective limit, is more suitable for measure-theoretic constructions. In this section, we will show that the algebraic dual of an infinite-dimensional vector space is indeed a measurable projective limit. Then we shall see how to make use of this important fact in scalar field theory. 

Let $\mathcal{V}$ be a real infinite-dimensional vector space and let $\Tilde{\mathcal{V}}$ be its algebraic dual, i.e. the space of all linear functionals on $\mathcal{V}$. Let $\mathcal{L}$ be the set of all finite-dimensional subspaces of $\mathcal{V}$. It is not difficult to verify that the binary operation `$\subset$' on the elements $\mathcal{L}$ is reflexive, transitive and anti-symmetric. Thus $\mathcal{L}$ is partially ordered, with 
\begin{equation}
    L' \geq L \Leftrightarrow L' \supset L, \quad \forall L', L \in\mathcal{L}. \label{eq3.73}
\end{equation}
Furthermore, there always exists a subspace greater than any other two subspaces. Therefore, $\mathcal{L}$ is a directed set. Next, let $\{\Tilde{L}: L\in\mathcal{L}\}$ be the collection of the algebraic duals of all $L \in \mathcal{L}$. For each $L', L \in \mathcal{L}$ such that $L' \geq L$, let $p_{L,L'}: \Tilde{L'}\to\Tilde{L}$ be the linear transformation such that every element in $\Tilde{L'}$ is mapped to its restriction to $\Tilde{L}$. Since any linear functional in $\Tilde{L}$ can be trivially extended to a linear functional in $\Tilde{L'}$ if $L' \geq L$, $p_{L,L'}$ are all surjective. Finally, it is straightforward to show that for $L'' \geq L' \geq L$,
\begin{equation}
    p_{L_,L''} = p_{L, L'}\circ p_{L', L''}. \label{eq3.74}
\end{equation}
Hence, by Definitions 3.1--3.3, the collection $\{\Tilde{L}, p_{L,L'}\}_{L,L'\in\mathcal{L}}$ forms a projective family. Let $\Tilde{V}_{\infty}$ be the corresponding projective family. We will now show that it is isomporphic to $\Tilde{V}$. 

Since all maps $p_{L,L'}$ are linear, $\Tilde{V}_{\infty}$ is a subspace of the direct product of all the subspaces $\Tilde{L}$, i.e. $\Tilde{V}_{\infty} \subset \prod_{L\in\mathcal{L}}\Tilde{L}$. Further, recall that an element in $\Tilde{V}_{\infty}$ is written as $(\phi_L \in \Tilde{L})_{L\in\mathcal{L}}$. Now for any $\phi \in \Tilde{V}$, let $\phi\restriction_L$ be its restriction to $\Tilde{L}$; clearly, $\phi\restriction_L = \phi_L$. Also, if $L'\geq L$, we have that 
\begin{equation*}
    \phi\restriction_L = p_{L,L'}(\phi\restriction_L'),
\end{equation*} 
from which it follows that the linear map
\begin{align}
    \omega:\, &\Tilde{V} \to \Tilde{V}_{\infty} \nonumber \\
    & \omega(\phi) = (\phi\restriction_L)_{L\in\mathcal{L}} \label{eq3.75}
\end{align}
is injective. Moreover, since $\Tilde{V}_{\infty}$ contains linear functionals by definition, the map above is also surjective. Therefore, $\Tilde{V}$ and $\Tilde{V}_{\infty}$ are isomorphic. 

If we have a measurable projective family $\{\Tilde{V}_{\infty}, \mathcal{B}_{\infty}\}$, $\mathcal{B}_{\infty}$ being the $\sigma$-algebra on $\Tilde{V}_{\infty}$, then the map $\omega$ also renders the dual space $\Tilde{V}$ measurable. We simply define the $\sigma$-algebra $\Tilde{\mathcal{B}}$ on $\Tilde{V}$ to be the collection of the preimages of the elements of $\mathcal{B}_{\infty}$ under $\omega$, i.e.
\begin{equation}
    \Tilde{\mathcal{B}} := \omega^{-1}\mathcal{B}_{\infty} \label{eq3.76}
\end{equation}
Thus the $(\Tilde{V}, \Tilde{\mathcal{B}})$ and $(\Tilde{V}_{\infty}, \mathcal{B}_{\infty})$ are also isomorphic as measurable spaces. 

We can also explicitly see how the $\sigma$-algebra $\Tilde{\mathcal{B}}$ (or $\mathcal{B}_{\infty}$) looks like. For any $\phi \in \mathcal{\Tilde{V}}$ and $v \in \mathcal{V}$, let $f_{v}:\Tilde{\mathcal{V}}\to\mathbb{R}$ be the function such that $f_{v}(\phi) = \phi(v)$. Then $\Tilde{\mathcal{B}}$ is the smallest $\sigma$-algebra
such that all the maps $f_v$ are measurable. In other words,
\begin{equation}
    \Tilde{B} = \mathbb{B}\left( \bigcup_{v\in\mathcal{V}}f^{-1}_{v}\mathcal{B}(\mathbb{R})\right), \label{eq3.77}
\end{equation}
where $\mathcal{B}(\mathbb{R})$ is the collection of all Borel sets in $\mathbb{R}$. 

With this, we are well-equipped to deal with the construction of measures on the dual space $\Tilde{\mathcal{V}}$. The fundamental reason that salvages the dual space as opposed to the original space lies in the following theorem, which, in general, does not exist for the latter. 

\begin{theorem}
    Any self-consistent family of finite Borel measures $\mu_L$ on the subspaces $\Tilde{L} \subset \Tilde{V}$ defines a countably additive measure on $(\Tilde{V}, \Tilde{\mathcal{B}})$. 
\end{theorem}
\begin{proof}
    Since a detailed proof uses concepts that have not been introduced, we just sketch the basic steps here. 

    Observe that since each subspace $\Tilde{L}$ is finite-dimensional, it is isomorphic to $\mathbb{R}^n$. Thus the $\sigma$-algebra $\mathcal{B}_L$ on each $\Tilde{L}$ is nothing but the Borel $\sigma$-algebra on $\mathbb{R}^n$. Now, $\mathbb{R}^n$ is a compact metric space, and thus so is $\Tilde{L}$. It can be shown \cite{yamasaki} that a family of self-consistent measures on a projective family of compact metric spaces can be uniquely extended to a countably additive measure on the projective limit. Since the projective limit here has been shown to be the dual space, the theorem is established. 
\end{proof}

On its own, Theorem 3.2 is not very useful, since it only guarantees the existence of a measure and does not reveal how to construct one. However, there exists a very useful characterisation of countably additive, finite measures on the dual space. It is called Bochner's theorem, which we now describe. 

Let $\mu$ be a finite and countably additive measure on $(\Tilde{\mathcal{V}}, \Tilde{\mathcal{B}})$. The \textit{Fourier transform}, or the \textit{generating functional}, of $\mu$ is a function $\chi: \mathcal{V} \to \mathbb{C}$ given by
\begin{equation}
    \chi(v) = \int_{\Tilde{\mathcal{V}}} d\mu(\phi)e^{i\phi(v)} \label{eq3.78}
\end{equation}
for every $v \in \mathcal{V}$. Furthermore, a function $f: \mathcal{V} \to \mathbb{C} $ is of \textit{positive type} if for all $c_1, \ldots, c_n \in \mathbb{C}$ and $v_1, \ldots, v_n \in \mathcal{V}$,
\begin{equation}
    \sum_{k, l =1}^{n}c_k \Bar{c}_l f(v_k - v_l) \geq 0. \label{eq3.79}
\end{equation}
We now present the Bochner theorem \cite{jose}. 

\begin{theorem}
    (Bochner theorem) A complex-valued function $\chi$ on an infinite-dimensional vector space $\mathcal{V}$ is the Fourier transform of a finite and countably additive measure on $(\Tilde{\mathcal{V}}, \Tilde{\mathcal{B}})$ if and only if it is of positive type and continuous on every finite-dimensional subspace of $\mathcal{V}$. Furthermore, the measure is normalised if and only if $\chi(0) = 1$. 
\end{theorem}

Let us now exemplify how to use Theorems 3.2 and 3.3 to construct measures on the algebraic dual of an infinite-dimensional vector space. 

\begin{mdframed}[style=testframe]
    \textbf{Gaussian measures}
    
    Let $\mathcal{V}$ be an infinite-dimensional vector space, and $\Tilde{\mathcal{V}}$ its algebraic dual. Let $\mathcal{L}$ denote the set of all finite-dimensional subspaces of $\mathcal{V}$ and for each $L \in \mathcal{L}$, let $\Tilde{L}$ be its (finite-dimensional) algebraic dual. 
    
    We begin by defining a self-consistent family of measures $\mu_L$ on finite-dimensional subspaces $\Tilde{L} \subset \Tilde{\mathcal{V}}$. Let us recall the self-consistency conditions on $\mu_L$. We require that if $L' \supset L$, then the measure $\mu_L$ on $\Tilde{L}$ is related to the measure $\mu_L'$ on $\Tilde{L}'$ by (cf. Eqs~(\ref{eq3.54}) and \ref{eq3.69}))
    \begin{equation}
        \mu_L = (p_{L,L'})_{*}\mu_{L'}, \label{eq3.80}
    \end{equation}
    where $p_{L,L'}$ is the projection map from $\Tilde{L}'$ to $\Tilde{L}$. 
    
    Now recall that if $\Tilde{L}$ is $n$-dimensional, then it is isomorphic to $\mathbb{R}^n$. Thus it suffices to define a measure on the latter. For this purpose, we generalise the (normalised) Gaussian measure in $\mathbb{R}$, 
    \begin{equation}
        d\mu(x) = \frac{1}{\sqrt{2\pi}}e^{-x^2}dx, \label{eq3.81}
    \end{equation}
    $dx$ being the usual Lebesgue measure on $\mathbb{R}$. Let $C$ be a positive-definite symmetric $n\times n$ matrix (i.e. all eigenvalues of $C$ are positive). We define the Gaussian measure $\mu^n_C$ of covariance $C$ on $\mathbb{R}^n$ to be
    \begin{equation}
        d\mu^n_C(x) = (2\pi)^{-n/2}\sqrt{\text{det}\,C}\exp\left({-\frac{1}{2}x_i C^{ij}x_j}\right)d^nx, \label{eq3.82}
    \end{equation}
    where $d^nx$ is the Lebesgue measure on $\mathbb{R}^n$ and $x = (x_1, \ldots, x_n)$. For computations, we can simplify Eq~(\ref{eq3.82}) as follows. Since $C$ is a symmetric matrix, its eigenvectors span $\mathbb{R}^n$. In particular, if all eigenvalues are distinct, the eigenvectors form an orthonormal basis of $\mathbb{R}^n$. If some of the eigenvalues are the same, then one can pick an orthonormal basis of each eigenspace, and apply the Gram-Schmidt decomposition to the resulting list of vectors to again obtain an orthonormal basis of $\mathbb{R}^n$. Thus, without loss of generality, we may assume that $C$ has $n$ distinct eigenvalues $\lambda_i$ and thus $n$ distinct eigenvectors $e^i$. Then $x = x_ie^i$ for any $x \in \mathbb{R}^n$. Since $\text{det}\,C = \lambda_1\cdots\lambda_n$, we get
    \begin{equation}
        d\mu_C^n(x) = (2\pi)^{-n/2}\prod_{i=1}^{n}\sqrt{\lambda_i}e^{-\frac{\lambda_i x_i^2}{2}}d^nx. \label{eq3.83}
    \end{equation}
    
    In terms of finite-dimensional Euclidean spaces, the self-consistency conditions on the measures mean that for $m < n$, the restriction of a Gaussian measure on $\mathbb{R}^n$ must be a Gaussian measure on $\mathbb{R}^m$. Indeed, supposing that $n - m = r$, Eq~(\ref{eq3.83}) readily yields
    \begin{equation}
        d\mu^m_C(x_1, \ldots, x_m) = \int_{\mathbb{R}^r}d\mu^n_C(x_1, \ldots, x_m, x_{m+1}, \ldots, x_n). \label{eq3.84} 
    \end{equation}
    To further confirm that the equation above is indeed equivalent to the self-consistency conditions (Eqs~(\ref{eq3.54}) and (\ref{eq3.69})), one can check that if $\Tilde{L} \subset \Tilde{L}'$, where $\Tilde{L} \cong \mathbb{R}^m$ and $\Tilde{L}' \cong \mathbb{R}^n$, the integral of a function on $\Tilde{\mathcal{V}}$ that is cylindrical with respect to both the subspaces is the same regardless of whether it is evaluated using $d\mu^m_C$ or $d\mu^n_C$. Therefore, Gaussian measures on finite-dimensional subspaces of $\Tilde{\mathcal{V}}$ constitute a self-consistent family of measures. Since they are also normalised (and thus finite), Theorem 3.2 guarantees the existence of a countably additive measure $\mu$ on $\Tilde{\mathcal{V}}$. 

    To see the precise form of the measure $\mu$ on $\Tilde{\mathcal{V}}$, we will first study the Fourier transform of finite-dimensional Gaussian measures and then make use of the Bochner theorem. Since the dual of $\mathbb{R}^n$ is $\mathbb{R}^n$ itself,  using Eqs~(\ref{eq3.78}) and (\ref{eq3.82}), for $x \in \mathbb{R}^n$, we get
    \begin{align}
        \chi(x) &= \int_{\mathbb{R}^n}d\mu^n_C(x) e^{ix^iy_i} \nonumber \\
        &= (2\pi)^{-n/2}\prod_{i=1}^{n}\int_{\mathbb{R}}d^ny\sqrt{\lambda_i}\exp\left({ix^iy_i}-\frac{1}{2}\lambda_iy_i^2\right) \quad \text{(no sum over $i$!)} \nonumber\\
        &= \prod_{i=1}^{n}e^{-\frac{(x^i)^2}{2\lambda_i}}, \label{eq3.85}
    \end{align}
    where we completed the square in the exponent in the second line and subsequently evaluated the resulting Gaussian integral. Now, $C$, being symmetric, has an inverse $C^{-1}$, which has eigenvalues $1/\lambda_i$. Therefore, $C^{-1}\phi = \sum_i(x^i/\lambda_i)e^i$ and so $x\cdot C^{-1}x = \sum_i((x^i)^2/\lambda_i)$, which yields
    \begin{equation}
        \chi(x) = \exp{\left(-\frac{1}{2}x\cdot C^{-1}x\right)} \label{eq3.86}
    \end{equation}
    We can further check that
    \begin{equation}
        \chi(0) = 1, \label{eq3.87}
    \end{equation}
    confirming that Gaussian measures are normalised. Furthermore, the function defined by Eq~(\ref{eq3.86}) is continuous on $\mathbb{R}^n$, and is always non-negative due to the positivity of $C$. Therefore, the conditions of the Bochner theorem are at least met on the finite-dimensional vector space $\mathbb{R}^n$. 

    We are now in a position to conjure a Fourier transform corresponding to the measure $\mu$ on the infinite-dimensional $\Tilde{\mathcal{V}}$. We observe that because $C$ is symmetric and positive definite, it actually defines an inner product on $\mathbb{R}^n$. Indeed, armed with Eq~(\ref{eq3.86}), we can recognise that every inner product on a finite-dimensional vector space determines a Gaussian measure on its dual space. With this insight and the Bochner theorem, we can characterise the measure $\mu$ on $\Tilde{\mathcal{V}}$. Fix an inner product $C: \mathcal{V} \times \mathcal{V} \to \mathbb{R}$, which we symbolically denote as
    \begin{equation}
        C(u,v) = \langle u, Cv \rangle \label{eq3.88}
    \end{equation}
    for every $u, v \in \mathcal{V}$. Then we assert that there is a unique Gaussian measure $\mu_C$ on $\Tilde{\mathcal{V}}$ with covariance $C$. Indeed, let 
    \begin{equation}
        \chi(v) = \exp{\left( -\frac{1}{2}x\cdot Cx \right)}. \label{eq3.89}
    \end{equation}
    As has been verified, this function is continuous on every finite-dimensional subspace of $\mathcal{V}$, and is of positive type in virtue of $C$ being positive definite. The conditions of the Bochner theorem are satisfied, and thus Eq~(\ref{eq3.89}) is the Fourier transform of a unique measure $\mu_C$ on $\Tilde{\mathcal{V}}$. This reveals the strength of the Bochner theorem: it lets us define a measure on the dual space without ever having to worry about its explicit form; insofar as computations are concerned, we have the explicit form of the Fourier transform at our disposal. 

    This last point about computations using the explicit form of the Fourier transform deserves some explanation. Eq~(\ref{eq3.89}) is far from being computationally explicit. But if we know what precisely is the vector space $\mathcal{V}$, we can be more vivid. For instance, suppose $\mathcal{V}$ is the Schwartz space $\mathcal{J}$ on $\mathbb{R}^D$. Then, of course,
    \begin{equation}
        \langle u, Cv\rangle = \int d^Dxd^Dx' u(x)\mathcal{C}(x,x')v(x'), \quad \forall u,v \in \mathcal{J}. \label{eq3.90}
    \end{equation}
    The fancy $\mathcal{C}$ bears its own name: the \textit{kernel} of $C$. It will soon become important\footnote{One might as well gasp at the hint of Green functions implicit in this equation.}. 
    
\end{mdframed}

Let us summarise the mathematical apparatus we have introduced so far in the context of scalar field theory. We started by identifying $\mathcal{P}$ as the classical configuration space. Detailed considerations revealed that quantum fields cannot be satisfactorily realised as elements of $\mathcal{P}$. Thus we were lead to consider the algebraic dual $\Tilde{\mathcal{P}}$ of $\mathcal{P}$, which is the space of all linear functionals on the Schwartz space $\mathcal{J}$, and is much larger than $\mathcal{P}$. Theorems 3.2 and 3.3 help us construct countably additive measures on $\Tilde{\mathcal{P}}$.

However, while $\mathcal{P}$ is too small to be considered as the configuration space of the quantum theory, the algebraic dual $\Tilde{\mathcal{P}}$ is too large for the same purpose. This is because physically, there must be at least some continuity conditions on our quantum fields, but $\Tilde{\mathcal{P}}$, being the space of \textit{all} linear functionals on $\mathcal{J}$, provides virtually no control over its elements. Therefore, one should expect a physical configuration $\Bar{\mathcal{P}}$ space to be somewhere between $\mathcal{P}$ and $\Tilde{\mathcal{P}}$, i.e. $\mathcal{P} \subset \Bar{\mathcal{P}} \subset \Tilde{\mathcal{P}}$. For infinite-dimensional topological vector spaces, there is indeed such a space, namely the \textit{topological dual}, which is the space of all \textit{continuous} functionals on the vector space. Thus $\Bar{\mathcal{P}}$ will be the topological dual of $\mathcal{J}$. It is called the space of \textit{tempered distributions}, and continuity of linear functionals in it is meant in the sense of being continuous in the topology defined by the sequence of seminorms\footnote{A seminorm is just like a usual norm, except that it need not be positive definite.} defined by Eq~(\ref{eq3.47}).

The question that we should then ask is whether the Bochner theorem is valid also for the topological dual. Intuitively, one should expect so, for in the Bochner theorem, the Fourier transform of the measure is required to be continuous on finite-dimensional subspaces of the vector space $\mathcal{V}$. This is equivalent to saying that the Fourier transform is continuous in the topology defined by finite-dimensional subspaces of $\mathcal{V}$. Thus, it is plausible to suggest that continuity of the Fourier transform in a weaker topology on $\mathcal{V}$ should shrink the support of the corresponding measure, i.e. a smoother Fourier transform should yield a measure supported on a proper subspace of the algebraic dual $\Tilde{\mathcal{P}}$. This intuitive expectation is indeed true, and is realised in the \textit{Bochner-Minlos} theorem, which in one of its myriad incarnations is as follows \cite{jose, ashtekar 3}.

\begin{theorem}
    (Bochner-Minlos theorem) Let $\mathcal{V}$ be a real nuclear space and $\mu$ be a measure on $(\Tilde{\mathcal{V}}, \Tilde{\mathcal{B}})$. If the Fourier transform of $\mu$ is continuous in the nuclear topology, then the measure is supported on the topological dual $\Bar{\mathcal{V}} \subset \Tilde{\mathcal{V}}$. Thus, every continuous and positive-type function on $\Bar{\mathcal{V}}$ defines a measure on $(\Bar{\mathcal{V}}, \Bar{\mathcal{B}})$, where $\Bar{\mathcal{B}} = \Tilde{\mathcal{B}} \cap \Bar{\mathcal{V}}$.  
\end{theorem}

We have broken character above by referring to new concepts (nuclear space, nuclear topology, etc.) without introducing them first. But provided that we restrict attention to the spaces in scalar field theory, we shall not need the full machinery of nuclear spaces. Accordingly, we need only focus on the Schwartz space $\mathcal{J}$ and its topological dual $\Bar{\mathcal{P}}$ introduced earlier. To apply the Bochner-Milnos theorem, we need only note that the nuclear topology on $\mathcal{J}$ coincides with the topology induced by the sequence of seminorms defined by Eq~(\ref{eq3.47}) \cite{jose}. 

The example of the Gaussian measure above easily adapts to the topological dual. In particular, the continuity of the Fourier transform (Eq~(\ref{eq3.89})) in the nuclear topology can be ensured by imposing conditions on the inner product $C$ (see \cite{jose} for example). 

Thus finally, we have picked the physical configuration space for scalar quantum field theory. It is the topological dual $\Bar{\mathcal{P}}$ of the Schwartz space $\mathcal{J}$. It is larger than the classical configuration space $\mathcal{P}$ and the existence of measures on it is ensured by the Bochner-Minlos theorem. 

\subsection{Quantum field theory regained}
It might seem at this point that the goal we had set ourselves up to, namely, making the path-integral formalism of free scalar field theory more rigorous, has been buried in the bewildering haystack of definitions, theorems and mathematical constructions presented above. However, we are now precisely in a position to discuss the rigorous construction of a free quantum scalar field. 

Let us for a moment forget about path-integral and canonical quantisation (and any other textbook quantisation scheme for that matter). We instead ask ourselves the question: what are the salient physical features we desire in a quantum field theory? We recall that the observable content of a quantum field theory is encoded in its correlation functions (Eqs~(\ref{eq3.43}) and \ref{eq3.44}). The physical principles underlying this observable content are those of special relativity and quantum mechanics. Accordingly, we require that our correlation functions be Lorentz covariant, our fields be local \footnote{That is, the commutator of field operators must vanish at spacelike separations.} and the vacuum state of our theory be unique. We can thus take these conditions to be the axioms of a theory and then find mathematical objects that satisfy these axioms. For Minkowskian free scalar field theory, these axioms are called the Wightman axioms and are realised by treating quantum field as functionals in the topological dual of the Schwartz space \cite{jaffe, vincent, wightman}. For Euclidean scalar field theory, there is an analogous list of axioms, called the Osterwalder-Schrader (OS) axioms, that yield a theory which upon analytic continuation yields a theory consistent with the Wightman axioms\footnote{The question of the equivalence of the two schemes is a delicate one. In general, the OS axioms are more general than the Wightman axioms. See \cite{jaffe, vincent} for details.}. It is here that are our involved constructions above become fruitful: the OS axioms are essentially conditions on measures on the topological dual of $\mathcal{J}$ \cite{jaffe}! 

For our purposes, it is not necessary to state the OS axioms in detail. Our essential motive was to understand the need for realising fields as functionals in the topological dual of the classical configuration space rather than as functions on that space itself. The preceding sections ought to have convinced us of that need. Here, it suffices to note that the Gaussian measure $\mu_C$ defined above satisfies the OS axioms if \cite{jaffe}
\begin{subequations}
\label{eq3.91}
    \begin{align}
        C &= (-\nabla^2 + m^2)^{-1}, \\
        \mu_C &= (-\nabla^2 + m^2)^{-1/2}, \\
        \mathcal{C}(x,x') &= \frac{1}{(2\pi)^D}\int d^Dp\, \frac{e^{ip\cdot(x-x')}}{m^2+p^2}, 
    \end{align}
\end{subequations}
where $p$ is the four-momentum. These equations ought to evoke some memories! Indeed, Eq~(\ref{eq3.91}c) is nothing but the Green function of scalar field theory! In fact, what is more, one can go on to calculate the correlation functions of the fields using Eq~(\ref{eq3.82}). We find \cite{jaffe, glimm} that for any covariance $C$ and functions $f_1,\ldots, f_n \in \mathcal{J}$, 
\begin{equation}
    \int\prod_{i=1}^N\phi(f_i)d\mu_C(\phi) = \sum_{\text{all pairs}}\prod_{i=1}^{N}\langle f_{m_i}, Cf_{n_i}\rangle, \label{eq3.92} 
\end{equation}
which, when $C = (-\nabla^2 + m^2)^{-1}$, is precisely the time-honoured Wick's theorem found in any standard textbook on quantum field theory (e.g. \cite{srednicki, peskin})! Very convincingly, the integral on the left-hand side above is as well-defined as anything can be -- the mathematical spectres haunting path integrals have all been exorcised. This is but a glimpse of constructive quantum field theory; more details can be found in the sources cited in the bibliography. 

\section{Quantum Configuration Space for Gravity}
Now that we have understood that the quantisation of a field theory requires an enlargement of the classical configuration space, we can return to gravity proper, and ask the question: what kind of an enlargement of the space $\mathcal{A/G}$ of (smooth) connections modulo gauge transformations will be required in quantum gravity? It is important to note here that quotienting $\mathcal{A}$ by $\mathcal{G}$ results in a nonlinear space, and therefore, a straightforward application of the results of the preceding section will not work. That is, we do not have access to a linear vector space, which is the essential ingredient in the Bochner and Bochner-Minlos theorems. Therefore, new concepts and techniques will need to be introduced to seek a suitable enlargement of $\mathcal{A/G}$. However, in achieving this goal and appreciating the inevitability of the route taken towards it, the rather lengthy discussions of the previous section will be essential.

Although $SU(2)$ is the group that primarily concerns us, most of the subsequent discussion would only make use of the fact that $SU(2)$ is a compact Lie group. Thus from now on, $\mathcal{G}$ refers to any compact Lie group, unless otherwise stated. 

\subsection{The holonomy algebra and its completion}
Since the original motivation for enlarging the classical configuration space is to obtain a Hilbert space on which the elementary $T$ variables have to be represented as operators, we proceed by harking back to the structure of the algebra of the $T$ variables. To this end, let us first focus on the $T^0$ variables. 

Let $\mathcal{HA}$ be the algebra generated by finite linear combinations of the $T^0$ variables with complex coefficients. We call this the \textit{holonomy algebra}. As demonstrated in Section 3.2.3, the Poisson brackets between the $T^0$ variables vanish -- thus $\mathcal{HA}$ is abelian. Furthermore, we also have closure under complex conjugation, and this implies that complex conjugation induces an involution $\star$ on $\mathcal{HA}$. In other words, $\mathcal{HA}$ is automatically a $\star$-algebra. Finally, we invoke the so-called \textit{Giles theorem} \cite{giles}. It states that one can construct any complex-valued, continuous, bounded function on $\mathcal{A/G}$ from the traces of holonomies. Therefore, stated precisely, $\mathcal{HA}$ is the abelian $\star$-subalgebra of the algebra of the set $C^0(\mathcal{A/G})$ of continuous and bounded complex-valued functions on $\mathcal{A/G}$.

We wish to find a representation of $\mathcal{HA}$ on a Hilbert space. Such a question is generally considered and answered in the representation theory of something called $C^{\star}$-algebras\footnote{$C^\star$-algebras are not only extremely powerful mathematical tools, but their use in physics offers a number of illuminating insights regarding otherwise obscure physical theories. For instance, the seemingly counterintuitive principles that underlie quantum mechanics, such as the uncertainty relation, the superposition principle, and so on, can be recast in a rather palatable and tangible form using $C^\star$-alebraic tools. See, for instance, Ref. \cite{gleason, strocchi}.}, studied, among others, by von Neumann, Gelfand, Mazur and Segal \cite{averson}. It will be worthwhile to introduce a bare minimum of concepts required to understand the results from the theory of $C^{\star}$-algebras that we will have to inevitably employ below. In what follows, all algebras will be assumed to be unital, i.e. containing the identity element. 

Let us first recall what a Banach space is, namely, a complete normed vector space, where by completeness we mean that every Cauchy sequence in the space converges to a point in the space, and the definition of convergence, of course, uses the metric induced by the norm. This allows us to introduce a 

\begin{definition}
    (Banach algebra) An associative algebra $\mathcal{K}$ that is also a Banach space and satisfies $||xy|| \leq ||x||\, ||y|| $ for every $x, y \in \mathcal{K}$, $||\cdot||$ being the norm on $\mathcal{K}$, 
\end{definition}
and thence a 
\begin{definition}
    ($C^{\star}$-algebra) A $\star$-algebra $\mathcal{K}$ that is also a Banach algebra and satisfies $||x^{\star}x|| = ||x||^2$ for every $x \in \mathcal{K}$. 
\end{definition}
Next, recall that a homomorphism is a map between two algebras that preserves vector addition and multiplication, scalar multiplication and the identity element, and that an isomorphism is a bijective homomorphism. If the algebras are $\star$-algebras, and we further impose the condition of $\star$-preservation, i.e. $\phi: \mathcal{K}_1 \to \mathcal{K}_2$ such that $\phi(x^{\star}) = \phi(x)^{\star}$, then the map is a $\star$-homomorphism or $\star$-isomorphism, as the case may be. Next, a homomorphism from an algebra $\mathcal{K}$ to linear operators on a Hilbert space $\mathcal{H}$ is called a representation of $\mathcal{K}$ on $\mathcal{H}$, and a $\star$-representation if the algebra is a $\star$-algebra and the map is $\star$-preserving. Finally, we introduce the notion of a cyclic representation. 

\begin{definition}
    (Cyclic representation) Given a $C^{\star}$-algebra $\mathcal{K}$ and a Hilbert space $\mathcal{H}$, a representation $\pi: \mathcal{K}\to B(\mathcal{H})$, where $B(\mathcal{H})$ is the set of bounded linear operators on $\mathcal{H}$, is called cyclic if there exists a cyclic vector, namely, a unit vector $v \in \mathcal{H}$  such that the set $\{\pi(x)v: x \in \mathcal{K}\}$ is dense\footnote{A subset is dense if its closure, or equivalently Cauchy completion, equals the parent set. Recall the example of rationals as a dense subset of reals.} in $\mathcal{H}$. 
\end{definition}

With this, we can introduce the following theorems \cite{averson, ashtekar 5, ashtekar 6}. 

\begin{theorem}
    (Gelfand-Naimark theorem) Every $C^{\star}$-algebra $\mathcal{K}$ is $\star$-isomorphic to the $C^{\star}$-algebra of all continuous bounded functions on a compact Hausdorff space called the \text{Gelfand spectrum} of $\mathcal{K}$. Furthermore, the Gelfand spectrum can be constructed directly from $\mathcal{K}$: it is the set $Hom(\mathcal{K}, \mathbb{C})$ of all $\star$-homomorphisms from $\mathcal{K}$ to the $\star$-algebra of complex numbers.   
\end{theorem}

\begin{theorem}
    (Gelfand-Naimark-Segal (GNS) construction) Let $\mathcal{K}$ be a $C^{\star}$-algebra and $\mathcal{H}$ a Hilbert space. Given a positive linear functional $\Gamma$ of unit norm on $\mathcal{K}$, there exists a cyclic representation $\pi: \mathcal{K}\to\mathcal{H}$ with a cyclic vector $\Omega$ such that 
    \begin{equation}
        \Gamma(x) = \langle \pi(x)\Omega, \Omega\rangle \label{eq3.93}
    \end{equation}
    for all $x \in \mathcal{K}$. 
\end{theorem}

\begin{theorem}
    (Riesz-Markov Theorem) Let $\mathcal{X}$ be a compact Hausdorff space and $C^0(\mathcal{X})$ the space of continuous bounded (complex-valued) functions on it. For every positive linear functional $\Psi$ on $\mathcal{X}$, there exists a unique regular Borel measure $\mu$ on $\mathcal{X}$ such that 
    \begin{equation}
        \Psi(f) = \int_{\mathcal{X}}d\mu(x)f(x) \label{eq3.94}
    \end{equation}
    for all $f \in C^0(\mathcal{X})$. 
\end{theorem}

To witness the power of these theorems, we need to convert the holonomy algebra $\mathcal{HA}$ into a $C^{\star}$-algebra. To this end, let us introduce a norm on $\mathcal{HA}$. We set 
\begin{equation}
    ||f|| = \sup_{[A]\in\mathcal{A/G}}|f([A])| \label{eq3.95}
\end{equation}
for all $f \in \mathcal{HA}$; $[A]$ is an equivalence class of connections that give rise to the same $T^0$ variable for all loops. The Cauchy completion $\overline{\mathcal{HA}}$ of $\mathcal{HA}$ with respect to this norm gives rise to an abelian $C^{\star}$-algebra \footnote{Here is a sketch of the proof. Recall that a Cauchy completion $\overline{M}$ of a normed space $M$ proceeds as follows. Two Cauchy sequences $(x_n), (y_n) \in M$ are said to be equivalent if $||x_n - y_n|| \to 0$. $\overline{M}$ is identified with the equivalence classes $[x_n]$ of all Cauchy sequences $(x_n)$ in $M$, and the norm on the latter induces a norm on the former, namely $||[x_n]||_{\overline{M}} := \lim_{n\to\infty} ||x_n||$. Now by Eq~(\ref{eq3.94}) and the properties of the absolute value function on complex numbers, it is easily verified that $\overline{\mathcal{HA}}$ is a $C^{\star}$-algebra. Note that the identity element in the algebra is simply the trace of a holonomy around the identity loop.}. 

Now let $\overline{\mathcal{A/G}}$ denote the Gelfand spectrum of $\overline{\mathcal{HA}}$. By the Gelfand-Naimark theorem, $\overline{\mathcal{HA}}$ is $\star$-isomorphic to the $C^\star$-algebra of $C^0(\overline{\mathcal{A/G}})$, i.e the set of continuous bounded functions on the Gelfand spectrum. The punchline is this. 
\begin{enumerate}[(1)]
    \item The theorems stated above allow us to construct a measure $\mu$ on $\overline{\mathcal{A/G}}$ and thus convert $C^0(\overline{\mathcal{A/G}})$ into the Hilbert space $L^2(\overline{\mathcal{A/G}}, \mu)$, which will be the home for general quantum states and the representation space for configuration operators. Hence, $\overline{\mathcal{A/G}}$ is the promised quantum configuration space.

    \item The classical configuration space $\mathcal{A/G}$ is densely embedded in $\overline{\mathcal{A/G}}$, which is, therefore, an enlargement of the former, as advertised.

    \item This enlargement is nontrivial, for it can be shown \cite{marolf} that $\mathcal{A/G}$ has measure zero, and in analogy with the classical configuration space of scalar field theory, additive measures on $\mathcal{A/G}$ cannot be extended to countably additive measures on $\overline{\mathcal{A/G}}$.
\end{enumerate}

We will now precisely see how Theorems 3.5--3.7 help us establish (1) and (2). The significantly involved mathematical constructions that are to follow are inevitable in view of the foregoing observations. 

\begin{mdframed}[style=testframe]
    \textbf{(2) $\mathcal{A/G}$ is densely embedded \cite{rendall} in $\overline{\mathcal{A/G}}$}

    For every $A \in \mathcal{A/G}$, define $\phi_A: \overline{\mathcal{HA}} \to \mathbb{C}$ such that $\phi_A(f) = f(A)$ for every $f \in \overline{\mathcal{HA}}$. As can be readily confirmed, every $\phi_A$ is a $\star$-homomorphism from $\overline{\mathcal{HA}}$ to the $\star$-algebra of $\mathbb{C}$ and thus, by the Gelfand-Naimark theorem, lies in the Gelfand spectrum $\overline{\mathcal{A/G}}$ of $\overline{\mathcal{HA}}$. 

    Define $j: \mathcal{A/G} \to \overline{\mathcal{A/G}}$ to be $j(A) = \phi_A$ for every $A \in \mathcal{A/G}$. Now given $A_1 \neq A_2$ in $\mathcal{A/G}$, there exists a loop $\gamma$ such that $T^0_\gamma(A_1) \neq T^0_\gamma(A_2)$. This is so because recall that $\mathcal{A/G}$ contains gauge equivalence classes of smooth connections, and gauge-equivalent connections give rise to a unique $T^0$ variable for any loop. Since the $T^0$ variables span $\overline{\mathcal{H/A}}$, it follows that $\phi_{A_1}(f) \neq \phi_{A_2}(f)$ for any $f$ in $\overline{\mathcal{H/A}}$. Thus $j$ is an injection, which entails that $\mathcal{A/G}$ is isomorphic to the image $j(\mathcal{A/G})$; in other words, $\mathcal{A/G} \subset \overline{\mathcal{A/G}}$. It will then suffice to show that $j(X)$ is dense in $\overline{\mathcal{A/G}}$.

    To this end, suppose $f$ is a continuous function on $\overline{\mathcal{A/G}}$ that vanishes on $j(\mathcal{A/G})$. By the Gelfand-Naimark theorem, there exists an isomorphism $I$ taking $f$ to some element $\Tilde{f}$ in $\overline{\mathcal{HA}}$. We thus have
    \begin{equation*}
        f(\phi_A) = I^{-1}(\Tilde{f})(\phi_A) = 0
    \end{equation*}
    for every $\phi_A = j(\mathcal{A/G})$. This means that $\Tilde{f}$ is vanishes everywhere on $\mathcal{A/G}$, and so $f$ is identically zero as well. 

    We have thus shown that every vanishing function on $j(\mathcal{A/G})$ vanishes on $\overline{\mathcal{A/G}}$ as well. This ensures that $j(\mathcal{A/G})$ is dense in $\overline{\mathcal{A/G}}$, for suppose that if it were not. Then since $\overline{\mathcal{A/G}}$ is compact Hausdorff (Gelfand-Naimark theorem), there would exist a continuous function on $\overline{\mathcal{A/G}}$ that vanished on the closure of $j(\mathcal{A/G})$ in $\overline{\mathcal{A/G}}$, but was nonzero on the complement of the closure. 

    \textbf{(1) Converting $C^0(\overline{\mathcal{A/G}})$ into a Hilbert space}

    Once again, we invoke the Gelfand-Naimark theorem. It guarantees that $\overline{\mathcal{A/G}}$ is compact. Thus by the Riesz-Markov theorem, there exists a unique regular Borel measure $\mu$ on $\overline{\mathcal{A/G}}$. Using this measure, we define an inner product on $C^0(\overline{\mathcal{A/G}})$,
    \begin{equation}
        \langle g, f \rangle := \int_{\overline{\mathcal{A/G}}}d\mu(A)\overline{g}(A)f(A),  \label{eq3.96}
    \end{equation}
    which in turn yields an $L^2$ norm $||f|| := \int d\mu |f|^2$. The Cauchy completion of $C^0(\overline{\mathcal{A/G}})$ with respect to this norm begets the Hilbert space $L^2(\overline{\mathcal{A/G}}, \mu)$. 

    Finally, we show that elements in $\overline{\mathcal{HA}}$ can be realised as bounded linear operators on $L^2(\overline{\mathcal{A/G}}, \mu)$. Now the GNS construction in Theorem 3.6 comes into play. It ensures that there exists a cyclic representation $\pi: \overline{\mathcal{A/G}} \to B(L^2(\overline{\mathcal{A/G}}, \mu))$ for every positive linear functional $\Gamma$ on $\overline{\mathcal{HA}}$ such that
    \begin{equation*}
        \Gamma(f) = \langle \pi(f)\Omega, \Omega \rangle
    \end{equation*}
    for any $f \in \overline{\mathcal{HA}}$, $\Omega$ being the cyclic vector. But since $\overline{\mathcal{HA}}$ is isomorphic to $C^0(\overline{\mathcal{A/G}})$ by the Gelfand-Naimark theorem, every positive linear functional on $\overline{\mathcal{HA}}$ can also be regarded as a positive linear functional on $C^0(\overline{\mathcal{A/G}})$. Now, by the Riesz-Markov theorem, positive linear functionals on $C^0(\overline{\mathcal{A/G}})$ are in one-to-one correspondence with regular Borel measures on $\overline{\mathcal{A/G}}$. This entails that
    \begin{equation}
        \Gamma(f) = \int_{\overline{\mathcal{A/G}}}d\mu(A)\Tilde{f}(A) \label{eq3.97}
    \end{equation}
    where $\Tilde{f}$ is the image of $f$ under the $\star$-isomorphism of the Gelfand-Naimark theorem. Comparing the preceding two equations, we see that for every $\psi\in L^2(\overline{\mathcal{A/G}}, \mu)$,
    \begin{equation}
        (\pi(f)\psi)(A) = \Tilde{f}(A)\psi(A). \label{eq3.98} 
    \end{equation}
    In other words, in every cyclic representation of the configuration operators, quantum states can be realised as square-integrable functions on $\overline{\mathcal{A/G}}$ on which the operators act by multiplication, as they should. This also entails that the configuration operators are self-adjoint, as they should be. 
\end{mdframed} 
  
To summarise, we wished to enlarge the classical configuration space $\mathcal{A/G}$ of smooth connections modulo gauge transformations. In order to look for the required enlargement, we studied the algebra of configuration variables. This we found was naturally extendible to an abelian $C^\star$-algebra, which in turn was amenable to the powerful results of Gelfand representation theory, which allowed us to enlarge $\mathcal{A/G}$ and thereby arrive at a suitable Hilbert space. 

Much remains to be done, however. In particular, we need to find the representation of the algebra of the momentum variables on $L^2(\overline{\mathcal{A/G}}, \mu)$ as well. Moreover, the measure $\mu$ so far is at most a mathematical curiosity without any physical significance. Physically, we require a measure that is invariant under diffeomorphisms of the spatial manifold $\Sigma$, since a change of coordinates should not change the Hilbert space of quantum states\footnote{This has nothing to do with solving the diffeomorphism constraint.}. To cater to the first concern, it is essential to develop differential calculus on $\overline{\mathcal{A/G}}$, whereas the second concern requires a proper understanding of functional integration on $\overline{\mathcal{A/G}}$. Either way, it is indispensable to broach the structure of $\overline{\mathcal{A/G}}$. This is the task we take up next.

\subsection{Characterisation of \texorpdfstring{$\overline{\mathcal{A/G}}$}{TEXT}}
There are a number of equivalent characterisations of $\overline{\mathcal{A/G}}$ in terms of projective limits. All have been explored and described in detail in the literature on loop quantum gravity \cite{ashtekar 2, ashtekar 3, ashtekar 4, ashtekar 5, ashtekar 6, ashtekar 7} (see \cite{ashtekar 5} in particular for a concise and relatively self-contained summary). To get a flavour for these rather lengthy constructions, we will describe one such characterisation \cite{ashtekar 7} in this section. It will be important in that it will allow us to construct a diffeomorphism-invariant measure on $\overline{\mathcal{A/G}}$ and represent the $T^1$ variables as ``momentum'' operators on $L^2(\overline{\mathcal{A/G}}, \mu)$.  

We have seen that $\mathcal{A/G}$ is a dense subset of $\overline{\mathcal{A/G}}$. Now, the former consists of \textit{smooth} connections. This suggests that $\overline{\mathcal{A/G}}$ contains connections that are not smooth, over and above those that are. This is reminiscent of the situation in scalar field theory, for instance. There, the classical configuration space comprised of smooth functions on spacetime, whereas the quantum configuration space consisted of distributions, which may not be smooth. 

These observations provide hints as to how $\overline{\mathcal{A/G}}$ may be realised as a projective limit. We begin by making precise the notion of a general connection which may or may not be smooth. For this purpose, it is instructive to abstract the properties one wishes to see in an object that can regarded as a connection. Recall that \textit{smooth} connections serve to parallel transport sections on (local regions of) $\Sigma \times \mathbb{R}^3$. Harking back to our excursion into parallel propagators in Section 3.2.1, we see that \textit{smooth} connections can be completely characterised as $SU(2)$-valued functions on $\Sigma$ that satisfy Eqs~(\ref{eq3.9}a, b, g, h), where the second two equations are smoothness conditions on the functions. Armed with this insight, we can define a generalised connection as a map from curves in $\Sigma$ to $SU(2)$ that satisfies only Eqs~(\ref{eq3.9}a, b). This strategy is essentially correct, as we will see. But before that, we need to introduce some definitions and notation.

\begin{figure}[ht]
    \centering

\tikzset{every picture/.style={line width=0.75pt}} 

\begin{tikzpicture}[x=0.75pt,y=0.75pt,yscale=-1,xscale=1]

\draw    (326,120) .. controls (353,135) and (374,90) .. (422,107) ;
\draw [shift={(422,107)}, rotate = 19.5] [color={rgb, 255:red, 0; green, 0; blue, 0 }  ][fill={rgb, 255:red, 0; green, 0; blue, 0 }  ][line width=0.75]      (0, 0) circle [x radius= 3.35, y radius= 3.35]   ;
\draw [shift={(378.63,107.23)}, rotate = 158.07] [color={rgb, 255:red, 0; green, 0; blue, 0 }  ][line width=0.75]    (10.93,-4.9) .. controls (6.95,-2.3) and (3.31,-0.67) .. (0,0) .. controls (3.31,0.67) and (6.95,2.3) .. (10.93,4.9)   ;
\draw [shift={(326,120)}, rotate = 29.05] [color={rgb, 255:red, 0; green, 0; blue, 0 }  ][fill={rgb, 255:red, 0; green, 0; blue, 0 }  ][line width=0.75]      (0, 0) circle [x radius= 3.35, y radius= 3.35]   ;
\draw    (422,107) .. controls (477,107) and (441,149) .. (507,118) ;
\draw [shift={(507,118)}, rotate = 334.84] [color={rgb, 255:red, 0; green, 0; blue, 0 }  ][fill={rgb, 255:red, 0; green, 0; blue, 0 }  ][line width=0.75]      (0, 0) circle [x radius= 3.35, y radius= 3.35]   ;
\draw [shift={(468.05,128.97)}, rotate = 206.33] [color={rgb, 255:red, 0; green, 0; blue, 0 }  ][line width=0.75]    (10.93,-4.9) .. controls (6.95,-2.3) and (3.31,-0.67) .. (0,0) .. controls (3.31,0.67) and (6.95,2.3) .. (10.93,4.9)   ;
\draw [shift={(422,107)}, rotate = 0] [color={rgb, 255:red, 0; green, 0; blue, 0 }  ][fill={rgb, 255:red, 0; green, 0; blue, 0 }  ][line width=0.75]      (0, 0) circle [x radius= 3.35, y radius= 3.35]   ;
\draw    (507,118) .. controls (509,85) and (453,62) .. (422,107) ;
\draw [shift={(422,107)}, rotate = 124.56] [color={rgb, 255:red, 0; green, 0; blue, 0 }  ][fill={rgb, 255:red, 0; green, 0; blue, 0 }  ][line width=0.75]      (0, 0) circle [x radius= 3.35, y radius= 3.35]   ;
\draw [shift={(463.62,81.98)}, rotate = 1.18] [color={rgb, 255:red, 0; green, 0; blue, 0 }  ][line width=0.75]    (10.93,-4.9) .. controls (6.95,-2.3) and (3.31,-0.67) .. (0,0) .. controls (3.31,0.67) and (6.95,2.3) .. (10.93,4.9)   ;
\draw [shift={(507,118)}, rotate = 273.47] [color={rgb, 255:red, 0; green, 0; blue, 0 }  ][fill={rgb, 255:red, 0; green, 0; blue, 0 }  ][line width=0.75]      (0, 0) circle [x radius= 3.35, y radius= 3.35]   ;
\draw    (388,166) .. controls (428,136) and (382,137) .. (422,107) ;
\draw [shift={(422,107)}, rotate = 323.13] [color={rgb, 255:red, 0; green, 0; blue, 0 }  ][fill={rgb, 255:red, 0; green, 0; blue, 0 }  ][line width=0.75]      (0, 0) circle [x radius= 3.35, y radius= 3.35]   ;
\draw [shift={(405.77,143.23)}, rotate = 264.25] [color={rgb, 255:red, 0; green, 0; blue, 0 }  ][line width=0.75]    (10.93,-4.9) .. controls (6.95,-2.3) and (3.31,-0.67) .. (0,0) .. controls (3.31,0.67) and (6.95,2.3) .. (10.93,4.9)   ;
\draw [shift={(388,166)}, rotate = 323.13] [color={rgb, 255:red, 0; green, 0; blue, 0 }  ][fill={rgb, 255:red, 0; green, 0; blue, 0 }  ][line width=0.75]      (0, 0) circle [x radius= 3.35, y radius= 3.35]   ;
\draw    (288,166) .. controls (315,181) and (345,143) .. (388,166) ;
\draw [shift={(388,166)}, rotate = 28.14] [color={rgb, 255:red, 0; green, 0; blue, 0 }  ][fill={rgb, 255:red, 0; green, 0; blue, 0 }  ][line width=0.75]      (0, 0) circle [x radius= 3.35, y radius= 3.35]   ;
\draw [shift={(331.51,163.14)}, rotate = 344.39] [color={rgb, 255:red, 0; green, 0; blue, 0 }  ][line width=0.75]    (10.93,-4.9) .. controls (6.95,-2.3) and (3.31,-0.67) .. (0,0) .. controls (3.31,0.67) and (6.95,2.3) .. (10.93,4.9)   ;
\draw [shift={(288,166)}, rotate = 29.05] [color={rgb, 255:red, 0; green, 0; blue, 0 }  ][fill={rgb, 255:red, 0; green, 0; blue, 0 }  ][line width=0.75]      (0, 0) circle [x radius= 3.35, y radius= 3.35]   ;
\draw    (203,155) .. controls (258,155) and (222,197) .. (288,166) ;
\draw [shift={(288,166)}, rotate = 334.84] [color={rgb, 255:red, 0; green, 0; blue, 0 }  ][fill={rgb, 255:red, 0; green, 0; blue, 0 }  ][line width=0.75]      (0, 0) circle [x radius= 3.35, y radius= 3.35]   ;
\draw [shift={(239,169.38)}, rotate = 47.65] [color={rgb, 255:red, 0; green, 0; blue, 0 }  ][line width=0.75]    (10.93,-4.9) .. controls (6.95,-2.3) and (3.31,-0.67) .. (0,0) .. controls (3.31,0.67) and (6.95,2.3) .. (10.93,4.9)   ;
\draw [shift={(203,155)}, rotate = 0] [color={rgb, 255:red, 0; green, 0; blue, 0 }  ][fill={rgb, 255:red, 0; green, 0; blue, 0 }  ][line width=0.75]      (0, 0) circle [x radius= 3.35, y radius= 3.35]   ;
\draw    (203,155) .. controls (243,125) and (197,126) .. (237,96) ;
\draw [shift={(237,96)}, rotate = 323.13] [color={rgb, 255:red, 0; green, 0; blue, 0 }  ][fill={rgb, 255:red, 0; green, 0; blue, 0 }  ][line width=0.75]      (0, 0) circle [x radius= 3.35, y radius= 3.35]   ;
\draw [shift={(219.21,119.47)}, rotate = 82.67] [color={rgb, 255:red, 0; green, 0; blue, 0 }  ][line width=0.75]    (10.93,-4.9) .. controls (6.95,-2.3) and (3.31,-0.67) .. (0,0) .. controls (3.31,0.67) and (6.95,2.3) .. (10.93,4.9)   ;
\draw [shift={(203,155)}, rotate = 323.13] [color={rgb, 255:red, 0; green, 0; blue, 0 }  ][fill={rgb, 255:red, 0; green, 0; blue, 0 }  ][line width=0.75]      (0, 0) circle [x radius= 3.35, y radius= 3.35]   ;
\draw    (323,106) .. controls (325,73) and (268,51) .. (237,96) ;
\draw [shift={(237,96)}, rotate = 124.56] [color={rgb, 255:red, 0; green, 0; blue, 0 }  ][fill={rgb, 255:red, 0; green, 0; blue, 0 }  ][line width=0.75]      (0, 0) circle [x radius= 3.35, y radius= 3.35]   ;
\draw [shift={(291.84,71.63)}, rotate = 188.36] [color={rgb, 255:red, 0; green, 0; blue, 0 }  ][line width=0.75]    (10.93,-4.9) .. controls (6.95,-2.3) and (3.31,-0.67) .. (0,0) .. controls (3.31,0.67) and (6.95,2.3) .. (10.93,4.9)   ;
\draw [shift={(323,106)}, rotate = 273.47] [color={rgb, 255:red, 0; green, 0; blue, 0 }  ][fill={rgb, 255:red, 0; green, 0; blue, 0 }  ][line width=0.75]      (0, 0) circle [x radius= 3.35, y radius= 3.35]   ;

\draw (373,80.4) node [anchor=north west][inner sep=0.75pt]    {$e_{1}$};
\draw (463,105.4) node [anchor=north west][inner sep=0.75pt]    {$e_{2}$};
\draw (470,62.4) node [anchor=north west][inner sep=0.75pt]    {$e_{3}$};
\draw (386,131.4) node [anchor=north west][inner sep=0.75pt]    {$e_{4}$};
\draw (335,163.4) node [anchor=north west][inner sep=0.75pt]    {$e_{1}^{-1}$};
\draw (259,45.4) node [anchor=north west][inner sep=0.75pt]    {$e_{3}^{-1}$};
\draw (189,120.4) node [anchor=north west][inner sep=0.75pt]    {$e_{4}^{-1}$};
\draw (216,175.4) node [anchor=north west][inner sep=0.75pt]    {$e_{2}^{-1}$};

\end{tikzpicture}
    \caption{An oriented graph with four edges.}
    \label{fig3.6}
\end{figure}
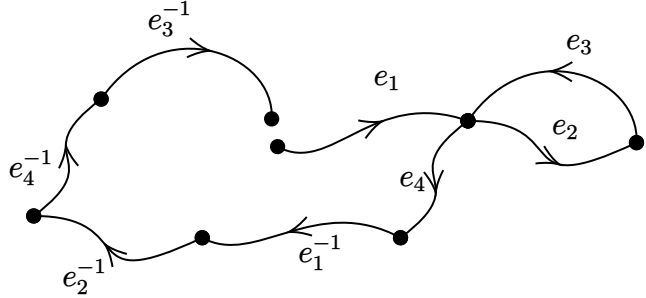

Two analytic\footnote{We had hinted earlier that analyticity is necessary in our constructions. We will precisely see in this section why that is so.} curves $e_1, e_2: [a,b] \to \Sigma$ are said to be equivalent if they are related by a reparametrisation. The equivalence class of all such curves will be called an \textit{analytic edge}, and an \textit{oriented} analytic edge if we demand the stronger condition that reparametrisations must be orientation-preserving. For notational simplicity, we shall use the same label $e$ for an equivalence class and its representatives. The composition of two edges, denoted by $e_1e_2$, is simply the composition of two representatives in each edge (such an operation is evidently well-defined).  The endpoints of an edge $e$ are called its vertices, and the set of all oriented analytic edges will be denoted by $\mathcal{E}$. Finally, a (oriented) \textit{graph} $\gamma$ is a finite subset of $\mathcal{E}$ containing (oriented) analytic edges satisfying the following properties (see Fig \ref{fig3.6}):
\begin{enumerate}[(i)]
    \item if $e \in \gamma$, then $e^{-1} \in \gamma$, $e^{-1}$ being the same edge but with opposite orientation;
    \item if $e_1 \neq e_2$ and $e_1 \neq e_2^{-1}$, then $e_1 \cap e_2$ is contained in the set of vertices of $e_1, e_2$;
    \item each vertex of an edge in $\gamma$ must be connected to a vertex of another edge in $\gamma$.
\end{enumerate}

The set of all graphs in $\Sigma$ will be denoted by $L$. We show that it is a directed set. To this end, we decree that given two graphs $\gamma', \gamma$,
\begin{equation}
    \gamma' \geq \gamma \label{eq3.99}
\end{equation}
if $\gamma'$ contains all the vertices of $\gamma$ and each edge in $\gamma$ can be written as composition of edges in $\gamma'$. Clearly, this is a partial order. Furthermore, since each edge is analytic, any two graphs $\gamma_1$ and $\gamma_2$ intersect at most at finitely many points\footnote{It is here that analyticity is essential. If we relax it to smoothness of edges, $L$ will fail to be a directed set, for smooth edges may intersect infinitely many times.}. Break the edges of the two graphs at these intersection points, and define $\gamma$ to be the graph containing the new edges thus obtained plus the nonintersecting edges of $\gamma_1$ and $\gamma_2$. Evidently, $\gamma \geq \gamma_1$ and $\gamma \geq \gamma_2$. Thus $L$ is a directed set.

Next, we introduce the space $\overline{\mathcal{A}}$ of \textit{generalised} connections on $\Sigma$ as the space of all maps $A: \mathcal{E}\to SU(2)$ from the set of all oriented analytic edges to the gauge group, such that
\begin{equation}
    A(e^{-1}) = [A(e)]^{-1} \quad \text{and} \quad A(e_1e_2) = A(e_1)A(e_2), \label{eq3.100}
\end{equation}
where $A(e_1e_2)$ is defined only if the composition $e_1e_2$ is another edge (i.e. is analytic). Eq~(\ref{eq3.100}) defines a general connection that may not be smooth. This is so because one could identify connections with the set of all parallel propagators associated to an edge in $\Sigma$. Now parallel propagators for smooth connections are essentially smooth maps from an edge to $SU(2)$ that satisfy Eq~(\ref{eq3.100}) or Eq~(\ref{eq3.9}a, b). Thus they can be regarded as a subset of all maps from an edge to $SU(2)$; such maps, by definition, are a subset of $\overline{\mathcal{A}}$. 

$\overline{\mathcal{A}}$ itself is essentially homemorhpic to a subset of infinitely many copies of $SU(2)$. To see this, for each edge $e \in \mathcal{E}$, let $\mathcal{A}_e$ denote the set of all maps from the one-point set $\{e\}$ to $SU(2)$. Clearly, $\overline{\mathcal{A}}\subset \prod_{e\in\mathcal{E}}\mathcal{A}_e$. However, $\mathcal{A}_e = \{e\}\times SU(2)$, i.e. $\mathcal{A}_e$ and $SU(2)$ are homeomorphic, hence topologically equivalent. This entails that
\begin{equation}
    \overline{\mathcal{A}} \sqsubset \prod_{e\in\mathcal{E}}SU(2), \label{eq3.101}
\end{equation}
where $\sqsubset$ denotes a topological embedding. 

Associated to $\overline{\mathcal{A}}$ we define a \textit{generalised} group $\overline{\mathcal{G}}$ of gauge transformations as the set of all maps $g: \Sigma\to SU(2)$, or equivalently as the Cartesian product
\begin{equation}
    \overline{\mathcal{G}} := \prod_{x\in\Sigma} SU(2). \label{eq3.102}
\end{equation}
For every $A \in \overline{\mathcal{A}}$, we define the action of a $g \in \overline{\mathcal{G}}$ on $A$ via
\begin{equation}
    [g(A)](e_{yx}) = (g_{y})^{-1}A(e_{yx})g_{x}, \label{eq3.103}
\end{equation}
where $e_{yx}$ is an edge from $x$ to $y$ in $\Sigma$, and $g_x$ is the group element assigned to $x$ by the map $g$. 

We will now show that the spaces $\overline{\mathcal{A}}$ and $\overline{\mathcal{G}}$ can be regarded as projective limits of some projective families. Given a graph $\gamma$, define $\mathcal{A}_{\gamma}$ to be the set of all maps $A_{\gamma}: \gamma \to SU(2)$ satisfying Eq~(\ref{eq3.100}). Clearly, $\mathcal{A}_{\gamma} \subset \overline{\mathcal{A}}$, and 
\begin{equation}
    \overline{\mathcal{A}} \subset \prod_{\gamma\in L}\mathcal{A}_\gamma. \label{eq3.104}
\end{equation}
Thus there exists a natural projection $p_{\gamma}: \overline{\mathcal{A}} \to \mathcal{A}_{\gamma}$ which sends each $A \in \overline{\mathcal{A}}$ to some $A_{\gamma} \in \mathcal{A}_{\gamma}$. By construction, this projection is surjective. Similarly, for each graph $\gamma$, we define 
\begin{equation}
    \mathcal{G}_{\gamma} = \{g_{\gamma}| g_\gamma: V_\gamma \to SU(2)\}, \label{eq3.105}
\end{equation}
where $V_\gamma$ is the set of vertices in $\gamma$. Again, we clearly have $\mathcal{G}_\gamma \subset \overline{\mathcal{G}} \subset \prod_{\gamma\in L}\mathcal{G}_\gamma$, and so there is a natural surjective projection\footnote{We use the same symbols for both projections since both are amenable to exactly the same manipulations and there is no risk of confusion.} $p_\gamma: \overline{\mathcal{G}}\to\mathcal{G}_\gamma$ that restricts $g \in \overline{\mathcal{G}}$ to some $g_\gamma \in \mathcal{G}_\gamma$. These projections allow us to render $\{\mathcal{A}_\gamma\}_{\gamma\in L}$ and $\{\mathcal{G}_\gamma\}_{\gamma\in L}$ into projective families whose projective limits are $\overline{\mathcal{A}}$ and $\overline{\mathcal{G}}$, respectively. We demonstrate below the essential steps for $\overline{\mathcal{A}}$; $\overline{\mathcal{G}}$ admits exactly the same treatment.

For any two graphs $\gamma' \geq \gamma$, define the map $p_{\gamma\gamma'}: \mathcal{A}_{\gamma'} \to \mathcal{A}_\gamma$ by 
\begin{equation}
    p_\gamma = p_{\gamma\gamma'}\circ p_{\gamma'} \label{eq3.106}
\end{equation}
This equation is precisely the condition that a projective limit should satisfy (see Eqs~(\ref{eq3.62}) and (\ref{eq3.67})). Thus it suffices to show that the maps $p_{\gamma\gamma'}$ are surjective and satisfy Eq~(\ref{eq3.60}), so that $\{\mathcal{A}_\gamma, p_{\gamma\gamma'}\}$ is a projective family. Surjectivity follows from the surjectivity of the projections $p_\gamma$, for let $A_\gamma \in \mathcal{A}_\gamma$. Then since $p_\gamma$ is surjective, there exists $A \in \overline{\mathcal{A}}$ such that $p_\gamma(A) = A_\gamma$. But then Eq~(\ref{eq3.105}) implies that $p_{\gamma}(A) = A_\gamma = p_{\gamma\gamma'}(p_{\gamma'}(A)) = p_{\gamma\gamma'}(A_{\gamma'})$ for some $A_{\gamma'} \in \mathcal{A}_{\gamma'}$. To verify Eq~(\ref{eq3.60}), pick three graphs $\gamma'' \geq \gamma' \geq \gamma$. From Eq~(\ref{eq3.105}), $p_{\gamma\gamma''}\circ p_{\gamma''} = p_{\gamma} = p_{\gamma\gamma'}\circ p_{\gamma'} = p_{\gamma\gamma'}\circ p_{\gamma'\gamma''}\circ p_{\gamma''}$, and so $p_{\gamma\gamma''} = p_{\gamma\gamma'}\circ p_{\gamma'\gamma''}$.

It is instructive to observe the structure of the spaces $\mathcal{A}_\gamma$ and $\mathcal{G}_\gamma$. From Eqs~(\ref{eq3.101}) and (\ref{eq3.102}), we see that $\mathcal{A}_\gamma \sqsubset \prod_{e\in\gamma}SU(2)$ and $\mathcal{G}_\gamma = \prod_{x\in V_\gamma}SU(2)$.
Since $\gamma$ and $V_\gamma$ are finite sets, these products contain finitely many copies of the three-dimensional group $SU(2)$. Therefore, the members of our projective family are finite-dimensional spaces. Hence, if $n$ is the number of edges in $\gamma$, and for each edge, we identify its two vertices with a two-point set, we may write
\begin{equation*}
    \mathcal{A}_\gamma \sqsubset SU(2)^n \quad \text{and} \quad \mathcal{G}_\gamma \cong SU(2)^n.
\end{equation*}
In fact, the finitude of $\gamma$ allows us to show further that $SU(2)^n \sqsubset \mathcal{A}_\gamma$. In other words, $\mathcal{A}_\gamma$ and $SU(2)^n$ are homeomorphic. This is easy to see. Denote by $\gamma_k$ a graph containing $k$ edges $e_1, \ldots, e_k$. We already know that $\mathcal{A}_{\gamma_1}\cong SU(2)$. But we also know that any $A_{\gamma_k} \in \mathcal{A}_{\gamma_k}$ can be written as $(A_{e_1}, \ldots, A_{e_k})$, where $A_{\gamma_i} \in \mathcal{A}_{e_i}$ for all $i \in \{1,\ldots, k\}$. Induction\footnote{Note that this induction only works works for finite graphs. It is on this account that Eq~(\ref{eq3.101}) only holds in one direction; its converse is false.} on $k$ then establishes the assertion. We thus conclude
\begin{equation}
    \mathcal{A}_\gamma \cong SU(2)^n \quad \text{and} \quad \mathcal{G}_\gamma \cong SU(2)^n \label{eq3.107}.
\end{equation}
An immediate consequence of this equation is that $\mathcal{A}_\gamma$ and $\mathcal{G}_\gamma$ are compact Hausdorff topological groups, since $SU(2)$ (and hence $SU(2)^n$) is compact Hausdorff.

The proof in the preceding paragraph provides a very useful insight. Note that for an edge $e$, the homeomorphism between $\mathcal{A}_e$ and $SU(2)$ holds even if we restrict the maps in $\mathcal{A}_e$ to be smooth $SU(2)$-valued functions (in the sense of Eq~(\ref{eq3.9}h, i)). But such smooth maps define the holonomy of smooth connections. We thus arrive at the following result.
\begin{lemma}
    Given any generalised connection $\overline{A} \in \overline{\mathcal{A}}$ and a graph $\gamma$, there exists a smooth connection $A$ whose holonomy around each edge $e \in \gamma$ equals $\overline{A}(e)$. That is,
    \begin{equation*}
        \overline{A}(e) = U_e[A] = \mathcal{P}e^{-\int_e A}
    \end{equation*}
    for every $e \in \gamma$.
\end{lemma}
This lemma provides a one-to-one correspondence between $\mathcal{A}_\gamma$ and \textit{smooth} connections. We shall have occasion to use this result later. Note that this result only holds for the edges of a particular graph and not for arbitrary edges. Note also that the $A$ above is not unique; one $\overline{A}$ may correspond to multiple smooth connections in this way. However, for all such smooth connections, their parallel propagators along each edge of the graph will, of course, be the same. Thus the use of this lemma would pose no ambiguities.  
\newline

We now have two projective limits, namely $\overline{\mathcal{A}}$ and $\overline{\mathcal{G}}$, at our disposal. How are they related to the quantum configuration space $\overline{\mathcal{A/G}}$? We will now answer this question, and will find that $\overline{\mathcal{A/G}}$, too, can be regarded as a projective limit. For this, we construct two new projective limits.

The first is simply the quotient $\overline{\mathcal{A}}/\overline{\mathcal{G}}$. Its elements are equivalence classes of generalised connections that are related by Eq~(\ref{eq3.103}) for some generalised gauge transformation. Such equivalence classes have a special name in mathematical parlance: \textit{conjugacy classes} of a group.  

Let us obtain the second promised projective limit. By construction, it is obvious that the group $\mathcal{G}_{\gamma}$ defined in Eq~(\ref{eq3.105}) naturally acts on $\mathcal{A}_\gamma$ through Eq~(\ref{eq3.103}), since it is defined on vertices of graphs, and gauge transformations act on vertices of edges in Eq~(\ref{eq3.103}). Consider now the space $\mathcal{A}_\gamma/\mathcal{G}_\gamma$. It contains equivalence classes of maps $A_\gamma: \gamma \to SU(2)$ that satisfy Eq~(\ref{eq3.100}) and are related by Eq~(\ref{eq3.103}) for some $g_\gamma \in \mathcal{G}_\gamma$ -- in other words, the conjugacy classes of $\mathcal{A}_\gamma$ with respect to $\mathcal{G}_\gamma$. Moreover, in view of Eq~(\ref{eq3.107}), we have
\begin{equation}
    \mathcal{A}_\gamma/\mathcal{G}_\gamma \cong SU(2)^n/Ad, \label{eq3.108} 
\end{equation}
where $n$ is the number of edges in $\gamma$. We note further that since gauge transformations act at vertices of edges in Eq~(\ref{eq3.103}), if $A \in \overline{\mathcal{A}}$ is restricted to a graph $\gamma$, then any gauge transformation $g \in \overline{\mathcal{G}}$ acting on $A$ will also be restricted to $\gamma$. In other words, for each $A \in \overline{\mathcal{A}}$ and $g \in \overline{\mathcal{G}}$, 
\begin{equation}
    p_{\gamma}(gA) = g_\gamma A_\gamma \label{eq3.109}
\end{equation}
for some $g_\gamma \in \mathcal{G}_\gamma$ and $A_\gamma \in \mathcal{A}_\gamma$. We say that the action of $\overline{\mathcal{G}}$ on $\overline{\mathcal{A}}$ is \textit{equivariant} with respect to the projections $p_\gamma$. This fact ensures that each projection $p_{\gamma\gamma'}$ defined above descends unambiguously to a projection
\begin{equation}
    p_{\gamma\gamma'}: \mathcal{A}_{\gamma'}/\mathcal{G}_{\gamma'} \to \mathcal{A}_\gamma/\mathcal{G}_\gamma, \label{eq3.110}
\end{equation}
which satisfies exactly the same properties. It follows that $\{\mathcal{A}_\gamma/\mathcal{G}_\gamma, p_{\gamma\gamma'}\}$ is a projective family. Let $\widetilde{\mathcal{A/G}}$ denote its projective limit. The fundamental result of this section is the fact that
\begin{equation}
    \overline{\mathcal{A}}/\overline{\mathcal{G}} \cong \widetilde{\mathcal{A/G}} \cong \overline{\mathcal{A/G}}. \label{eq3.111}
\end{equation}
In words, the quotient of projective limits, the projective limits of quotients, and the Gelfand spectrum of the holonomy $C^\star$-algebra are all equivalent. This is a very rich characterisation of the quantum configuration space, and will be indispensable for the tools developed in the rest of this chapter. 

That Eq~(\ref{eq3.111}) holds is a highly nontrivial fact. It does not hold for any general spaces. The essential ingredient that makes it possible is the fact that all three spaces in the equation are compact Hausdorff topological spaces. We already know that $\overline{\mathcal{A/G}}$ is compact Hausdroff. The compact-Hausdorff-ness of the other two spaces deserves to be thoroughly investigated. It essentially depends on the fact that the projective limits at our disposal arise from projective families of compact Hausdorff spaces. In particular, we have the following theorem \cite{ribes}.

\begin{theorem}
    The projective limit $\widetilde{\mathcal{X}}$ of a projective family $\{\mathcal{X}_\gamma, p_{\gamma, \gamma'}\}_{\gamma, \gamma' \in L}$ of compact Hausdorff spaces is compact. 
\end{theorem}
\begin{proof}
    We will have to make use of the following two well-known results from topology \cite{munkres}. 

    \begin{lemma}
    An arbitrary product of Hausdorff spaces is Hausdorff.
\end{lemma}
\begin{lemma}
    (Tychonoff theorem\footnote{That this theorem has a name and the preceding one has not is because of its highly nontrivial character and proof. See any standard text on topology, for instance \cite{munkres}.}) An arbitrary product of compact spaces is compact. 
\end{lemma}

We have that
\begin{equation*}
    \widetilde{\mathcal{X}} \subset \prod_{\gamma\in L}\mathcal{X}_\gamma,
\end{equation*}
which, together with Lemma 3.4 and the fact that $\mathcal{X}_\gamma$ are all Hausdorff, entails that $\widetilde{\mathcal{X}}$ is Hausdorff. 

Next, we show that $\widetilde{\mathcal{X}}$ is, in fact, a closed subset of $\prod_{\gamma\in L}\mathcal{X}_\gamma$, which being a product of compact spaces, is compact by the Tychonoff theorem. Since every closed subset of a compact space is compact \cite{munkres}, $\widetilde{\mathcal{X}}$ is compact as well. 

As for the fact that $\widetilde{\mathcal{X}}$ is closed, assume that it is not. Let $(X_\gamma)_{\gamma\in L} \in \prod \mathcal{X}_\gamma\setminus\widetilde{\mathcal{X}}$. Then since $\widetilde{\mathcal{X}}$ is the projective limit, there exist $\gamma_1, \gamma_2$ with $\gamma_2 \geq \gamma_1$ such that $p_{\gamma_1\gamma_2}(X_{\gamma_2}) \neq X_{\gamma_1}$ for some $X_{\gamma_1}, X_{\gamma_2} \in \mathcal{X}_{\gamma_1}, \mathcal{X}_{\gamma_2}$. Since $\mathcal{X}_{\gamma_1}$ is Hausdorff, we can pick in it disjoint neighbourhoods $U$ and $V$ of $X_{\gamma_1}$ and $p_{\gamma_1\gamma_2}(X_{\gamma_2})$, respectively, and also arrange for them to have empty intersection with $\widetilde{\mathcal{X}}$. Furthermore, let $V'$ be a neighbourhood of $X_{\gamma_2}$ such that $p_{\gamma_1\gamma_2}(V') \subset V$. Then the open set $\prod_\gamma B_\gamma \subset \prod_\gamma \mathcal{X}_\gamma$ with $B_{\gamma_1} = U, B_{\gamma_2} = V'$ and $B_{\gamma} = \mathcal{X}_{\gamma}$ for all $\gamma \neq \gamma_1, \gamma_2$ is a neighbourhood of $(X_{\gamma})$ that is disjoint from $\widetilde{\mathcal{X}}$. Thus the complement of $\widetilde{\mathcal{X}}$ is open, and so the set is closed.  
\end{proof}

Let us now establish the correspondence in Eq~(\ref{eq3.111}). We will proceed in three steps. 
\begin{enumerate}[(1)]
    \item First, we will show that the projective limit $\widetilde{\mathcal{A/G}}$ is the Gelfand spectrum of the $C^\star$-algebra of cylindrical functions on $\widetilde{\mathcal{A/G}}$.
    \item Second, we will find that this $C^\star$-algebra is isomorphic to the holonomy $C^\star$-algebra $\overline{\mathcal{HA}}$, and so $\widetilde{\mathcal{A/G}}$ is nothing but the Gelfand spectrum $\overline{\mathcal{A/G}}$ of $\overline{\mathcal{HA}}$.
    \item Finally, we will see that $\overline{\mathcal{A}}/\overline{\mathcal{G}}$ and $\widetilde{\mathcal{A/G}}$ are homeomorphic as topological spaces. 
\end{enumerate}
As with Theorem 3.8, in proving (1) and (3), the only relevant property of the projective families under consideration will be their being compact Hausdorff. Thus, where possible, we will be more general and work with any projective family of compact Hausdorff spaces.

\begin{mdframed}[style=testframe]
    \begin{enumerate}[(1)]
        \item Let $\{\mathcal{X}_\gamma, p_{\gamma\gamma'}\}_{\gamma,\gamma'\in L}$ be a projective family with projective limit $\widetilde{\mathcal{X}}$ and $C^0(\mathcal{X}_\gamma)$ be the set of all continuous bounded functions on $\mathcal{X}_\gamma$. Then, as will be recalled from Section 3.3.3, the set $Cyl(\widetilde{\mathcal{X}})$ of cylindrical functions on $\widetilde{X}$ contains functions $f$ such that $f = p^*_\gamma(f_\gamma)$ for some $\gamma\in L$ and $f_\gamma\in C^0(\mathcal{X}_\gamma)$, where $p_\gamma: \widetilde{\mathcal{X}}\to\mathcal{X}_\gamma$.  

        $Cyl(\widetilde{\mathcal{X}})$ gives rise to a $C^\star$-algebra. To see this, we first note that the representation of a cylindrical function is independent of $\gamma$. This is so because if $\gamma' \geq \gamma$, then since a function $f \in Cyl(\Tilde{\mathcal{X}})$ cylindrical with respect to $\gamma$ is also cylindrical with respect to $\gamma$, it follows from the definition of a projective limit that
        \begin{equation}
            f_{\gamma'} = p^*_{\gamma\gamma'}f_\gamma, \label{eq3.112}
        \end{equation}
        provided we restrict attention to only those $X_{\gamma'} \in \mathcal{X}_{\gamma'}$ that come from some $(X_\gamma)_{\gamma\in L} \in \widetilde{\mathcal{X}}$ (and this will always be the case so long as we remain in the space $\widetilde{\mathcal{X}}$). This in turn implies that if $\gamma''\geq\gamma'\geq\gamma$, then
        \begin{equation}
            p^*_{\gamma\gamma''}f_\gamma = p^*_{\gamma'\gamma''}f_{\gamma'}, \label{eq3.113}
        \end{equation}
        which allows us to define addition and multiplication on $Cyl(\widetilde{\mathcal{X}})$, as follows. Let $f, g \in(\widetilde{\mathcal{X}})$. By definition, there exist some $\gamma_1, \gamma_2 \in L$ such that $f = p^*_{\gamma_1}f_{\gamma_1}$ and $g = p^*_{\gamma_1}g_{\gamma_1}$. But since $L$ is a directed set, we can find some $\gamma \geq \gamma_1, \gamma_2$. This, in view of Eq~(\ref{eq3.113}), means that we can always arrange for any $f$ and $g$ to be represented on the same space $\mathcal{X}_\gamma$. Then it is clearly well-defined to set 
        \begin{subequations}
        \label{eq3.114}
            \begin{align}
                 f + g = &p^*_\gamma (f_\gamma + g_\gamma), \quad fg = p^*_\gamma (f_\gamma g_\gamma), \\
                af &= p^*_\gamma(af_\gamma), \quad f^* = p^*_\gamma(f^*_\gamma).   
            \end{align}
            Here, $a \in \mathbb{C}$, and the star on $f$ indicates complex conjugation. These equations render $Cyl(\widetilde{\mathcal{X}})$ into a $\star$-algebra, upon which we can put a norm by
            \begin{align}
                ||f|| = \sup_{x_\gamma\in\mathcal{X}_\gamma} |f_\gamma(x_\gamma)|,
            \end{align}
            which is again well-defined due to Eq~(\ref{eq3.113}). Since $||f^*||=||f||$, the completion of $Cyl(\widetilde{\mathcal{X}})$ under this norm yields a $C^\star$-algebra, which we denote by $\overline{Cyl}(\widetilde{\mathcal{X}})$.
        \end{subequations}
        Next, we show that $\overline{Cyl}(\widetilde{\mathcal{X}})$ is isomorphic to the set $C^0(\widetilde{\mathcal{X}})$ of continuous bounded functions on the projective limit. Then by the Gelfand-Naimark theorem, $\widetilde{\mathcal{X}}$ is the Gelfand spectrum of $\overline{Cyl}(\widetilde{\mathcal{X}})$. 

            To establish the required isomorphism, we note that any cylindrical function is essentially a continuous bounded function on some member of our projective family. Thus any cylindrical function is by definition a continuous bounded function on the projective limit. Moreover, $C^0(\widetilde{\mathcal{X}})$ is complete in its sup norm, which in turn coincides with Eq~(\ref{eq3.114}c) for cylindrical functions. Therefore, any function in the closure $\overline{Cyl}(\widetilde{\mathcal{X}})$ also belongs to $C^0(\widetilde{\mathcal{X}})$. This shows that $\overline{Cyl}(\widetilde{\mathcal{X}}) \subset C^0(\widetilde{\mathcal{X}})$. All we need to do then is to show the converse, i.e. that every continuous bounded function on $\widetilde{\mathcal{X}}$ can be realised as the pullback of a continuous bounded function on $\mathcal{X}_\gamma$ for some $\gamma\in L$. This we now do in the specific case for which $\mathcal{X}_\gamma = \mathcal{A}_\gamma$ and $\widetilde{\mathcal{X}} = \overline{\mathcal{A}}$; the other two projective families admit a similar treatment \footnote{The same result can be shown, using a different, albeit more technical, argument for any compact Hausdorff projective family; see \cite{ashtekar 6}}. 
            
            (Could not be completed due to lack of time)

        \item (Could not be completed due to lack of time)
            
        \item (Could not be completed due to lack of time)
        
    \end{enumerate}
\end{mdframed}

\section{Quantum Kinematics}
We have probed the structure of the quantum configuration space $\overline{\mathcal{A/G}}$ in detail. We are now in a position to answer the concerns raised at the end of Section 3.4.1. 

\subsection{The Ashtekar-Lewandowski measure}
As we have seen in Section 3.4.1, one can realise quantum states as functions on $\overline{\mathcal{A/G}}$ that are square-integrable with respect to a measure $\mu$ whose existence is ensured by the Riesz-Markov theorem. We were also then able to realise the $T^0$ variables as operators on $L^2(\overline{\mathcal{A/G}}, \mu)$. In particular, we found that one could define  
\begin{equation}
    (\hat{T}^0_{\alpha})\psi(A) := T^0_\alpha(A)\psi(A), \label{eq3.115}
\end{equation}
where $\psi \in L^2(\overline{\mathcal{A/G}}, \mu)$ and $T^0_\alpha(A)$ is the trace of the holonomy of $A$ around the loop $\alpha$. By construction, these operators are self-adjoint. In the language of standard quantum mechanics, we have constructed configuration/position operators. Note that in this construction, only the existence of the measure $\mu$ was required; we did not need to choose a particular one. But of course, to carry out computations, such as calculations of inner products, one needs to specify a particular measure on $\overline{\mathcal{A}}$. In this section, we will explicitly construct a specific measure on $\overline{\mathcal{A/G}}$ called the \textit{Ashtekar-Lewandowski} measure. Its virtue lies in the fact that it is invariant with respect to diffeomorphisms of the spatial manifold $\Sigma$. Furthermore, there exists a theorem, called the LOST theorem \cite{lewandowski}, which states that under mild assumptions, the Ashtekar-Lewandowski representation of loop quantum gravity is unique upto unitary equivalences; this is quite reminiscent of the highly important Stone-von-Neumann theorem in quantum mechanics. 

We are now squarely in the situation we encountered when constructing a measure on the quantum configuration space of scalar field theory in Section 3.3. There, we saw that the configuration space can be realised as a projective limit of its finite-dimensional subspaces, on which we know how to construct explicit measures (e.g. the Gaussian measure), and Theorem 3.2 guaranteed the extension of these measures to the configuration space. Here, we have also realised the quantum configuration space $\overline{\mathcal{A/G}}$ as a projective limit, albeit of finite-dimensional compact Hausdorff topological spaces $\mathcal{A}_\gamma/\mathcal{G}_\gamma$. Does some extension theorem exist for such spaces as well? The answer, as hinted towards the end of Section 3.3.3, is yes. In particular, we have the following theorem \cite{ashtekar 6, jose, yamasaki}.    
\begin{theorem}
    Any self-consistent family of regular Borel measures on a projective family of compact Hausdorff spaces can be extended to a unique regular Borel measure on the projective limit. 
\end{theorem}
\begin{proof}
    Let $\{\mu_\gamma\}$ be a family of regular Borel measures on $\{\mathcal{X}_\gamma\}$ satisfying (c.f. Eq~(\ref{eq3.69}))
    \begin{equation}
        \mu_\gamma = (p_{\gamma\gamma'})_*\mu_{\gamma'}\quad \forall \gamma'\geq\gamma. \label{eq3.116}
    \end{equation}
    Let $f\in C^0(\widetilde{\mathcal{X}})$. As shown in the previous section, $f = p^*_\gamma f_\gamma$ for some $\gamma\in L$ and $f_\gamma\in C^0(\mathcal{X}_\gamma)$. Then it follows from Eqs~(\ref{eq3.116}) and (\ref{eq3.113}) that the following functional on $C^0(\widetilde{\mathcal{X}})$ is well-defined:
    \begin{equation}
        \Gamma(f) = \int_{\mathcal{X}_\gamma}d\mu_\gamma f_\gamma. \label{eq3.117}
    \end{equation}
    If $f \geq 0$, then $\Gamma(f) \geq 0$. Thus the functional is positive on $C^0(\widetilde{\mathcal{X}})$. It is also linear. Since $\widetilde{\mathcal{X}}$ is compact Hausdorff, by the Riesz-Markov theorem, there exists a unique regular Borel measure $\mu$ on $\widetilde{\mathcal{X}}$ such that
    \begin{equation}
        \Gamma(f) = \int_{\widetilde{\mathcal{X}}}d\mu f. \label{eq3.118}
    \end{equation}
\end{proof}

Comparing the preceding two equations, we further see that
    \begin{equation}
        \mu = p^*_\gamma \mu_\gamma. \label{eq3.119}
    \end{equation}
That is, $\mu$ is nothing but the pullback of the measure $\mu_\gamma$ by the projection $p_\gamma$. Thus the moral of Theorem 3.9 is that we can virtually forget about $\widetilde{\mathcal{X}}$ and instead define measures, functions, integrals, and so on, on finite-dimensional spaces $\mathcal{X}_\gamma$ and just work in them. 

Let us now focus on our particular spaces $\{\mathcal{A}_\gamma\}$ and $\{\mathcal{A}_\gamma/\mathcal{G}_\gamma\}$ and their associated projective limits. Physically, only the latter are relevant for us. However, as we will repeatedly see, it is often convenient to introduce gauge-invariant objects on $\mathcal{A}_\gamma$ ($\overline{\mathcal{A}}$); gauge-invariance would then ensure that such objects project down unambiguously to $\mathcal{A}_\gamma/\mathcal{G}_\gamma$ ($\widetilde{\mathcal{A/G}}$). The first instance of this fortuitousness will be seen in our derivation of the diffeomorphism-invariant Ashtekar-Lewandowski measure on our projective limit(s). 

Recall that in scalar field theory, the Gaussian measure emerged as the natural measure to be defined on the finite-dimensional subspaces of the quantum configuration space. Physical considerations such as Lorentz invariance then fixed the explicit form of the covariance of that measure. Here the situation is even better, for the choice of a natural measure on the relevant finite-dimensional spaces automatically satisfies our physical constraints, namely diffeomorphism and gauge invariance. The finite-dimensional spaces $\mathcal{A}_\gamma$ are essentially homeomorphic to finitely many copies of $SU(2)$ (see Eq~(\ref{eq3.107})), which are compact Lie groups. Now, there exists the so-called Haar measure on any compact Lie group. It is both left- and right-invariant with respect to the group action. Therefore, it is gauge-invariant. Finally, since no background structures (e.g. metrics, connections, volume forms, etc.) are required in its definition (see the next Section), it is diffeomorphism-invariant (for free!). By Eq~(\ref{eq3.119}), it thus extends uniquely to a diffeomorphism-invariant measure $\mu_{AL}$ -- the sought-after Ashtekar-Lewandowski measure -- on our projective limits. 

A little more explicitly, let $\gamma$ be a graph with $n$ edges, and let $\mu_H$ be the Haar measure on $SU(2)$. Then we define a measure $\mu_H^{(n)}$ on $\mathcal{A}_\gamma \sqsubset SU(2)^n$ by
\begin{equation}
    \mu_H^{(n)} := \mu_H\otimes \cdots \otimes \mu_H \label{eq3.120}
\end{equation}
Let $q: SU(2)^n \to SU(2)^n/Ad$ be the quotient map between the mentioned spaces. This means that a measure on $\tilde{\mu}_H^{(n)}$ induces a measure on $\mathcal{A}_\gamma/\mathcal{G}_\gamma \subset SU(2)^n/Ad$:
\begin{equation}
    \tilde{\mu}_H^{(n)} = \Tilde{q}\mu_H\otimes\cdots\mu_H, \label{eq3.121}
\end{equation}
where $\Tilde{q}$ is the induced map, i.e. $\Tilde{q} f = f\circ \Tilde{q}$. Finally, in accordance with Eq~(\ref{eq3.118}), we obtain the Ashtekar-Lewandowski measure $\mu_{AL}$ on $\widetilde{\mathcal{A/G}}$ or $\overline{\mathcal{A}}/\overline{\mathcal{G}}$ by a pullback:
\begin{equation}
    \mu_{AL} = p^*_\gamma \tilde{\mu}_H^{(n)}. \label{eq3.122}
\end{equation}
We shall refer to $L^2(\overline{\mathcal{A/G}},d\mu_{AL})$ as $\mathcal{H}_{kin}$, the \textit{kinematical Hilbert space} of a loop quantum gravity.

Now that we have an explicit measure on the configuration space, we can begin asking how to perform explicit calculations with quantum states, such as computing integrals on our function spaces. For this, we first need to study the as-yet elusive Haar measure in more detail. This we now do. 

\subsection{The Haar measure}
As we briskly mentioned in the preceding section, the Haar measure is a left- and right-invariant measure that exists on any compact Lie group; it is also normalised. In this section, we shall establish these claims, and see the explicit form of the measure for some simple groups, such as $U(1)$ and $SU(2)$. 

Let $G$ be a compact Lie group. We wish to prove the existence and uniqueness of a measure $dg$ on $G$ such that for all $h, g \in G$,
\begin{subequations}
    \label{eq3.123}
    \begin{align}
        \int_G dg &= 1, \\
        \int_G dg f(hg) = \int_G f(g), &\quad \int_G dg f(gh) = \int_G dg f(g),
    \end{align}
\end{subequations}
where $f$ is a complex-valued function on $G$. The first equation equation is the condition of normalisation, which we can always achieve because $G$ is compact. The crux of the matter lies in demonstrating left and right invariance, as demanded in the next two equations. We shall follow the informal route taken in Ref.\cite{creutz} by deriving an expression for the measure under the assumption of existence, and then showing that the expression works. 

Let $n$ be the dimension of $G$, and consider an arbitrary parametrisation of the group elements of $G$ in terms of a set of parameters $\alpha_i$, where $\alpha_i$ belongs to some subset $D$ of $\mathbb{R}^n$ and $i \in \{1,\cdots, n\}$. For instance, $SU(2)$ is diffeomorphic to the $\partial S_3$ boundary of a unit three-sphere, and can thus can be parametrised by $\{\theta, \phi, \chi| -\pi\leq \theta \leq \pi, 0\leq \phi, \chi \leq 2\pi\}$. Thus we can write $G = \{g(\alpha) \in G| \alpha \in D\}$, and represent the multipilication of two group elements as $g(\beta)g(\gamma) = g(\alpha(\beta,\gamma))$ for some $\alpha, \beta, \gamma \in D$, $\alpha$ being a function of the other two parameters. The advantage of working with a parametrisation is that one can now write integrals of functions on the group as ordinary $n$-dimensional integrals over $D$, 
\begin{equation}
    \int_G dg f(g) = \int_D d\alpha_1\cdots d\alpha_n J(\alpha) f(g(\alpha)), \label{eq3.124}
\end{equation}
where $J(\alpha)$ is a parametric representation of the measure we seek. Our task is thus reduced to finding a $J(\alpha)$ that meets the constraints in Eq~(\ref{eq3.123}). To achieve this goal, we first note that the right-invariance property of the measure now looks like
\begin{equation}
    \int_D d\beta J(\beta)f(g(\beta)) = \int_D d\beta J(\beta) f(g(\alpha(\beta, \gamma))) \label{eq3.125}
\end{equation}
where $\gamma$ is the parameter on which $h$ in Eq~(\ref{eq3.123}b) depends. Since $\alpha, \beta, \gamma \in \mathbb{R}^n$, we can perform a change of variables from $\beta$ to $\alpha(\beta,\gamma)$, obtaining
\begin{equation}
    \int_D d\beta J(\beta)f(g(\beta)) = \int_D d\alpha \left|\frac{\partial\alpha(\beta,\gamma)}{\partial\beta} \right|^{-1}J(\beta)f(g(\alpha)), \label{eq3.126}
\end{equation}
where $|\partial\alpha/\partial\beta|$ is the Jacobian of the variable transformation. Since the preceding eqution is true for any $f$, it follows that
\begin{equation}
    J(\alpha) = \left|\frac{\partial\alpha(\beta,\gamma)}{\partial\beta} \right|^{-1} J(\beta). \label{eq3.127}
\end{equation}
Now if $\beta$ tends towards the parameter corresponding to the identity element $e$, then $\alpha(\beta, \gamma)$ tends to $\gamma$ up to a trivial multiplicative factor which can be absorbed into $K$. We then arrive at
\begin{equation}
    J(\gamma) = K\left|\frac{\partial\alpha(\beta,\gamma)}{\partial\beta} \right|^{-1}_{\beta = e}, \label{eq3.128}
\end{equation}
where $K = J(e)$ is a trivial normalisation factor to be determined by Eq~(\ref{eq3.123}a), and $\beta = e$ symbolically stands for the value of the parameter corresponding to the identity element. Thus, a left-invariant measure, if it exists, is determined by a Jacobian (\ref{eq3.128}). Hence, assuming the existence of the relevant transformations in the parameter space, a left-invariant measure on a compact group exists. 

We now show uniqueness of this measure. For this, we ought to show that Eq~(\ref{eq3.127}) holds for any $\beta$ and $\gamma$. Accordingly, 
let $\eta(\beta, \gamma)$ be an arbitrary function of $\beta$ and $\gamma$. In view of Eqs~(\ref{eq3.127}) and (\ref{eq3.128}), we wish that 
\begin{equation}
    J(\eta(\beta,\gamma)) = K\left| \frac{\partial\alpha(\delta, \eta(\beta,\gamma))}{\partial\delta}\right|^{-1}_{\delta=e} = K\left|\frac{\partial\eta(\beta,\gamma)}{\partial\beta} \right|^{-1}\left|\frac{\partial\alpha(\beta,\gamma)}{\partial\delta} \right|^{-1}_{\delta=e}. \label{eq3.129}
\end{equation}
But this is indeed the case, because the associativity of group multiplication warrants 
\begin{equation}
    \alpha(\delta, \eta(\beta, \gamma)) = \alpha(\eta(\delta, \beta), \gamma), \label{eq3.130}
\end{equation}
which upon differentiation with respect to $\delta$ and an application of the chain rule yields Eq~(\ref{eq3.129}). This establishes the uniqueness of the left-invariant measure.

Finally, it remains to be seen that our measure is also right-invariance. To see this, let $Dg$ be a measure on $G$ which satisfies
\begin{equation*}
    \int_G Dg f(g) = \int_G dg f(g_o^{-1}gg_o)
\end{equation*}
for some fixed but arbitrary $g_o \in G$; $dg$ is the unique left-invariant measure constructed above. Then it follows that
\begin{equation*}
    \int Dg f(hg) = \int dg f(hg_o^{-1}gg_o) = \int dg f(g_o^{-1}gg_o) = \int Dg f(g),
\end{equation*}
where the second equality follows from left invariance. Thus $Dg$ is left invariant. But since the left-invariant measure is unique, $Dg = dg$. Now the left invariance of $Dg$ implies that
\begin{equation*}
    \int Dg f(g_o^{-1}gg_o) = \int Dg f(gg_o) = \int Dg f(g).
\end{equation*}
That is, $Dg$ and hence $dg$ is right-invariant as well. 

To make the integrals with respect to the Haar measure computationally tractable, we introduce a metric on $G$, namely
\begin{equation}
    M_{ij} = \text{tr}\left(g^{-1}\frac{\partial g}{\partial\alpha_i}g^{-1}\frac{\partial g}{\partial\alpha_j}\right) \label{eq3.131}
\end{equation}
for all $g \in G$. Then differential geometric wisdom decrees that
\begin{equation}
    \int_G dg f(g) = K\int_D d\alpha \,\sqrt{\text{det}\,M}f(g(\alpha)). \label{eq3.132} 
\end{equation}

Let us now discuss specific examples. Suppose $G = U(1)$. A simple parametrisation of this group is
\begin{equation}
    U(1) = \{e^{i\theta}|0 \leq \theta < 2\pi\}. \label{eq3.133}
\end{equation}
We find using Eqs~(\ref{eq3.131}) and (\ref{eq3.132}) that
\begin{equation}
    \int_{U(1)}dg f(g) = \frac{1}{2\pi}\int_{0}^{2\pi}d\theta f(e^{i\theta}). \label{eq3.134}
\end{equation}
More relevant to our purposes is $SU(2)$, which it will be recalled, has the following parametrisation in the fundamental representation:
\begin{equation}
    SU(2) = \{a_o + a_i\tau^i| a^2 := a_o^2 + \delta^{ij}a_ia_j = 1\}. \label{eq3.135}
\end{equation}
We find
\begin{equation}
    \int_{SU(2)}dg f(g) = \frac{1}{\pi^2}\int d^4a\,\delta(a^2 - 1)f(g). \label{eq3.136}
\end{equation}
For further details as to how to efficiently compute integrals of explicit functions on $SU(2)$, see Ref. \cite{creutz}. 

\subsection{Construction of momentum operators}
So far, we have rigorously constructed a Hilbert space $\mathcal{H}_{kin}$ for quantum states, and have represented the configuration variables $T^0$ as operators on this space. It is now time to promote the momentum variables $T^1$ to operators on $\mathcal{H}_{kin}$. We now turn to this task. As we shall see, the fairly complicated characterisation of the quantum configuration space presented in the previous section will be immensely useful.

It will be instructive to recall how momentum variables are promoted to operators in standard quantum mechanics. For a particle moving in three dimensions, for instance, we have $\hat{p} = -i\hbar\nabla$. This is a vector field\footnote{Recall that a (smooth) vector field on a manifold $M$ is a linear map $X: C^\infty(M)\to C^\infty(M)$ that satisfies the Leibniz rule: for every $f, g \in M$, $X(fg) = X(f)g + fX(g)$. Also, at every $x\in M$, $X$ defines a tangent vector by $v_x(f)$ for every $f\in C^\infty(M)$, where $v_x \in T_xM$, thought of as a map from $C^\infty(M)$ to $\mathbb{R}$.} on $\mathbb{R}^3$, the domain space of quantum states. This suggests that the $T^1$ variables are to be represented as vector fields on $\overline{\mathcal{A/G}}$, the domain space of quantum states in loop quantum gravity. In addition, we will want the commutator between the configuration and momentum operators to agree with the Poisson brackets between their classical counterparts (see Eq~(\ref{eq3.28})). To accomplish this task, we will have to cleverly construct suitable vector fields on $\overline{\mathcal{A/G}}$, as we shall see in detail below. 

As should be evident by now, we will first introduce a consistent family of relevant objects on the finite-dimensional spaces $\mathcal{A}_\gamma$ and consistency will ensure their extension to $\overline{\mathcal{A}}$; furthermore, we will ensure that the objects are gauge invariant, whence they will consistently project down to $\overline{\mathcal{A/G}}$. Note that this is possible only because of the detailed characterisation of $\overline{\mathcal{A/G}}$ in terms of projective limits, condensed, as it were, in Eq~(\ref{eq3.111}).

\subsubsection{Vector fields on a Lie group}
Let us begin with a simple problem. Consider an edge $e$ and the associated space of maps $\mathcal{A}_e$. From the results of the previous section, we know that $\mathcal{A}_e \cong SU(2)$. We wish to obtain a simple characterisation of vector fields on $SU(2)$. In fact, we can be more general, and study vector fields on any Lie group. Associated to every Lie group $G$ is its Lie algebra $\mathfrak{g}$, which, recall, is the tangent space of $G$ at the identity element. We will show that in a certain sense, any vector field on $G$ can be completely determined by its action ``near'' the identity, i.e. in terms of the Lie algebra $\mathfrak{g}$. 

Before proceeding, we will require some definitions and notation. Adding another recalling to our endless list of recallings, recall that a diffeomorphism between smooth manifolds is a bijection whose inverse is differentiable as well.  For any diffeomorphism $\phi: G \to G$, we define the differential $d\phi_g$ of $\phi$ at every $g\in G$ to be the map from $T_gG$ to $T_{\phi(g)}G$ such that for every $f \in C^\infty(G)$, $d\phi_g(v)(f) = v(f\circ\phi)$ for all $v \in T_gG$; this definition makes sense because $f\circ\phi \in C^\infty(G)$. A differential satisfies the Leibniz rule -- hence its name. Note further that a differential can also be thought of as the pushforward $\phi_*$ of $v\in T_gG$ by $\phi$.

Now let Vect$(G)$ be the space of all vector fields on $G$. We say that two vector fields $X, Y \in \text{Vect}(G)$ are $\phi$-related if for every $g\in G$, $d\phi_g(X_g) = Y_{\phi(g)}$. In terms of pushforwards, we can write $Y = \phi_*(X)$, which is related to the definition in terms of differentials via $(\phi_*X)_h = d\phi_{\phi^{-1}(h)}(X_{\phi^{-1}(h)})$ for every $h\in G$ (it will be helpful to draw some pictures).    

For a given $g \in G$, the \textit{left translation} of $G$ by $g$ is a diffeomorphism $L_g: G \to G$ such that $L_g(h) = gh$ for all $h \in G$. Analogously, we have the notion of right translation $R_g$. We say that a vector field $X \in \text{Vect}(G)$ is \textit{left invariant} if it is invariant under all left translations. In other words, it is $L_g$-related to itself for every $g \in G$, 
\begin{equation}
    d(L_g)_{h}(X_{h}) = X_{gh}, \quad \forall g, h \in G. \label{eq3.137} 
\end{equation}
More succinctly, $(L_g)_*(X) = X$. A similar definition, \textit{mutatis mutandis}, holds for right-invariant vector fields too. Since $(L_g)_*(aX + bY) = a(L_g)_*(X) + b(L_g)_*(Y)$, the set $LG$ of left-invariant vector fields on $G$ is a linear subspace of Vect$(G)$. Again, a similar statement holds for right-invariant vector fields. 

We are now ready to drop the punchlines: (1) any vector field can be written as a linear functional combination of left-invariant (or right-invariant) vector fields, and (2) $LG$ is isomorphic (as a vector space) to the Lie algebra $\mathfrak{g}$ of $G$. We shall now prove these claims \cite{lee}.

Define a map $\varepsilon: LG \to T_oG$ such that $\varepsilon(X) = X_o$ for all $X \in LG$, $o$ being the identity element in $G$. Clearly, $\varepsilon$ is linear over $\mathbb{R}$. We show that it is also bijective and hence, an isomorphism between $LG$ and $T_eG$, or equivalently between $LG$ and $\mathfrak{g}$.

To show injectivity, suppose $\varepsilon(X) = \varepsilon(Y)$ for some $X, Y \in LG$. Then linearity implies $\varepsilon(X - Y) =  (X-Y)_o = 0$, and the left-invariance of $X - Y$ entails $(X-Y)_g = d(L_g)_o(X-Y)_o = 0$ for all $g \in G$. Thus $X - Y = 0$, yielding injectivity.

Next, we show surjectivity. For this, define a vector field $V^L$ on $G$ via
\begin{equation}
    V^L_g = d(L_g)_o(V) \label{eq3.138}
\end{equation}
for every $V \in T_oG$. $V^L$ so defined is a left-invariant vector field, since 
\begin{equation*}
    d(L_h)_g(V^L_g) = d(L_h)_g(d(L_g)_e(V)) = d(L_{hg})_o(V) = V^L_{hg}
\end{equation*}
where the second-last equality follows from $L_h\circ L_g = L_{hg}$. Now observe that $\varepsilon(V^L) = V^L_o = d(L_o)_o(V) = V \in T_oG$, which means that $\varepsilon$ is surjective. This completes the proof that $LG \cong T_oG$. 

Next, we establish that any vector field can be written as a linear combination of left-invariant vector fields. This is straightforward to exhibit. Let $\{e_1, \ldots, e_n\}$ be a basis of $T_oG$. Then $\{d(L_g)_o(e_1), \ldots, d(L_g)_o(e_n)\}$
is a basis of $T_gG$ for any $g \in G$, but as already noted, the vector fields defined by $d(L_g)_o(e_i)$ are all left invariant.   

Finally, let us describe a perhaps more intuitive way of writing left- and right-invariant vector fields. Given any $\Lambda \in \mathfrak{g}$, we know that $e^{t\Lambda} \in G$. Then in view of Eq~(\ref{eq3.138}), given any function $f: G \to \mathbb{R}$ we can write
\begin{equation*}
    \Lambda^L_g(f) = d(L_g)_o(\Lambda)(f) = \Lambda(f\circ L_g) = \left.\frac{d}{dt}(f(L_g(e^{\Lambda t})))\right|_{t=0} = \left.\frac{d}{dt}(f(ge^{\Lambda t}))\right|_{t=0}.
\end{equation*}
 In this way, we can associate a left-invariant vector field to any element of the Lie algebra. Furthermore, since it is immaterial whether we choose $e^{t\Lambda}$ or $e^{-t\Lambda}$, we can choose one for left-invariant fields and the other for right-invariant fields. That is, for any $\Lambda \in \mathfrak{g}$, we associate the following left- and right-invariant vector fields, respectively:
 \begin{equation}
     \Lambda^L(f(g)) = \left. \frac{d}{dt}f(ge^{\Lambda t}) \right|_{t=0}, \quad  \Lambda^R(f(g)) = \left. \frac{d}{dt}f(e^{-\Lambda t}g) \right|_{t=0}. \label{eq3.139}
 \end{equation}
For what it's worth, this equation can also be taken as the definition of left- and right-invariant vector fields. Ugly as it may seem, it offers the advantage of seeing that these two fields are close cousins, in the sense that pushforward of a left-invariant field by the map $i: G \to G$, $i(g) = g^{-1}$ yields a right-invariant field. For,
\begin{equation}
    i_*(\Lambda^L(f(g))) = \left. \frac{d}{dt}f(i(ge^{\Lambda t})) \right|_{t=0} = \left. \frac{d}{dt}f(e^{-\Lambda t}g^{-1}) \right|_{t=0} = \Lambda^R(f(g^{-1})). \label{eq3.140}
\end{equation}

We are now ready to tackle the problem of constructing suitable vector fields on the quantum configuration space.

\subsubsection{Enter momentum operators}
Recall that the $T^1$ variables were defined as integrals of the $T^a$ variables over a surface $S$ embedded in the spatial manifold $\Sigma$, and we chose to foliate $S$ by a one-parameter family of loops (see Eq~(\ref{eq3.10})). Thus, it will be convenient to first introduce an operator corresponding to a single loop in the foliation. In other words, we shall first find a representation of $T^a_\alpha(t) = \text{tr}\,(U_\alpha(t) \Tilde{E}^a(t))$ (Eq~(\ref{eq3.10}), where $\alpha$ is an arbitrary loop in the foliation. 

To this end, consider an analytic loop $\alpha$ in $\Sigma$. Note that $\alpha$ can be thought of as a graph with one edge. Thus, it is well-defined to consider a space $\mathcal{A}_\gamma$ of general connections with $\gamma \geq \alpha$. We wish to define vector fields on $\mathcal{A}_\gamma$. These will, of course, act on functions $f_\gamma \in C^\infty(\mathcal{A}_\gamma)$. And we want, in particular, vector fields associated with $\text{tr}\,(U_\alpha(t) \Tilde{E}^a(t))$, where $U_\alpha(t)$ is the holonomy of a connection and $\Tilde{E}^a(t)$ is a triad. Let us break our task into two parts. First we will focus on promoting $U_\alpha(t)$ to things that act on functions $f_\gamma \in C^\infty(\mathcal{A}_\gamma)$, and then we will do the same for $\Tilde{E}^a(t)$. 

As for $U_\alpha(t)$, we can promote it to a map $U_\alpha: C^\infty(\mathcal{A}_\gamma) \to C^\infty(\mathcal{A}_\gamma)$ such that for every $f_\gamma\in C^\infty(\mathcal{A}_\gamma)$ and $A_\gamma \in \mathcal{A}_\gamma$,
\begin{equation}
    U_\alpha(f_\gamma)(A_\gamma) := U_\alpha[A] = \text{tr}\,\left( \mathcal{P}\exp{\int_\alpha A} \right), \label{eq3.141}
\end{equation} 
where $A$ is a smooth connection whose parallel propagator along each edge $e$ $\gamma$ equals $A_\gamma(e)$ (cf. Lemma 3.4).

Next, consider $\Tilde{E}^a$. We invoke three facts. First, at $x\in \Sigma$, $\Tilde{E}^a = iB^a_i(x)\tau^i$, where $B^a_i(x)\in T_x\Sigma$ (recall that as originally conceived, triads were the spatial projection of tetrads, which are essentially vector fields on the spatial manifold $\Sigma$). Second, $\mathcal{A}_\gamma\cong SU(2)^n$, where $n$ is the number of edges $\{e_1,\ldots, e_n\}$ in $\gamma$. Third, a triad is inserted at $\alpha$ at a particular point. The first and second facts suggest that we represent $B^a_i(x)$ as a collection of vector fields on $SU(2)$, where each field in the collection corresponds to the group associated with a particular edge in $\gamma$. Furthermore, the $i\tau^i$ in the decomposition of $\Tilde{E}^a$ form a basis for $\mathfrak{su}(2)$ and thus naturally invite us to use the left- and right-invariant fields corresponding to $i\tau_i$. The third fact provides us with the hint that one ought to get nonzero contributions from the action of our fields only when the edges intersect $\alpha$ at the point where a triad is inserted. This further gives rise to a number of subtleties. First, the only meaningful points to consider on an edge are its endpoints. Second, since $\alpha$ and the edges are all oriented, their relative orientations at the intersection points also matter. This in turn means that it matters whether an intersection occurs at the beginning point of an edge or its ending point, for one could have a closed edge, in which case it will depend on the particular endpoint under consideration whether the corresponding edge is ``incoming'' or ``outgoing'' at an intersection point. To cater to this, we can associate left-invariant vector fields to the beginning points of edges, and right-invariant fields to the ending points. Finally, we also need to take into account the orientations of $\alpha$ and the edges relative to the tangent spaces from which the $B^a_i$ come. Now these tangent spaces are three-dimensional, and since $\Sigma$ is also three-dimensional, no extrinsic orientation can be conferred upon them. It will also not do to consider the two-dimensional subspaces of these tangent spaces, since in the end, the momentum variables are defined on two-dimensional surfaces. Thus we will consider one-dimensional subspaces of the tangent spaces. With these considerations in mind, we conjure a clever guess for the vector fields that will represent the $T^a_\alpha$ variables as operators.

\begin{figure}[htbp]
    \centering

\tikzset{every picture/.style={line width=0.75pt}} 

\begin{tikzpicture}[x=0.75pt,y=0.75pt,yscale=-1,xscale=1]

\draw    (159,74) -- (303,64) ;
\draw [shift={(231,69)}, rotate = 356.03] [color={rgb, 255:red, 0; green, 0; blue, 0 }  ][fill={rgb, 255:red, 0; green, 0; blue, 0 }  ][line width=0.75]      (0, 0) circle [x radius= 3.35, y radius= 3.35]   ;
\draw  [color={rgb, 255:red, 189; green, 16; blue, 224 }  ,draw opacity=1 ] (185.5,114.5) .. controls (185.5,89.37) and (205.87,69) .. (231,69) .. controls (256.13,69) and (276.5,89.37) .. (276.5,114.5) .. controls (276.5,139.63) and (256.13,160) .. (231,160) .. controls (205.87,160) and (185.5,139.63) .. (185.5,114.5) -- cycle ;

\draw   (175,68) .. controls (179.22,70.39) and (183.23,71.56) .. (187.02,71.51) .. controls (183.45,72.78) and (180.1,75.27) .. (176.96,78.97) ;
\draw [color={rgb, 255:red, 208; green, 2; blue, 27 }  ,draw opacity=1 ]   (231,69) .. controls (252,94) and (218,124) .. (254,131) ;
\draw [shift={(238.05,97.39)}, rotate = 97.06] [color={rgb, 255:red, 208; green, 2; blue, 27 }  ,draw opacity=1 ][line width=0.75]    (10.93,-4.9) .. controls (6.95,-2.3) and (3.31,-0.67) .. (0,0) .. controls (3.31,0.67) and (6.95,2.3) .. (10.93,4.9)   ;
\draw   (263,62) .. controls (267.22,64.39) and (271.23,65.56) .. (275.02,65.51) .. controls (271.45,66.78) and (268.1,69.27) .. (264.96,72.97) ;

\draw  [color={rgb, 255:red, 189; green, 16; blue, 224 }  ,draw opacity=1 ] (183.35,97.87) .. controls (187.53,95.41) and (190.55,92.53) .. (192.41,89.23) .. controls (191.72,92.95) and (192.19,97.1) .. (193.83,101.67) ;
\draw  [color={rgb, 255:red, 189; green, 16; blue, 224 }  ,draw opacity=1 ] (273.71,84.28) .. controls (272.15,88.87) and (271.75,93.03) .. (272.5,96.74) .. controls (270.59,93.47) and (267.52,90.64) .. (263.29,88.25) ;

\draw    (317,74) -- (461,64) ;
\draw [shift={(389,69)}, rotate = 356.03] [color={rgb, 255:red, 0; green, 0; blue, 0 }  ][fill={rgb, 255:red, 0; green, 0; blue, 0 }  ][line width=0.75]      (0, 0) circle [x radius= 3.35, y radius= 3.35]   ;
\draw  [color={rgb, 255:red, 189; green, 16; blue, 224 }  ,draw opacity=1 ] (343.5,114.5) .. controls (343.5,89.37) and (363.87,69) .. (389,69) .. controls (414.13,69) and (434.5,89.37) .. (434.5,114.5) .. controls (434.5,139.63) and (414.13,160) .. (389,160) .. controls (363.87,160) and (343.5,139.63) .. (343.5,114.5) -- cycle ;

\draw   (333,68) .. controls (337.22,70.39) and (341.23,71.56) .. (345.02,71.51) .. controls (341.45,72.78) and (338.1,75.27) .. (334.96,78.97) ;
\draw [color={rgb, 255:red, 208; green, 2; blue, 27 }  ,draw opacity=1 ]   (366,7) .. controls (387,32) and (353,62) .. (389,69) ;
\draw [shift={(371.53,48.13)}, rotate = 276.38] [color={rgb, 255:red, 208; green, 2; blue, 27 }  ,draw opacity=1 ][line width=0.75]    (10.93,-4.9) .. controls (6.95,-2.3) and (3.31,-0.67) .. (0,0) .. controls (3.31,0.67) and (6.95,2.3) .. (10.93,4.9)   ;
\draw   (421,62) .. controls (425.22,64.39) and (429.23,65.56) .. (433.02,65.51) .. controls (429.45,66.78) and (426.1,69.27) .. (422.96,72.97) ;
\draw  [color={rgb, 255:red, 189; green, 16; blue, 224 }  ,draw opacity=1 ] (341.35,97.87) .. controls (345.53,95.41) and (348.55,92.53) .. (350.41,89.23) .. controls (349.72,92.95) and (350.19,97.1) .. (351.83,101.67) ;
\draw  [color={rgb, 255:red, 189; green, 16; blue, 224 }  ,draw opacity=1 ] (431.71,84.28) .. controls (430.15,88.87) and (429.75,93.03) .. (430.5,96.74) .. controls (428.59,93.47) and (425.52,90.64) .. (421.29,88.25) ;

\draw    (161,307) -- (305,297) ;
\draw [shift={(233,302)}, rotate = 356.03] [color={rgb, 255:red, 0; green, 0; blue, 0 }  ][fill={rgb, 255:red, 0; green, 0; blue, 0 }  ][line width=0.75]      (0, 0) circle [x radius= 3.35, y radius= 3.35]   ;
\draw  [color={rgb, 255:red, 189; green, 16; blue, 224 }  ,draw opacity=1 ] (187.5,347.5) .. controls (187.5,322.37) and (207.87,302) .. (233,302) .. controls (258.13,302) and (278.5,322.37) .. (278.5,347.5) .. controls (278.5,372.63) and (258.13,393) .. (233,393) .. controls (207.87,393) and (187.5,372.63) .. (187.5,347.5) -- cycle ;

\draw [color={rgb, 255:red, 208; green, 2; blue, 27 }  ,draw opacity=1 ]   (233,302) .. controls (254,327) and (220,357) .. (256,364) ;
\draw [shift={(238.53,343.13)}, rotate = 276.38] [color={rgb, 255:red, 208; green, 2; blue, 27 }  ,draw opacity=1 ][line width=0.75]    (10.93,-4.9) .. controls (6.95,-2.3) and (3.31,-0.67) .. (0,0) .. controls (3.31,0.67) and (6.95,2.3) .. (10.93,4.9)   ;
\draw  [color={rgb, 255:red, 189; green, 16; blue, 224 }  ,draw opacity=1 ] (185.35,330.87) .. controls (189.53,328.41) and (192.55,325.53) .. (194.41,322.23) .. controls (193.72,325.95) and (194.19,330.1) .. (195.83,334.67) ;
\draw  [color={rgb, 255:red, 189; green, 16; blue, 224 }  ,draw opacity=1 ] (275.71,317.28) .. controls (274.15,321.87) and (273.75,326.03) .. (274.5,329.74) .. controls (272.59,326.47) and (269.52,323.64) .. (265.29,321.25) ;
\draw    (319,308) -- (463,298) ;
\draw [shift={(391,303)}, rotate = 356.03] [color={rgb, 255:red, 0; green, 0; blue, 0 }  ][fill={rgb, 255:red, 0; green, 0; blue, 0 }  ][line width=0.75]      (0, 0) circle [x radius= 3.35, y radius= 3.35]   ;
\draw  [color={rgb, 255:red, 189; green, 16; blue, 224 }  ,draw opacity=1 ] (345.5,348.5) .. controls (345.5,323.37) and (365.87,303) .. (391,303) .. controls (416.13,303) and (436.5,323.37) .. (436.5,348.5) .. controls (436.5,373.63) and (416.13,394) .. (391,394) .. controls (365.87,394) and (345.5,373.63) .. (345.5,348.5) -- cycle ;

\draw [color={rgb, 255:red, 208; green, 2; blue, 27 }  ,draw opacity=1 ]   (368,241) .. controls (389,266) and (355,296) .. (391,303) ;
\draw [shift={(375.05,269.39)}, rotate = 97.06] [color={rgb, 255:red, 208; green, 2; blue, 27 }  ,draw opacity=1 ][line width=0.75]    (10.93,-4.9) .. controls (6.95,-2.3) and (3.31,-0.67) .. (0,0) .. controls (3.31,0.67) and (6.95,2.3) .. (10.93,4.9)   ;
\draw  [color={rgb, 255:red, 189; green, 16; blue, 224 }  ,draw opacity=1 ] (343.35,331.87) .. controls (347.53,329.41) and (350.55,326.53) .. (352.41,323.23) .. controls (351.72,326.95) and (352.19,331.1) .. (353.83,335.67) ;

\draw   (177,301) .. controls (181.22,303.39) and (185.23,304.56) .. (189.02,304.51) .. controls (185.45,305.78) and (182.1,308.27) .. (178.96,311.97) ;
\draw   (265,295) .. controls (269.22,297.39) and (273.23,298.56) .. (277.02,298.51) .. controls (273.45,299.78) and (270.1,302.27) .. (266.96,305.97) ;
\draw   (335,302) .. controls (339.22,304.39) and (343.23,305.56) .. (347.02,305.51) .. controls (343.45,306.78) and (340.1,309.27) .. (336.96,312.97) ;
\draw   (423,296) .. controls (427.22,298.39) and (431.23,299.56) .. (435.02,299.51) .. controls (431.45,300.78) and (428.1,303.27) .. (424.96,306.97) ;
\draw  [color={rgb, 255:red, 189; green, 16; blue, 224 }  ,draw opacity=1 ] (433.71,318.28) .. controls (432.15,322.87) and (431.75,327.03) .. (432.5,330.74) .. controls (430.59,327.47) and (427.52,324.64) .. (423.29,322.25) ;

\draw (282,44.4) node [anchor=north west][inner sep=0.75pt]    {$W$};
\draw (165,85.4) node [anchor=north west][inner sep=0.75pt]    {$\textcolor[rgb]{0.74,0.06,0.88}{\alpha ^{-}}$};
\draw (276,76.4) node [anchor=north west][inner sep=0.75pt]    {$\textcolor[rgb]{0.74,0.06,0.88}{\alpha ^{+}}$};
\draw (434,76.4) node [anchor=north west][inner sep=0.75pt]    {$\textcolor[rgb]{0.74,0.06,0.88}{\alpha }\textcolor[rgb]{0.74,0.06,0.88}{^{+}}$};
\draw (323,85.4) node [anchor=north west][inner sep=0.75pt]    {$\textcolor[rgb]{0.74,0.06,0.88}{\alpha }\textcolor[rgb]{0.74,0.06,0.88}{^{-}}$};
\draw (440,44.4) node [anchor=north west][inner sep=0.75pt]    {$W$};
\draw (278,309.4) node [anchor=north west][inner sep=0.75pt]    {$\textcolor[rgb]{0.74,0.06,0.88}{\alpha }\textcolor[rgb]{0.74,0.06,0.88}{^{+}}$};
\draw (167,318.4) node [anchor=north west][inner sep=0.75pt]    {$\textcolor[rgb]{0.74,0.06,0.88}{\alpha }\textcolor[rgb]{0.74,0.06,0.88}{^{-}}$};
\draw (284,277.4) node [anchor=north west][inner sep=0.75pt]    {$W$};
\draw (436,310.4) node [anchor=north west][inner sep=0.75pt]    {$\textcolor[rgb]{0.74,0.06,0.88}{\alpha }\textcolor[rgb]{0.74,0.06,0.88}{^{+}}$};
\draw (325,319.4) node [anchor=north west][inner sep=0.75pt]    {$\textcolor[rgb]{0.74,0.06,0.88}{\alpha }\textcolor[rgb]{0.74,0.06,0.88}{^{-}}$};
\draw (442,278.4) node [anchor=north west][inner sep=0.75pt]    {$W$};
\draw (218,279.4) node [anchor=north west][inner sep=0.75pt]    {$\textcolor[rgb]{0.82,0.01,0.11}{e^{+}}$};
\draw (205,211.4) node [anchor=north west][inner sep=0.75pt]    {$( +)( -)( -) \ +\ ( +)( +)( +) \ =\ 2$};
\draw (220,171.4) node [anchor=north west][inner sep=0.75pt]    {$( a)$};
\draw (377,170.4) node [anchor=north west][inner sep=0.75pt]    {$( b)$};
\draw (219,45.4) node [anchor=north west][inner sep=0.75pt]    {$\textcolor[rgb]{0.82,0.01,0.11}{e}\textcolor[rgb]{0.82,0.01,0.11}{^{-}}$};
\draw (380,70.4) node [anchor=north west][inner sep=0.75pt]    {$\textcolor[rgb]{0.82,0.01,0.11}{e}\textcolor[rgb]{0.82,0.01,0.11}{^{-}}$};
\draw (381,308.4) node [anchor=north west][inner sep=0.75pt]    {$\textcolor[rgb]{0.82,0.01,0.11}{e}\textcolor[rgb]{0.82,0.01,0.11}{^{+}}$};
\draw (210,438.4) node [anchor=north west][inner sep=0.75pt]    {$( -)( -)( -) \ +\ ( -)( +)( +) \ =\ -2$};
\draw (222,403.4) node [anchor=north west][inner sep=0.75pt]    {$( c)$};
\draw (378,404.4) node [anchor=north west][inner sep=0.75pt]    {$( d)$};

\end{tikzpicture}
    \caption{If an edge lies entirely inside $\alpha$, then $k(e) = 0$. Otherwise, there are four possible relative orientations, as shown above. For each configuration, we move down along $e$, and pass into $W$ and $\alpha$. If the orientation of a component encountered during the traversal is the same as the direction of traversal, we assign a positive sign to that component; otherwise, we assign a negative sign. The first term in each sum corresponds to traversal along $\alpha^-$ and the second to that along $\alpha^+$. It is easy to see that if one starts traversing along some component other than $e$, the same results are obtained. Thus the assignments above are well-defined.}
    \label{fig3.7}
\end{figure}
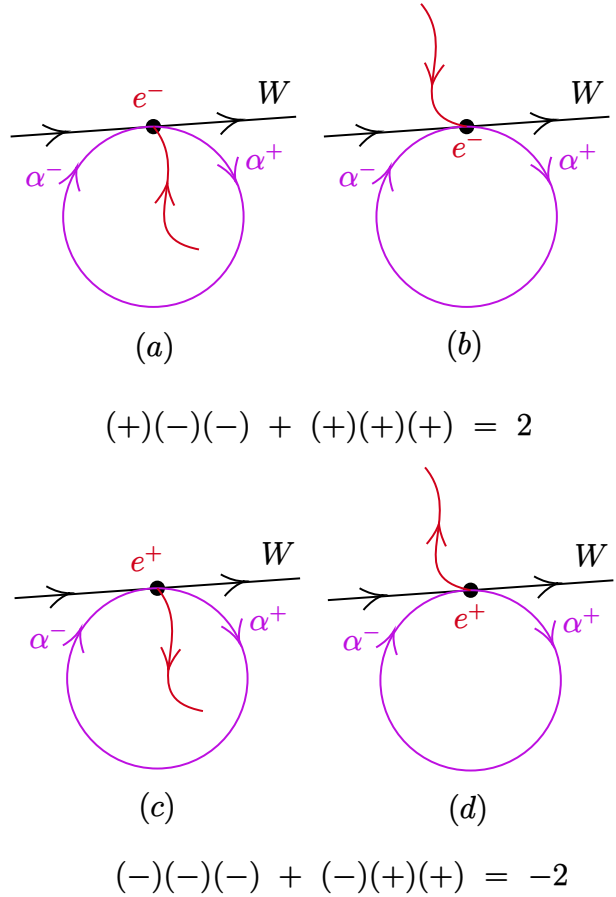

Fix a point $p$ on $\alpha$, and consider a one-dimensional subspace $W$ of the tangent space at $p$. We define \cite{ashtekar 5} the vector field $X_{\alpha, W}: C^\infty(\mathcal{A}_\gamma)\to C^\infty(\mathcal{A}_\gamma)$ such that for all $f_\gamma \in C^\infty({\mathcal{A}_\gamma})$, 
\begin{subequations}
\label{eq3.142}
    \begin{align}
        X_{\alpha,W}(f_\gamma) = -i\text{tr}\,(U_\alpha(f_\gamma)\tau^i)\sum_{e\in\gamma}\left[ k^-(e)X^L_{e,i}(f_\gamma) + k^+(e)X^R_{e,i}(f_\gamma) \right],
    \end{align}
    \begin{align}
    k^{\pm}(e) = \begin{cases}
 0& \text{ if } e^{\pm}\neq p, \\ 
 \frac{1}{4}\left[ \text{sgn}(\dot{e}^{\pm}, \dot{\alpha}^+, W) +  \text{sgn}(\dot{e}^{\pm}, \dot{\alpha}^-, W)\right]& \text{ if } e^{\pm}=p. 
\end{cases}
    \end{align}
\end{subequations}
Let us subdue this monstrous equation. $U_\alpha$ is the vector field defined by Eq~(\ref{eq3.141}). $\tau^i$ are the Pauli matrices. $X^L_{e,i}$ and $X^R_{e,i}$ are the left- and right-invariant vector fields that come from the Pauli matrices via Eq~(\ref{eq3.139}), defined on the copy of $SU(2)$ associated with the edge $e$. $e^+$ and $e^-$, respectively, are the beginning and ending points of the edge $e$, and $\alpha^+$ and $\alpha^-$, respectively are the ingoing and outgoing segments of $\alpha$ at $p$. The overdots indicate tangent vectors to the respective curves, and serve to specify their orientations. Finally, $\text{sgn}(\dot{e}^{\pm}, \dot{\alpha}^{\pm}, W) = 0, \pm 1$, depending on the relative orientations of $\dot{e}^{\pm}, \dot{\alpha}^{\pm}$ and $W$ -- the precise prescription is given in Fig~\ref{fig3.7}. We call $k^{\pm}(e)$ the orientation factor associated to the edge $e$.

So far, we have promoted the $T^a$ variables to vector fields on $\mathcal{A}_\gamma$. The transition to the $T^1$ variables is now straightforward. Given a surface $S\in \Sigma$ foliated by a one-parameter family loops, we set
\begin{equation}
    \hat{T}^1_S = \sum_{x\in S}X_{\alpha_x, W_x}, \label{eq3.143}
\end{equation}
where $\alpha_x$ is a loop in $S$ passing through $x$, and $W_x$ is a one-dimensional subspace of the tangent space $\Sigma$ at $x$. The uncountable number of loops in the sum above may appear uncomfortable, but appearances can be illusory! Since any graph $\gamma$ has finitely many edges, which will thus intersect $S$ only finitely many times, according to Eq~(\ref{eq3.142}), the action of $\hat{T}^1_S$ on any $f_\gamma \in C^\infty(\mathcal{A}_\gamma)$ will have only finitely many nonzero terms. 

We now have vector field representations of the momentum variables on the projective family $\{\mathcal{A}_\gamma\}$ labelled by graphs in $\Sigma$. Let us now see how one can consistently extend these vector fields to vector fields on $\overline{\mathcal{A}}$. In other words, we know how our fields act on $f_\gamma \in C^\infty(\mathcal{A}_\gamma)$. We must now ask whether those fields can be consistently extended to act on functions in $C^\infty({\overline{\mathcal{A}}})$. To begin with, note that the results of the previous section entail that $C^\infty(\overline{\mathcal{A}})$ is isomorphic to $Cyl^\infty(\overline{\mathcal{A}})$, i.e. the space of \textit{smooth} cylindrical functions on $\overline{\mathcal{A}}$. This lets us precisely say what it means to consistently extend vector fields defined on a projective family to the corresponding projective limit. Given two graphs $\gamma' \geq \gamma$, and two vector fields $X_\gamma$ and $X_{\gamma'}$ defined, respectively, on $\mathcal{A}_\gamma$ and $\mathcal{A}_{\gamma'}$, if two functions $f_\gamma \in C^\infty{\mathcal{A}_\gamma}$ and $f_{\gamma'} \in C^\infty{\mathcal{A}_{\gamma'}}$ represent the same cylindrical function $f\in Cyl^\infty(\overline{\mathcal{A}})$, we want the action of the two fields on the respective functions to coincide. That is, $X_{\gamma}(f_\gamma) = X_{\gamma'}(f_{\gamma'})$. Now if $X_\gamma = X_\alpha$ for all $\gamma$, this consistency condition of ours is always met, for suppose $A_\gamma, A_{\gamma'}$ comes from a particular $(A_{\gamma})_{\gamma\in L} \in \overline{\mathcal{A}}$. Then by Eq~(\ref{eq3.112}), 
\begin{equation*}
    X_{\alpha}(f_{\gamma'}(A_\gamma')) = X_{\alpha}f_{\gamma}(p_{\gamma\gamma'}(A_{\gamma'})) = X_\alpha(f_\gamma(A_\gamma)). 
\end{equation*}
Thus, the momentum vector fields are well-defined on $\overline{\mathcal{A}}$.

Finally, we must extend these fields to $\overline{\mathcal{A/G}}$. For this, they must be invariant under $\overline{\mathcal{G}}$. But this is true by construction, since gauge transformations do not change the location of the vertices of a graph, and the factor of $\text{tr}\,(U_\alpha\tau^i)$ involved in the definition (\ref{eq3.142}) is manifestly gauge invariant. Therefore, we have consistently promoted the momentum variables to operators on $\mathcal{H}_{kin} = L^2(\overline{\mathcal{A/G}}, d\mu_{AL})$. Finally, it is possible to show \cite{ashtekar 5, ashtekar 7} that all the operators constructed above are self-adjoint on their respective domains; thus step 5 of the quantisation scheme sketched in Section 3.1 is complete. In fact, all the steps except for the last two are almost complete -- what remains to be done is to ensure that the commutator algebra of the configuration and momentum operators mimics their classical Poisson algebra, as demanded in step 2. This we shall verify in the next subsection.  

A remark is in order. The discussion in this section has been much less rigorous than that in the previous sections. In particular, most of the arguments presented above for the consistent extension of vector fields from the projective family to its projective limit are rather heuristic. For a more thorough treatment, we refer the reader to Ref. \cite{ashtekar 7}.

\subsection{Canonical commutation relations}
We have clearly ventured deep into the waters of mathematical niceties. Lest the density of our abstractions suffocates us, we must find a way back to the surface and breathe in the air of physics. Thus, in this subsection, we shall test the faithfulness of our toilsome construction of the configuration and momentum operators by investigating whether or not these operators satisfy commutation relations that mimic the Poisson brackets between their classical counterparts.

The operators $\hat{T}^0_\alpha$ are abelian by construction, and hence their commutator vanishes, as expected. 

Next we consider the commutation relations between the configuration and momentum operators. Let $\alpha$ and $\beta$ be two loops in $\Sigma$. For any $\psi \in \mathcal{H}_{kin}$ and $A \in \overline{\mathcal{A/G}}$, we find
\begin{align}
    [X_{\alpha,W}, \hat{T}^0_\beta]\psi(A) &= X_{\alpha, W}(T^0_\beta(A)\psi(A)) - T^0_\beta(A)X_{\alpha,W}(\psi(A)) \nonumber \\
    &= X_{\alpha, W}(T^0_\beta(A))\psi(A) + T^0_\beta(A)X_{\alpha, W}(\psi(A)) - T^0_\beta(A) X_{\alpha, W}(\psi(A)) \nonumber \\
    &= X_{\alpha, W}(T^0_\beta(A))\psi(A), \label{eq3.144}
\end{align}
where we made use of the product rule. Thus all we need to do is to evaluate the action of $X_{\alpha, W}$ on $T^0_\beta(A)$, trace of the holonomy of $A$ around $\beta$. This is fairly straightforward. 
\begin{align}
    X_{\alpha, W}(T^0_\beta(A)) &= -i\text{tr}\,(U_\alpha(A)\tau^i)\left[ k^-(\beta)\left.\frac{d}{dt}\text{tr}\,(U_\beta(A)e^{it\tau_i})\right|_{t=0}\right.   \nonumber \\ 
    &\qquad\qquad\qquad + \left.k^+(\beta)\left.\frac{d}{dt}\text{tr}\,(e^{-it\tau_i}U_\beta(A))\right|_{t=0} \right] \nonumber \\
    &= (k^-(\beta) - k^+(\beta))\text{tr}\,(U_\alpha(A)\tau^i)\text{tr}\,(\tau_i U_\beta(A))\nonumber \\
    &= (k^-(\beta) - k^+(\beta))U_\alpha(A)^{AB}\tensor{\tau}{^i_{BA}}\tensor{\tau}{_i^{CD}}U_\beta(A)_{DC} \nonumber \\
    &= -2(k^-(\beta) - k^+(\beta))U_\alpha(A)^{AB}\delta^{(C}_B\delta^{D)}_A U_\beta(A)_{DC} \nonumber \\
    &= -(k^-(\beta) - k^+(\beta))(U_\alpha(A)^{DC}U_\beta(A)_{DC} + U_\alpha(A)^{CD}U_\beta(A)_{DC}) \nonumber \\
    &=  -(k^-(\beta) - k^+(\beta))(-T^0_{\alpha \# \beta^{-1}} +  T^0_{\alpha \# \beta}). \label{eq3.145}
\end{align}
According to Fig~\ref{fig3.7}, whenever the factor $(k^-(\beta) - k^+(\beta))$ is nonzero, it has the value $+1$. Thus, except for the factor of the delta function, Eq~(\ref{eq3.145}) reduces to $-i$ times Eq~(\ref{eq3.23}), and if we substitute Eq~(\ref{eq3.145}) into Eq~(\ref{eq3.143}), we get exactly $-i$ times Eq~(\ref{eq3.28}) -- precisely what we wish. 

Finally, we turn towards the commutator of the momentum operators. Let $\psi_\gamma$ be the representation of $\psi \in \mathcal{H}_{kin}$ on a graph $\gamma$. For two loops $\alpha$ and $\beta$, we have
\begin{align}
    [X_{\alpha, V}, X_{\beta, W}]\psi(A) &= X_{\alpha, V}[-i\text{tr}(U_\beta(A)\tau^i)\sum_{e\in\gamma}\underbrace{\{k^-(e)X^L_{e,i}(\psi_\gamma(A)) + k^+(e)X^R_{e,i}(\psi_\gamma(A))\}}_{K_i(e, \beta, W)}]\nonumber \\
    &\qquad\qquad\qquad\qquad +\binom{\alpha\leftrightarrow\beta}{V\leftrightarrow W} \nonumber \\
    &= -\text{tr}(U_\alpha(A)\tau^j)\sum_{e'\in\gamma}\left[k^-(e')X^L_{e',j}\left(\text{tr}(U_\beta(A)\tau^i)\sum_{e\in\gamma}K_i(e,\beta,W)\right)\right. \nonumber \\
    &\left.\qquad\qquad\qquad\qquad + k^+(e')X^R_{e',j}\left(\text{tr}(U_\beta(A)\tau^i)\sum_{e\in\gamma}K_i(e,\beta,W)\right)\right] \nonumber \\
    &\qquad -\text{tr}(U_\beta(A)\tau^i)\text{tr}(U_\alpha(A)\tau^j)\sum_{e,e'\in\gamma}\left[k^-(e')X^L_{e',j}(K_i(e,\beta, W))\right. \nonumber \\
    &\qquad\qquad\qquad\qquad\qquad\qquad\qquad +\left. k^+(e')X^R_{e',j}(K_i(e,\beta, W)) \right] \nonumber \\
    & \qquad\qquad\qquad\qquad +\binom{\alpha\leftrightarrow\beta}{V\leftrightarrow W} \nonumber \\
    &= -i\text{tr}(U_\alpha(A)\tau^i)\text{tr}(U_\alpha(A)\tau_j\tau^i)\underbrace{\sum_{e,e'\in\gamma}(k^-(e')-k^+(e'))K_i(e,\beta,W)}_{M_i(e,\alpha, \beta, V, W)} \nonumber \\
    &= -i\left[\text{tr}(U_{\alpha^{-1}\#\beta}(A)\tau^i) - \text{tr}(U_{\beta^{-1}\#\alpha}(A)\tau^i)\right]M_i(e,\alpha,\beta,V,W). \label{eq3.146}
\end{align}
The factor $M_i(e,\alpha,\beta, V, W)$ is nonzero only if $\alpha$, $\beta$, $V$, $W$ and $e$ all intersect. Thus, substituting the preceding equation into Eq~(\ref{eq3.143}), we see that the commutator of two momentum operators is nonzero only when their corresponding surfaces intersect -- once again, precisely our expectation. 

This shows that the commutator algebra of the configuration and momentum operators mimics the classical Poisson algebra. We are in safe territory.

\section{Spin Networks}
A fastidious perusal of the preceding sections ought to have revealed that until now, we have completed the first five steps of the quantisation program outlined in Section 3.1. Moreover, the Gauss constraint introduced in the previous chapter has also been dealt with, since the states in $\mathcal{H}_{kin} = L^2(\overline{\mathcal{A/G}}, d\mu_{AL})$ are by construction gauge invariant. What remains to be done is the implementation of the diffeomorphism and Hamiltonian constraints at the quantum level. But to do so, we first need to see how the states in $\mathcal{H}_{kin}$ look like. Thus in this section, we will derive \cite{ashtekar 5, ashtekar 6, giesel} an orthonormal basis of $\mathcal{H}_{kin}$. States written in this basis are called \textit{spin network} states. We will find this basis to be extremely convenient for explicit calculations, and for finding solutions to the diffeomorhpism and Hamiltonian constraints in the next chapter. 

For any graph $\gamma$, let $\mathcal{H}_\gamma = L^2(\mathcal{A_\gamma/G_\gamma}, d\mu_H)$, i.e. the space of square-integrable functions on $\mathcal{A_\gamma/G_\gamma}$. Since the union of $\mathcal{H}_\gamma$ over $\gamma$ is dense\footnote{This follows from the results of Section 3.4. Recall that the $\mu_{AL}$-completion of the sup-norm completion of $Cyl(\overline{\mathcal{A/G}})$ is $L^2(\overline{\mathcal{A/G}}, d\mu_{AL})$. On the other hand, the $\mu_H$-completion of $C^0(\mathcal{A_\gamma/G_\gamma})$ is $L^2(\mathcal{A_\gamma/G_\gamma}, d\mu_H)$, and the union of $C^0(\mathcal{A_\gamma/G_\gamma})$ over $\gamma$ is $Cyl(\mathcal{A_\gamma/G_\gamma})$.} in $\mathcal{H}_{kin} = L^2(\overline{\mathcal{A/G}}, d\mu_{AL})$, a basis of the former is also a (not necessarily orthonormal) basis of the latter. We will, therefore, derive an orthonormal basis for $\mathcal{H}_\gamma$ and then show how it can be converted into an orthonormal basis for $\mathcal{H}_{kin}$. 

\subsection{Peter-Weyl theorem}
Let us start with the simplest kind of problem that is similar to the behemoth that currently stands in front of us. Consider a Lie group $\mathcal{G}$ and the space $L^2(\mathcal{G}, d\mu_H)$ of complex-valued functions on $\mathcal{G}$ that are square-integrable with respect to the Haar measure. How to construct an orthonormal basis for $L^2(\mathcal{G})$ (the Haar measure is implied, so we won't bother writing it any longer)? The answer is provided by the celebrated Peter-Weyl theorem, proven by Hermann Weyl and his student Fritz Peter. To understand the import of this theorem, an important digression into representation theory of Lie groups is in order. 

Recall that a representation of a group $G$ on a vector space $V$ is a group homomorphism from $G$ to $GL(V)$, i.e. $\rho: G\to GL(V)$ such that $\rho(gh) = \rho(g)\rho(h)$ for all $h, g \in G$. Here, $GL(V)$ is the set of all bijective linear transformations from $V$ to itself, i.e. invertible operators on $V$, which form a group under functional multiplication. If $V$ is finite-dimensional, then $GL(V)$ is simply $GL(n)$, the set of all invertible $n\times n$ matrices, where $n$ is the dimension of $V$; since we shall only be concerned with finite-dimensional representations, we will make this identification once and for all. 

A subspace $W\subset V$ is said to be $G$-invariant if $\rho(g)w \in W$ for all $g\in G$ and $w\in W$. The restriction $\rho|_W$ of $\rho$ to $W$ is called a subrepresentation. Note that $V$ itself and the empty set are trivial $G$-invariant subspaces. If a representation admits no nontrivial subrepresentation, it is called \textit{irreducible}. Conversely, if there is at least one nontrivial subrepresentation, then we have a \textit{reducible} representation. In terms of matrix representations, reducibility of a representation $\rho$ translates to there being a similarity transformation $S$ that transforms every matrix $\rho(g)$ into upper-triangular form, i.e. $\rho'(g) = S^{-1}\rho(g)S$ is upper triangular for every $g \in G$. 

Now let $H$ be a Hilbert space. Then a representation $\rho: G\to H$ is called unitary if $\rho(g)$ is a unitary operator on $H$ for all $g\in G$. The Peter-Weyl theorem is concerned with the unitary representations of a compact Lie group $\mathcal{G}$ on a complex Hilbert space $H$. It has multiple parts \cite{knapp}. The parts that we will be interested in pertain to the unitary representations of $\mathcal{G}$ on $L^2(\mathcal{G})$. In particular, they give a recipe for constructing an orthonormal basis for $L^2(\mathcal{G})$ in terms of finite-dimensional unitary representations of $\mathcal{G}$. 

This is how it works. To begin with, the theorem asserts that every irreducible unitary representation of $\mathcal{G}$ is finite-dimensional. Now let $\Gamma$ be the set of all irreducible unitary representations of $\mathcal{G}$; for convenience, we may write $\Gamma = \{\rho^j: j\in \Lambda\}$, where $\Lambda$ is some indexing set. For each $\rho^j \in \Gamma$, let the associated (finite-dimensional) vector space be denoted by $V_j$. The theorem further says that $L^2(\mathcal{G})$ can be decomposed as the (closure of) the direct over all inequivalent irreducible unitary representations of $\mathcal{G}$ . More precisely, we have
\begin{equation}
    L^2(\mathcal{G}) = \overline{\bigoplus}_{j\in\Lambda}V_j\otimes V^*_j, \label{eq3.147} 
\end{equation}
where $V^*_j$ is the dual of $V_j$. Finally, given this decomposition, the theorem constructs an explicit orthonormal basis of $L^2(\mathcal{G})$; it is essentially a mutually orthonormal collection of orthonormal bases of $V_j\otimes V^*_j$. 

To see how this works in practice, let $d_j$ be the dimension of $V_j$. For each $j\in\Lambda$, choose an orthonormal basis $\{e_m: m\in (1, \ldots, d_j)\}$ of $V_i$, and let $ \phi_m$ denote the corresponding dual basis of $V^*_j$. Then for every $g\in\mathcal{G}$, $\rho^j(g)$ can be thought of as a matrix with elements given by
\begin{equation}
    \rho^j_{mn}(g) := \phi_m(\rho^j(g)e_n). \label{eq3.148}
\end{equation}
Notice that $\rho^j_{mn}$ are complex-valued functions on $\mathcal{G}$. They form a basis of $V_j\otimes V^*_j$. We thus define\footnote{The use of bras and kets here is purely for notational convenience, and is offered to assuage the alienating effects upon the reader of the intensely mathematical journey she must have undertaken while ruminating over the contents of this chapter -- the sight of a tool from the physicist's toolkit may well be refreshing.} functions $|a^j_{mn}\rangle: \mathcal{G} \to \mathbb{C}$ such that
\begin{equation}
    \langle g|a^j_{mn}\rangle := \sqrt{d_j}\rho^j_{mn}(g). \label{eq3.149}
\end{equation}
The $|a^j_{mn}\rangle$ form an orthonormal basis for $V_j\otimes V^*_j$, and by Eq~(\ref{eq3.147}), $\{|a^j_{mn}\rangle: j\in \Lambda\}$ forms a collection of orthonormal vectors in $L^2(\mathcal{G})$. The utility of the Peter-Weyl theorem consists in its showing that this collection also forms a \textit{basis}\footnote{One may worry here about $\Lambda$ being possibly uncountable. But this may be avoided (see \href{https://math.stackexchange.com/questions/524203/uncountable-orthonormal-system-in-hilbert-spaces}{here}) by the very fact that we have an orthonormal basis for our Hilbert space. This is because for any $|h\rangle\in L^2(\mathcal{G})$, the set $B = \{j\in\Lambda: \langle a^j_{mn}|h\rangle \neq 0\}$ is countable, and we may write $|h\rangle = \sum_{j\in B} \langle a^j_{mn}|h\rangle| a^j_{mn}\rangle$, for otherwise, there would be infinitely many nonzero terms in $|\langle h|h\rangle|^2 = \sum_{j\in\Lambda} |\langle h|a^j_{mn}\rangle|^2$. Therefore, without loss of generality, we may identify $\Lambda = \mathbb{Z}_{+}$.} of $L^2(\mathcal{G})$:
\begin{equation}
    \langle a^j_{pq}| a^k_{rs}\rangle = \int_{\mathcal{G}}d\mu_H(g)\sqrt{d_j}\sqrt{d_k}\,\overline{\rho^j_{pq}(g)}\rho^{k}_{rs}(g) = \delta^{jk}\delta_{pr}\delta_{qs}, \label{eq3.150}
\end{equation}
where $\mu_H$ is, of course, the Haar measure and the overline indicates complex conjugation. In summary, the Peter-Weyl theorem gives us an orthonormal basis of $L^2(\mathcal{G})$ in terms of the matrix elements of finite-dimensional irreducible unitary representations of $\mathcal{G}$. This, as we shall see, will be of immense use to us. 

\subsection{Spin network functions}
Consider now\footnote{Here and in future, we will omit explicit reference to the measures on the spaces we are dealing with. They will be clear from the context: for a graph, we will always have copies of the Haar measure, and for $\overline{\mathcal{A/G}}$, the Ashtekar-Lewandowski measure is always implied.} $L^2(\mathcal{A}_\gamma)$ for an arbitrary graph $\gamma$ with, say, $N$ edges. It is obvious from the results of Sections 3.4 and 3.5 that
\begin{equation}
    L^2(\mathcal{A}_\gamma) \cong L^2(SU(2)^N)\cong  \bigotimes_{e\in \gamma}L^2(SU(2)). \label{eq3.151}
\end{equation}
We can thus use the results of the previous subsection to construct an orthonormal basis of $L^2(\mathcal{A}_\gamma)$. Since now we have a tensor product of Hilbert spaces over $SU(2)$, we can decompose each space into irreducible unitary representations of $SU(2)$. Then we can use the associativity of the tensor product to rewrite the whole $L^2(\mathcal{A}_\gamma)$ as a direct sum of tensor products of Hilbert spaces. More precisely, let $\{\rho^j\}$ be the set of irreducible unitary representations of $SU(2)$. To each edge $e$ in $\gamma$, we associate a set of representations $\rho^{j_e}$, and for each value of $j_e$, we define an orthonormal basis $\rho^{j_e}_{m_en_e}$ of $L^2(SU(2))$ in the manner of Eq~(\ref{eq3.149}). Collecting all the labels $\{j_{e_1}, \ldots, j_{e_N}\} := \mathbf{j}$, $\{m_{e_1}, \ldots, m_{e_N}\} := \mathbf{m}$ and $\{n_{e_1}, \ldots, n_{e_N}\} := \mathbf{n}$, we introduce the so-called spin-network basis $|s^{\mathbf{j}}_{\gamma, \mathbf{m}, \mathbf{n}}\rangle: \mathcal{A}_\gamma \to \mathbb{C}$ such that for all $A\in \mathcal{A}_\gamma$
\begin{equation}
    \langle A|s^{\mathbf{j}}_{\gamma, \mathbf{m}, \mathbf{n}}\rangle = \sqrt{d^{j_{e_1}}}\cdots\sqrt{d^{j_{e_N}}}\rho^{j_{e_1}}_{m_{e_1}n_{e_1}}(A(e_1))\cdots\rho^{j_{e_N}}_{m_{e_N}n_{e_N}}(A(e_N)). \label{eq3.152} 
\end{equation}
From Eq~(\ref{eq3.147}) and (\ref{eq3.151}), and the associativity of the tensor product, it follows that these spin-network states are a basis for $L^2(\mathcal{A}_\gamma)$. Indeed, we can write
\begin{equation}
    L^2(\mathcal{A}_\gamma) \cong \bigoplus_{\mathbf{j}}\bigotimes_{i=1}^{N}V_{j_{e_i}}\otimes V^*_{j_{e_i}}. \label{eq3.153}
\end{equation}
To make some contact with familiar territory, recall that the irreducible unitary representations of $SU(2)$ are typically labelled by their `spin' $j = 0, 1/2, 1, 3/2, \ldots$, the dimension of the spin-$j$ representation being $d^j = 2j + 1$. The $2j+1$ basis states of each representation are labelled by $m$, where $m \in (-j, -j+1, \ldots, j-1, j)$. Thus we are doing nothing but labelling each edge of $\gamma$ with spins $j$ and in effect, generalising the concept of spin representations to finitely many copies of $SU(2)$; this is where the spin-network basis gets its name from. 

However, this is barely half the work required of us. We have constructed an orthonormal basis of $L^2(\mathcal{A}_\gamma)$, but our eventual goal is to do the same for $L^2(\mathcal{A}_\gamma/\mathcal{G}_\gamma)$ and subsequently for $L^2(\overline{\mathcal{A/G}})$. Now recall that gauge transformations in $\mathcal{G}_\gamma$ act only at the vertices of edges in $\gamma$ (see Eq~(\ref{eq3.103})). To see how this action translates to $L^2(\mathcal{A}_\gamma)$ as decomposed in Eq~(\ref{eq3.153}), we must actually decompose it further. To this end, for a graph $\gamma$ with $N$ edges, let us define
\begin{equation}
    \mathcal{H}_{\gamma, \mathbf{j}} := \bigotimes_{i=1}^{N}V_{j_{e_i}}\otimes V^*_{j_{e_i}}, \label{eq3.154} 
\end{equation}
so that $L^2(\mathcal{A}_\gamma) \cong \bigoplus_{\mathbf{j}}\mathcal{H}_{\gamma, \mathbf{j}}$. Focus on one particular set of representations $\mathbf{j}$. Note that we label each edge $e$ in $\gamma$ with a particular representation $j_e \in \mathbf{j}$. In other words, we identify different edges of a graph with reference to the representations they correspond to. Now, different edges begin and end at different vertices in the graph, and so, we may alternatively think of labelling the vertices of a graph with representations. Thus, let $V(\gamma)$ be the vertices of the graph $\gamma$. For each vertex $v_i \in V(\gamma)$, let $E(v_i)$ be the set of edges intersecting at $v_i$. Then the tensor product over edges in Eq~(\ref{eq3.154}) may be decomposed into two parts, the first part being a product over vertices rather than edges, and the second part being a product over edges intersecting at a vertex -- this will evidently be an equivalent way of keeping track of all the edges in the decomposition. Therefore, we arrive at
\begin{subequations}
\label{eq3.155}
    \begin{align}
    \mathcal{H}_{\gamma, \mathbf{j}} &= \bigotimes_{i=1}^{M}\mathcal{H}_{v_i}, \\
    \mathcal{H}_{v_i} := &\bigotimes_{e\in E(v_i)}V_{j_{e}}\otimes V^*_{j_{e}}, 
\end{align}
\end{subequations}
where $M$ is the number of vertices in $\gamma$. Thus, the spin-network basis in Eq~(\ref{eq3.152}) can be rewritten as
\begin{equation}
    \langle A| s^{\mathbf{j}}_{\gamma, \mathbf{m}, \mathbf{n}}\rangle = \prod_{v\in V(\gamma)}\prod_{e\in E(v)}\sqrt{2j_e+1}\rho^{j_e}_{m_en_e}(A(e)). \label{eq3.156}
\end{equation}

What is the advantage of this convoluted reformulation? As we shall now see, decomposition along vertices allows us to construct an orthonormal basis of $L^2(\mathcal{A}_\gamma/\mathcal{G}_\gamma)$. Recall again that gauge transformations in $\mathcal{G}_\gamma$ act on $\mathcal{A}_\gamma$ along the vertices in $\gamma$. That is, given $A\in \mathcal{A}_\gamma$ and $g\in \mathcal{G}_\gamma$, we have (see Eq~(\ref{eq3.103}))
\begin{equation}
    gA(e) = g(e(a))A(e)g^{-1}(e(b)) \label{eq3.157}
\end{equation}
for every edge $e\in \gamma$, $e(a)$ and $e(b)$ being the beginning and ending points of $e$, respectively. Since group representations $\rho^{j_e}$ are homomorphic, their matrix elements transform as
\begin{equation}
    g\rho^{j_e}_{m_en_e}(A(e)) = \sum_{k_e, l_e} \rho^{j_e}_{m_ek_e}(g(e(a)))\rho^{j_e}_{k_el_e}(A(e))\rho^{j_e}_{l_en_e}(g^{-1}(e(b))). \label{eq3.158} 
\end{equation}
Clearly, finding an orthonormal basis for $L^2(\mathcal{A}_\gamma/\mathcal{G}_\gamma)$ amounts to finding a $\mathcal{G}_\gamma$-invariant subset of the spin-network basis states $\{| s^{\mathbf{j}}_{\gamma, \mathbf{m}, \mathbf{n}}\rangle\}$. In view of the preceding considerations, this is equivalent to finding subspaces of $\mathcal{H}_{v_i}$ for every vertex $v_i \in V(\gamma)$ such that they are invariant under the transformation (\ref{eq3.158}). To facilitate this task, we define bases of the space $\mathcal{H}_{v_i}$. Since each $\mathcal{H}_{v_i}$ is a tensor product over, say, $k$ spin representations, one for each edge coming in or going out at $v_i$, we may denote an arbitrary basis of $\mathcal{H}_{v_i}$ as a $(0, k)$ tensor $t_{n_1\cdots n_k}^i$. Similarly, we define bases $\Tilde{t}_i^{n_1\cdots n_k}$ of the dual spaces $\mathcal{H}^*_{v_i}$. A dual space carries the dual $\Tilde{\rho}$ of a representation $\rho$, where $\Tilde{\rho}$ is defined by $\Tilde{\rho}(g) := \rho(g^{-1})^{T}$ for all group elements $g$. In particular, we pick $\Tilde{t}_i$ to be dual bases:
\begin{equation}
    \Tilde{t}_i(t^j) = \Tilde{t}_i^{n_1\cdots n_k}t_{n_1\cdots n_k}^i = \delta^i_j, \label{eq3.159}
\end{equation}
where summation over repeated indices is implied. Being tensors, these bases have a natural action of $\mathcal{G}_\gamma$ on them:
\begin{subequations}
    \label{eq3.160}
    \begin{align}
        g\cdot \Tilde{t}^{n_1\cdots n_k}_i &= \rho^{j_{e_1}}\tensor{(g(v_i))}{^{n_1}_{m_1}}\cdots \rho^{j_{e_k}}\tensor{(g(v_i))}{^{n_k}_{m_k}}t^{m_1\cdots m_k}_i, \\
        g\cdot t_{n_1\cdots n_k}^i &= \rho^{j_{e_1}}\tensor{(g^{-1}(v_i))}{_{n_1}^{m_1}}\cdots \rho^{j_{e_k}}\tensor{(g^{-1}(v_i))}{_{n_k}^{m_k}}t_{m_1\cdots m_k}^i.
    \end{align}
\end{subequations}
Here we have defined $\rho^j_{mn}(g(v)) := \rho^j\tensor{(g(v))}{^m_n}$ to make use of the summation convention. Using these bases, we can construct arbitrary tensors in $\mathcal{H}_{v_i}$. For instance, a $(1,0)$ tensor $V$ will be given by $V= V^nt^i_n$. 

Next, we substitute Eq~(\ref{eq3.158}) into Eq~(\ref{eq3.156}) to find the transformation properties of the spin-network basis:
\begin{equation}
    \langle gA|s^{\mathbf{j}}_{\gamma, \mathbf{m}, \mathbf{n}}\rangle = \prod_{v\in V(\gamma)}\prod_{e\in E(v)}\sqrt{2j_e+1} \rho^{j_e}\tensor{(g(e(a)))}{^{m_e}_{q_e}}\rho^{j_e}\tensor{(A(e))}{^{q_e}_{r_e}}\rho^{j_e}\tensor{(g^{-1}(e(b)))}{^{r_e}_{n_e}} \label{eq3.161}
\end{equation}
This can be massaged into a more illuminating form. Observe that for a particular vertex $v$, the second product above contains edges that intersect at $v$. Thus for each edge in the product, $v$ is either its starting point, ending point or both. Therefore, either one or both of the factors among $\rho(g(e(a)))$ and $\rho(g^{-1}(e(b)))$ can be replaced with $\rho(g(v))$. If only one of the factors, say $\rho(g(e(a)))$, is so replaced, the other factor can be replaced by $\rho(g^{-1}(v'))$, where $v'$ is some other vertex which is the beginning or ending point of the edge in question; and we are eventually going to consider the products over edges intersecting at $v'$ as well, since the first product above runs over all vertices. This suggests writing Eq~(\ref{eq3.156}) in a different manner. Notice that if an edge begins at $v$, it can be thought of as going out at $v$, and if it ends at $v$, we can take it to be coming in at $v$. Thus, for each vertex $v$, we break the set $E(v)$ into two parts, namely $E^+(v)$ and $E^-(v)$, which contain, respectively, the outgoing and ingoing edges at $v$. Then we may write
\begin{equation}
    \langle A|s^{\mathbf{j}}_{\gamma, \mathbf{m}, \mathbf{n}}\rangle = \prod_{v\in V(\gamma)}\prod_{e+\in E^{+}(v)}\prod_{e'\in E^{-}(v)}\sqrt{2j_e+1}\sqrt{2j_{e'}+1}\rho^{j_e}\tensor{(A(e))}{^{m_e}_{n_e}}\rho^{j_{e'}}\tensor{(A(e'))}{^{m_{e'}}_{n_{e'}}}. \label{eq3.162}
\end{equation}
Substituting Eq~(\ref{eq3.161}) into the preceding one reveals transformations occurring at every outgoing and ingoing edge at every vertex:
\begin{align}
    \langle A|s^{\mathbf{j}}_{\gamma, \mathbf{m}, \mathbf{n}}\rangle = &\prod_{v\in V(\gamma)}\prod_{e+\in E^{+}(v)}\prod_{e'\in E^{-}(v)}\sqrt{2j_e+1}\sqrt{2j_{e'}+1} \nonumber \\
    & \times \rho^{j_e}\tensor{(g(v))}{^{m_e}_{q_e}}\rho^{j_e}\tensor{(A(e))}{^{q_e}_{r_e}}\rho^{j_e}\tensor{(g^{-1}(v))}{^{r_e}_{n_e}}\nonumber \\
    & \times \rho^{j_{e'}}\tensor{(g(v))}{^{m_{e'}}_{q_{e'}}}\rho^{j_{e'}}\tensor{(A({e'}))}{^{q_{e'}}_{r_{e'}}}\rho^{j_{e'}}\tensor{(g^{-1}(v))}{^{r_{e'}}_{n_{e'}}}. \label{eq3.163}
\end{align}
To obtain a gauge-invariant state, we ought to cancel the factors of $\rho(g(v))$ and $\rho(g^{-1}(v))$ occurring above at each vertex $v$. This can be done by constructing appropriate $(k,l)$ tensors using the bases and dual bases of the spaces $\mathcal{H}_{v_i}$ introduced above. Suppose that the vertex $v_i\in V(\gamma)$ has $k$ outgoing edges and $l$ ingoing edges. Consider an arbitrary $(k,l)$ tensor at $v_i$, i.e. a tensor $I$ in 
\begin{equation}
    \bigotimes_{p=1}^{l}(V_{j_p}\otimes V^*_{j_p})^*\otimes \bigotimes_{q=1}^{k}(V_{j_q}\otimes V^*_{j_q}). \label{eq3.164}
\end{equation}
By construction, $I$ is invariant under gauge transformations, its components $I^{n_1\cdots n_k}_{i,m_1\cdots m_l}$ transforming inversely as the bases $t^i$ and $\Tilde{t}_i$ transform (Eq~\ref{eq3.160}). In other words, each upper index of $I^{n_1\cdots n_k}_{i,m_1\cdots m_l}$ transforms with factors of $\rho(g(v_i))$ and each lower index transforms with factors of $\rho(g^{-1}(v_i))$. Therefore, we can contract the corresponding indices in $I^{n_1\cdots n_k}_{i,m_1\cdots m_l}$ and the factors of $\rho(A(e))$ and $\rho(A(e'))$ in the product over edges at the vertex $v_i$. If we do this for all vertices, we will have obtained a gauge-invariant state. Explicitly, we define the gauge-invariant spin-network basis $|s^{\mathbf{j}}_{\gamma, \mathbf{I}}\rangle$, where $\mathbf{I} = \{I_v: v\in V(\gamma)\}$, as
\begin{align}
    \langle A| s^{\mathbf{j}}_{\gamma, \mathbf{I}}\rangle &= \prod_{v\in V(\gamma)}I_v \cdot \prod_{e\in E(v)}\sqrt{2j_e+1}\rho^{j_e}(A(e)) \nonumber \\
    &= \prod_{v\in V(\gamma)}\sqrt{2j_{e_1}+1}\cdots\sqrt{2j_{e_k}+1}\cdots\sqrt{2j_{e_{k+l}}+1}\nonumber \\
    &\qquad \times I^{n_1\cdots n_k}_{v, m_{k+1}\cdots m_{k+l}}\rho^{j_{e_1}}\tensor{(A(e_1))}{^{m_{1}}_{n_1}}\cdots \rho^{j_{e_k}}\tensor{(A(e_k))}{^{m_{k}}_{n_k}}\nonumber \\
    &\qquad\times \rho^{j_{e_{k+1}}}\tensor{(A(e_{k+1}))}{^{m_{k+1}}_{n_{{k+1}}}}\cdots \rho^{j_{e_{k+l}}}\tensor{(A(e_{k+l}))}{^{m_{k+l}}_{n_{{k+l}}}}, \label{eq3.165}
\end{align}
where we have split the product over edges at a vertex $v$ into outgoing and ingoing edges in such a way that $e_1,\ldots, e_k$ are outgoing edges and $e_{k+1}, \ldots, e_{k+l}$ are ingoing edges at $v$. Thus, we have constructed an orthonormal basis of $L^2(\mathcal{A}_\gamma/\mathcal{G}_\gamma)$. The tensors $I$ are called \textit{intertwiners}. We will write
\begin{equation}
    L^2(\mathcal{A}_\gamma/\mathcal{G}_\gamma) \cong \bigoplus_{\mathbf{j}}\bigotimes_{v\in V(\gamma)}\text{Inv}\left(\bigotimes_{e\in E(v)}V_{j_{e}}\otimes V^*_{j_{e}}\right), \label{eq3.166}
\end{equation}
where $\text{Inv}\left(\bigotimes_{e\in E(v)}V_{j_{e}}\otimes V^*_{j_{e}}\right)$ denotes the gauge-invariant part of $\bigotimes_{e\in E(v)}V_{j_{e}}\otimes V^*_{j_{e}}$, obtained via the intertwining procedure described above.  
\newline

Finally, it remains to be seen how to extend the $\mathcal{G}_\gamma$-invariant orthonormal basis  $|s^{\mathbf{j}}_{\gamma, \mathbf{I}}\rangle$ to an orthonormal basis of $\mathcal{H}_{kin} = L^2(\overline{\mathcal{A/G}})$. At first sight, it might appear that one could obtain an orthonormal basis for $L^2(\overline{\mathcal{A/G}})$ from simply the direct sum over all graphs, each term in the sum being an orthonormal basis of a particular graph. However, recall that $L^2(\overline{\mathcal{A/G}})$ contains cylindrical functions, and a function cylindrical with respect to a graph is also cylindrical with respect to every larger graph. Therefore, the spaces $L^2(\mathcal{A}_\gamma/\mathcal{G}_\gamma)$ and $L^2(\mathcal{A}_{\gamma'}/\mathcal{G}_{\gamma'})$ belonging to different graphs $\gamma$ and $\gamma'$ may not be orthogonal. Thus, even though one may extend the orthonormal bases of $\{L^2(\mathcal{A}_\gamma/\mathcal{G}_\gamma)\}$ to a basis of $L^2(\overline{\mathcal{A/G}})$, but the basis so obtained will not be orthogonal. To resolve this difficulty, we introduce some new Hilbert spaces corresponding to each graph \cite{giesel, ashtekar 2}. To each graph $\gamma$, let $\mathcal{H}^{'}_\gamma$ be the subspace of $L^2(\mathcal{A}_\gamma/\mathcal{G}_\gamma) := \mathcal{H}_\gamma$ that is orthogonal to the space $\mathcal{H}_{\Tilde{\gamma}}$ associated with every graph $\gamma$ which is \textit{strictly} contained in $\gamma$. That is, every vertex of $\Tilde{\gamma}$ is contained in $\gamma$, and every edge in $\Tilde{\gamma}$ can be written as a composition of edges in $\gamma$, \textit{and} the converse of these two statements is \textit{false}. As can at once be verified, this definition ensures that if $f \in \mathcal{H}^{'}_\gamma$, then $f \notin \mathcal{H}^{'}_\lambda$ for any $\lambda$ distinct from $\gamma$. Therefore, we may write 
\begin{equation}
    \mathcal{H}_{kin} = \bigoplus_{\gamma \in L}\mathcal{H}^{'}_\gamma. \label{eq3.167}
\end{equation}

\chapter{Demise of the Continuum}
In this chapter, we will finally use the machinery constructed in Chapter 3 to obtain a prediction of loop quantum gravity, namely, that spatial geometry is discrete at the Plank scale. This prediction, if it is true, is truly paradigm-shifting, since it rings a death knell for our preconceived notion of the space we inhabit as a continuum. 

But what exactly does it mean to say that spatial geometry is discrete? Recall that in classical general relativity, certain geometric quantities, such as the area of a two-surface or the volume of a three-surface, are physical observables. Since such quantities are functions of the phase-space variables, for which we now have unambiguous operators in the quantum theory, it makes sense to define appropriate operators corresponding to these functions as well, and study their spectra. We will perform precisely this task for the area of a two-surface, and shall find that the spectrum of the resulting operator in the quantum theory comes out to be discrete. This is the sense in which spatial geometry is understood to be discrete. Thus, for instance, if one could measure the area of a surface to an extraordinary precision (i.e. at the Plank scale), in principle, one should find that the values of area are discretely spaced, in analogy to the discretely spaced energy levels of, say, a quantum harmonic oscillator. 

This chapter is primarily based on the rigorous construction of the area operator presented in Ref. \cite{ashtekar 8}. The derivation of the area functional in terms of the triads is adapted from Ref. \cite{giesel}.

\section{Area Operator Derived}
Let $S$ be a finite, analytic two-surface embedded in the three-dimensional spatial manifold $\Sigma$. Let $X^a$, $a \in \{1, 2, 3\}$, be the coordinates of $S$ in $\Sigma$, and let $u_i$, $i\in \{1,2\}$, be coordinates intrinsic to $S$. The ADM three-metric $q_{ab}$ on $\Sigma$ induces a two-metric $h_{ij}$ on $S$:
\begin{align}
    ds^2 &= q_{ab}dX^a dX^b \nonumber \\
    &= q_{ab}\frac{\partial X^a}{\partial u^i}\frac{\partial X^b}{\partial u^j}du^i du^j := h_{ij}du^i du^j, \label{eq4.1}
\end{align}
where 
\begin{equation}
    X^a_{,i} := \frac{\partial X^a}{\partial u^i} \label{eq4.2}
\end{equation}
are vector fields tangent to $S$. Denoting the determinant of the induced metric $h_{ij}$ by $h$, we can express the area of $S$ as 
\begin{equation}
    A_S = \int_S d^2u \sqrt{h}. \label{eq4.3}
\end{equation}
This is the classical area observable in general relativity. We wish to quantise it. In order to do that, we first need to re-express it in terms of the new (real) variables introduced in Chapter 2. In particular, $h$ depends on $q$, which in turn depends on the triads (cf. Eq~(\ref{eq2.30})). Thus Eq~(\ref{eq4.2}) is to be written in terms of the triads. 

To this end, define an everywhere normal one-form $n^a$ on $S$:
\begin{equation}
    n_a := \epsilon_{abc}X^b_{,1}X^c_{,2}, \label{eq4.4}
\end{equation}
and recall that the inverse $q^{ab}$ of $q_{ab}$ can be written as
\begin{equation}
    q^{ab} = \frac{1}{2q}\epsilon^{acd}\epsilon^{bef}q_{ce}q_{df}, \label{eq4.5}
\end{equation}
whence
\begin{align}
    q\, n_a n_b q^{ab} &= q\, n_a n_b \frac{1}{2q}\epsilon^{acd}\epsilon^{bef}q_{ce}q_{df} \nonumber \\
    &= \frac{1}{2}\epsilon_{akl}X^k_{,1}X^l_{,2}\epsilon_{bmn}X^m_{,1}X^n_{,2}\epsilon^{acd}\epsilon^{bef}q_{ce}q_{df} \nonumber \\
    &= h_{11}h_{22} - h_{12}h_{21} = h, \label{eq4.6}
\end{align}
where the last line follows from properties of the Levi-Civita symbol and Eq~(\ref{eq4.1}). Finally, comparing Eqs~(\ref{eq4.1}) and (\ref{eq2.30}), we see that
\begin{equation}
    A_S = \int_S d^2u \sqrt{n_an_b \Tilde{E}^a_i(u)\Tilde{E}^{bi}(u)}. \label{eq4.7}
\end{equation}

We now wish to promote Eq~(\ref{eq4.7}) to an operator on the Hilbert space $\mathcal{H}_{kin}$. At first sight, this seems to be a formidable task, since there is a square-root involved, and it is not always possible to give any sense to the notion of the square-root of an operator; special cases in which this can be done are possible, but may be too restrictive for us, such as the fact that if an operator is normal, its square root exists. But this does not mean that we are in completely uncharted territory. Inherited field-theoretic wisdom tells us that when transition to the quantum theory involves the risk of treading ill-defined paths, it is best to regularise one's classical expressions to obtain better control over them. Regularisation consists in smearing one's classical variables over a region, promoting the smeared variables to operators and then take a limit to remove the smearing. More precisely, since the $\Tilde{E}^a_i$ in the integrand above depend on points $u$ in a two-dimensional surface, we use a two-dimensional smearing function to define smeared triads. Accordingly, let $f_\epsilon(x, y)$ be a one-parameter family of functions on $S$ that tend to $\delta^2(x,y)$ as $\epsilon\to 0$. Then we can write
\begin{equation}
    [\Tilde{E}_i]_f(u) = \int_S d^2v f_\epsilon(u, v)n_a\Tilde{E}^a_i(v), \label{eq4.8}
\end{equation}
and thus
\begin{align}
    [A_S]_f &= \int_S d^2u\left[ \int d^2v f_\epsilon(u, v)n_a\Tilde{E}^a_i(v) \int_S d^2w f_\epsilon(u,w)n_b\Tilde{E}^{bi}(w)\right]^{1/2} \nonumber \\
    &= \int_S d^2u\, \left[[\Tilde{E}_i]_f(u)[\Tilde{E}^{i}]_f(u)\right]^{1/2}. \label{eq4.9}
\end{align}
These are the \textit{regularised} triads and area functionals, and the smearing fields are called \textit{regulators}. 

The advantage of smearing lies in our hope that if one promotes the non-smeared variables (in this case the triads) to operators on the quantum mechanical Hilbert space, substitutes the operator versions of those variables into the smeared equations (\ref{eq4.9}) that contain products and square-roots of operators, and then finally takes the limit $\epsilon \to 0$, one obtains a well-defined operator. Experience with Minkowskian quantum field theory tells us that this expectation is usually realised, and this is the reason we employ the smearing trick here. 

However, there is a potential caveat that is present in quantum gravity but not in Minkowskian quantum field theory. Notice from Eq~(\ref{eq4.8}) that since $\Tilde{E}^a_i$ are vector densities of weight 1 in the variable $v$, and the integral is two-dimensional, whence the integrand must be of density weight 1, it follows that the regulator $f_\epsilon(u,v)$ must be a vector density of weight 1 in the variable $u$. But the definition of a vector density involves factors of the determinant of the metric on the spatial manifold, and thus the smearing fields are background dependent\footnote{The triads are not background-dependent in this way, even though they are also vector densities of nonzero weight. This is so because they are themselves the fundamental canonical variables from which other objects, including the metric, have to constructed.}. On the other hand, we want operators in loop quantum gravity to be background-independent, in virtue of our regarding the background-independence of general relativity as a fundamental physical fact about gravity that must be preserved in a correct theory of quantum gravity. On this account, we should require further that once the regulators are removed after passage to the quantum theory, the resulting operators must not depend on the smearing fields or any other background-dependent fields introduced in any intermediate step of the regularisation procedure. This is not a requirement in Minkowskian field theory, since it is by definition a background-dependent theory, the geometry of spacetime being assumed to be flat Minkwoskian. 

With these considerations in mind, let us proceed to promote the regularised triads and area functional to operators on $\mathcal{H}_{kin}$. In principle, it is possible to follow a construction along the lines of Section 3.5.3 to represent $\Tilde{E}^a_i$ as vector fields on the space of cylindrical functions of generalised connections, and thence to obtain the operator analogues of the regularised variables. However, here we will use the fact that since the triads are conjugate momenta of smooth connections, one can perhaps more intuitively replace the triads with functional derivatives with respect to smooth connections, just as we do in ordinary quantum mechanics? That is, we set 
\begin{equation}
    \hat{\Tilde{E}}^a_i(x) = i\frac{\delta}{\delta A^i_a(x)}, \label{eq4.10}
\end{equation}
Since we seek operators on $\mathcal{H}_{kin}$, which essentially contains cylindrical functions, we will begin by considering the action of Eq~(\ref{eq4.10}) on cylindrical functions. For convenience, we will first consider cylindrical functions on $\overline{\mathcal{A}}$ and only at the end promote all our constructions to the physically relevant $\overline{\mathcal{A/G}}$. Thus let $\gamma$ be a fixed graph with $N$ edges, and consider a cylindrical function $\psi$ on $\overline{\mathcal{A}}$. For all $\overline{A} \in \overline{\mathcal{A}}$, we have that
\begin{equation}
    \psi(\overline{A}) = \psi_\gamma(\overline{A}(e_1), \ldots, \overline{A}(e_N)), \label{eq4.11}
\end{equation}
where $e_1, \ldots e_N$ are the edges in $\gamma$, and $\psi_\gamma$ is a smooth complex-valued function on $SU(2)^N$; here we have exploited the isomorphism between $\mathcal{A}_\gamma$ and $SU(2)^N$. Now, it seems a rather silly idea to act Eq~(\ref{eq4.10}) on such a function as above, for the former contains \textit{smooth} connections, while the latter has \textit{generalised} connections. Indeed, one might even be tempted to regard this seemingly ill-define procedure as the reason behind our not using Eq~(\ref{eq4.10}) in Section 3.5.3. However, this is not really a difficulty, for we could invoke Lemma 3.4 to find a \textit{smooth} connection $A$ whose parallel propagator along each edge $e$ of $\gamma$ equals $\overline{A}(e)$, 
\begin{equation}
    \overline{A}(e_I) = \mathcal{P}\exp{\left(-\int_{e_I}A\right)} := U_{I}(1,0)[A], \label{eq4.12}
\end{equation}
where we have parametrised the edges using the interval $[0,1]$. Then, we have
\begin{equation}
    \hat{\Tilde{E}}^a_i(x)\cdot \psi(\overline{A}) = i\frac{\delta}{\delta A^i_a(x)}\psi_\gamma(U_1, \ldots, U_N) = i\sum_{I=1}^{N} \frac{\delta U_I}{\delta A^i_a(x)}\frac{\partial \psi_\gamma}{\partial U_I}, \label{eq4.13}
\end{equation}
where we have made use of the chain rule. To evaluate the right-hand side above, we note that using Eqs~(\ref{eq3.17}) and (\ref{eqs2.39}), it is not difficult to obtain
\begin{equation}
    \frac{\delta U_I}{\delta A^i_a(x)} = \frac{1}{2}\int_0^1 ds\,\delta^3(e_I(s), x)\dot{e}^a(s)U_I(1,s)\tau_i U_I(s,0), \label{eq4.14}
\end{equation}
whence
\begin{equation}
    \hat{\Tilde{E}}^a_i(x)\cdot \psi(\overline{A}) = \frac{1}{2}\sum_{I}\int_0^1 ds\,\delta^3(e_I(s), x)\dot{e}^a_I(s)U_I(1,s)\tau_i U_I(s,0)\frac{\partial\psi_\gamma}{\partial U_I}. \label{eq4.15}
\end{equation}
Owing to the presence of the Dirac delta functions, this is a badly divergent quantity. But this is where regularisation helps. Substituting the preceding equation into Eq~(\ref{eq4.8}), we have
\begin{equation}
    [\widehat{\Tilde{E}}_i]_f(u)\cdot \psi(\overline{A}) = \frac{1}{2}\sum_{I}\int_S d^2v\, f_\epsilon(u,v)\int_0^1 ds\,\delta^3(e_I(s), v)n_a\dot{e}^a_I(s)U_I(1,s)\tau_i U_I(s,0)\frac{\partial\psi_\gamma}{\partial U_I}. \label{eq4.16}
\end{equation}
Is this well-defined? It indeed is, but to verify this, we must massage the equation above into a more convenient form. To this end, we make the following assumptions \cite{ashtekar 8} on the graph $\gamma$:
\begin{enumerate}[(i)]
    \item If an edge of $\gamma$ contains a segment lying in the surface $S$, then the entire edge lies in the closure of $S$.
    \item Otherwise, each edge of $\gamma$ intersects $S$ at most once, and we call such an intersection point an isolated point.
    \item The isolated points of intersecting edges and the surface $S$ must all be at the vertices of $\gamma$. 
\end{enumerate}
These conditions engender no loss of generality, for we can replace a graph $\gamma$ that does not satisfy them with one that does by subdividing the edges of $\gamma$ to add more vertices until the conditions are met, and the new graph so obtained will be greater than $\gamma$, whence $\psi$ will also be cylindrical with respect to the new graph. We can now significantly simplify Eq~(\ref{eq4.16}). If an edge does not intersect $S$, the delta function is obviously zero, and so there is no contribution to the sum. If an edge lies in $S$, then its tangent vector $\dot{e}^a$ is orthogonal to $n_a$, and thus again, there is no contribution to the sum. Therefore, only edges intersecting $S$ at isolated points give nonzero contributions. Such edges are of two types: those that are outgoing at the intersection point, and those that are ingoing. For outgoing (ingoing) edges, the intersection points are their beginning (ending) points, and so the parameter $s=0$ ($s=1$), yielding $U_I(s,0) = \mathds{1}$ ($U_I(1,s) = \mathds{1}$). Furthermore, supposing that $n_a$ points outwards from $S$, and both $n_a$ and $\dot{e}^a_I$ have unit norm, for outgoing (ingoing) edges, $n_a\dot{e}^a_I = +1$ if $\dot{e}^a_I$ also points outwards, which is to say that $e_I$ lies above (below) $S$; on the other hand, $n_a\dot{e}^a_I = -1$ if $e_I$ lies below (above) $S$. In other words, the contributions to the sum of outgoing and ingoing edges are of opposite parity. These considerations entail that 
\begin{subequations}
    \label{eq4.17}
    \begin{equation}
        [\widehat{\Tilde{E}}_i]_f(u)\cdot\psi(\overline{A}) = \frac{1}{2}\left[\sum_{I=1}^{N}k_If_\epsilon(u, v_I)X_{I,i} \right]\cdot \psi(\overline{A}(e_1), \ldots, \overline{A}(e_N)),
    \end{equation}
    where 
    \begin{equation}
        k_I =     \begin{cases}
 \,\,\,0& \text{ if $e_I$ lies inside $S$ or does not intersect it} \\ 
 +1& \text{ if $e_I$ has an isolated intersection with $S$ and lies above it} \\ 
 -1& \text{ if $e_I$ has an isolated intersection with $S$ and lies below it }  
\end{cases}
    \end{equation}
    and $X_{I,i}$ are vector fields assigned to a vertex $v_I$ of $\gamma$ by the following formula:
    \begin{equation}
    X_{I,i} = \left\{\begin{matrix}
\tensor{(\overline{A}(e_I)\tau_i)}{_A^B}\frac{\partial\psi_\gamma}{\partial\tensor{(\overline{A}(e_I))}{_B^A}} \quad \text{ when $e_I$ is outgoing}\\

-\tensor{(\tau_i\overline{A}(e_I))}{_A^B}\frac{\partial\psi_\gamma}{\partial\tensor{(\overline{A}(e_I))}{_B^A}} \quad \text{ when $e_I$ is outgoing.}
\end{matrix}\right.
    \end{equation}
\end{subequations}
It is interesting to observe that 
\begin{align*}
    \tensor{(\overline{A}(e_I)\tau_i)}{_A^B}\frac{\partial\psi_\gamma}{\partial\tensor{(\overline{A}(e_I))}{_B^A}} &= -i\left. \frac{d}{dt}\psi_\gamma(\overline{A}(e_1),\ldots, \overline{A}(e_I)e^{it\tau_i}, \ldots, \overline{A}(e_N)) \right|_{t=0},\\
    \tensor{(\overline{A}(e_I)\tau_i)}{_A^B}\frac{\partial\psi_\gamma}{\partial\tensor{(\overline{A}(e_I))}{_B^A}} &= i\left. \frac{d}{dt}\psi_\gamma(\overline{A}(e_1),\ldots, e^{-it\tau_i}\overline{A}(e_I), \ldots, \overline{A}(e_N)) \right|_{t=0}.
\end{align*}
That is, these are nothing but left- and right-invariant vector fields on the space of smooth cylindrical functions! As a bonus, we have derived another way of deriving the momentum operators in loop quantum gravity; perhaps this alternative construction is more intuitive, since Eq~(\ref{eq4.10}) is what we do in ordinary quantum mechanics (compare this with the construction presented in Section 3.5.3). 

We have successfully promoted the regularised triads (\ref{eq4.8}) to operators on $L^2(\mathcal{A}_\gamma)$. We now do the same to the area functional (\ref{eq4.9}). Let the integrand in the functional be denoted by $g_f$. Eqs~(\ref{eq4.9}) and (\ref{eq4.17}) yield
\begin{equation}
    \hat{g}_f(u)\cdot\psi_\gamma = \frac{1}{4}\left[ \sum_{I,J}k_Ik_Jf_\epsilon(u, v_I)f_\epsilon(u,v_J)X_{I,i}X^i_J\right]\cdot \psi_\gamma \label{eq4.18} 
\end{equation}
for all $\psi_\gamma \in L^2(\mathcal{A}_\gamma)$. This sum can simplified by choosing $\epsilon$ small enough to ensure $f_\epsilon(u, v_I)f_\epsilon(u,v_J) = 0$ unless $v_I = v_J$. Then we can regroup the sum with respect to the vertices of $\gamma$ that lie in $S$:
\begin{equation}
    \hat{g}_f(u)\cdot\psi_\gamma = \frac{1}{4}\left[ \sum_{v\in V(\gamma)} (f_\epsilon(u, v))^2\sum_{I_v, J_v}k_{I_v}k_{J_v}X_{I_v, i}X^i_{J_v} \right]\cdot\psi_\gamma, \label{eq4.19}
\end{equation}
where $V(\gamma)$ is the set of vertices in $S$ and $I_v$, $J_v$ denote the edges arriving at or leaving the vertex $v$. Next, we take the square root (which we are allowed to do since the operators $X_{I,i}$ are self-adjoint on $L^2(\mathcal{A}_\gamma)$ (see \cite{ashtekar 8})) of the expression above and choose $\epsilon$ to be sufficiently small so that $f_\epsilon(u, v)$ is nonzero for at most one vertex at each point $u \in S$, whence we can take the sum over $v$ outside the square root. Hence, 
\begin{equation}
    [\hat{A}_S]_f(u)\cdot\psi_\gamma = \frac{1}{2}\sum_{v\in V(\gamma)}f_\epsilon(u,v)\left[\sum_{I_v, J_v} k_{I_v}k_{J_v}X_{I_v, i}X^i_{J_v}\right]^{1/2}\cdot \psi_\gamma.  \label{eq4.20}
\end{equation}
This is a cute expression, for the dependence on the coordinates of $S$ lies entirely outside the square root -- a phenomenon referred to as \textit{point-splitting}. This is fortunate because taking the limit $\epsilon\to 0$, we see that
\begin{equation}
    \lim_{\epsilon\to 0}\,[\hat{A}_S]_f(u)\cdot\psi_\gamma = \frac{1}{2}\sum_{v\in V(\gamma)}\delta^2(u,v)\left[\sum_{I_v, J_v} k_{I_v}k_{J_v}X_{I_v, i}X^i_{J_v}\right]^{1/2}\cdot \psi_\gamma,  \label{eq4.21}
\end{equation}
from where we can finally read off the area operator:
\begin{equation}
    \hat{A}_S\cdot\psi_\gamma = \frac{1}{2}\sum_{v\in V(\gamma)}\left[\sum_{I_v, J_v} k_{I_v}k_{J_v}X_{I_v, i}X^i_{J_v}\right]^{1/2}\cdot\psi_\gamma. \label{eq4.22}
\end{equation}
This operator is consistent in the sense described in Section 3.5.3, and can thus be extended unambiguously to $L^2(\overline{\mathcal{A}})$. Since it is also manifestly gauge invariant (contracted internal indices), it also projects down to $\mathcal{H}_{kin} = L^2(\overline{\mathcal{A/G}})$. 

It is worth pointing out that the area operator is not (spatially) diffeomorphism invariant, because diffeomorphisms of $\Sigma$ will change the intersection points of the surface $S$ and a graph. However, this does not mean that diffeomorphism-invariant geometric operators do not exist. The operator corresponding to the volume of a region in $\Sigma$ is, in particular, diffeomorphism covariant. But we will not construct the volume operator in this thesis; the interested reader is referred to \cite{ashtekar 9}. It is also important to note that if one includes matter fields in the theory and uses them to define areas of surfaces, one does actually get diffeomorphism-invariant area operators \cite{ashtekar 5}. 

\section{Spectrum of the Area Operator}
In this section, we calculate the eigenvalues of the area operator. To begin with, let us recast Eq~(\ref{eq4.22}) in a more elegant form. Observe that the structure of this equation is of a single sum of operators, each of which is associated to a vertex of a graph $\gamma$. In other words, we may write 
\begin{equation}
    \hat{A}_S\cdot\psi_\gamma = \frac{1}{2}\sum_{v\in V(\gamma)}\sqrt{\Delta_{S, v}}\cdot\psi_\gamma, \label{eq4.23}
\end{equation}
where
\begin{equation}
    \Delta_{S, v} := \sum_{I_v, J_v}k_{I_v}k_{J_v}X_{I_v, i}X^i_{J_v} \label{eq4.24}
\end{equation}
is an operator corresponding to the vertex $v$ of $\gamma$ lying in the surface $S$; we call it a \textit{vertex operator}. Since we have assumed that each edge of $\gamma$ intersects $S$ at most once, the edges arriving at or leaving two different vertices must be distinct. Therefore, vertex operators corresponding to two different vertices must commute. It then suffices to work out the spectrum of one vertex operator at a time. 

Consider then a particular vertex operator $\Delta_{S,v}$. Motivated from Eq~(\ref{eq4.17}b), we can divide the edges of $\gamma$ that intersect $S$ at $v$ into three categories: (1) the edges $e_1, \ldots, e_d$ that lie below $S$; (2) the edges $e_{d+1}, \dots, e_u$ that lie above $S$, and (3) the edges $e_{u+1}, \ldots, e_t$ that are tangential to $S$. For want of a better name, let us call the edges in (1) and (2) edges of type `down' and `up', respectively. We then define the following operators:
\begin{subequations}
\label{eq4.25}
    \begin{align}
            J_{S,v}^{(d)i} &:= -i(X^i_1 + \cdots + X^i_d), \qquad J_{S,v}^{(u)i} := -i(X^i_{d+1} + \cdots + X^i_{u}), \\
             J_{S,v}^{(t)i} &:= -i(X^i_{u+1} + \cdots + X^i_{t}), \qquad J_{S,v}^{(d+u)i} := J_{S,v}^{(d)} + J_{S,v}^{(u)},
    \end{align}
\end{subequations}
where $X^i_1, X^i_2$, etc. are the vector fields (\ref{eq4.17}c) in $\mathfrak{su}(2)$ assigned to vertices $v_1, v_2$, etc. of $\gamma$. The next step is to specify the operators $X^i_1, X^i_2$, etc. Since they belong to the Lie algebra of $SU(2)$ and are mutually commuting, we can identify them with distinct angular momentum operators. That is, for each vertex $v$, fix a representation of $\mathfrak{su}(2)$ and identify $-iX^i_I$ with the $i$th component of the angular momentum operators associated with that representation. Then $J_{S, v}^{(d)i}$, $J_{S, v}^{(u)i}$ and $J_{S, v}^{(t)i}$ can be thought of as \textit{total} `down', `up' and `tangential' angular momentum operators in the $i$th direction at the vertex $v$. For instance, suppose we work in the fundamental representation, which is given in terms of the Pauli matrices $\tau^i$. Then 
\begin{equation}
    X^i_n = \mathds{1}_{SU(2)}\otimes\cdots\otimes\mathds{1}_{SU(2)}\otimes\tau^i\otimes\mathds{1}_{SU(2)}\otimes\cdots\mathds{1}_{SU(2)}, \label{eq4.26}
\end{equation}
where $\tau^i$ occurs at the $n$th position in the tensor product, which is over all the edges intersecting at $v$. 

In terms of the $J$ operators (Eq~(\ref{eq4.25})), we can express the vertex operators as
\begin{align}
    \Delta_{S,v} &= (J_{S, v}^{(d)i} - J_{S, v}^{(d)i})((J_{S, v, i}^{(d)} - J_{S, v, i}^{(d)})) \nonumber \\
    &= 2(J_{S, v}^{(d)})^2 + 2(J_{S, v}^{(u)})^2 - (J_{S, v}^{(d+u)})^2, \label{eq4.27}
\end{align}
where $J^2 := J^iJ_i$, and  the second line follows from the commutativity of $J_{S,v}^{(u)}$ and $J_{S,v}^{(d)}$; note that the tangential operators do not occur in the preceding equations by virtue of Eq~(\ref{eq4.17}b). Elementary quantum mechanics now immediately yields the eigenvalues of $\Delta_{S,v}$:
\begin{subequations}
    \label{eq4.28}
    \begin{align}
        \lambda_{S,v} = 2j^{(d)}(j^{(d)} + 1) + 2j^{(u)}(j^{(u)} + 1) - j^{(d+u)}(j^{(d+u)} + 1), 
    \end{align}
    where $j^{(d)}$, $j^{(u)}$ and $j^{(d+u)}$ are half-integers subject to the condition
    \begin{align}
        j^{(d+u)} \in \{|j^{(d)} - j^{(u)}|, |j^{(d)} - j^{(u)}| + 1, \ldots, j^{(d)} + j^{(u)}\}.
    \end{align}
\end{subequations}
Thus the eigenvalues $a_S$ of the area operator, which is a sum over (square roots of) commuting vertex operators, are given by
\begin{equation}
    a_S = \frac{1}{2}\sum_{v\in V(\gamma)}\sqrt{2j_v^{(d)}(j^{(d)} + 1) + 2j_v^{(u)}(j_v^{(u)} + 1) - j_v^{(d+u)}(j_v^{(d+u)} + 1)}, \label{eq4.29}
\end{equation}
where $j_v$ denotes the spin representation associated with the vertex $v$. As advertised, the spectrum of the area operator is manifestly discrete. 

Let us consider the special case of there being only one vertex at which $\gamma$ and $S$ intersect. Inspecting Eq~(\ref{eq4.29}), we see that the smallest eigenvalue $a_p$ occurs either when $j^{(u)_v} = 0$ and $j^{(d)_v} = 1/2$ or vice versa. Restoring SI units, we find that
\begin{equation}
    a_P = l^2_p\frac{\sqrt{3}}{4}, \label{eq4.30}
\end{equation}
where $l^2_p$ is the Plank length\footnote{We have been working in natural units so far. Thus restoring them means that we have to multiply by appropriate factors of $\hbar$, $c$ and $G$ so as to obtain a quantity with the SI units of area, namely m$^2$. The Plank length is the only such combination.}. This is the smallest quantum of area as predicted by loop quantum gravity. The appearance of the Plank length indicates that the discreteness of spatial geometry manifests only at extremely small scales. 

\chapter{The Apocalypse of Constraints}
At long last, we squarely confront the constraints of general relativity at the quantum level. The $\mathcal{H}_{kin}$ constructed in the previous chapter is by construction gauge invariant. Thus all we now need to do is to solve the other two constraints of general relativity. But as we shall see, this task is easier said than done. In this chapter, we will provide a full solution to the diffeomorphism constraint, and a partial, unsatisfactory solution to the Hamiltonian constraint. In fact, there is no completely non-controversial way of solving the Hamiltonian constraint in loop quantum gravity to this day. 

Our treatment of the diffeomorphism constraint follows that of Ref. \cite{ashtekar 5, alok}, while that of the Hamiltonian constraint is adapted from Ref. \cite{ashtekar 2}.

\section{Diffemorphism Constraint}
\subsection{Finite vs. infinitesimal diffeomorphisms}
For convenience, let us recall the expression of the spatial diffeomorphism constraint (Eq~(\ref{eq2.3.2}b)):
\begin{equation}
    V_a = F^i_{ab}\Tilde{E}^b_i \approx 0. \label{eq5.1}
\end{equation}
For reasons that must now be obvious, we will be interested in the smeared version of this constraint:
\begin{equation}
    V_{\Vec{N}} = \int_\Sigma d^3x N^a F^i_{ab}\Tilde{E}^b_i(x) \approx 0, \label{eq5.2}
\end{equation}
where $N^a$ is a vector field on the spatial manifold $\Sigma$. As explained in Chapter 1, physically, the diffeomosphism constraint (infinitesimally) deforms $\Sigma$ along the one-parameter family of (spatial) diffeomorphisms $\phi_t$ generated by $N^a$.

The first task that we wish to undertake is to promote $V_{\Vec{N}}$ to an operator on the kinematical Hilbert space $\mathcal{H}_{kin} = L^2(\overline{\mathcal{A/G}}, d\mu_{AL})$. To this end, consider first smooth complex-valued functions on $\mathcal{A/G}$. On $\mathcal{A/G}$, we can define the operator analogue of $V_{\Vec{N}}$ via (\ref{eq4.10}). We then find that
\begin{align}
    \hat{V}_{\Vec{N}}\cdot \psi &= \int_\Sigma d^3x N^a F^i_{ab}\frac{\delta\psi}{\delta A^i_b} = -\{\psi, V_{\Vec{N}}\} = \mathcal{L}_{\Vec{N}}\psi \label{eq5.3}
\end{align}
for all $\psi \in C^{\infty}(\mathcal{A/G})$ (the curly braces denote Poisson brackets). We now ask: can we extend the operation of $\hat{V}_{\Vec{N}}$ from $C^\infty(\mathcal{A/G})$ to $\mathcal{H}_{kin}$? Unfortunately, the answer to this question is in the negative, as we now show \cite{ashtekar 5}. Let us restrict our attention to traces of holonomies around a loop $\alpha$. From Eq~(\ref{eq5.3}), we have 
\begin{equation}
    \hat{V}_{\Vec{N}}\cdot(T_\alpha)(A) = \lim_{t\to 0}\frac{T_{\phi_t(\alpha)}-T_\alpha}{t}(A), \label{eq5.4}
\end{equation}
where $\phi_t(\alpha)$ is the image of $\alpha$ under the family $\phi_t$ of diffeomorphisms generated by $N^a$, and the limit is taken pointwise in $\mathcal{A/G}$. Since $\overline{\mathcal{A/G}}$ is obtained by the $\mu_{AL}$-completion of the holonomy algebra, the preceding equation suggests that for the action of $\hat{V}_{\Vec{N}}$ to be well-defined on $\overline{\mathcal{A/G}}$, we must require that
\begin{equation}
    \lim_{t\to 0}||T_{\phi_t(\alpha)}-T_\alpha ||^2 = \lim_{t\to 0}\int_{\overline{\mathcal{A/G}}}d\mu_{AL}(T_{\phi_t(\alpha)}-T_\alpha)^2 = 0. \label{eq5.5}
\end{equation}
Let $\phi_t$ be such that it leaves $\alpha$ invariant and $\phi_t(\phi_s(\alpha)) = \alpha_{u(s,t)}$ for all $s> 0$. Then since $\mu_{AL}$ is diffeomorphism-invariant, it follows that
\begin{equation}
    \int_{\overline{\mathcal{A/G}}}d\mu_{AL}T^2_\alpha = \int_{\overline{\mathcal{A/G}}}d\mu_{AL}T^2_{\phi_t(\alpha)} \label{eq5.6} 
\end{equation}
for all $t$, and that there exists $t_o$ such that for all $0 < t < t_o$
\begin{equation}
    \int_{\overline{\mathcal{A/G}}}d\mu_{AL}T_\alpha T_{\phi_t(\alpha)} = k, \label{eq5.7}
\end{equation}
where $k$ is a constant. Substituting Eqs~(\ref{eq5.6}) and (\ref{eq5.7}) into Eq~(\ref{eq5.5}) reveals that $k = \int T^2_\alpha$, and thus,
\begin{equation}
    T_{\phi_t(\alpha)} = T_\alpha
\end{equation}
for all $0 < t < t_o$. This means that if we represent $T_\alpha$ as an operator on $\mathcal{H}_{kin}$, then $\hat{T}_{\phi_t(\alpha)} = \hat{T}_\alpha$ for all $0 < t < t_o$. However, since $\phi_t(\alpha)$ is a loop distinct from $\alpha$ if $t \neq 0$, as elements of the holonomy algebra, $T_\alpha \neq T_{\phi_t(\alpha)}$. This means that one cannot represent the traces of holonomies as faithful operators on $\mathcal{H}_{kin}$. Since we do not want this pathological behaviour, we cannot extend the action of $\hat{V}_{\Vec{N}}$ from $\mathcal{A/G}$ to $\overline{\mathcal{A/G}}$. 

We seem to have hit a roadblock. Since we cannot represent the diffeomorphism constraint as an operator on $\mathcal{H}_{kin}$, the first method of dealing with constraints outlined in Section 1.3.2 becomes unavailable. However, this does not mean that all hope is lost; we still have the group-theoretic approach at our disposal, and we shall see that it saves the day. 

Let us begin by recalling that $V_{\Vec{N}}$ is the generator of \textit{infinitesimal} diffeomorphisms, and so the preceding arguments at most establish the impossibility of representing infinitesimal diffeomorphisms as operators in the quantum theory. We still have the choice of considering \textit{finite} diffeomorphisms. These are furnished by the flow $\phi_t$ generated by $\Vec{N}$. Accordingly, for each vector field $\Vec{N}$, we define an operator $\hat{U}_{\Vec{N}}(t)$ on $C^\infty(\mathcal{A/G})$ given by
\begin{equation}
    \hat{U}_{\Vec{N}}(t)\cdot\psi := (\phi_t)_{\star}\cdot\psi = \psi\circ\phi_t \label{eq5.9} 
\end{equation}
for every $\psi \in C^\infty(\mathcal{A/G})$. Now we ask: can this operator be extended to $\mathcal{H}_{kin}$? Fortunately, the very property of $\mu_{AL}$ -- namely, diffeomorphism invariance -- that was our bane in the infinitesimal case now becomes our salvation, for we find that for all $\psi_1, \psi_2 \in \mathcal{H}_{kin}$
\begin{align}
    \langle \hat{U}_{\Vec{N}}(t)\psi_1, \hat{U}_{\Vec{N}}(t)\psi_2\rangle &= \int_{\overline{\mathcal{A/G}}}d\mu_{AL} \overline{\psi_1(\phi_t(A))}\psi_2(\phi_t(A)) \nonumber \\
    &= \int_{\overline{\mathcal{A/G}}}d\mu_{AL} \overline{\psi_1(A)}\psi_2(A) = \langle \psi_1, \psi_2\rangle.  
\end{align}
That is, $\hat{U}_{\Vec{N}}(t)$ is a unitary operator, whence it can be uniquely\footnote{Here is some elaboration of this point. Unitarity of an operator on a dense subspace of a Hilbert space can be used to obtain a unitary extension on the whole Hilbert space. Furthermore, a unitary operator is by definition bounded. Now, every bounded linear operator on a normed vector space is continuous. Finally, two continuous functions agreeing on a dense subspace of a topological space agree on the whole space, and so we have uniqueness.} extended to $\mathcal{H}_{kin}$, since $\mathcal{A/G}$ is dense in $\overline{\mathcal{A/G}}$. Thus we can henceforth focus on finite spatial diffeomorphisms, represented as unitary operators on the kinematical Hilbert space. Notice that the set of all one-parameter family of finite diffeomorphisms $\phi_t$ forms a group. Therefore, we are in the group-theoretic framework of solving the constraints; we shall refer to the spatial diffeomorphism group as $Diff(\Sigma)$.

\subsection{Group averaging}
Now that we have an operator representing spatial diffeomorphisms, the next step is to find quantum states that are invariant under the action of this operator. These will be the diffeomorphism-invariant states of the theory, spanning the diffeomorphism-invariant Hilbert space.

The first avenue to look for diffeomorphism-invariant states is some suitable subspace of $\mathcal{H}_{kin}$ itself. However it is not difficult to see that no such nontrivial subspace exists. Pick a cylindrical function in $\mathcal{H}_{kin}$. It is determined by its value on a particular graph in the spatial manifold $\Sigma$. Now, since spatial diffeomorphisms move points of the manifold around, they must transform between different graphs by moving their edges and vertices around. This entails that under a diffeomorphism, the value of a cylindrical function on a particular graph may change. Since we allow all possible diffeomorphisms, and thus all possible transformations between graphs, the only function left invariant under the action of all diffeomorphisms is the constant function. Thus, unless one can perform the Herculean task of gleaning some physical insight from a constant function, we are left with no choice but to abandon $\mathcal{H}_{kin}$ in our search for a diffeomorphism-invariant Hilbert space.

In view of the preceding considerations, it is clear that we require a space larger than $\mathcal{H}_{kin}$ to find a home for diffeomorphism-invariant Hilbert states. The dual $\mathcal{H}^{\star}_{kin}$ of $\mathcal{H}_{kin}$ naturally lends itself to this task. The idea is to average over all the diffeomorphic images of a given state to obtain a nontrivial diffeomorphism-invariant state. The resulting state cannot lie in $\mathcal{H}_{kin}$, for otherwise, it would not be nontrivial and diffeomorphism invariant. Our expectation is that it will instead lie in $\mathcal{H}^{\star}_{kin}$. As we shall see, this is indeed the case.

Thus, what we are after is a map $\eta$ from $\mathcal{H}_{kin}$ to $\mathcal{H}^{\star}_{kin}$ that outputs diffeomorphism-invariant \textit{dual} states. In this context, diffeomorphism invariance translates to the condition that for all $\psi\in\mathcal{H}_{kin}$ the action of $\eta(\psi) \in \mathcal{H}^{\star}_{kin}$ is invariant under diffeomorphisms. That is, 
\begin{equation}
    \eta(\psi)[\hat{U}(g)\cdot\psi'] = \eta(\psi)[\psi'] \quad \forall g \in Diff(\Sigma), \, \psi, \psi' \in \mathcal{H}_{kin}, \label{eq5.10}
\end{equation}
where $\hat{U}(g)$ is the unitary representation of $g$ on $\mathcal{H}_{kin}$. By linearity of dual vectors, the range $\mathcal{V}$ of the map $\eta$ will be a vector subspace of $\mathcal{H}^{\star}_{kin}$, but not necessarily a Hilbert space. To ensure the latter, we can restrict $\eta$ to be real and positive, i.e.
\begin{equation}
    \eta(\psi)[\psi'] = \overline{\eta(\psi')[\psi]} \text{ and } \eta(\psi)[\psi] \geq 0 \quad \forall \psi, \psi'\in\mathcal{H}_{kin}. \label{eq5.11}
\end{equation}
Then it is easily verified that
\begin{equation}
    \langle\eta(\psi), \eta(\psi')\rangle_{\mathcal{V}} := \eta(\psi')[\psi] \label{eq5.12}
\end{equation}
provides an inner product\footnote{The positions of $\psi$ and $\psi'$ are switched on both sides of Eq~(\ref{eq5.12}) because $\eta$ is an anti-linear map.} on $\mathcal{V}/\sim$, where the quotient is taken over states with zero norm. Completion of $\mathcal{V}/\sim$ under this inner product thus gives a diffeomorphism-invariant Hilbert space, which we shall denote by $\mathcal{H}_{diff}$. 

The map $\eta$ is called a \textit{group averaging map}. To understand the rationale behind this name, let us explicitly construct such a map. It will now be convenient to work in a specific basis of $\mathcal{H}_{kin}$. Let us thus pick the gauge-invariant spin-network functions constructed in Chapter 3. We can identify a spin-network state $|s\rangle \in \mathcal{H}_{kin}$ with its projection onto $\mathcal{H}_\gamma$, where $\gamma$ is the smallest graph with respect to which $\ket{s}$ is cylindrical. We can thus write $\ket{s} \cong \ket{s^{\mathbf{j}}_{\gamma, \mathbf{I}}}$, where $\mathbf{j} = \{j_{e_1}, \ldots, j_{e_n}\}$ is the set of spin-$j$ representations associated to the edges $e_1, \ldots, e_n$ of $\gamma$, and $\mathbf{I} = \{I_v : v\in V(\gamma)\}$ is the set of intertwiners corresponding to the vertices $v\in V(\gamma)$ of $\gamma$. This simplification enables us to see the unitary action of $Diff(\Sigma)$ on spin-network states. A diffeomorphism $\phi \in Diff(\Sigma)$ moves the edges and vertices of $\gamma$ around in $\Sigma$, yielding a different graph $\phi(\gamma)$. We thus simply have
\begin{equation}
    \hat{U}(\phi)\ket{s^{\mathbf{j}}_{\gamma, \mathbf{I}}} = \ket{s^{\phi(\mathbf{j})}_{\gamma, \phi(\mathbf{I})}}, \label{eq5.13}
\end{equation}
where\footnote{One might worry here about the possibility of there being an increase in the number of edges and vertices in a graph under a diffeomorphism as a result of different edges crossing each other. But this would never happen, for a diffeomorphism is by definition invertible, whence two different points of the spatial manifold cannot be mapped to the same point.} $\phi(\mathbf{j}) := \{j_{\phi(e)}: e\in\gamma\}$ and $\phi(\mathbf{I}) := \{I_{\phi(v)}: v\in\gamma\}$ denote the spin and intertwiner assignments to the edges and vertices of the graph $\phi(\gamma)$, respectively. Now let $[s]$ be the set of all distinct diffeomorphic images of $\ket{s}$ under $Diff(\Sigma)$; we call it the \textit{orbit} of $\ket{s}$. We define \cite{alok} a map $\eta: \mathcal{H}_{kin}\to\mathcal{H}^{\star}_{kin}$ such that
\begin{equation}
    \eta(\ket{s}) = \eta_{[s]}\sum_{\ket{s'}\in[s]}\bra{s'}, \label{eq5.14}
\end{equation}
where $\eta_{[s]}$ is a positive parameter, which we shall fix later. Since spin-network functions corresponding to distinct graphs are orthonormal, the elements of $[s]$ are all mutually orthonormal. Thus the action of $\eta(\ket{s})$ on an arbitrary spin-network state either vanishes or equals $\eta_{[s]}$. Combined with the fact that the sum is over \textit{all} distinct diffeomorphic images of $\ket{s}$, this entails that the map $\eta$ satisfies Eq~(\ref{eq5.10}). Furthermore, the positivity of $\eta_{[s]}$ implies that Eq~(\ref{eq5.11}) is also satisfied. Consequently, $\eta$ is a group averaging map. The reason for the name is clear now: we sum over all diffeomorphic relatives of a state in $\mathcal{H}_{kin}$ to obtain a diffeomorphism-invariant state in $\mathcal{H}^{\star}_{kin}$. 

While the preceding treatment may seem satisfactory at first sight, for we have successfully constructed a set of diffeomorphism states, a closer inspection reveals a glaring ambiguity, which arises from a subtle fact about how observables in the quantum theory act on spin-network states corresponding to unequal graphs. Recall that in the classical theory, phase-space functions that have vanishing Poisson brackets with the constraints of the theory are identified with observables. When we promote these observables and constraints to operators in the quantum theory, their commutators with each other must vanish. More precisely, in this context, for every diffeomorphism $\phi$, and every observable operator $\hat{O}$, 
\begin{equation}
    \hat{O}\hat{U}(\phi) = \hat{U}(\phi)\hat{O}. \label{eq5.15}
\end{equation}
Now let $\ket{s_1}$ and $\ket{s_2}$ be two spin-network states corresponding to two graphs $\gamma_1$ and $\gamma_2$, respectively, such that\footnote{Since we are considering the smallest graphs corresponding to each spin network, inequality in this context excludes the possibility that one graph is greater or smaller than the other; both are simply not comparable.} $\gamma_1 \neq \gamma_2$. Consider a vector field on $\Sigma$ that vanishes everywhere on $\gamma$ but is transverse to an open subset of $\gamma_2$. The diffeomorphisms generated by this vector field change $\ket{s_1}$ but leave $\ket{s_2}$ invariant. Thus there are infinitely many diffeomorphisms that affect $\ket{s_1}$ but do nothing to $\ket{s_2}$; let us denote such diffeomorphisms by $\phi_1$. We then find that
\begin{equation}
    \bra{s_2}\hat{O}\ket{s_1} = \bra{s_2}\hat{O}\hat{U}^{\dagger}(\phi_1)\ket{s_1} = \bra{s_2}\hat{U}^{\dagger}(\phi_1)\hat{O}\ket{s_1}, \label{eq5.16} 
\end{equation}
where the second equality follows from Eq~(\ref{eq5.15}). In other words, the projections of the state $\hat{O}\ket{s_1}$ along the infinitely many states $\hat{U}(\phi_1)\ket{s_1}$ are all equal. Therefore, $\bra{s_2}\hat{O}\ket{s_1} = 0$. That is, an observable cannot map between spin networks belonging to unequal graphs. We say that such spin networks lie in different \textit{superselection sectors}. 

We can describe the ambiguity in our construction of diffeomorphism-invariant states. If two spin networks are superselected, their group-averaged images are also superselected unless their underlying graphs are diffeomorphic. Thus there exist infinitely many sectors of diffeomorphism-invariant states that cannot be mapped to each other by any observable. Which one of these infinitely many sectors is physically relevant? One answer to this question is that we treat each distinct sector as a physically different realisation of diffeomorphism-invariant quantum gravitational systems, leaving the specification of the sector which we inhabit up to experiment. However, in order to adopt this interpretation consistently, we must fix the value of the parameter $\eta_{[s]}$ in Eq~(\ref{eq5.14}) in a diffeomorphism-invariant manner. This we shall now do. We will show that the imposition of another condition on a group averaging map fixes $\eta_{[s]}$ in the desired manner. The condition is the following:
\begin{equation}
    \eta(\psi)[\hat{O}\psi'] = \eta(\hat{O}^{\dagger}\psi)[\psi'], \quad \forall \psi, \psi' \in \mathcal{H}_{kin}. \label{eq5.17}
\end{equation}
That is, we require a group averaging map to commute with all observables (again, the appearance of the dagger on the right-hand side above is a consequence of the anti-linearity of $\eta$). 

To make use of this condition, let us first introduce some essential concepts and definitions. Given a group $G$, an element $g\in G$, and a subgroup $H$ of $G$, the left (right) coset of $H$ in $G$ by $g$ is the subgroup $gH := \{gh: h\in H\}$ ($Hg := \{hg: h\in H\}$). If all the left and right cosets of $H$ equal, $H$ is called a normal subgroup. Then the binary relation on $G$ defined by identifying elements belonging to a normal subgroup $N$ is an equivalence relation. We denote the set of all equivalence classes under this relation as $G/N$, and can check that it is also a group in its own right. It contains the left (or right) cosets of $N$ in $G$. For example, consider the additive group of integers $\mathbb{Z} = \{0, \pm 1, \pm 2, \ldots\}$. All subgroups in it normal, since the group is abelian. For illustration, consider the subgroup $3\mathbb{Z} := \{0, \pm 3, \pm 6, \ldots\}$. The distinct (left) cosets of this subgroup are $3\mathbb{Z}$, $1 + 3\mathbb{Z} = \{1,\pm 3 + 1, \pm 6 + 1, \ldots\}$, $2 + 3\mathbb{Z} = \{2,\pm 3 + 2, \pm 6 + 2, \ldots\}$. Thus $\mathbb{Z}/3\mathbb{Z} = \{3\mathbb{Z}, 1+3\mathbb{Z}, 2+3\mathbb{Z}\}$. 

Now consider a fixed spin-network state $\ket{s}$ with the smallest underlying graph being $\gamma$. Let $Sym_s$ be the subset of $Diff(\Sigma)$ that leaves $\ket{s}$ invariant. Since a diffeomorphic image of $\ket{s}$ must also be invariant under $Sym_s$, it follows\footnote{$\hat{U}(\chi)\hat{U}(\phi)\ket{s} = \hat{U}(\phi)\hat{U}(\chi)\ket{s}$ for all $\chi \in Sym_s$ and $\phi\in Diff(\Sigma)$.} that $Sym_s$ is a normal subrgoup of $Diff(\Sigma)$. Furthermore, since the states in the orbit $[s]$ of $\ket{s}$ are related by distinct diffeomorphisms, we can identify these states with the cosets of $Sym_s$ in $Diff(\Sigma)$. We can then rewrite Eq~(\ref{eq5.14}) as 
\begin{equation}
    \eta(\ket{s}) = \eta_{[s]}\sum_{\phi\in Diff(\Sigma)/Sym_s}\bra{s}\hat{U}^{\dagger}(\phi). \label{eq5.18}
\end{equation}
This was warm-up for what is ahead. We next define $Sym^0_s$ to be the subset of $Sym_s$ that preserves the edges of $\gamma$, i.e.
\begin{equation}
    Sym^0_s = \{\phi\in Diff(\Sigma): \phi(e) = e \quad  \forall e\in\gamma\}. \label{eq5.19}
\end{equation}
Evidently, $Sym^0_s$ is a normal subgroup of $Sym_s$. Moreover, the group $D_s := Sym_s/Sym^0_s$ of cosets of $Sym^0_s$ in $Sym_s$ is \textit{finite}, since it contains diffeomorphisms that involve only permutations of the edges of $\gamma$. Let $|D_s|$ denote the cardinality of $D_s$; It is worth noting that since $|D_s| = |D_{s'}|$ for all $s' \in [s]$, $|D_s|$ is a diffeomorphism-invariant number. 

We are now ready to evaluate \cite{alok} the parameter $\eta_{[s]}$ for the group averaging map $\eta$ defined in Eq~(\ref{eq5.14}). Let $\ket{s_1}$ and $\ket{s_2}$ be two spin networks corresponding to graphs $\gamma_1$ and $\gamma_2$ that are diffeomorphic. Given an observable $\hat{O}$, we impose Eq~(\ref{eq5.17}) on $\eta$:
\begin{equation}
    \eta(\hat{O}\ket{s_1})[s_2] = \eta(\ket{s_1})[\hat{O}^{\dagger}\ket{s_2}]. \label{eq5.20}
\end{equation}
We will evaluate both sides of this equation to determine $\eta_{[s]}$. We begin by observing that we can expand the state $\hat{O}\ket{s_1}$ as
\begin{equation}
    \hat{O}\ket{s_1} = \sum_{i=1}^{N}\lambda_i\hat{U}(\phi_i)\ket{s_2} + \ket{\chi}, \quad \bra{s_2}\hat{U}^{\dagger}(\phi)\ket{\chi} = 0\, \forall \phi\in Diff(\Sigma). \label{eq5.21}
\end{equation}
The sum is over all the components of $\hat{O}\ket{s_1}$ along the orbit of $\ket{s_2}$; without loss of generality, we may take the terms in the sum to be mutually orthogonal. $\ket{\chi}$, on the other hand, represents the components of $\hat{O}\ket{s_1}$ that are orthogonal to $[s_2]$. Thus, substituting Eq~(\ref{eq5.21}) into Eq~(\ref{eq5.14}) yields the left-hand side of Eq~(\ref{eq5.20}):
\begin{equation}
     \eta(\hat{O}\ket{s_1})[s_2] = \eta_{[s_2]}\sum_{i=1}^{N}\lambda_i. \label{eq5.22}
\end{equation}
Next, let us focus on the right-hand side. Define a map $\eta^0: \mathcal{H}_{kin}\to\mathcal{H}^{\star}_{min}$ such that
\begin{equation}
    \eta^0(\ket{s}) = \eta_{[s]}\sum_{\phi\in Diff(\Sigma)/Sym^0_s}\bra{s}\hat{U}^{\dagger}(\phi). \label{eq5.23}
\end{equation}
Comparing this equation with Eq~(\ref{eq5.18}) and recalling the relationship between $Sym_s$ and $Sym^0_s$, we find that
\begin{equation}
    \eta^0(\ket{s}) = |D_s|\,\eta(\ket{s}), \label{eq5.24}
\end{equation}
which, combined with Eq~(\ref{eq5.15}), entails that
\begin{equation}
    \eta(\ket{s_1})[\hat{O}^{\dagger}\ket{s_2}] = \eta_{[s_1]}\sum_{i=1}^{N}\overline{\lambda}_ix_i, \label{eq5.25} 
\end{equation}
where 
\begin{equation}
    x_i = \frac{1}{|D_{s_1}|}\sum_{\phi\in Diff(\Sigma)/Sym^0_{s_1}}\overline{\bra{s_2}\hat{U}(\phi)\hat{U}(\phi_i)\ket{s_2}}. \label{eq5.26}
\end{equation}
It can be shown \cite{alok} that there only $|D_{s_2}|$ number of nonzero terms in the preceding sum, and thus $x_i = |D_{s_2}|/|D_{s_1}|$. Thus, equating Eqs~(\ref{eq5.22}) and \ref{eq5.25} reveals that
\begin{equation}
    \eta_{[s]} = C|D_s|, \label{eq5.27}
\end{equation}
where $C$ is some unimportant positive constant. As promised, $\eta_{[s]}$ is indeed diffeomorphism 
invariant. 

\section{Hamiltonian Constraint}
Finally, we confront the mother of all constraints, without which there is no sense in any talk about quantum \textit{dynamics}. Unlike the Gauss and diffeomorphism constraints, we will not be able to solve the Hamiltonian constraint in a fully background-independent manner. Nonetheless, as will become abundantly clear, what we will achieve shall be a significant improvement over the situation encountered in the canonical quantisation of the ADM variables, where the Wheeler-DeWitt equation remains, at best, a formal device and thus the dream of understanding quantum dynamics stays ever so elusive. 

\subsection{Taming the constraint}
Let us start by observing that, unlike the diffeomorphism constraint, a group-theoretic approach is hard to imagine in order to deal with the Hamiltonian constraint. This is because, as hinted in Section 1.2, the finite canonical transformations generated by the Hamiltonian constraint are not very well understood -- the essential difficulty can be traced back to the appearance of the nontrivial $(\cdots)$ terms in Eq~(\ref{eq1.33*}b), which prohibit one to describe the finite transformations generated by the Hamiltonian constraint as a one-parameter group of maps that are integral curves of a differential equation like Eq~(\ref{eq1.33**}). Therefore, we have to proceed in the footsteps of Dirac (see Section 1.3.2) by promoting the Hamiltonian constraint to an operator on $\mathcal{H}_{kin}$ (more ideally, on $\mathcal{H}_{diff}$) and then seeking its null space. This is the method we shall currently employ. 

Let us countenance our constraint afresh (cf. Eq~(\ref{eq2.3.2}a)):
\begin{equation}
    S = \frac{\epsilon^{ijk}}{\Tilde{E}}\Tilde{E}^a_i\Tilde{E}^b_jF_{abk} - \frac{4}{\Tilde{E}}\Tilde{E}^a_{[i}\Tilde{E}^b_{j]}K^i_{[a}K^j_{b]}. \label{eq5.28}
\end{equation}
Remember that we are keeping the Barbero-Immirzi parameter $\beta$ to be 1. We shall only solve the first term in the constraint above, since it will suffice to elucidate the main ideas and both the merits and demerits of the full approach. Accordingly, we define the smeared version of the constraint:
\begin{equation}
    S_N = \int_\Sigma d^3x N\epsilon^{ijk}\frac{\Tilde{E}^a_i\Tilde{E}^b_j}{\Tilde{E}}F_{abk}, \label{eq5.29}
\end{equation}
where $N$ is the lapse function, introduced in Chapters 1 and 2. It might appear that the decision to not absorb $\Tilde{E}$ into $N$ (as we did in the Palatini and Ashtekar formulations in Chapter 2) may pose problems, but on the contrary, it will allow us to recast the constraint in a form that is much more tractable in terms of being promoted to an operator. To see this, we recall that the volume $V$ of $\Sigma$ (if assumed to be compact) is given by 
\begin{equation}
    V = \int_\Sigma d^3x \sqrt{q} = \int_\Sigma d^3x E = \int_\Sigma d^3x \sqrt{\Tilde{E}}, \label{eq5.30}
\end{equation}
where $q$ is the determinant of the three-metric on $\Sigma$ and $E$ is the determinant of the de-densitised triads $E^a_i$. Next, observe that 
\begin{equation}
    \epsilon^{ijk}\frac{\Tilde{E}^b_j\Tilde{E}^c_k}{\Tilde{E}} = \Tilde{\epsilon}^{abc} E E^i_a, \label{eq5.31}
\end{equation}
where $\Tilde{\epsilon}^{abc}$ is a Levi-Civita density of weight 1. Finally, we note that
\begin{equation}
    \frac{\delta V}{\delta \Tilde{E}^a_i} = \frac{1}{2}\sqrt{\Tilde{E}}\Tilde{E}^i_a = \frac{1}{2}E^2 E^i_a. \label{eq5.32}
\end{equation}
Combined, these equations yield \textit{Thiemann's identity}:
\begin{equation}
    2\Tilde{\epsilon}^{abc}\{A^i_a, V\} =  \epsilon^{ijk}\frac{\Tilde{E}^b_j\Tilde{E}^c_k}{\Tilde{E}}, \label{eq5.33}
\end{equation}
whence the smeared constraint becomes 
\begin{equation}
    S_N = 2\int_\Sigma d^3x \,N(x)\,\Tilde{\epsilon}^{abc}\,\text{tr}\left(F_{ab}(x)\{A_c(x), V\} \right). \label{eq5.34}
\end{equation}
As we will see, this expression is much easier to promote to an operator in the quantum theory. 

\subsection{Taming it more}
Passage to the quantum theory is facilitated by thinking about the operator analogue of the curvature $F_{ab}$. It turns out that $F_{ab}$ can be approximated by connections up to first order. More precisely, if $\alpha$ is a loop that bounds a coordinate $P$ place of area $\epsilon^2$, then 
\begin{equation}
    \int_P F = \frac{1}{2}\left[A(\alpha^{-1}) - A(\alpha) \right] + O(\epsilon^2), \label{eq5.35}
\end{equation}
where $A(\alpha)$ denotes the holonomy of the connection $A$ around $\alpha$. Similarly, if $s$ is a line segment of length $\epsilon$, then one can show that
\begin{equation}
    \left\{\int_s A, V\right\} = -A(s)^{-1}\{A(s), V\} + O(\epsilon), \label{eq5.36}
\end{equation}
$A(s)$ being the parallel propagator of $A$ along $s$. Now, connections have well-defined quantum analogues in terms of $SU(2)$-valued maps. As for the volume functional $V$, we briskly mentioned in Chapter 4 that their operator analogues also exist. Let us see how these look like. Any region $R$ in $\Sigma$ has a volume operator $\hat{V}_R$ that acts on an arbitrary cylindrical function $\psi_\gamma$ as
\begin{subequations}
    \label{eq5.37}
    \begin{equation}
        \hat{V}_{R}\cdot\psi_\gamma = \sum_{v\in V(\gamma)}\sqrt{|\hat{q}_v|}\cdot\psi_\gamma,
    \end{equation}
    where
    \begin{equation}
        \hat{q}_v\cdot\psi_\gamma = \frac{1}{48}\epsilon_{ijk}\sum_{I_v, J_v, K_v}k(I_v, J_v, K_v)X^i_{I_v}X^j_{J_v}X^k_{K_v}\cdot\psi_\gamma.  
    \end{equation}
\end{subequations}
All the terms above are familiar from the previous chapter, except $k(I_v, J_v, K_v)$, which is an orientation factor that is 0 if the tangent vectors of the three edges $e_{I_v}$, $e_{J_v}$ and $e_{K_v}$ arriving at or leaving $v$ are linearly dependent or do not intersect $R$, and $\pm 1$ if they intersect $R$ at $v$, and are linearly independent and oriented positively or negatively with respect to a fixed orientation on $\Sigma$ \cite{ashtekar 9}. One could now exploit the preceding three equations in extending $S_N$ to an operator by approximating the integral in Eq~(\ref{eq5.34}) by a sum over small regions of dimensions $\epsilon$ and replacing the curvature, the Poisson brackets and the volume functionals in each region by those equations. This is the general idea which we shall now implement in detail. 

Partition the spatial manifold $\Sigma$ into cells $\triangle$ of arbitrary shape and of spatial extent not exceeding $\epsilon$ in any direction -- such a partition is called a \textit{triangulation}.  In each cell, fix a point $v_{\triangle}$ and define edges $e_I$, $I = 1, \ldots, n_e$, and loops $\alpha_i$, $i = 1, \ldots, n_\alpha$, that all start at $v_{\triangle}$ and lie entirely within $\triangle$. We shall denote this whole structure of cells and the objects they contain by $R_\epsilon$, and call it a regulator. From Eqs~(\ref{eq5.35}) and (\ref{eq5.36}), we see that within each cell $\triangle$, the Hamiltonian constraint (\ref{eq5.34}) can be approximated as
\begin{equation}
    S_{N}^{R_\epsilon, \triangle} = N(v_{\triangle})\sum_{i,I}C^{iI}\text{tr}\left(\left[A(\alpha_i) - A(\alpha_i^{-1}) \right]A(e_I)^{-1}\{A(e_I), V_{\triangle}\} \right), \label{eq5.38}
\end{equation}
where $C^{iI}$ is a fixed constant that depends on the shape of the cell but not on $\epsilon$. Furthermore, depending on the regulator, the sum
\begin{equation}
    S^{R_\epsilon}_N = \sum_{\triangle}S^{R_\epsilon, \triangle}_N \label{eq5.39}
\end{equation}
approaches $S_N$ as $\epsilon \to 0$:
\begin{equation}
    \lim_{\epsilon\to 0}S^{R_\epsilon}_N = S_N. \label{eq5.40}
\end{equation}
Regulators that satisfy this condition are called \textit{permissible} regulators. We shall promote such permissibly regulated to operators and then take the limit $\epsilon\to 0$ to obtain the quantum version of the Hamiltonian constraint. This is quite analogous to the what we did with area functionals in the previous chapter, the difference being that now, the more complicated structure of the Hamiltonian constraint calls for a more involved regularisation procedure. 

As an example of a permissible regulator, consider a triangulation of $\Sigma$ by cubic cells of side length $\epsilon$. Let $v_{\triangle}$ be a corner of the cube $\triangle$, take the three sides of the cube meeting at this corner to be the edges associated with $\triangle$, and let the loops associated with $\triangle$ be the boundaries of the three faces shared by these three sides. Then the constants $C^{iI}$ are all 1, and we indeed recover $S_N$ in the limit $\epsilon\to 0$. 

Transition to the quantum theory now seems straightforward. We simply replace the holonomies in Eq~(\ref{eq5.38}) with generalised connections applied to the edges and loops, $V_{\triangle}$ with the operator $\hat{V}_{\triangle}$ defined in Eq~(\ref{eq5.37}), and the Poisson brackets with $-i$ times the corresponding commutator. We thus obtain
\begin{subequations}
    \label{eq5.41}
    \begin{equation}
        \hat{S}^{R_\epsilon}_N = \sum_{\triangle}\hat{S}^{R_\epsilon,\triangle}_N,
    \end{equation}
    \begin{equation}
        \hat{S}^{R_\epsilon,\triangle}_N = -iN(v_{\triangle})\sum_{iI}C^{iI}\text{tr}\left(\left[\pi(\Bar{A}(\alpha_i)) - \pi(\Bar{A}(\alpha_i^{-1})) \right]\pi(\Bar{A}(e_I)^{-1})[\pi(\Bar{A}(e_I)), \hat{V}_{\triangle}] \right),
    \end{equation}
\end{subequations}
where $\Bar{A}$ denotes a generalised connection, and $\pi(\Bar{A})$ is the corresponding $SU(2)$ element in the representation $\pi$. This yields a well-defined operator on the space of cylindrical functions in $\mathcal{H}_{kin}$, and it is self-adjoint, since $\hat{V}$ is self-adjoint. However, here we encounter a nontrivial issue. In order that the Hamiltonian constraint operator be physically meaningful, we must require it to be diffeomorphism covariant, in the sense that under diffeomorphisms of a graph, the action of the constraint operator on cylindrical functions should transform covariantly. This in turn amounts to finding an appropriate transformation condition on the regulators. We will now formulate this condition. 

Notice first that the definition of the volume operator (\ref{eq5.37}) entails that the action of $\hat{S}^{R_\epsilon,\triangle}_N$ on a cylindrical function $\psi_\gamma$ is nonzero only when one of the edges $e_I$ in $\triangle$ intersects a vertex of the graph $\gamma$. This means that we need only determine how those cells of a regulator that contain vertices of $\gamma$ transform under diffeomorphisms. Moreover, for a fixed graph, we can always choose $\epsilon$ small enough that every vertex is contained in exactly one cell -- we will call such regulators \textit{refined}. We can now formulate the transformation condition on regulators. To each graph $\gamma$, we associate a refined and permissible regulator $R_{\epsilon, \gamma}$. We then define a \textit{diffeomorphism-covariant quantum regulator} to be a family $\{R_{\epsilon, \gamma} : \epsilon > 0, \gamma \in L\}$ of refined and permissible regulators that satisfy the following condition: if $(\gamma, v)$ is diffeomorphic to $(\gamma', v')$, where $v, v'$ are vertices of $\gamma, \gamma'$, then for every $\epsilon$ and $\epsilon'$, $(\gamma, v, \triangle, (e_I), (\alpha_i))$ is diffeomorphic to $(\gamma', v, \triangle', (e'_I), (\alpha'_i))$, where $\triangle$ and $\triangle'$ are the cells corresponding to $R_{\epsilon, \gamma}$ and $R_{\epsilon', \gamma'}$, and containing $v$ and $v'$, respectively. Such diffeomorphism-covariant regulators exist (see \cite{ashtekar 2} and the references therein for examples). 

Diffeomorphism-covariant quantum regulators give rise to diffeomorphism-covariant regulated constraint operators (\ref{eq5.41}). The final step now consists in removing the regulators to obtain well-defined Hamiltonian constraint operators on $\mathcal{H}_{kin}$. It turns out, however, that the operators $\hat{S}^{R_{\epsilon, \gamma}, \Delta}_N$ do not converge in $\mathcal{H}_{kin}$ as $\epsilon \to 0$. To see this, consider the action of these operators on spin-network states. Since $\hat{S}^{R_{\epsilon, \gamma}, \Delta}_N$ are linear operators on $L^2(\mathcal{A}_\gamma)$, they map a spin-network state on $\gamma$ to a different spin-network state on $\gamma$. Now, if $\epsilon\neq\epsilon'$, then a regulator of dimensions $\epsilon$ must, in general, contain a different set of edges in each cell than a regulator of dimensions $\epsilon'$. These two observations imply that the actions of $\hat{S}^{R_{\epsilon, \gamma}, \Delta}_N$ and $\hat{S}^{R_{\epsilon', \gamma}, \Delta}_N$ on the same spin-network state $\ket{s_\gamma}$ should yield distinct spin-network states. But by construction, distinct spin-network states on the same graph are orthonormal, whence $\bra{s_\gamma}\hat{S}^{R_{\epsilon, \gamma}, \Delta}_N\hat{S}^{R_{\epsilon', \gamma}, \Delta}_N\ket{s_\gamma} = 0$, which implies that $\lim_{\epsilon\to 0}\hat{S}^{R_{\epsilon, \gamma}, \Delta}_N$ does not exist.

Have we hit a roadblock? Certainly not, for just as in the case of the diffeomorphism constraint, we have recourse to the dual space $\mathcal{H}^{\star}_{kin}$. We extend the action of regulated constraint operators (\ref{eq5.41}) to $\mathcal{H}^{\star}_{kin}$ in the obvious fashion. For each $\rho\in \mathcal{H}^{\star}_{kin}$, we define
\begin{equation}
    (\hat{S}^{R_{\epsilon}}_N\cdot\rho)(\psi) := \rho(\hat{S}^{R_{\epsilon}}_N\cdot\psi) \label{eq5.42} 
\end{equation}
for every $\psi\in\mathcal{H}$. This allows us to remove the regulator by defining
\begin{equation}
    \hat{S}_N = \lim_{\epsilon\to 0}\hat{C}^{R_\epsilon}_N \label{eq5.43}
\end{equation}
via
\begin{equation}
    (\hat{S}_N\cdot\rho)(\psi) := \lim_{\epsilon\to 0} \rho(\hat{S}^{R_{\epsilon}}_N\cdot\psi), \label{eq5.44}
\end{equation}
and identify the domain of $\hat{S}_N$ in $\mathcal{H}^{\star}_{kin}$ as the set of states $\rho$ for which the limit above exists; it can be shown that this domain is nonempty (see \cite{ashtekar 2, alok}). Thus, we have successfully constructed a quantum version of the Hamiltonian constraint as an operator on $\mathcal{H}^{\star}_{kin}$. Physical states are elements of $\mathcal{H}^{\star}_{kin}$ that lie in the null space of this operator. But before ambition gets the better of us, let us pause here to highlight some problems with the foregoing constructions. 

To begin with, it is obvious that the action of the constraint operators, even after the regulators are removed, depends on the choice of the regulators. This is because, as already explained, owing to the different arrangements and numbers of edges and loops per cell in different regulators, the action of differently regulated constraints on the same state in $\mathcal{H}_{kin}$ yields possibly different states. Since the limit in Eq~(\ref{eq5.44}) is taken pointwise, this implies that different regulators give rise to different states in $\mathcal{H}^{\star}_{kin}$ for the same state in $\mathcal{H}_{kin}$. Therefore, the final constraint operator carries a memory of the classical background space on which the regulators were constructed. Not only does this leave open the question of which regulator ought to be considered physical, but it also means that our constructions run afoul of full background independence, contrary to one of the central aims of non-perturbative quantum gravity. 

Another ambiguity associated with our approach is the one that features in any approach to implementing the constraints on the quantum level via Dirac quantisation (see Section 1.3.2). This is the problem of operator ordering, which manifests itself in our choice of writing the curvature, connection and the volume functional in the specific order in which they appear in Eq~(\ref{eq5.34}). How can we be sure that this is the correct ordering? As we discussed in Section 1.3.2, modulo empirical input, there is no fully satisfactory answer to this question, except perhaps concerns pertaining to simplicity.

Nevertheless, there is a justification for our approach that offers some immunity against the foregoing concerns. To access it, let us probe the domain of the Hamiltonian constraint operator in $\mathcal{H}^{\star}_{kin}$. Generally, this domain will depend on the choice of regulators, but there is a distinguished subspace of it that is independent of that choice: it is precisely the diffeomorphism-invariant Hilbert space $\mathcal{H}_{diff}\subset\mathcal{H}^{\star}_{kin}$ constructed in the preceding section! To see this, notice that for a fixed regulator $R_\epsilon$, cells corresponding to different, unequal choices of $\epsilon$ differ in terms of their shape. In other words, the regulators $R_\epsilon$ corresponding to different values of $\epsilon$ are related by a diffeomorphism. This implies that up to diffeomorphisms, for every regulator $R_\epsilon$, the operator $\hat{S}^{R_\epsilon}_N$ is independent of the choice of $\epsilon$, i.e. for every $\epsilon, \epsilon'$, there exists a diffeomorphism $\phi$ such that
\begin{equation}
    \hat{S}^{R_\epsilon}_N\cdot\psi = \hat{U}(\phi)\cdot[\hat{S}^{R_{\epsilon'}}_N\cdot\psi] \label{eq5.45}
\end{equation}
for every $\psi \in \mathcal{H}_{kin}$ and every lapse function $N$. Now suppose $\rho \in \mathcal{H}_{diff}$. Then Eqs~(\ref{eq5.10}) and (\ref{eq5.45}) imply that 
\begin{equation}
    \hat{S}^{R_\epsilon}_N\cdot\rho = \hat{S}^{R_{\epsilon'}}_N\cdot\rho \label{eq5.46}
\end{equation}
for all $\epsilon, \epsilon'$ and $N$. Therefore, the action of $\hat{S}^{R_\epsilon}_N$ on $\mathcal{H}_{diff}$ is independent of $\epsilon$, whence the limit in Eq~(\ref{eq5.44}) becomes trivial. We conclude that irrespective of the choice of regulators, all diffeomorphism-invariant states lie in the domain of $\hat{S}_N$. This fact provides some justification for according at least some physical relevance to our partially background-dependent construction, and offers hope for finding the solutions to all the constraints of general relativity simultaneously. Recalling the shortcomings of the ADM quantisation program from Chapter 1, this is an improvement by leaps and bounds. 

\section{Quantum Gravitational States}
In this section, following Ref. \cite{ashtekar 2}, we will provide an algorithm to find solutions to both the diffeomorphism and Hamiltonian constraints simultaneously, i.e. diffeomorphism-invariant states that are also annihilated by the Hamiltonian constraint operator constructed in the previous section.

As we explained earlier, obtaining explicit solutions to the Hamiltonian constraint requires that we make a choice of diffeomorphism-covariant quantum regulators. We will make the following choice, due originally to Thomas Thiemann \cite{ashtekar 2}. Let $R_{\epsilon, \gamma}$ be a regulator associated with a graph $\gamma$. We restrict the edges $(s_I)$ and loops $(\alpha_i)$ assigned to each cell $\triangle$ of $R_{\epsilon, \gamma}$ containing a vertex $v$ of $\gamma$ as follows. First, every edge $s_I$ must be a proper segment of an edge of $\gamma$ that is outgoing at $v$. Next, to every pair of graph edges $e_I$, $e_J$, assign a triangular loop $\alpha_i$ subject to the following conditions: (1) $\alpha_i$ contains $v$ but no other point of $\gamma$; (2) $\alpha_i$ lies in a 2-plane that contains $e_I$, $e_J$ and is defined up to diffeomorphisms that preserve the edges of $\gamma$, and (3) $\alpha_i$ is oriented clockwise with respect to the orientation defined in the 2-plane by the ordered pair of segments $(s_I, s_J)$. We call loops such as $\alpha_i$ \textit{extraordinary}. Finally, the constants $C^{iJ}$ in the regulated operators (\ref{eq5.41}) are $0$ or $\pm k$, depending on the relative orientation of the tangent vectors to the segments $s_I$, $s_J$ and $s_K$ at $v$ with respect to the background orientation on $\Sigma$, $k$ being a fixed constant.

Now let $\eta(\ket{s^{\mathbf{j}}_{\gamma, \mathbf{I}}})$ be a group-averaged diffeomorphism-invariant state in $\mathcal{H}^{\star}_{kin}$. There can be two possibilities: either $\gamma$ itself contains an extraordinary loop or it does not. In the former case, $\hat{S}_{N}\eta(\ket{s^{\mathbf{j}}_{\gamma, \mathbf{I}}})) = 0$, so that any group-averaged spin networks corresponding graphs that do not have any extraordinary loops are solutions to the Hamiltonian constraint. But these are not very interesting. Let us, therefore, focus on graphs containing extraordinary loops. 

Let $\gamma$ be a fixed graph with extraordinary loops, and let $\mathbf{j}$ be the set of spin-j representations of $SU(2)$ associated with the edges of $\gamma$. Consider the set $\Gamma^{(n)}_{(\gamma, \mathbf{j})}$ of all spin-labelled graphs $(\gamma', \mathbf{j}')$ that can be obtained from $(\gamma, \mathbf{j})$ by diffeomorphisms and by creating $n$ extraordinary loops labelled by $j(\pi)$, where $\pi$ is the $SU(2)$ representation used in the definition of the constraint operator (\ref{eq5.41}). Let $\mathcal{D}^{(n)}_{(\gamma, \mathbf{j})}$ be the set of of all group-averaged spin-network states coming from group-averaging the elements of $\Gamma^{(n)}_{(\gamma, \mathbf{j})}$. The spaces $\mathcal{D}^{(n)}_{(\gamma, \mathbf{j})}$ are finite-dimensional and satisfy the following property:
\begin{equation}
    \Gamma^{(n)}_{(\gamma, \mathbf{j})} \neq \Gamma^{(n')}_{(\gamma', \mathbf{j}')} \Rightarrow \mathcal{D}^{(n)}_{(\gamma, \mathbf{j})} \cap \mathcal{D}^{(n')}_{(\gamma', \mathbf{j'})} = \varnothing, \label{eq5.47}
\end{equation}
which implies that every $\psi \in \mathcal{H}_{diff} \subset \mathcal{H}^{\star}_{kin}$ can be uniquely decomposed as
\begin{equation}
    \psi = \sum_{\gamma, \mathbf{j}, n} \psi^{(n)}_{(\gamma, \mathbf{j})}, \quad \text{where} \quad \psi^{(n)}_{(\gamma, \mathbf{j})} \in \mathcal{D}^{(n)}_{(\gamma, \mathbf{j})}. \label{eq5.48}
\end{equation}
We can thus conclude that
\begin{equation}
    \hat{S}_N\cdot\psi = 0 \Leftrightarrow \hat{S}_N\cdot\psi^{(n)}_{(\gamma, \mathbf{j})} = 0 \quad \forall n, (\gamma, \mathbf{j}). \label{eq5.49}
\end{equation}
In other words, any general solution to the Hamiltonian constraint can be found by finding solutions in finite-dimensional subspaces of $\mathcal{H}_{diff}$, a task which amounts to solving a finite set of linear equations -- a computer-programmable problem! Analogous results hold for the part of the Hamiltonian constraint (\ref{eq5.29}) that we have not solved.  

\chapter*{\center Epilogue}
\addcontentsline{toc}{chapter}{\protect\numberline{}Epilogue}
What are we to make of our coddiwomple through quantum gravity? Compared with the situation described in Chapter 1 with respect to the geometrodynamical variables, we indeed have made astonishing progress. To begin with, we have obtained a fully consistent version of general relativity quantised via a well-defined prescription in an almost (modulo a fully satisfactory treatment of the Hamiltonian constraint) \textit{background-independent manner}. In the final picture, there is no memory of any metric-dependent objects defined on the spacetime manifold. In fact, through the area operator of Chapter 4, the quantum theory ends up punching holes in the spacetime continuum that we began with in the classical theory:

\begin{verse}
    Oh these scissors from the quantum attic,\\
    Ripping to shreds the spacetime fabric,\\
    Ruining ten years of Einstein's painful toil,\\
    What a spoil, cries the continuum, what a spoil!
\end{verse}

But this comes at a price. Having reached the lofty heights of the divine quantum theory, how are we to descend to the mundane classical world we left behind? Why is that important, one might ask? Well, in the absence of empirical input, all we have at our disposal to physically justify the bold and terrifying mathematical hitchhiking that we have committed ourselves to is the certainty that under a suitable approximation, we recover the classical theory that fueled our journey in the first place. But a fabric so severely mutilated as we have butchered the spacetime fabric is hardly capable of being mended back to its original condition. Indeed, semi-classical and classical approximations of loop quantum gravity and whether or not they yield the correct physical theories at the relevant scales are still open problems. Some progress has been made, but are still a long way from achieving the end goal (see, for instance, Ref.~\cite{ashtekar 11} for some idea of how one may attack the problem in the connection representation of loop quantum gravity). A very active area of research in this context is the spinfoam formalism of loop quantum gravity \cite{rovelli 3, ashtekar 10}.

In addition to this, one of our ultimate goals was to find a home for physical quantum states, which are states satisfying a suitably quantised version of general relativistic constraints. We have not fully achieved this goal. In particular, the Hamiltonian constraint has not been quantised in a fully background-independent manner. Nonetheless, both a completely satisfactory solution to the diffeomorphism constraint and the tight matching between diffeomorphism-invariant states and the support of the Hamiltonian constraint operator are highly nontrivial facts. Furthermore, there is also the so-called ``master constraint'' approach to the constraints that was pioneered by Thomas Thiemann \cite{thiemann} and that we have not been able to visit. These considerations hold promising hopes for future progress. 

Another area that we have not explored is the application of the theory developed here to quantum cosmology. This is also a very active area of research in which loop quantum gravity has a number of achievements to its name, such as a derivation of the blackhole entropy and the removal of spacetime singularities. Again, these are enviable virtues of the formalism developed in this thesis, and provide justification for our central theme, namely that connections of a compact ``gauge group'' constitute perhaps a much better arena to explore the terrain of canonical quantum gravity. For further details about loop quantum cosmology, we refer the reader to Ref. \cite{ashtekar 2, ashtekar 10, ashtekar 12}. 

%

\end{document}